%% file: ms.tex
\documentclass[A4,12pt]{article}

\usepackage{fullpage}
\usepackage{color}
\usepackage{url}
\usepackage{graphicx}
\usepackage{epstopdf}
\usepackage{xspace}
\usepackage[cmex10]{amsmath}
\usepackage{amssymb}
\usepackage{amsthm}
\usepackage{algorithmicx}
\usepackage[noend]{algpseudocode}
\usepackage{listings}

\usepackage{pgfplots}
\pgfplotsset{compat=newest}
\usepackage{pgfplotstable}

\usepackage{fancyvrb}
\usepackage{ifthen}
\usepackage{tikz}
\usepackage[caption=false,font=footnotesize]{subfig}

\usepackage{titlesec}
\titleformat{\paragraph}
{\normalfont\normalsize\bfseries}{\theparagraph}{1em}{}
\titlespacing*{\paragraph}
{0pt}{3.25ex plus 1ex minus .2ex}{1.5ex plus .2ex}

\makeatletter
\newcounter{subsubparagraph}[subparagraph]
\renewcommand\thesubsubparagraph{%
  \thesubparagraph.\@arabic\c@subsubparagraph}
\newcommand\subsubparagraph{%
  \@startsection{subsubparagraph}    
    {6}                              
    {\parindent}                     
    {3.25ex \@plus 1ex \@minus .2ex} 
    {-1em}                           
    {\normalfont\normalsize\bfseries}}
\newcommand\l@subsubparagraph{\@dottedtocline{6}{10em}{5em}}
\newcommand{\subsubparagraphmark}[1]{}
\makeatother

\newcommand{\FINDATE}{29.02.2016} 
\newcommand{\DNUM}{D2.3}
\newcommand{\DNAME}{Power models, energy models and libraries for energy-efficient concurrent data structures and algorithms}
\newcommand{\DFMTNAME}{Power models, energy models and libraries \\ for energy-efficient concurrent \\ data structures and algorithms}
\newcommand{\DSHORTNAME}{Power models, energy models and libraries}

\usepackage{siunitx}

\newcommand\UB{\mathit{UB}}

\algrenewcommand\alglinenumber[1]{\scriptsize #1:}

\algnewcommand{\LComment}[1]{\Statex  \(\triangleright\) #1 \hfill~}

\newcounter{tempEquationCounter} 
\newcounter{thisEquationNumber}
\newenvironment{floatEq}
{\setcounter{thisEquationNumber}{\value{equation}}\addtocounter{equation}{1}
\begin{figure*}[!t]
\normalsize\setcounter{tempEquationCounter}{\value{equation}}
\setcounter{equation}{\value{thisEquationNumber}}
}
{\setcounter{equation}{\value{tempEquationCounter}}
\hrulefill\vspace*{4pt}
\end{figure*}
}

\algdef{SE}[DOWHILE]{Do}{doWhile}{\algorithmicdo}[1]{\algorithmicwhile\ #1}%

\algnewcommand\EMPTY{\textbf{EMPTY}}

\algblockdefx[NAME]{Start}{End}%
	[1]{\textbf{Struct} #1:}%
	{EndStruct}
\algnotext[NAME]{End}

\algblockdefx[Name]{START}{END}%
	[1]{#1}%
	{Ending}
\algnotext[Name]{END}

\usepackage{amsmath,amsthm,amssymb}
\usepackage{fixltx2e}
\usepackage{tikz}
\usetikzlibrary{calc,positioning}
\usetikzlibrary{shapes}
\usetikzlibrary{matrix}
\usetikzlibrary{decorations}
\usetikzlibrary{decorations.pathmorphing}
\usetikzlibrary{decorations.markings}
\usetikzlibrary{backgrounds}

\usepackage[linesnumbered,vlined,noend,ruled,nofillcomment]{algorithm2e}

\usepackage{ifthen}
\usepackage{arrayjobx}

\usepackage{stmaryrd}

\usepackage{verbatim}  

\newcommand{\inte}[2][0]
{ \{ #1,\dots,#2 \} }
\newcommand{\kth}[1]{\ema{#1^{\mathrm{th}}}}
\newcommand{\kst}[1]{\ema{#1^{\mathrm{st}}}}

\newcommand{\pro}[1]{\ema{(\mathcal{P}_{#1}})}

\newtheorem{theorem}{Theorem}

\newtheorem{lemma}{Lemma}
\newtheorem{corollary}{Corollary}
\newtheorem{property}{Property}

\theoremstyle{definition}
\newtheorem{definition}{Definition}
\newtheorem{remark}{Remark}

\newcommand{\ct}{\ema{P}}

\newcommand{\ema}[1]{\ensuremath{#1}\xspace}
\newcommand{\trl}{\ema{\ct_{rl}}}
\newcommand{\scas}{\ema{\mathit{cc}}}
\newcommand{\fcas}{\ema{\mathit{cc}}}
\newcommand{\mem}{\ema{\mathit{rc}}}

\newcommand{\calrl}{\ema{\mathit{cw}}}

\newcommand{\rint}[4]{\ema{\int_{#1}^{#2} #3 \, \mathit{d#4}}}
\newcommand{\expansion}[1]{\ema{e\left(#1\right)}}

\newcommand{\expa}{\ema{e}}

\newcommand{\expansionp}[1]{\ema{e'\left(#1\right)}}
\newcommand{\trlo}{\ema{\trl^{(0)}}}

\newcommand{\exppl}{(\texttt{+})}
\newcommand{\expmi}{(\texttt{-})}

\newcommand{\cw}{\ema{\mathit{cw}}}
\newcommand{\pw}{\ema{\mathit{pw}}}
\newcommand{\rc}{\ema{\mathit{rc}}}
\newcommand{\cc}{\ema{\mathit{cc}}}
\newcommand{\rlw}{\ema{\mathit{rlw}}}
\newcommand{\rlwp}{\ema{\rlw^{\expmi}}}
\newcommand{\psiz}{\ema{\pw^{\exppl}}}
\newcommand{\rlsiz}{\ema{\rlw^{\exppl}}}

\newcommand{\thru}{\ema{T}}

\newcommand{\ctot}{\ema{P}}
\newcommand{\thr}[1]{\ema{\mathcal{T}_{#1}}}

\newcommand{\shiftf}{\ema{\mathit{d}}}
\newcommand{\shift}[1]{\ema{\mathit{d}\left(#1\right)}}

\newcommand{\rl}{retry loop\xspace}
\newcommand{\rls}{retry loops\xspace}

\newcommand{\RLs}{Retry Loops\xspace}

\newcommand{\re}{retry\xspace}
\newcommand{\res}{retries\xspace}
\newcommand{\REs}{Retries\xspace}

\newcommand{\ps}{parallel section\xspace}
\newcommand{\pss}{parallel sections\xspace}
\newcommand{\ds}{data structure\xspace}
\newcommand{\dss}{data structures\xspace}
\newcommand{\casexp}{{\it Compare-And-Swap}\xspace}
\newcommand{\cas}{{\it CAS}\xspace}

\newcommand{\rf}{{\it Read}\xspace}
\newcommand{\faa}{{\it Fetch-and-Increment}\xspace}

\newcommand{\delmin}{\FuncSty{DeleteMin}\xspace}
\newcommand{\enqop}{\FuncSty{Enqueue}\xspace}
\newcommand{\deqop}{\FuncSty{Dequeue}\xspace}
\newcommand{\popop}{\FuncSty{Pop}\xspace}
\newcommand{\pushop}{\FuncSty{Push}\xspace}
\newcommand{\incop}{\FuncSty{Increment}\xspace}
\newcommand{\decop}{\FuncSty{Decrement}\xspace}

\newcommand{\casop}[1]{\FuncSty{CAS\textsubscript{#1}}\xspace}

\newcommand{\caca}{wasted \re}
\newcommand{\cacas}{wasted \res}

\newcommand{\flo}{\ema{f^{\exppl}}}
\newcommand{\fup}{\ema{f^{\expmi}}}

\newcommand{\tlo}{\ema{\thru^{\exppl}}}
\newcommand{\tup}{\ema{\thru^{\expmi}}}

\newcommand{\ghz}[1]{\ema{#1\,\mathrm{GHz}}}
\newcommand{\megb}[1]{\ema{#1\,\mathrm{MB}}}

\newcommand{\ie}{\textit{i.e.}\xspace}
\newcommand{\etal}{\textit{et al.}\xspace}
\newcommand{\eg}{\textit{e.g.}\xspace}
\newcommand{\etc}{\textit{etc.}\xspace}

\newcommand\rr[1]{}

\newcommand\pp[1]{#1}

\newcommand\posrem[2]{#1}

\newcommand\tra[1]{}

\newcommand\falseparagraph[1]{{\bf #1:}\xspace}

\newcommand\hardcon{hardware conflict\xspace}

\newcommand\hardcons{hardware conflicts\xspace}
\newcommand\Hardcons{Hardware conflicts\xspace}

\newcommand\logcons{logical conflicts\xspace}
\newcommand\Logcons{Logical conflicts\xspace}

\graphicspath{{figs-chalmers/disc/},{figs-chalmers/aggregate/},{figs-uit/},{figs-uit-model/figures/},{figs-uit-model/diagrams/}}

\begin{document}

\thispagestyle{empty}

\vspace{-3cm}
\begin{center}
\textbf{SEVENTH FRAMEWORK PROGRAMME}\\
\textbf{THEME ICT-2013.3.4}\\
Advanced Computing, Embedded and Control Systems
\end{center}
\bigskip

\begin{center}
\includegraphics[width=\textwidth]{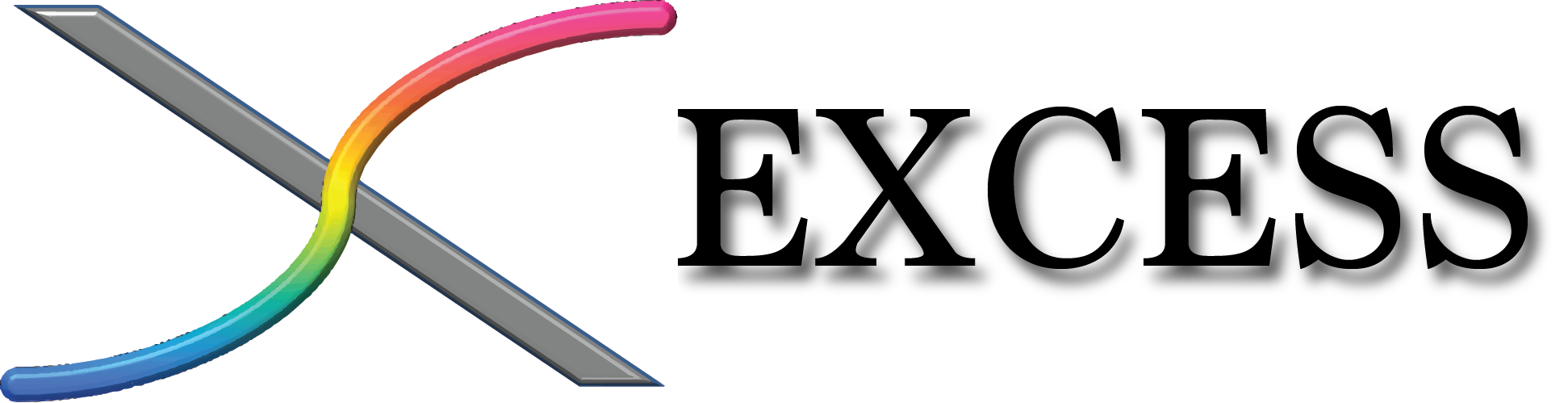}
\end{center}
\bigskip

\begin{center}
Execution Models for Energy-Efficient
Computing Systems\\
Project ID: 611183
\end{center}
\bigskip

\begin{center}
\Large
\textbf{\DNUM} \\
\textbf{\DNAME}
\end{center}
\bigskip

\begin{center}
\large
Phuong Ha, Vi Tran, Ibrahim Umar,
Aras Atalar, Anders Gidenstam, 
Paul Renaud-Goud, Philippas Tsigas, 
Ivan Walulya
\end{center}

\vfill

\begin{center}
\includegraphics[width=3cm]{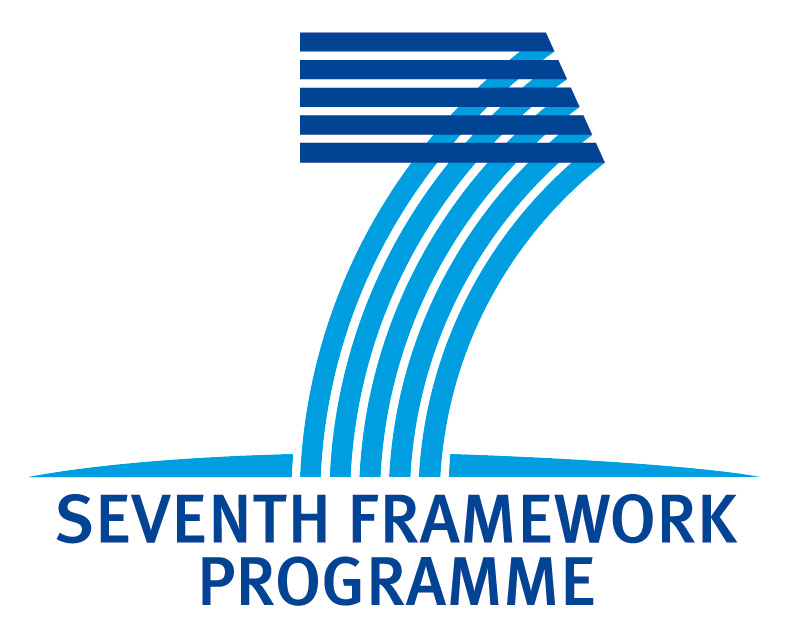}\\
Date of preparation (latest version): \FINDATE \\
Copyright\copyright\ 2013 -- 2016 The EXCESS Consortium \\
\hrulefill \\
The opinions of the authors expressed in this document do not
necessarily reflect the official opinion of EXCESS partners or of
the European Commission.
\end{center}

\newpage

\section*{DOCUMENT INFORMATION}

\vspace{1cm}

\begin{center}
\begin{tabular}{ll}
\textbf{Deliverable Number} & \DNUM \\
\textbf{Deliverable Name} & \begin{minipage}{10cm}{\DFMTNAME}\end{minipage} \\
\textbf{Authors}

& Phuong Ha \\
& Vi Tran \\
& Ibrahim Umar \\
& Aras Atalar\\
& Anders Gidenstam \\
& Paul Renaud-Goud\\
& Philippas Tsigas \\
& Ivan Walulya \\

\textbf{Responsible Author} & Phuong Ha\\
& e-mail: \url{phuong.hoai.ha@uit.no} \\
& Phone: +47 776 44032 \\
\textbf{Keywords} & High Performance Computing; \\
& Energy Efficiency \\
\textbf{WP/Task} & WP2/Task 2.1, 2.2, 2.3 \\
\textbf{Nature} & R \\
\textbf{Dissemination Level} & PU \\
\textbf{Planned Date} &  29.02.2016\\
\textbf{Final Version Date} & 28.02.2016 \\
\textbf{Reviewed by} &  \\
\textbf{MGT Board Approval} & \\ 
\end{tabular}
\end{center}

\newpage

\section*{DOCUMENT HISTORY}

\vspace{1cm}

\begin{center}
\begin{tabular}{llll}
\textbf{Partner} &
\textbf{Date} &
\textbf{Comment} &
\textbf{Version} \\
UiT (P.\ Ha, V.\ Tran) & 14.12.2015 & Deliverable skeleton & 0.1 \\
Chalmers (P.\ Renaud-Goud) & 22.1.2016 & Input - energy model and energy evaluation & 0.2 \\
UiT (P.\ Ha, V.\ Tran, I .\ Umar) & 29.1.2016 & Input - energy/power models and libraries & 0.3 \\
UiT (I .\ Umar) & 19.2.2016 & Revised chapter 6 & 0.4 \\
UiT (V.\ Tran) & 19.2.2016 & Revised chapter 1, 2, 3 & 0.5 \\
Chalmers (P.\ Renaud-Goud) & 22.2.2016 & Revised chapter 1, 4, 5 & 0.6 \\
UiT (P.\ Ha, V.\ Tran) & 26.2.2016 & Minor fixes & 0.7 \\
\end{tabular}
\end{center}

\newpage

\begin{abstract}
\input{Abstract.tex}
\end{abstract}

\newpage

\section*{Executive Summary}
\input{executive-summary.tex}





\newpage

\tableofcontents

\newpage


\section{Introduction} \label{sec:introduction}
\subsection{Purpose}

\input{intro-purpose.tex}

\subsection{Power and Energy Models}
\label{power-energy-models}
\input{motivation-models.tex}

\subsubsection{Power Models for Ultra-low Power Embedded Systems}  
\input{motivation-ULP-power.tex}

\subsubsection{Energy Complexity Model for Multithreaded Algorithms}
\input{motivation-EPEM.tex}

\subsubsection{Energy Model on CPU for Lock-free Data-structures with Low-level of Disjoint-access Parallelism}
\input{chalmers-motiv-disc}

\subsubsection{Energy Evaluation on Myriad2 for Multiway Aggregation on Streaming Applications}
\input{chalmers-motiv-aggr}

\subsection{Energy-efficient and Concurrent Data Structures and Algorithms}
\input{uit-temp/motiv-uit-tree}

\subsection{Contributions}
\input{intro-contribution.tex}

\newpage
\section{Power Models for Ultra-low Power Embedded Systems}  
\label{sec:Myriad-power-models}
\input{Myriad-power-models}

\newpage
\section{Energy Complexity Model for Multithreaded Algorithms}
\label{sec:EPEM}
\input{EPEM}

\newpage
\section{Energy Model for Lock-free Data-structures with Low-level of Disjoint-access Parallelism}
\label{sec:chalmers-disc}
\input{chalmers-disc}
\newpage
\section{Energy Evaluation on Myriad2 for Multiway Aggregation on Streaming Applications}
\label{sec:chalmers-aggregate}
\input{chalmers-aggregate}
\newpage
\section{Libraries of Energy-efficient and Concurrent Data Structures}  \label{sec:Concurrent-Data-Structures}
In this section, we describe our study on libraries of concurrent  search trees.

\subsection{Concurrent Search Trees} 
\input{uit-temp/ibrahim-uit-myriad2.tex}

\newpage
\clearpage



\section{Conclusions} \label{sec:Conclusion}
\input{conclusion.tex}

\newpage

\bibliographystyle{plain}
\bibliography{../WP6-bibtex/longhead,../D2.1/D2.1_related_papers,../WP6-bibtex/excess,../WP6-bibtex/related-papers,../WP6-bibtex/peppher-related,UiT-Vi-D2-3,bib-chalmers/biblio-disc,bib-chalmers/biblio-aggr,UiT-Ibrahim}

\section*{Glossary}
\input{glossary.tex}

\end{document}

%% file: Abstract.tex
This deliverable reports the results of the power models, energy models and libraries for energy-efficient concurrent data structures and algorithms as available by project month 30 of Work Package 2 (WP2). It reports i) the latest results of Task 2.2-2.4 on providing programming abstractions and libraries for developing energy-efficient data structures and algorithms and ii) the improved results of Task 2.1 on investigating and modeling the trade-off between energy and performance of concurrent data structures and algorithms. The work has been conducted on two main EXCESS platforms: Intel platforms with recent Intel multicore CPUs and Movidius Myriad platforms.

Regarding modeling the trade-off between energy-efficiency and performance of concurrent data structures and algorithms, we report in this deliverable four energy/power model evaluation studies, including: an improved power model for EXCESS platforms compared to Deliverable D2.2 (i.e., Movidius Myriad1), a new energy complexity model (namely, EPEM - Energy-aware Parallel External Memory) for multi-threaded algorithms, the modeling of the performance and the energy consumption of \dss on a CPU platform, as well as an investigation on the optimization of streaming applications on Myriad2, from three points of view, which are
performance, energy consumption, and space. 

Regarding developing novel programming abstractions and libraries, we have collected additional performance profiles using CPU counters and a cycle accurate simulator (i.e., GEM5), which are useful to gain insight into the relation between reduced data movements and energy efficiency. Moreover, we have implemented DeltaTree, the locality-aware data structures, and a fast concurrent B-Tree on Myriad2 platform, and have shown that a specialized ultra low-power embedded platform such as Movidius Myriad2 can also benefit from the locality-aware data structures.

%% file: executive-summary.tex




Work package 2 (WP2) aims to develop libraries for energy-efficient inter-process communication and data sharing on the EXCESS platforms. In order to set the stage for these tasks, WP2 needs to investigate and model the trade-offs between energy consumption and performance of data structures and algorithms for inter-process communication. WP2 also provides concurrent data structures and algorithms that support energy-efficient massive parallelism while minimizing inter-component communication. The developed data structures and algorithms are locality- and heterogeneity-aware.

This Deliverable D2.3 includes the results of four on-going tasks (i.e., Task 2.1-Task 2.4). In summary, Task 2.1 (PM1 - PM36) aims for modeling the trade-off between energy and performance in concurrent data structures and algorithms. Task 2.2 (PM7 - PM36) objective is to identify essential concurrent data structures and algorithms for inter-process communication. Task 2.3 (PM7 - PM36) develops locality- and heterogeneity-aware concurrent data structures and Task 2.4 (PM7 - PM36) develops locality- and heterogeneity-aware memory access algorithms. 

The latest results of Task 2.1 on investigating and modeling the trade-off between energy and performance \cite{HaTUTGRWA:2014} are presented in the Sections \ref{sec:Myriad-power-models}, \ref{sec:EPEM}, \ref{sec:chalmers-disc}, \ref{sec:chalmers-aggregate} (power/energy model studies). The latest results of Tasks 2.2-2.4 on energy-efficient and concurrent programming abstractions and libraries \cite{HaTUAGRT15} available by project month 30 are also summarized in this report in Section \ref{sec:Concurrent-Data-Structures}. 

\subsubsection*{Power/Energy models} 
The studies on proposing new power and energy models help to investigate the trade-off between energy-efficiency and performance of concurrent data structures and algorithms. The new power and energy models provide the understanding of energy consumption of concurrent data structures and algorithms.
\begin{itemize}
\item We have improved the power models for EXCESS platforms (i.e., Movidius Myriad1) from the power models presented in Deliverable D2.2. The latest RTHpower models have modeled the power consumption of Myriad1 both when computation and data transfer are performed in parallel and separately.
\item We have proposed EPEM, a new energy complexity model for multithreaded algorithms. This new general and validated energy complexity model for parallel (multithreaded) algorithms abstracts away possible multicore platforms by their static and dynamic energy of a computational operation and data access, and derives the energy complexity of a given algorithm from its {\em work}, {\em span} and {\em I/O} complexity. The new model is validated by different sparse matrix vector multiplication (SpMV) algorithms (e.g., compressed sparse column (CSC) and compressed sparse block (CSB)) running on high performance computing (HPC) platforms (e.g., Intel Xeon and Xeon Phi) and using nine sparse matrix types from Florida Matrix Collection. The new energy complexity model is able to characterize and compare the energy consumption of SpMV kernels according to three aspects: different algorithms, different input matrix types and different platforms. 
\item We have continued the modelling of the performance and the
energy consumption of \dss on a CPU platform. In Deliverable D2.2, we
have successfully modelled lock-free queues: we have been able to
predict the throughput, the power dissipated by the chip and the
energy per operation of six different implementations of lock-free
queues by measuring only a very few points of the studied domain. In
the present deliverable, we target the same metrics, which are
performance- and/or energy-related, and take some more steps: we
present the prediction of those metrics on a larger set of \dss,
namely stack, shared counter, queue and priority queue on a
single-socket processor. To obtain these estimates of the metrics, we
need even less measurement points than previously.
\item We have investigated on the optimization of streaming
applications on Myriad2, from three points of view, which are
performance, energy consumption, and space. To do this, we have
focused our evaluation on an operator that is widely used in data
streaming applications: the multiway aggregator. Several
implementations of this aggregator has been tested, employing several
queue implementations.
\end{itemize}
\subsubsection*{Libraries of concurrent data structures and algorithms}
We describe a set of implemented concurrent search trees as well as their energy and performance analyses.
\begin{itemize}
\item In the previous deliverable, we have shown that locality-aware concurrent search trees were able to consume less energy while able to maintain better throughput than the locality-oblivious concurrent search trees. Recently, we have collected additional performance profiles using CPU counters and a cycle accurate simulator (i.e., GEM5), which are useful to gain insight on the relation between reduced data movements and energy efficiency.

\item We have implemented DeltaTree and a fast concurrent B-Tree on Myriad2 platform, and have shown that a specialized ultra low-power embedded platform such as Movidius Myriad2 can also benefit from the fine-grained locality data structures. The experimental results show that GreenBST has up to 100\% better energy efficiency and more operations/second on the x86, ARM, and Xeon Phi platforms.
\end{itemize}
This report is organized as follows. Section \ref{sec:introduction} provides the background and motivations of the work presented in this deliverable. Sections \ref{sec:Myriad-power-models}, \ref{sec:EPEM}, \ref{sec:chalmers-disc}, \ref{sec:chalmers-aggregate} discuss power and energy models studies.
Section \ref{sec:Concurrent-Data-Structures} describes the second prototype with latest updates of EXCESS libraries including numerous concurrent search tree implementations as well as their performance and energy analyses. Section ~\ref{sec:Conclusion} concludes the report by summarizing the latest results and future works. 

%% file: intro-purpose.tex
The goal of Work package 2 (WP2) is to develop programming abstraction and libraries for inter-process communication and data sharing on EXCESS platforms, along with investigating and modeling the trade-offs between energy consumption and performance of data structures and algorithms for inter-process communication. WP2 also concerns supporting energy-efficient massive parallelism through scalable concurrent data structures and algorithms that strive for the energy limit, and minimizing inter-component communication through locality- and heterogeneity-aware data structures and algorithms.  


This report summarizes i) the latest results of Task 2.1 on investigating and modeling the trade-off between energy and performance of concurrent data structures and algorithms ii) the improved results of Task 2.2 on providing essential concurrent data structures and algorithms for inter-process communication and well as results of Task 2.3 on developing novel concurrent data structures that are locality- and heterogeneity-aware.

%% file: motivation-models.tex
This section explains the motivations of four energy/power model studies, including: an improved power model for EXCESS platforms compared to Deliverable D2.2 (i.e., Movidius Myriad1), a new energy complexity model (namely, EPEM - Energy-aware Parallel External Memory) for multi-threaded algorithms, the modeling of the performance and the energy consumption of \dss on a CPU platform, as well as an investigation on the optimization of streaming applications on Myriad2.

%% file: motivation-ULP-power.tex
Devising accurate power models is crucial to gain insights into how a computer system consumes power and energy. Significant efforts have been devoted to devising power and energy models, resulting in several seminal papers in the literature such as \cite{Alonso2014, jacobson2011, Choi2013, Choi2014, Korthikanti2009, Korthikanti2010, 7108419, Mishra:2015, Snowdon:2009}. The models are either platform specific \cite{jacobson2011} or application specific \cite{Alonso2014}.
Jacobson et al. \cite{jacobson2011} provided an insightful analysis in power modeling methodologies via a range of abstraction levels. They also proposed accurate power modeling methodologies for POWER-family processors. Alonso et al. \cite{Alonso2014} proposed energy models for three key dense-matrix factorizations.  
Roofline model of energy \cite{Choi2013, Choi2014} considers both algorithmic and platform properties. However, the Roofline model does not consider the number of cores running applications as a model parameter (i.e., coarse-grained models). Theoretical models by Korthikanti et al. \cite{Korthikanti2010, Korthikanti2009} are based on strong theoretical assumptions and are not yet validated on real platforms. Imes et al. \cite{7108419} provided a portable approach to make real-time decision and run the chosen configuration to minimize energy consumption. However, the approach requires systems supporting hardware resources (e.g., model-specific register) to expose energy data to the software during run-time. Mishra et al. \cite{Mishra:2015} used a probabilistic approach to find the most energy-efficient configuration by combining online and offline machine-learning approaches. However, this approach collects a significant amount of data to feed to its probabilistic network. RTHpower models proposed in this study are lightweight and applicable to any ultra-low power embedded systems as long as there is a means to measure the energy consumption of micro-benchmarks on the targeted platform. 

Recently, ultra-low power (ULP) embedded systems have become popular in the scientific community and industry, especially in media and wearable computing. ULP systems can achieve low energy per instruction down to a few pJ \cite{Alioto2012}. Alioto \cite{Alioto2012} mentioned that techniques such as pipe-lining, hardware replication, ultra-low-voltage memory design and leakage-reducing make a system ultra-low power. In order to model ULP systems where energy per instruction can be as low as few pJ, more accurate fine-grained approaches are needed. 
However, to the best of our knowledge, there are no application-general, fine-grained and validated models yet that provide insights into how an application running on an ULP embedded system consumes energy and, particularly, whether the race-to-halt (RTH) strategy (i.e, system is run as fast as possible, and then switched to idle state to save energy) that is widely used in high-performance computing (HPC) systems is still applicable to ULP embedded systems.

%% file: motivation-EPEM.tex
Like time complexity models that have significantly contributed to the analysis and development of fast algorithms, energy complexity models for parallel algorithms are desired as crucial means to develop energy efficient algorithms for ubiquitous multicore platforms. Ideal energy complexity models should be validated on real multicore platforms and applicable to a wide range of parallel algorithms. However, existing energy complexity models \cite{Alonso2014, Korthikanti2009, Korthikanti2010} for parallel algorithms are either theoretical without model validation or algorithm-specific without being applicable to a wide range of algorithms. This work presents a new general validated energy complexity model for parallel (multithreaded) algorithms.

%% file: chalmers-motiv-disc.tex
We consider the modeling and the analysis of the performance of
lock-free concurrent data structures. Lock-free designs employ an
optimistic conflict control mechanism, allowing several threads to
access the shared data object at the same time. They guarantee that at
least one concurrent operation finishes in a finite number of its own
steps regardless of the state of the operations.  Our analysis
considers such lock-free data structures that can be represented as
linear combinations of fixed size \rls.

Our main contribution is a new way of modeling and analyzing a general
class of lock-free algorithms (including stack, shared counter, queue,
priority queue), achieving predictions of throughput that are close to
what we observe in practice.  We emphasize two kinds of conflicts that
shape the performance: (i) \hardcons, due to concurrent calls to
atomic primitives; (ii) \logcons, caused by simultaneous operations on
the shared \ds.
We show how to deal with these hardware and logical conflicts
separately, and how to combine them, so as to calculate the throughput
of lock-free algorithms.

We propose also a common framework that, in addition to providing a
better understanding of the performance impacting factors, enables a
fair comparison between lock-free implementations by covering the
whole contention domain. This comparison translates into the ability
to choose the best implementation at hand, with respect to the
actual application that uses the \ds.
This part of our analysis comes with a method for calculating a good
back-off strategy to finely tune the performance of a lock-free
algorithm. Our experimental results, based on a set of widely used
concurrent data structures and on abstract lock-free designs, show
that our analysis follows closely the actual code behavior.

%% file: chalmers-motiv-aggr.tex
The transition from uniprocessor to multiprocessor designs in embedded
systems is not as straightforward as in general purpose machines due
to the fact that limitations in terms of space availability and energy
consumption are introduced depending on the application's purpose and
target environment along with the hardware used.

Developing this kind of systems in a way that it uses memory and
energy in an optimal way, and such that performance is not affected at
a big percentage, is of utmost importance and data structures used
within an application play a significant role towards this goal.

The work presented here aims to exhibit data structures that are
suited for embedded systems, and to investigate trade-offs between
different implementations in terms of energy consumption, memory
utilization and performance. Through this investigation the focus will
be on data streaming applications implementing multiway aggregation of
the received data.

Although efficient data structures for a concurrent environment have
been studied extensively, the issue of appropriate data structures for
data streaming applications has been neglected. Concurrent data
structures play a major role between aggregation stages, because they
are in charge of the regulation of the parallelism and the load
balancing in this streaming applications.
For this reason we have developed such a streaming application, which
is based on the research conducted on concurrent data structures. An
efficient solution providing lower latency, bigger throughput and
energy efficiency at the data aggregation function of the application
is then achieved.

%% file: uit-temp/motiv-uit-tree.tex
Recent research has suggested that improving fine-grained data-locality is one of
the main approaches to improving energy efficiency and performance. 
However, no previous research has investigated the effect of the approach on these metrices in the case of concurrent data structures. 

Our study investigates how fine-grained data locality influences energy efficiency and performance in concurrent search trees, a crucial data structure that is widely used in several important systems. We conduct a set of experiments on three lock-based concurrent search trees: GreenBST, a portable fine-grained locality-aware concurrent search tree; CBTree, a coarse-grained locality-aware concurrent B+tree; LFBST, a locality-oblivious non-blocking binary search tree; and citrus, a locality-oblivious read-copy-update (RCU) concurrent search tree. We run the experiments on a commodity x86 platform, an embedded ARM platform, and an accelerator board based on the Intel Xeon Phi. The experimental results show that GreenBST has up to 100\% better energy efficiency and more operations/second on the x86, ARM, and Xeon Phi platforms.
The results also confirm that portable fine-grained locality can improve energy efficiency and performance in concurrent search trees.

%% file: intro-contribution.tex
The main achievements in this report are summarized as follows.

\subsubsection{Power/Energy models}
\begin{itemize}

\item We have improved the power model for one EXCESS platform (Movidius Myriad1) initially described in Deliverable D2.2. The power
models characterize applications by their operational intensity that
can be extracted from any application. The models are validated with
three application kernels, namely dense matrix multiplication, sparse
matrix vector multiplication and breadth first search. Based on the
models, we propose a framework to predict whether it is
energy-efficient to apply race-to-halt (RTH) strategy (i.e. running an
application with a maximum number of cores). For the application
kernels, the proposed framework is able to predict when to use RTH and
when not to use RTH precisely. The experimental results show that we
can save up to 60\% energy for dense matrix multiplication, 61\%
energy for sparse matrix vector multiplication by using RTH and 5\%
energy for breadth first search by \emph{not} using RTH.
\item We have proposed a new energy complexity model for multithreaded
algorithms. This new general and validated energy complexity model for
parallel (multithreaded) algorithms abstracts away possible multicore
platforms by their static and dynamic energy of a computational
operation and data access, and derives the energy complexity of a
given algorithm from its {\em work}, {\em span} and {\em I/O}
complexity. The new model is validated by different sparse matrix
vector multiplication (SpMV) algorithms (e.g., compressed sparse
column (CSC) and compressed sparse block (CSB)) running on high
performance computing (HPC) platforms (e.g., Intel Xeon and Xeon Phi)
and using nine sparse matrix types from Florida Matrix Collection \cite{Davis:2011}. The
new energy complexity model is able to characterize and compare the
energy consumption of SpMV kernels according to three aspects:
different algorithms, different input matrix types and different
platforms.
\item We have continued the modelling of the performance and the
energy consumption of \dss on a CPU platform. In Deliverable D2.2, we
have successfully modelled lock-free queues: we have been able to
predict the throughput, the power dissipated by the chip and the
energy per operation of six different implementations of lock-free
queues by measuring only a very few points of the studied domain. In
the present deliverable, we targeted the same metrics, which are
performance- and/or energy-related, and take some more steps: we
present the prediction of those metrics on a larger set of \dss,
namely stack, shared counter, queue and priority queue on a
single-socket processor. To obtain these estimates of the metrics, we
need even less measurements points than previously.
\item We have investigated on the optimization of streaming
applications on Myriad2, from three points of view, which are
performance, energy consumption, and space. To do this, we have
focused our evaluation on an operator that is widely used in data
streaming applications: the multiway aggregator. Several
implementations of this aggregator has been tested, employing several
queue implementations.



\end{itemize}
\subsubsection{Libraries of concurrent data structures and algorithms}
\begin{itemize}
\input{uit-temp/ibrahim-uit-intro.tex}
\end{itemize}  

%% file: uit-temp/ibrahim-uit-intro.tex
\item We have added two state-of-the art concurrent search trees into the concurrent search tree library. In the previous deliverable, we have shown that locality-aware concurrent search trees were able to consume less energy lead while able to maintain better throughput than the locality-oblivious concurrent search trees. Recently, we have collected additional performance profiles using CPU counters and a cycle accurate simulator (i.e., GEM5), which are useful to gain insight on the relation between reduced data movements and energy efficiency.

\item We have implemented DeltaTree and a fast concurrent B-Tree on Myriad2 platform, and have shown that a specialized ultra low-power embedded platform such as Movidius Myriad2 can also benefit from the fine-grained locality data structures.

%% file: Myriad-power-models.tex
This section presents the improvements on power models for Myriad1 platform from the model proposed in EXCESS Deliverable D2.2 \cite{HaTUAGRT15}.
The study is summarized to three main works as follows.
\begin{itemize}
\item We propose new application-general fine-grained power models (namely, RTHpower) that provide insights into how a given application consumes power when running on an ultra-low power embedded system \cite{Tran2016}. The RTHpower models consider three parameter groups: platform properties, application properties (e.g. operational intensity and scalability) and execution settings (e.g., the number of cores executing a given application). 
The models consider both platform and application properties, which give more insights into how to design applications to achieve better energy efficiency. (cf. Section \ref{sec:PowerModel}) 
\item We train the new RTHpower models on an ultra-low power embedded system, namely Movidius Myriad using different sets of micro-benchmarks and validate the models using two computation kernels from Berkeley dwarfs \cite{Asa06} and one data-intensive kernel from Graph500 benchmarks \cite{6114175}. The three chosen application kernels are dense matrix multiplication (Matmul), sparse matrix vector multiplication (SpMV) and breadth first search (BFS). The model fitting has percentage error at most 7\% for  micro-benchmarks and 10\% for application benchmarks (cf. Section \ref{sec:Validation}). 
\item We investigate the RTH strategy on an ultra-low power embedded platform using the new RTHpower models. We propose a framework that is able to predict precisely when to and when not to apply the RTH strategy in order to minimize energy consumption. We validate the framework using micro-benchmarks and application kernels. From our experiments, we show real-world scenarios when to use RTH and when not to use RTH. We can save up to 61\% energy for dense matrix multiplication, 59\% energy for SpMV by using RTH and up to 5\% energy for BFS by not using RTH. (cf. Section \ref{sec:Framework})
\end{itemize}

\subsection{RTHpower - Analytical Power Models}
\label{sec:PowerModel}
For the flow of reading, we first summarize a power model for operation units described in Deliverable D.2.2 and then develop it to the improved RTHpower models considering application properties in this Deliverable.
\subsubsection{A Power Model for Operation Units}
The experimental results of the micro-benchmarks suite for operation units show that the power consumption of Movidius Myriad1 platform is ruled by Equation \ref{eq:Pfirst}. In the equation, the static power $P^{sta}$ is the required power when the Myriad chip is on, including memory storage power; the active power $P^{act}$ is the power consumed when a SHAVE core is on and actively performing computational work; the dynamic power $P^{dyn}(op)$ is the power consumed by each operation unit such as arithmetic units (e.g., IAU, VAU, SAU, CMU) or load/store units (e.g., LSU0, LSU1) in one SHAVE.  The experimental results show that different operation units have different $P^{dyn}(op)$ values as listed in Table~\ref{table:Punits}. The total dynamic power of a SHAVE core is the sum of all dynamic power from involved units. If benchmarks or programs are executed with $n$ SHAVE cores, the active and dynamic power needs to be multiplied with the number of used SHAVE cores. By using regression fitting techniques, the average value of $P^{sta}$ and $P^{act}$ from all micro-benchmarks are computed in Equation \ref{eq:Pstat1} and Equation \ref{eq:Pstat2}. Table \ref{table:Parameters} provides the description of parameters in the proposed models. 
\begin{equation} \label{eq:Pfirst}
	P^{units}=P^{sta}+ n \times \left(P^{act} + \sum_i{P^{dyn}_i(op)} \right)
\end{equation}
\begin{equation} \label{eq:Pstat1}
	P^{sta}=61.81 \text{ mW}
\end{equation}
\begin{equation} \label{eq:Pstat2}
	P^{act}=29.33 \text{ mW}
\end{equation}
\begin{table}
\caption{\textbf{$P^{dyn}(op)$} of SHAVE Operation Units}
\label{table:Punits}
\begin{center}
\begin{tabular}{lll}
\hline\noalign{\smallskip}
Operation & Description & $P^{dyn}$ (mW)\\ 
\noalign{\smallskip}\hline\noalign{\smallskip}
SAUXOR  &  Perform bitwise exclusive-OR on scalar              & 14.68          \\ 
SAUMUL  &  Perform scalar multiplication               & 17.69          \\ 
VAUXOR  &  Perform bitwise exclusive-OR on vector               & 34.34         \\
VAUMUL  &  Perform vector multiplication             & 51.98         \\ 
IAUXOR  &  Perform bitwise exclusive-OR on integer              & 15.91          \\ 
IAUMUL  &  Perform integer multiplication              & 18.48          \\ 
CMUCPSS &  Copy scalar to scalar             & 12.62          \\ 
CMUCPIVR&  Copy integer to vector              & 18.84         \\ 
LSULOAD &  Load from a memory address to a register            & 29.87         \\ 
LSUSTORE&  Store from a register to a memory address              & 37.49          \\
\noalign{\smallskip}\hline
\end{tabular}
\end{center}
\end{table}
\begin{table}
\caption{Model Parameter List}
\label{table:Parameters}
\begin{center}
\begin{tabular}{ll}
\hline\noalign{\smallskip}
Parameter & Description \\
\noalign{\smallskip}\hline\noalign{\smallskip}
$P^{sta}$                & Static power of a whole chip         \\ 
$P^{act}$                 & Active power of a core        \\ 
$P^{dyn}(op)$                  & Dynamic power of an operation unit        \\ 
$P^{LSU}$                  & Dynamic power of Load Store Unit          \\ 
$P^{ctn}$                  & Contention power of a core waiting for data          \\ 
$m$                 & Average number of active cores accessing data       \\ 
$n$                & Number of assigned cores to the program      \\ 
$I$              & Operational intensity of an application         \\ 
$\alpha$                & Time ratio of data transfer to computation       \\
$\beta$ & Tuning parameter of an application\\ 
\noalign{\smallskip}\hline\noalign{\smallskip}
\end{tabular}
\end{center}
\end{table}
\subsubsection{RTHpower Models for Applications}
\label{sec:ModelEquation}
Since typical applications require both computation and data movement, we use the concept of operational intensity proposed by Williams et al.\cite{Williams2009} to characterize applications. An application can be characterized by the amount of computational work $W$ and data transfer $Q$. $W$ is the number of operations performed by an application. $Q$ is the number of transferred bytes required during the program execution. Both $W$ and $Q$ define the operational intensity $I$ of applications as in Equation \ref{eq:PIWQ}. 
\begin{equation} \label{eq:PIWQ}
	I= \frac{W}{Q}
\end{equation}
As the time required to perform one operation is different from the time required to transfer one byte of data, we introduce a parameter to the models: time ratio {$\alpha$} of transferring one byte of data to performing one arithmetic operation. Ratio {$\alpha$} is the property of an application on a specific platform and its value depends on the application. 
Since the time to access data and time to perform computation work can be overlapped, during a program execution, the SHAVE core can be in one of the three states: performing computation, performing data transfer or performing both computation and data transfer in parallel. An application either has data transfer time longer than computation time or vice versa. Therefore, there are two models for the two cases for higher accuracy (as compared to Deliverable D2.2, only one model represents both cases).
\begin{itemize}
\item If data transfer time is longer than computation time, the model follows Equation \ref{eq:PalbiWQ}. The execution can be modeled as two (composed) periods: one is when computation and data transfer are performed in parallel and the other is when only data transfer is performed. Fraction $\frac{W}{\alpha \times Q}$ represents the overlapped time of computation and data transfer. Fraction $\frac{\alpha \times Q-W}{\alpha \times Q}$ represents the remaining time for data transfer. 
\begin{equation} 
\label{eq:PalbiWQ} 
\begin{split}
	P&= P^{comp||data} \times \frac{W}{\alpha \times Q}+  P^{data} \times \frac{\alpha \times Q-W}{\alpha \times Q}
\end{split}
\end{equation}
\item If computation time is longer than data transfer time, estimated power follows Equation \ref{eq:PalleWQ}. The execution can be modeled as two periods: one is when computation and data transfer are performed in parallel and the other is when only computation is performed.
\begin{equation} 
\label{eq:PalleWQ} 
\begin{split}
	P&= P^{comp||data} \times \frac{\alpha \times Q}{W} + P^{comp} \times \frac{W-\alpha \times Q}{W}
\end{split}
\end{equation}
\end{itemize}

After converting W and Q to I by using Equation ~\ref{eq:PIWQ}, the final models are simplified as Equation \ref{eq:PalbiI} and Equation \ref{eq:PalleI},
\begin{equation} 
\label{eq:PalbiI} 
\begin{split}
	P&= P^{comp||data} \times \frac{I}{\alpha}+  P^{data} \times \frac{\alpha-I}{\alpha}
\end{split}
\end{equation}
\begin{equation} 
\label{eq:PalleI} 
\begin{split}
	P&= P^{comp||data} \times \frac{\alpha}{I} + P^{comp} \times \frac{I-\alpha}{I}
\end{split}
\end{equation}
where $P^{data}$, $P^{comp}$ and $P^{comp||data}$ are explained below:
\paragraph{Data transfer power} $P^{data}$ is the power consumed by the whole chip when only data transfer is performed. $P^{data}$ is computed by Equation \ref{eq:Pdata}. 
In Equation \ref{eq:Pdata}, $P^{sta}$ is the static power; $P^{act}$ is the active power; $n$ is the number of active cores assigned to run the application;  $m$ is the average number of SHAVE cores accessing data in parallel during the application execution; contention power $P^{ctn}$ is the power overhead occurring when a SHAVE core waits for accessing data because of the limited memory ports (or bandwidth) or cache size in the platform architecture. Therefore, $n-m$ is the average number of SHAVE cores waiting for memory access during the application execution. 
\begin{equation} \label{eq:Pdata}
\begin{split}
	P^{data} &= P^{sta} + min(m,n) \times (P^{act}+ P^{LSU})\\
	&+ max(n-m, 0) \times P^{ctn} 
\end{split}
\end{equation}
\paragraph{Computation power} $P^{comp}$ is the power consumed by the whole chip when only computation is performed. $P^{comp}$ is computed by Equation \ref{eq:Pcomp}. Each core runs its arithmetic units (e.g. IAU, SAU, VAU) to perform computation work. There is no contention power due to no memory access. Therefore, all assigned cores are active and contribute to the power consumption.
\begin{equation} 
\label{eq:Pcomp} 
\begin{split}
	P^{comp}&= P^{sta} + n \times (P^{act}+\sum_i{P^{dyn}_i(op)})
\end{split}
\end{equation}
\paragraph{Computation and data transfer power} $P^{comp||data}$ is the power consumed by the whole chip when computation and data transfer are performed in parallel. $P^{comp||data}$ is computed by Equation \ref{eq:Pcompdata}. In this case, there is contention power due to data waiting. $P^{comp||data}$ is different from $P^{data}$ in the aspect that the active cores also run arithmetic units that contribute to total power as $\sum_i{P^{dyn}_i(op)}$.
\begin{equation} 
\label{eq:Pcompdata}
\begin{split}
	P^{comp||data} &= P^{sta} \\
	&+ min(m,n) \times (P^{act}+ P^{LSU}+\sum_i{P^{dyn}_i(op)}) \\
	&+ max(n-m, 0) \times P^{ctn}
\end{split}
\end{equation}
\subsection{Model Training and Validation}
\label{sec:Validation}
This section presents the experimental results including two sets of micro-benchmarks and three application kernels (i.e., {\it matmul}, SpMV and BFS)  that are used for training and validating the models .
\subsubsection{Model Training with Micro-benchmarks}
\label{sec:model-validation-micro}
\begin{figure}[!t] \centering
\small
\begin{tikzpicture}
\begin{axis}[
title=Percentage Errors of Micro-benchmarks for Operational Units,
xlabel=Number of Cores,
ylabel=Percentage Error,
width=7cm, height=4cm,
legend style={at={(0.5,-0.5)},
anchor=north,legend columns=-1},
xtick={1,2,4,8},
ytick={-10,-5,0,5,10},
yticklabel={\pgfmathprintnumber\tick\%}
]

\addplot+[
error bars/.cd,
x dir=both, x explicit,
y dir=both, y explicit,
]
table[x=NumberOfCores, y=VauXor]
{
NumberOfCores VauXor	SauXorVauXor	SauXorCmuCpssIauXor
1	3	-4	2
2	3	-5	3
4	5	-5	5
8	6	-4	5
};

\addplot+[
error bars/.cd,
x dir=both, x explicit,
y dir=both, y explicit,
]
table[x=NumberOfCores, y=SauXorVauXor]
{
NumberOfCores VauXor	SauXorVauXor	SauXorCmuCpssIauXor
1	3	-4	2
2	3	-5	3
4	5	-5	5
8	6	-4	5
};

\addplot+[
error bars/.cd,
x dir=both, x explicit,
y dir=both, y explicit,
]
table[x=NumberOfCores, y=SauXorCmuCpssIauXor]
{
NumberOfCores VauXor	SauXorVauXor	SauXorCmuCpssIauXor
1	3	-4	2
2	3	-5	3
4	5	-5	5
8	6	-4	5
};
\legend{VauXor, SauXorVauXor, SauXorCmuCpssIauXor} 
\end{axis}
\end{tikzpicture}
\caption{The upper-/ lower-bounds on percentage errors of the model fitting for \emph{unit-suite} shown by the worst cases of three categories: one unit (e.g., SauXor), two units (e.g., IauXorVauXor) and three units (e.g., SauXorCmuIauXor). The absolute percentage errors of micro-benchmarks for operation units are at most 6\%.}
\label{fig:OperationUnitError}
\end{figure}
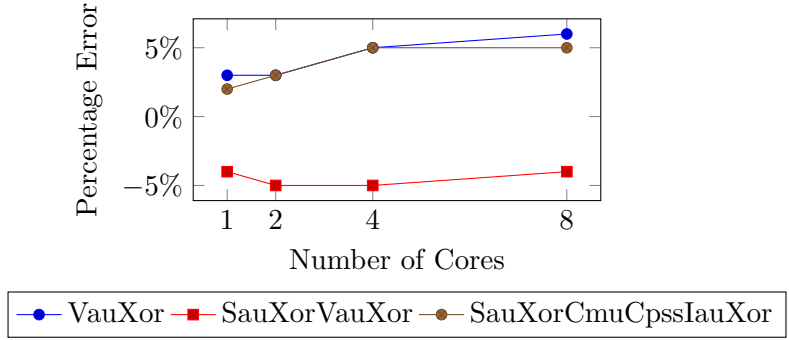

Analyses of experimental results are performed based on two sets of micro-benchmarks: 22 micro-benchmarks for operation units called \emph{unit-suite}  and 9 micro-benchmarks for different operational intensities called \emph{intensity-suite}. Each micro-benchmark is executed with different numbers of SHAVE cores to measure its power consumption.
\paragraph{Micro-benchmarks for Operation Units}
We assess the fitting of the power model for operation units (Equation \ref{eq:Pfirst}) using data from \emph{unit-suite}. The micro-benchmarks of \emph{unit-suite} are listed in Table \ref{table:Micro-benchmarks}. We calculate the percentage errors of the model fitting and plot them in Figure \ref{fig:OperationUnitError}. Percentage error is calculated as $PE=\frac{measurement - estimation}{measurement}$. The {\em absolute} percentage error is the absolute value of the percentage error. For this model, the absolute percentage errors are at most 6\%. Figure \ref{fig:OperationUnitError} shows the percentage error of the worst cases in all three categories: one unit, two pipe-lined units and three pipe-lined units. These results prove that the model is applicable to micro-benchmarks using either a single (e.g., performing bit-wise exclusive-OR on scalar unit: SauXor) or pipe-lined arithmetic units in parallel (e.g., performing Xor on scalar and integer units, in parallel with copying from scalar to scalar unit: SauXorCmuCpssIauXor). The model also shows the compositionality of the power consumption not only for multiple SHAVE cores but also for multiple operation units within a SHAVE core.
\begin{table*}
\caption{Micro-benchmarks for Operation Units}
\label{table:Micro-benchmarks}
\centering
\resizebox{\columnwidth}{!}{%
\begin{tabular}{ll}
\hline\noalign{\smallskip}
Description & Micro-benchmark Name \\
\noalign{\smallskip}\hline\noalign{\smallskip}
 10 micro-benchmarks using one unit                 & SAUXOR, SAUMUL, IAUXOR, IAUMUL, VAUXOR, VAUMUL, CMUCPSS,    \\
    (cf. Table 2)            &   CMUCPIVR, LSULOAD, LSUSTORE       \\ 
11 micro-benchmarks using two units    &   SAUXOR-CMUCPSS, SAUXOR-CMUCPIVR, SAUXOR-IAUMUL,  \\ 
  &   SAUXOR-IAUXOR, SAUXOR-VAUMUL, SAUXOR-VAUXOR, SAUMUL-IAUXOR,   \\ 
      & IAUXOR-VAUXOR, IAUXOR-VAUMUL, IAUXOR-CMUCPSS, LSULOAD-STORE   \\ 
1 micro-benchmark using three units & SAUXOR-IAUXOR-CMUCPSS \\ 
\noalign{\smallskip}\hline\noalign{\smallskip}
\end{tabular}
}
\end{table*}

\paragraph{Micro-benchmarks for Application Intensities}
Since any application requires both computation and data movement, we design intensity-based micro-benchmarks which execute both arithmetic units (e.g., SAU) and two data transfer units (e.g., LSU0, LSU1) in a parallel manner. They are implemented with parallel instruction pipeline supported by the platform. In order to validate the RTHpower models, this \emph{intensity-suite} indicates different values of operation intensities (from 0.25 to 64). Operational intensity $I$ is retrieved from the assembly code by counting the number of arithmetic instructions and the number of load/store instructions. 

\begin{figure}[!t] \centering
\small
\begin{tikzpicture}
\begin{axis}[
title=Percentage Errors of Intensity-based Micro-benchmarks,
xlabel=Operational Intensity,
ylabel=Percentage Error,
width=7cm, height=4cm,
legend style={at={(0.5,-0.5)},
anchor=north,legend columns=-1},
xtick={1,4,8,16,32,64},
ytick={-10,-5,0,5,10},
yticklabel={\pgfmathprintnumber\tick\%}
]

\addplot+[
error bars/.cd,
x dir=both, x explicit,
y dir=both, y explicit,
]
table[x=intensity, y=deviation]
{
intensity	1-core	deviation	4-cores	 6-cores	8-cores
0.25	15.15	3.46	-4.74	20.31	1.11
0.5	7.25	3.01	-3.69	13.16	0.84
1	7.75	2.56	-2.40	-7.04	0.79
2	1.37	1.78	-2.04	-2.01	0.50
4	2.72	1.44	-1.54	-1.74	0.37
8	-0.57	0.55	-0.59	35.18	0.10
16	-22.89	-1.65	5.14	89.87	-1.27
32	-22.66	0.64	4.16	98.80	-1.20
64	-18.98	7.13	-3.13	111.22	1.95
};

\addplot+[
error bars/.cd,
x dir=both, x explicit,
y dir=both, y explicit,
]
table[x=intensity, y=deviation]
{
intensity	1-core	2-cores	deviation	6-cores	8-cores
0.25	15.15	3.46	-4.74	20.31	1.11
0.5	7.25	3.01	-3.69	13.16	0.84
1	7.75	2.56	-2.40	-7.04	0.79
2	1.37	1.78	-2.04	-2.01	0.50
4	2.72	1.44	-1.54	-1.74	0.37
8	-0.57	0.55	-0.59	35.18	0.10
16	-22.89	-1.65	5.14	89.87	-1.27
32	-22.66	0.64	4.16	98.80	-1.20
64	-18.98	7.13	-3.13	111.22	1.95
};

\addplot+[
error bars/.cd,
x dir=both, x explicit,
y dir=both, y explicit,
]
table[x=intensity, y=deviation]
{
intensity	1-core	2-cores	4-cores	 6-cores	deviation
0.25	15.15	3.46	-4.74	20.31	1.11
0.5	7.25	3.01	-3.69	13.16	0.84
1	7.75	2.56	-2.40	-7.04	0.79
2	1.37	1.78	-2.04	-2.01	0.50
4	2.72	1.44	-1.54	-1.74	0.37
8	-0.57	0.55	-0.59	35.18	0.10
16	-22.89	-1.65	5.14	89.87	-1.27
32	-22.66	0.64	4.16	98.80	-1.20
64	-18.98	7.13	-3.13	111.22	1.95
};
\legend{2-cores, 4-cores, 8-cores} 
\end{axis}
\end{tikzpicture}
\caption{The absolute percentage errors of RTHpower model fitting for \emph{intensity-suite} (operational intensity $I$ from 0.25 to 64) are at most 7\%.}
\label{fig:Intensity-Deviation}
\end{figure}
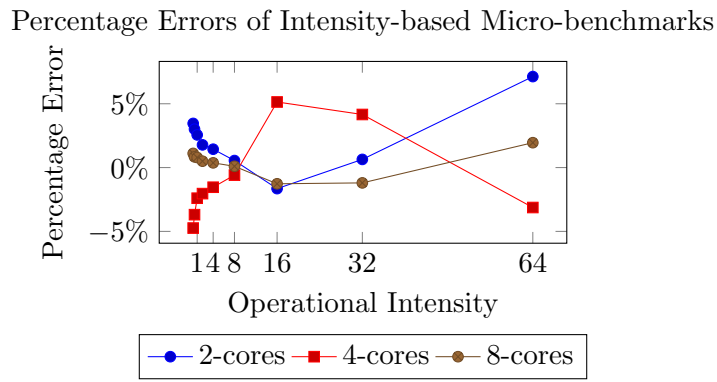

\begin{figure}[!t] \centering
\resizebox{0.6\columnwidth}{!}{ \includegraphics{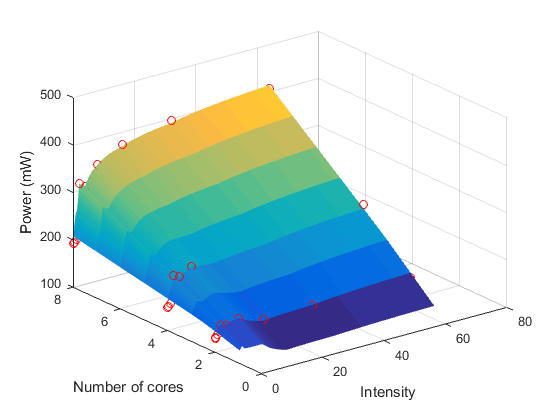}}
\caption{The estimated power range of varied intensities and numbers of cores from RTHpower models. The dots in the figure represent measurement data.}
\label{fig:TSKEstimation}
\end{figure}

In the models, there are platform-dependent parameters such as $\alpha$, $m$ and $P^{ctn}$. The parameter values for each application operational intensity are derived from experimental results by using Matlab function $lsqcurvefit$ . For the application intensities from 0.25 to 1, $\alpha$ is found bigger than operational intensity $I$ meaning that data transfer time is longer than computation time. The estimated power model follows Equation \ref{eq:PalbiI}. For operational intensity from 2 to 64, $\alpha$ is less than $I$ meaning that data transfer time is less than computation time. The estimated power follows Equation \ref{eq:PalleI}.
We plot the percentage errors of the model fitting for intensity-based micro-benchmarks in Figure \ref{fig:Intensity-Deviation}. 
In order to obtain a  full range of estimated power with any values of intensities and numbers of cores, a fuzzy logic approach, namely Takagi Sugeno Kang (TSK) mechanism \cite{6313399}, is applied to the RTHpower models. Each intensity has a parameter set, including $\alpha$, $P^{ctn}$ and $m$. Based on the RTHpower models, each parameter set provides an individual function to estimate the power of an application based on its intensity value and a number of cores. The TSK mechanism considers the individual functions as membership functions and combines them into a general function that can be used for any input (i.e., intensity $I$ and number of cores $n$). The membership functions of the fuzzy sets are triangular\cite{Pedrycz:1994}. After implemented the approach using Matlab Fuzzy Logic toolbox, the full range of estimated power is obtained and presented in Figure \ref{fig:TSKEstimation}. It is observed that when the intensity value increases, the power-up (i.e., the power consumption ratio of the application executed with $n$ cores to the application executed with 1 core) is also increased. The small dip in the diagram is due to the switch from Equation \ref{eq:PalbiI} to Equation \ref{eq:PalleI} at the intensity $I=2$.
\subsubsection{Model Validation with Application Kernels}
\begin{figure}[!t] \centering
\resizebox{0.5\columnwidth}{!}{ \includegraphics{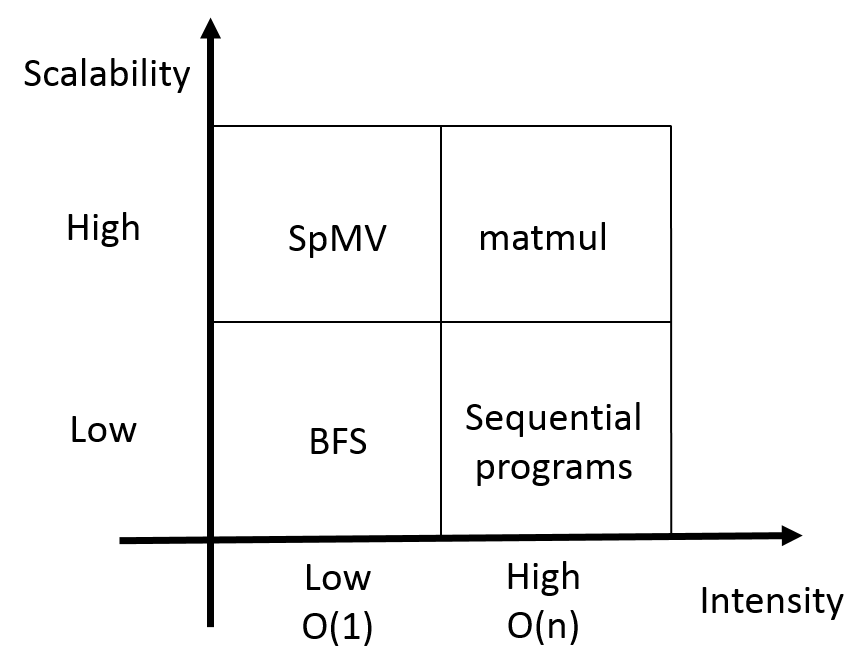}}
\caption{Application Categories}
\label{fig:App-Cate}
\end{figure}

The following application kernels have been chosen to implement and validate the RTHpower models on Myriad1: {\it matmul} (a computation-intensive kernel), SpMV (a kernel with dynamic access patterns), and BFS (a data-intensive kernel of Graph500 benchmarks\cite{6114175}). All three kernels belong to the list of Berkeley dwarfs \cite{Asa06} and are able to cover the two dimensions of operational intensity and performance speed-up as shown in Figure \ref{fig:App-Cate}.
{\it matmul} is proved to have high intensity and scalability \cite{Ofenbeck:2014}. SpMV has low operational intensity and high speed-up due to its parallel scalability \cite{William2007}. BFS, on the other hand, has low operational intensity and saturated low scalability \cite{Cook:2013}. Since the available benchmark suites in literature are not executable on Myriad1 platform, the three mentioned kernels have been implemented by the authors using the Movidius Development Kit for Myriad1. As the RTHpower models will be used to predict whether the RTH strategy is an energy efficient approach for an given application, we focus mainly on two settings: the 8-core setting representing the RTH strategy (i.e., using all available cores of Myriad1) and the 1-core setting  representing the other extreme (i.e., using a minimum number of cores).
\paragraph{Dense Matrix Multiplication}
{\it Matmul} has been implemented on Myriad1 by using both C and assembly languages. The {\it matmul} algorithm computes matrix C based on two input matrices A and B $C=A \times B$. All three matrices in this benchmark are stored in DDR RAM. Matrix elements are stored with float type equivalent to four bytes. The number of operations and accessed data are calculated based on matrix size $n$ as: $W=2\times N^3$ and $Q=16\times N^2$  \cite{Ofenbeck:2014}. Intensity of {\it matmul} is also varied with matrix size as: $I = \frac{W}{Q} = \frac{N}{8}$. 
The experiments are conducted until matrix size 1024x1024, the largest size that Myriad1 RAM memory can accommodate. Figure \ref{fig:MatmulDeviation} shows that the percentage error of {\it matmul} estimated power compared to measured power is -25\% on average for 1 core and 12\% on average for 8 cores. 

We observe that operational intensity is not enough to capture other factors such as the communication pattern and  potential performance/power overheads due to the implementation. E.g., although a sequential version and a parallel version of a {\it matmul} algorithm have the same intensity, it is obvious that they have different communication pattern (intuitively, the sequential version doesn't have communication between cores). Since different parallel versions for different number of cores have different communication patterns (e.g., sequential version vs. 8-core version), ignoring the mentioned factors contributes to the percentage errors. Therefore, we improve the models by introducing a tunning parameter $\beta$ to the models in Equation \ref{eq:Pimp1} and Equation \ref{eq:Pimp2}, where $\beta$ is computed in Equation \ref{eq:beta}. Note that the tuning parameter $\beta$ for each sequential/parallel version (e.g., 1-core version or 8-core version) is fixed across problem sizes and therefore it can be obtained during kernel installation and then saved as meta-data for each version in practice. E.g., with 1 core, $\beta=\frac{1}{1-25\%}$. With 8 cores, $\beta = \frac{1}{1+12\%}$. After the improvement, the percentage errors are at most 10\% as shown in Figure \ref{fig:MatmulDeviation}. 
\begin{equation} 
\label{eq:beta} 
\begin{split}
\beta = \frac{1}{1 + \overline{PE}}.
\end{split} 
\end{equation}
\begin{figure}[!t] \centering
\small
\begin{tikzpicture}
\begin{axis}[
title=Model Percentage Errors of Dense Matrix Multiplication (matmul),
xlabel=Matrix Size,
ylabel=Percentage Error,
width=7cm, height=4cm,
legend style={at={(0.5,-0.5)},
anchor=north,legend columns=-1},
legend columns=2,
symbolic x coords={
128x128,
256x256,
512x512,
1024x1024,
},
x tick label style={rotate=35,anchor=east},
ytick={-20,-10,0,10,20},
yticklabel={\pgfmathprintnumber\tick\%}
]

\addplot+[
error bars/.cd,
x dir=both, x explicit,
y dir=both, y explicit,
]
table[x=size, y=deviation]
{
size	deviation 8-cores 1-core-after-improved 8-cores-after-improved
128x128	-31.34	2.29	-8.03	-9.50
256x256	-29.06	14.78	-4.56	2.42
512x512	-25.13	22.82	0.94	8.81
1024x1024	-17.78	20.72	9.79	7.22
};

\addplot+[
error bars/.cd,
x dir=both, x explicit,
y dir=both, y explicit,
]
table[x=size, y=deviation]
{
size	1-core deviation 1-core after improved 8-cores after improved
128x128	-31.34	2.29	-8.03	-9.50
256x256	-29.06	14.78	-4.56	2.42
512x512	-25.13	22.82	0.94	8.81
1024x1024	-17.78	20.72	9.79	7.22
};

\addplot+[
error bars/.cd,
x dir=both, x explicit,
y dir=both, y explicit,
]
table[x=size, y=deviation]
{
size	1-core 8-cores deviation 8-cores-after-improved
128x128	-31.34	2.29	-8.03	-9.50
256x256	-29.06	14.78	-4.56	2.42
512x512	-25.13	22.82	0.94	8.81
1024x1024	-17.78	20.72	9.79	7.22
};

\addplot+[
error bars/.cd,
x dir=both, x explicit,
y dir=both, y explicit,
]
table[x=size, y=deviation]
{
size	1-core 8-cores 1-core-after-improved deviation
128x128	-31.34	2.29	-8.03	-9.50
256x256	-29.06	14.78	-4.56	2.42
512x512	-25.13	22.82	0.94	8.81
1024x1024	-17.78	20.72	9.79	7.22
};
\legend{1-core-original,8-cores-original,1-core-improved,8-cores-improved} 
\end{axis}
\end{tikzpicture}
\caption{Absolute percentage errors of estimated power from measured power of {\it matmul}. After incorporating the tuning parameter, the absolute percentage errors of {\it matmul} are at most 10\%.}
\label{fig:MatmulDeviation}
\end{figure}
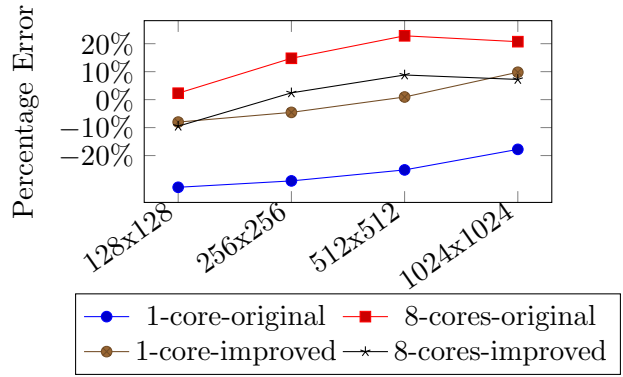
\begin{equation} 
\label{eq:Pimp1} 
\begin{split}
	P_{improved}= (P^{comp||data} \times \frac{I}{\alpha}+  P^{data} \times \frac{\alpha-I}{\alpha}) \times \beta
\end{split} 
\end{equation}
\begin{equation} 
\label{eq:Pimp2} 
\begin{split}
	P_{improved}= (P^{comp||data} \times \frac{\alpha}{I} + P^{comp} \times \frac{I-\alpha}{I})\times \beta
\end{split} 
\end{equation}
\paragraph{Sparse Matrix Vector Multiplication}
SpMV implementation on Myriad1 is written in C language. All input matrix and vector of this benchmark reside in DDR RAM. This implementation uses the common data layout of SpMV which is compressed sparse row (csr) format \cite{Saad:2003}. There is no random generator supported in the RISC core so a 5 non-zero elements per row is fixed in all experiments. 
Each element of matrix and vector is stored with float type of four bytes. 
From our implementation analysis, the number of operations and accessed data are proportional to the size of a matrix dimension $N$ as: $W=5 \times 2\times N$ and $Q=5 \times 2 \times 4 \times N$. Operational intensity of SpMV  therefore, does not depend on matrix size and is a fixed value: $I = \frac{W}{Q}= 0.25$.

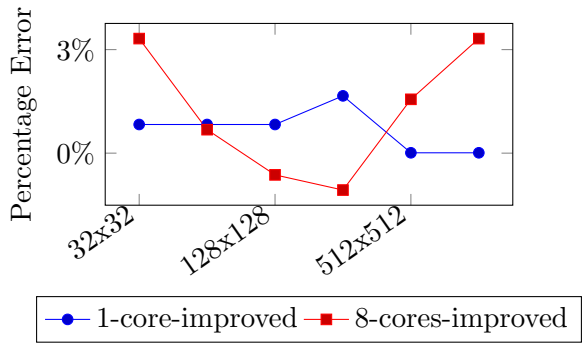
\begin{figure}[!t] \centering
\small
\begin{tikzpicture}
\begin{axis}[
title=Model Percentage Errors of Sparse Matrix Vector Multiplication,
xlabel=Matrix Size,
ylabel=Percentage Error,
width=7cm, height=4cm,
legend style={at={(0.5,-0.5)},
anchor=north,legend columns=-1},
legend columns=2,
symbolic x coords={32x32,
64x64,
128x128,
256x256,
512x512,
1024x1024,
},
x tick label style={rotate=35,anchor=east},
ytick={-6,-3,0,3,6},
yticklabel={\pgfmathprintnumber\tick\%}
]

\addplot+[
error bars/.cd,
x dir=both, x explicit,
y dir=both, y explicit,
]
table[x=size, y=deviation]
{
size	1-core 8-cores deviation 8-cores-after-improved
32x32	14.19	-6.25	0.83	3.32
64x64	14.19	-8.74	0.83	0.68
128x128	14.19	-9.93	0.83	-0.63
256x256	15.15	-10.32	1.66	-1.07
512x512	13.25	-7.92	0.01	1.56
1024x1024	13.25	-6.25	0.01	3.32
};

\addplot+[
error bars/.cd,
x dir=both, x explicit,
y dir=both, y explicit,
]
table[x=size, y=deviation]
{
size	1-core 8-cores 1-core-after-improved deviation
32x32	14.19	-6.25	0.83	3.32
64x64	14.19	-8.74	0.83	0.68
128x128	14.19	-9.93	0.83	-0.63
256x256	15.15	-10.32	1.66	-1.07
512x512	13.25	-7.92	0.01	1.56
1024x1024	13.25	-6.25	0.01	3.32
};
\legend{1-core-improved,8-cores-improved} 
\end{axis}
\end{tikzpicture}
\caption{Absolute percentage errors of estimated power from measured power of SpMV. After incorporating the tuning parameter, the absolute percentage errors of SpMV are at most 4\%.}
\label{fig:SpMVDeviation}
\end{figure}

Figure \ref{fig:SpMVDeviation} shows the percentage error of SpMV estimated power using Equation \ref{eq:Pimp1} and \ref{eq:Pimp2} compared to measured power. The $\beta$ values for 1-core and 8-core versions of SpMV are $\frac{1}{1 + 14\%}$ and $\frac{1}{1 - 9\%}$, respectively. The absolute percentage errors are at most 4\% as shown in Figure \ref{fig:SpMVDeviation}. SpMV has lower modeling errors than {\it matmul} since SpMV has a fixed intensity value on different matrix sizes.
\paragraph{Breadth First Search}
We also implemented BFS - a data-intensive Graph500 kernel, on Myriad1. BFS is the graph kernel to explore the vertices and edges of a directed graph from a starting vertex. We use the implementation of current Graph500 benchmark (omp-csr) and port it to Myriad1. The output BFS graphs after running BFS implementation on Myriad1 are verified by the verification step of original Graph500 code to ensure the output graphs are correct.

The size of a graph is defined by its scale and edgefactor. In our experiments, we mostly use the default edgefactor of 16 from the Graph500 so that each vertex of the graph has $16$ edges in average. The graph scales are varied from 14 to 17 and the graphs has from $2^{14}$ to $2^{17}$ vertices. It is noted that graph scale 17 is the largest scale that the DDR RAM of Myriad1 can accommodate. From the implementation analysis, the operational intensity of BFS is a fixed value: $I = \frac{W}{Q} = 0.257$ and does not depend on edgefactor or scale. 

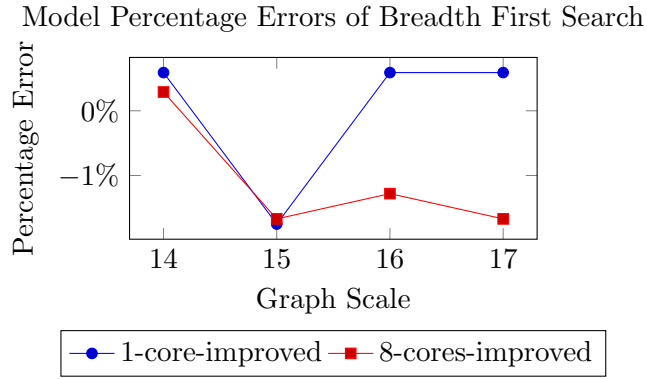
\begin{figure}[!t] \centering
\small
\begin{tikzpicture}
\begin{axis}[
title=Model Percentage Errors of Breadth First Search,
xlabel=Graph Scale,
ylabel=Percentage Error,
width=7cm, height=4cm,
legend style={at={(0.5,-0.5)},
anchor=north,legend columns=-1},
legend columns=2,
symbolic x coords={14,15,16,17},
ytick={-2,-1,0,1,2},
yticklabel={\pgfmathprintnumber\tick\%},
]

\addplot+[
error bars/.cd,
x dir=both, x explicit,
y dir=both, y explicit,
]
table[x=scale, y=deviation]
{
scale	deviation	8-cores-after	1-core-before	8-cores-before
14	0.59	0.29	8.81	-17.79
15	-1.75	-1.67	6.28	-19.40
16	0.59	-1.28	8.81	-19.08
17	0.59	-1.67	8.81	-19.40
};

\addplot+[
error bars/.cd,
x dir=both, x explicit,
y dir=both, y explicit,
]
table[x=scale, y=deviation]
{
scale	1-core-after	deviation	1-core-before	8-cores-before
14	0.59	0.29	8.81	-17.79
15	-1.75	-1.67	6.28	-19.40
16	0.59	-1.28	8.81	-19.08
17	0.59	-1.67	8.81	-19.40
};
\legend{1-core-improved,8-cores-improved} 
\end{axis}
\end{tikzpicture}
\caption{Absolute percentage errors of estimated power from measured power of BFS. After incorporating the tuning parameter, the absolute percentage errors are at most 2\%.}
\label{fig:BFSDeviation}
\end{figure}

Figure \ref{fig:BFSDeviation} shows the percentage error of BFS estimated power using Equation \ref{eq:Pimp1} and \ref{eq:Pimp2} compared to measured power . The $\beta$ values for 1-core and 8-core versions of BFS are $\frac{1}{1+8\%}$ and $\frac{1}{1-19\%}$, respectively. The absolute percentage errors are at most 2\% as shown in Figure \ref{fig:BFSDeviation}.
\subsection{Race-to-halt Prediction Framework}
\label{sec:Framework}
With the RTHpower models, we want to identify how many cores the system should use to run an application to achieve the least energy consumption. 
In order to answer the question, we need to consider the performance 
speed-up and power-up of an application on a specific platform. 

From Amdahl's Law~\cite{4563876} the theoretical maximum speed-up of an application running on a multicore system is derived as Equation \ref{eq:Amdahl-law}, where $p$ denotes the fraction of the application that can be parallelized and $n$ is the number of cores:
\begin{equation} 
\label{eq:Amdahl-law} 
	\mbox{\it speed-up} \leq \frac{1}{(1-p)+ \frac{p}{n}}
\end{equation}
\subsubsection{Framework Description}
The purpose of this framework is to identify when to and when not to use RTH for a given application. The two required inputs for making decision are power-up and performance speed-up of the application executed with $n$ cores, where $n$ is the maximum number of cores. 
\begin{itemize}
\item Step 1: Identify meta-data, including speed-up and operational intensity, of a given application by one of the three main approaches listed: i) doing theoretical analysis to find the amount of computation work $W$, data transfer $Q$ and operational intensity $I$ as well as identify the maximum speed-up of a given application; ii) executing the application on a targeted platform (e.g., Myriad1) to measure its performance speed-up and extract its operational intensity $I$; iii) using profiling tools \cite{Liu2014} to extract the number of operations $W$ and the amount of data transferred $Q$ as well as the performance speed-up of an application on a common platform (e.g., Intel platform).
\item Step 2: Compute power consumption of an application running with one core and with a maximum number of cores by the RTHpower models. Note that the RTHpower models are able to estimate power consumption for any number of cores by changing parameter $n$ in the models. For verifying the RTH strategy, we only need to apply the model for a single core and all cores. 
\item Step 3: Compare the energy consumption of the application between using 1 core and using a maximum number of cores to identify whether running a maximum number of cores is the most energy-efficient.
\end{itemize}
\subsubsection{Framework Validation}
The framework is validated with three micro-benchmarks and three application kernels. In this validation, the values of operational intensity $I$ are extracted from theoretical analysis of the implementations and performance speed-up is identified by executing the micro-benchmarks or application kernels with different numbers of cores.
\paragraph{Race-to-halt for Micro-benchmarks}
We first validate the framework with micro-benchmarks. In this validation, we measure the power-up and performance speed-up of three micro-benchmarks: one with 60\% parallel code, one with 100\% parallel code and a small-size micro-benchmarks which has high overhead. All three micro-benchmarks have operational intensity $I=0.25$. Namely, in the micro-benchmarks, each SauXor instruction is followed by a LsuLoad instruction which loads 4 bytes.

All three micro-benchmarks have the same assembly code wrapped inside a loop. The number of iterations to repeat the code are the difference among them. 
We run the micro-benchmarks on one SHAVE for 1 000 000 times. If the micro-benchmark has 100\% parallel code, running it on $n$ SHAVEs requires each SHAVE performing $\frac{1}{n}$ of the amount of work (e.g., if performing the micro-benchmark on 8 SHAVEs, each SHAVE needs to run 125 000 times). Similarly, if the micro-benchmark has a parallel fraction of 60\%, then running the program on $n$ SHAVEs requires each SHAVE to perform $(1-0.6)+(\frac{0.6}{n})$ of the amount of work (e.g.,if performing the micro-benchmark on 8 SHAVEs, each SHAVE needs to run 475 000 times). For small-size micro-benchmark, the code is executed 8 times with 1 core and once with 8 cores. Since the amount of computation is small, the relative overhead of initializing the platform and executing the small-size micro-benchmark is high. 

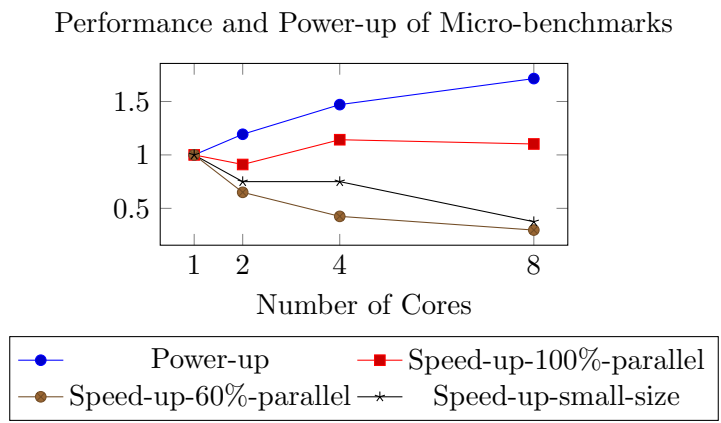
\begin{figure}[!t] \centering
\small
\begin{tikzpicture}
\begin{axis}[
title=Performance and Power-up of Micro-benchmarks,
xlabel=Number of Cores,
width=7cm, height=4cm,
legend style={at={(0.5,-0.5)},
anchor=north,legend columns=-1},
legend columns=2,
xtick={1, 2, 4, 8},
]

\addplot+[
error bars/.cd,
x dir=both, x explicit,
y dir=both, y explicit,
]
table[x=cores, y=powerscalability]
{
cores	powerscalability
1	1
2	1.193277311
4	1.470588235
8	1.714285714
};

\addplot+[
error bars/.cd,
x dir=both, x explicit,
y dir=both, y explicit,
]
table[x=cores, y=speedup]
{
cores speedup 
1	1
2	0.910371319
4	1.142168675
8	1.102325581
};

\addplot+[
error bars/.cd,
x dir=both, x explicit,
y dir=both, y explicit,
]
table[x=cores, y=speedup]
{
cores speedup 
1	1
2	0.649908592
4	0.424985057
8	0.296806512
};

\addplot+[
error bars/.cd,
x dir=both, x explicit,
y dir=both, y explicit,
]
table[x=cores, y=speedup]
{
cores speedup 
1	1
2	0.75
4	0.75
8	0.375
};

\legend{Power-up, Speed-up-100\%-parallel, Speed-up-60\%-parallel, Speed-up-small-size} 
\end{axis}
\end{tikzpicture}
\caption{Performance and power-up of micro-benchmarks with operational intensity $I=0.25$. All three reported micro-benchmarks have performance speed-up less than platform power-up.}
\label{fig:Micro-Speed-up}
\end{figure}

\begin{figure}[!t] \centering
\small
\begin{tikzpicture}
\begin{axis}[
title=Energy Consumption of Micro-benchmarks,
xlabel=Number of Cores,
ylabel=Energy (mJ),
width=7cm, height=4cm,
legend style={at={(0.5,-0.5)},
anchor=north,legend columns=-1},
legend columns=2,
xtick={1, 2, 4, 8},
]

\addplot+[
error bars/.cd,
x dir=both, x explicit,
y dir=both, y explicit,
]
table[]
{
x	y
1	169.218
2	221.804
4	217.875
8	263.16

};

\addplot+[
error bars/.cd,
x dir=both, x explicit,
y dir=both, y explicit,
]
table[]
{
x	y
1	194.8528068
2	229.4848075
4	207.5381265
8	266.0719045
};

\addplot+[
error bars/.cd,
x dir=both, x explicit,
y dir=both, y explicit,
]
table[]
{
x	y
1	169.218
2	310.696
4	585.55
8	977.364
};

\addplot+[
error bars/.cd,
x dir=both, x explicit,
y dir=both, y explicit,
]
table[]
{
x	y
1	194.8528068
2	321.4550312
4	557.7691334
8	988.1786779
};

\addplot+[
error bars/.cd,
x dir=both, x explicit,
y dir=both, y explicit,
]
table[]
{
x	y
1	342
2	500
4	600
8	1616
};

\addplot+[
error bars/.cd,
x dir=both, x explicit,
y dir=both, y explicit,
]
table[]
{
x	y
1	357
2	568
4	700
8	1632
};

\legend{100\%-parallel-measurement,100\%-parallel-estimation,60\%-parallel-measurement,60\%-parallel-estimation, small-size-measurement,small-size-estimation} 
\end{axis}
\end{tikzpicture}
\caption{Energy consumption of micro-benchmarks with operational intensity $I=0.25$.
For all three reported micro-benchmarks, the programs executed with 1 core consume the least energy, compared to 2, 4, 8 cores, from both measured data and estimated data.}
\label{fig:Micro-Energy}
\end{figure}
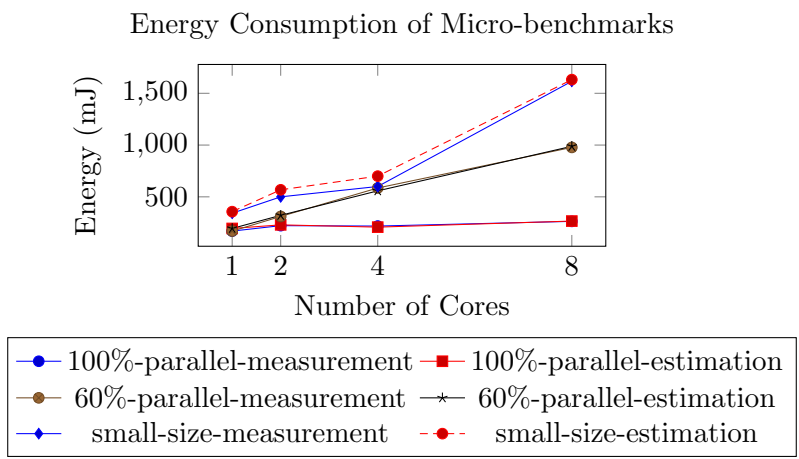

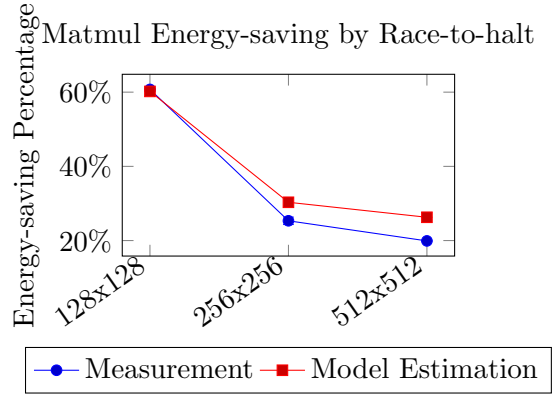
\begin{figure}[!t] \centering
\small
\begin{tikzpicture}
\begin{axis}[
title=Matmul Energy-saving by Race-to-halt,
xlabel=Matrix Size,
ylabel=Energy-saving Percentage,
width=6cm, height=4cm,
legend style={at={(0.5,-0.5)},
anchor=north,legend columns=-1},
symbolic x coords={
128x128,
256x256,
512x512,
},
x tick label style={rotate=35,anchor=east},
yticklabel={\pgfmathprintnumber\tick\%}
]
\addplot+[
error bars/.cd,
x dir=both, x explicit,
y dir=both, y explicit,
]
table[y error=error]
{
x y error
128x128	60.73 0
256x256	25.33 1
512x512	19.91 0
};

\addplot+[
error bars/.cd,
x dir=both, x explicit,
y dir=both, y explicit,
]
table[y error=error]
{
x y error
128x128	60.20 0
256x256	30.32 1
512x512	26.28 0
};

\legend{Measurement, Model Estimation} 
\end{axis}
\end{tikzpicture}
\caption{{\it matmul} Energy-saving by Race-to-halt. This diagram shows how many percentages of energy-saving if executing {\it matmul} with 8 cores instead of 1 core. Since the energy-saving percentage is positive over different matrix sizes, RTH is an energy-saving strategy for {\it matmul}. Energy-saving percentage from model estimation for {\it matmul} has standard deviation less than 3\%.}
\label{fig:Matmul-RTH}
\end{figure}

\begin{figure}[!t] \centering
\small
\begin{tikzpicture}
\begin{axis}[
title=SpMV Energy-saving by Race-to-halt,
xlabel=Matrix Size,
ylabel=Energy-saving Percentage,
width=8cm, height=4cm,
legend style={at={(0.5,-0.5)},
anchor=north,legend columns=-1},
symbolic x coords={32x32,
64x64,
128x128,
256x256,
512x512,
1024x1024,
},
x tick label style={rotate=35,anchor=east},
yticklabel={\pgfmathprintnumber\tick\%}
]
\addplot+[
error bars/.cd,
x dir=both, x explicit,
y dir=both, y explicit,
]
table[y error=error]
{
x y error
32x32	46.46	0.00
64x64	53.76	0.98
128x128	59.28	0.55
256x256	58.14	3.00
512x512	55.46	1.66
1024x1024	49.06	1.34
};

\addplot+[
error bars/.cd,
x dir=both, x explicit,
y dir=both, y explicit,
]
table[y error=error]
{
x y error
32x32	44.99	0.00
64x64	53.75	0.98
128x128	59.81	0.56
256x256	59.21	3.05
512x512	54.68	1.63
1024x1024	47.22	1.29

};

\legend{Measurement, Model Estimation} 
\end{axis}
\end{tikzpicture}
\caption{SpMV Energy-saving by Race-to-halt. This diagram shows how many percentages of energy-saving if execute SpMV with 8 cores instead of 1 core. Since the energy-saving percentage is positive over different matrix sizes, RTH is a energy-saving strategy for SpMV.}
\label{fig:SpMV-RTH}
\end{figure}
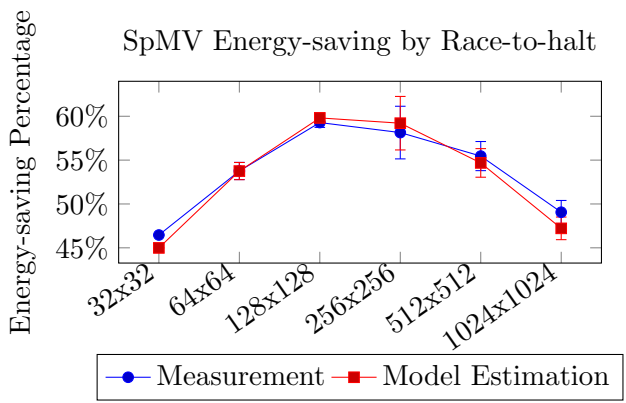

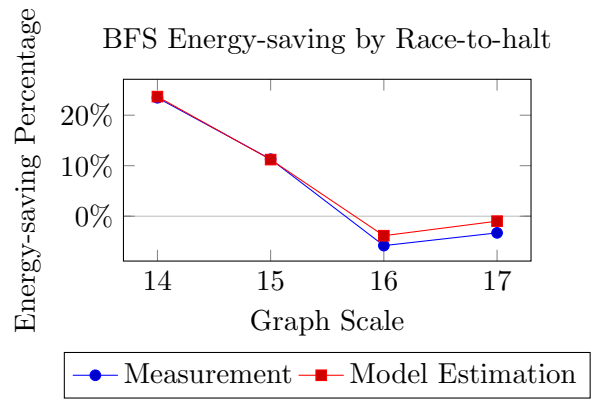
\begin{figure}[!t] \centering
\small
\begin{tikzpicture}
\begin{axis}[
title=BFS Energy-saving by Race-to-halt,
xlabel=Graph Scale,
ylabel=Energy-saving Percentage,
width=7cm, height=4cm,
legend style={at={(0.5,-0.5)},
anchor=north,legend columns=-1},
xtick=data,
minor ytick={0},grid=minor,
yticklabel={\pgfmathprintnumber\tick\%}
]
\addplot+[
error bars/.cd,
x dir=both, x explicit,
y dir=both, y explicit,
]
table[y error=error]
{
x y error
14	23.45	0.66
15	11.29	0.30
16	-5.82	-0.06
17	-3.30	-0.04
};

\addplot+[
error bars/.cd,
x dir=both, x explicit,
y dir=both, y explicit,
]
table[y error=error]
{
x y error
14	23.68 0
15	11.21 0
16	-3.86 0
17	-0.98 0

};

\legend{Measurement,Model Estimation} 
\end{axis}
\end{tikzpicture}
\caption{BFS Energy-saving by Race-to-halt. This diagram shows how many percentages of energy-saving if executing BFS with 8 cores instead of 1 core. The positive percentages at scale 14 and 15 mean that RTH should be applied. The negative percentages at scale 16 and 17 mean that RTH should not be applied. The standard deviation of BFS energy-saving percentage is less than 3\%, from scale 14 to 17.}
\label{fig:BFS-RTH}
\end{figure}

Figure~\ref{fig:Micro-Speed-up} shows that the power-up of running $n$ SHAVEs to the program running 1 SHAVE varies from 1 (1 core) to 1.71 (8 cores) for operational intensity $I=0.25$. If the performance speed-up is bigger than the power-up, RTH is an energy-saving strategy. If the speed-up is less than the power-up, running the program with the maximum number of cores consumes more energy than running it with 1 core. Note that when this happens, assigning one core to run the program is more energy-efficient and race-to-halt is no longer applicable for saving energy. For all three micro-benchmarks in this validation, the performance speed-up is identified by running them over different numbers of cores.
The energy consumption of the three micro-benchmarks is shown in Figure \ref{fig:Micro-Energy}. All three micro-benchmarks achieve the least energy consumption when executed with one core, from both measured and estimated data. The model estimation and actual measurement show that RTH is not applicable to the three micro-benchmarks.
\paragraph{Race-to-halt for Dense Matrix Multiplication}
The {\it matmul} application has increasing values of operational intensity over input sizes and its performance speed-up is higher than its power-up on Myriad1. 
Therefore, running {\it matmul} with the 8 cores is more energy-efficient than running it with one core. Figure ~\ref{fig:Matmul-RTH} shows percentages of energy-saving if executing {\it matmul} with 8 cores instead of 1 core, from both measured and estimated data. The energy saving percentage is computed based on the energy gap of running 1 core and 8 cores divided by energy consumed by running 1 core as in Equation \ref{eq:ES}.
\begin{equation} 
\label{eq:ES} 
	 ES = \frac{E^{1core} - E^{8cores}}{E^{1core}}
\end{equation}
The framework predict that RTH should be applied to {\it matmul} over different matrix sizes. By using RTH for {\it matmul}, we can save from 20\% to 61\% of {\it matmul} energy consumption. RTH is a good strategy for {\it matmul}. We observe that the energy saving reduces when matrix size increases due to the decrease of performance speed-up from size 128x128. The reason is that a matrix size bigger than 128x128 makes the data set no longer fit in the last level cache (or L2 cache of 64KB) and thereby lowers performance (in flops).  
\paragraph{Race-to-halt for Sparse Matrix Vector Multiplication}
SpMV has a fixed value of operational intensity over input sizes. From the RTHpower models as well as measurement data, the power-up of SpMV is relatively constant. However, SpMV has performance speed-up higher than its power-up. 
Therefore, running SpMV with the maximum number of cores is more energy-efficient than running it with one core. Figure ~\ref{fig:SpMV-RTH} shows how many percentages of energy-saving if executing SpMV with 8 cores instead of 1 core, from both measured and estimated data. The framework can predict that RTH should be applied to SpMV over different matrix sizes. By using RTH for SpMV, we can save from 45\% to 59\% of SpMV energy consumption. RTH is a good strategy for SpMV. The energy saving increases from size 32x32 to 128x128 since the data fits in L1 cache. 
\paragraph{Race-to-halt for Breadth First Search}
In our set of application kernels implemented on Myriad1, 
BFS is the application kernel able to prove that running with a
maximum number of cores does not always give the least energy consumption. From both measured data and estimated data, the total energy consumed by running with one core is less than the total energy by running with 8 cores at scale 16 and 17. There are negative values of -5\% and -3\% in Figure~\ref{fig:BFS-RTH} if running BFS with 8 cores instead of 1 cores at scale 16 and 17, respectively. The RTHpower models can predict when to apply RTH precisely for different scales. 

The result can be explained by the relation between BFS power-up and performance speed-up. Since BFS has a fixed value of operational intensities across graph scale, from RTHpower models (cf. Equation \ref{eq:Pimp1} and \ref{eq:Pimp2}), it is understood that BFS power consumption does not depend on graph scales and its power-up is a fixed value. From the measurement results, we also observe that BFS power-up is relatively constant over the graph scales. However, BFS speed-up in our experiments decreases when scale increases. The reason is that with the same graph degree, when scale increases, the graph becomes more sparse and disconnected. Compared to the Graph500 implementation, BFS search on Myriad1 are performed from a chosen subset of source nodes. The speed-up then, becomes less than power-up at scale 16 and 17. Therefore, running BFS with 8 cores at bigger graph scales (i.e., 16 and 17 in our experiments) consumes more energy than running BFS with one core.
\subsection{Conclusion}
In this study, new fine-grained power models have been proposed to provide insights into how a given application consumes energy when executing on an ultra-low power embedded system. In the models, applications are represented by their operational intensity. The models have been validated on Movidius Myriad1, an ultra-low power embedded platform. Experimental results on 31 micro-benchmarks and three application kernels have shown high accuracy of estimated data by the model. 
Based on the models, we have devised a new framework to predict whether RTH is applicable to a given application. The framework has been validated by both micro-benchmarks and real application kernels, showing a prediction accuracy that is good enough for the purpose of deciding about RTH. Improving and applying the models and framework to other embedded platforms (e.g. ARM) and application kernels (e.g. other Berkeley dwarfs) are parts of our future work.    

%% file: EPEM.tex
\subsection{Introduction}
\input{EPEM/introduction}

\subsection{Related Work - Overview of energy models}
\label{related-work}
\input{EPEM/related-work}

\subsection{EPEM Shared Memory Machine Model}
\label{EPEM-model}
\input{EPEM/EPEM-memory-model.tex}

\subsection{Energy Complexity in EPEM model}
\label{energy-model}
\input{EPEM/EPEM-energy-complexity-model} 

\subsection{A Case Study - SpMV Energy Complexity}
\label{SpMV-energy}
\input{EPEM/SpMV-energy-complexity}

\subsection{Validation of EPEM Model}
\label{validation}
\input{EPEM/SpMV-implementation}
\input{EPEM/validation}

\subsection{Conclusion}
\input{EPEM/conclusion}

%% file: EPEM/introduction.tex
Understanding the energy complexity of algorithms is crucially important to improve the energy efficiency of algorithms and reducing energy consumption of computing systems. By knowing the energy complexity of algorithms, the algorithm designers can choose one algorithm over the others to achieve energy optimization. Devising energy models is one of the main approaches to characterize the energy complexity of algorithms on computing systems. Energy models help to investigate the trade-offs between energy consumption and performance of algorithms as well as their inter-process communication. 
 
Significant efforts have been devoted to develop power and energy models in literature \cite{Alonso2014, Choi2013, Choi2014, Korthikanti2009, Korthikanti2010, 7108419, Mishra:2015, Snowdon:2009}. However, there are no analytic models for multithreaded algorithms that are both applicable to every algorithms and comprehensively validated yet. The existing {\em parallel} energy models are either theoretical studies without validation or only applicable for specific algorithms. Modeling energy consumption of {\em parallel} algorithms is difficult since the energy models need to model both algorithm characteristics and platform properties. Algorithm characteristics include computational workload, memory workload, data-accessing patterns and scalable parallelism. Platform properties include static and dynamic energy of memory accesses and computational operations. The previous energy model studies have not considered all the mentioned algorithm and platform properties.

The existing models and their classification are summarized in Table \ref{table:energy-model-summary}. The previous studies have not covered all listed aspects: ability to analyze the energy complexity of parallel algorithms (i.e. Energy complexity analysis for parallel algorithms), whether applicable to general algorithm (i.e., Algorithm Generality), whether the model is validated (i.e., Validation). Table \ref{table:energy-model-summary} also shows how this work is different from the other studies. 
\begin{table*}
\caption{Energy Model Summary}
\label{table:energy-model-summary}
\begin{center}
\begin{tabular}{lllll}
\hline\noalign{\smallskip}
\textbf{Study} 	 &\textbf{Energy complexity} 	&\textbf{Algorithm}  	 &\textbf{Validation}  	 \\
	 &\textbf{analysis for} 	&\textbf{ generality}  		&  	\\
	 &\textbf{parallel algorithms} && 				\\
\noalign{\smallskip}\hline\noalign{\smallskip}
LEO \cite{Mishra:2015} 	&No 	&General	&Yes 	\\
POET \cite{7108419} 	&No 	&General	&Yes	\\
Koala \cite{Snowdon:2009} 	&No 	&General	&Yes	\\
Roofline \cite{Choi2013,Choi2014}	&No	&General	&Yes	\\  
Energy scalability \cite{Korthikanti2009, Korthikanti2010}	&Yes	&General	&No	\\ 
Sequential energy complexity \cite{Roy2013}   	&No	&General	&Yes\\
Alonso  et al. \cite{Alonso2014}	&Yes	&Algorithm-specific	&Yes	\\
Malossi  et al. \cite{Malossi2015}	&Yes	&Algorithm-specific	&Yes	\\
\textbf{EPEM model}       	&\textbf{Yes}	&\textbf{General}	&\textbf{Yes}	\\
\noalign{\smallskip}\hline\noalign{\smallskip}
\end{tabular}
\end{center}
\end{table*}

The energy complexity model EPEM (Energy-aware Parallel External Memory) proposed in this study is for general multithreaded algorithms and validated on three aspects: different algorithms for a given problem, different input types types on different platforms. The proposed model is an analytic model which characterizes both algorithms (e.g., representing algorithm by {\em work}, {\em span} and {\em I/O} complexity) and platform properties (e.g., representing platforms with static and dynamic energy of memory accesses and computational operations). By considering {\em work} and {\em span} complexity, the new energy model can applied to any multithreaded algorithms. 
 
Since the new EPEM energy model focuses on analyzing the energy complexity of algorithms, the model does not give the estimation of absolute energy consumption. The new model, however, provides the algorithms designers the understanding of how an algorithms consumes energy and give the insight on how to choose one algorithms over the others on different input types and platforms. This methodology has been validated for two SpMV algorithms running on two high performance platforms (Intel Xeon and Xeon Phi), computing nine matrix input types from Florida matrix collection \cite{Davis:2011}. The validation results prove the practicability and applicability the EPEM energy complexity model when comparing the energy consumption of different algorithms as well as comparing different input types on different platforms.
In this work, the following contributions have been made.
\begin{itemize}
\item Devise a new energy model EPEM for analyzing the energy complexity of multithreaded algorithms based on their {\em work}, {\em span} and {\em I/O} complexity (cf. Section \ref{energy-model}). 
\item Conduct a case study to demonstrate the methodology how to apply the EPEM model to find energy complexity of three sparse matrix vector multiplication (SpMV) algorithms (i.e., Compressed Sparse Column(CSC) and Compressed Sparse Block(CSB) and Compressed Sparse Row(CSR))(cf. Section \ref{SpMV-energy}). 
\item Validate the EPEM energy complexity model according to three aspects: different algorithms, different input types and different platforms. The results show the precise prediction on which validated SpMV algorithm (i.e., CSB or CSC) consumes more energy when computing different matrix input types from Florida matrix collection \cite{Davis:2011} (cf. Section \ref{validation}). The model platform-related parameters for 11 platforms, including x86, ARM and GPU, are also provided for further uses of the proposed energy complexity model.  
\end{itemize}

%% file: EPEM/related-work.tex
\begin{table*}
\caption{Energy Model Details}
\label{table:energy-model-details}
\centering
\resizebox{\columnwidth}{!}{%
\begin{tabular}{llllllll}
\hline\noalign{\smallskip}
\textbf{Study} 	 &\textbf{Parallel-} 	&\textbf{Applicability}  	 &\textbf{Validation} 	&\textbf{Communication}  	 &\textbf{Pre-run} 	 	 &\textbf{Application} \\
 	& \textbf{Algorithm } 	&\textbf{}  	  	&\textbf{} &\textbf{model}  	 &\textbf{Overhead} 	&\textbf{properties} \\ 	  
	 &\textbf{Support} &&&&&&\\						
\noalign{\smallskip}\hline\noalign{\smallskip}						
LEO \cite{Mishra:2015} 	&parallel 	&Yes 	     &Yes 	&No	&Yes	       	&None\\
\noalign{\smallskip}\hline\noalign{\smallskip}	
POET \cite{7108419} 	&parallel 	&Yes	    & Yes  	&No	&No	 	&None\\
\noalign{\smallskip}\hline\noalign{\smallskip}	
Koala \cite{Snowdon:2009} 	&parallel 	&Yes	     &Yes  	&No	&Yes&None \\
\noalign{\smallskip}\hline\noalign{\smallskip}	
Roofline        	&sequential	&Yes	     &Yes  	&Von Neumann	&No  	 &Operational  \\    
 \cite{Choi2013,Choi2014}		       	&&&       	&shared cached		 &&intensity	       \\
\noalign{\smallskip}\hline\noalign{\smallskip}
Energy	&parallel	&Yes	&No	&Message passing	&No 	&No. of messages\\
scalability \cite{Korthikanti2009}			&&&&&&				No. of computations\\
\noalign{\smallskip}\hline\noalign{\smallskip}				
Energy	&parallel	&Yes	&No	&CREW PEM	&No 	&No. of mem-accesses\\
scalability \cite{Korthikanti2010}							&&&&&&No. of computations\\
\noalign{\smallskip}\hline\noalign{\smallskip}	
Sequential 	&sequential	&Yes	     &Yes	&Uni-processor	&No	&Work complexity      \\
energy   			&&&	& with parallel	&	 	&I/O complexity\\
complexity \cite{Roy2013}   			&&	&	&memory-bank&	 	&\\
\noalign{\smallskip}\hline\noalign{\smallskip}	
Alonso	&parallel	&No(Dense matrix	     &Yes  	&No	&Yes	& Application tasks     \\
 et al. \cite{Alonso2014}		&&factorization)	&&&&		\\		
\noalign{\smallskip}\hline\noalign{\smallskip}	
Malossi 	&parallel	&No(Algebraic 	     &Yes  	&Shared memory	&Yes    	&No. of arithmetic, barrier      \\
et al. \cite{Malossi2015}	  	&& kernels)   					 &&&&mem-accesses, reduction\\
\noalign{\smallskip}\hline\noalign{\smallskip}	
EPEM        	&parallel	&Yes	&Yes	&EPEM	&No   	&Work, Span, I/O\\
model	     		&&&				&&&Input types    \\  
\noalign{\smallskip}\hline\noalign{\smallskip}	
\end{tabular}
}
\end{table*}
Devising accurate power models is crucial to gain insights into how a computer system consumes energy. Significant efforts have been devoted to predict energy, resulting in several energy model studies in the literature including analytic models \cite{Alonso2014, Choi2013, Choi2014, Korthikanti2009, Korthikanti2010} and energy models to find energy-optimized system configurations  \cite{7108419, Mishra:2015, Snowdon:2009}. We present the summary of existing modeling studies in Table \ref{table:energy-model-summary}. The characteristics of each approaches are extracted as the list of categories, including: whether the models support parallel algorithms (i.e., Parallel Algorithm Support), whether the model is applicable to general algorithms (i.e., Applicability), whether the model is validated (i.e., Validation), the communication model (i.e., Communication model), whether there is pre-run overhead before estimating energy consumption of applications (i.e., Pre-run Overhead) and how the model represents applications (i.e., App-properties). This summary is not an exhaustive survey on the topic of energy models. However, we believe the Table \ref{table:energy-model-details} represents the most current studies on energy models.

Energy models for finding energy-optimized system configurations for a given application have been recently reported [12, 16, 19]. Imes et al. \cite{7108419} used controller theory and linear programming to find energy-optimized configurations for an application with soft real-time constraints at runtime. Mishra et al. \cite{Mishra:2015} used hierarchical Bayesian model in machine learning to find  energy-optimized configurations. They used offline learning to train the Bayesian model with a training set of applications with different patterns, and used online learning to quickly estimate the optimal configuration for a given application. Snowdon et al. \cite{Snowdon:2009} developed a power management framework called Koala which models the energy consumption of the platform and monitors an application' energy behavior. By matching an application's behavior with the system policy, an energy-optimized configuration is determined at runtime. Although the energy models for finding energy-optimized system configurations have resulted in energy saving in practice, they focus on characterizing system platforms rather than applications and therefore are not appropriate for analyzing the energy complexity of application algorithms. 

Another direction of energy modeling study is to predict the energy consumption of applications by analyzing applications without actual execution on real platforms which we classify as analytic models. The analytic modeling approaches also provide algorithm designers the understanding about how their applications consume energy, helping them to improve the energy complexity of their algorithms.

Energy roofline models \cite{Choi2013, Choi2014} are some of the comprehensive energy models that abstract away possible algorithms in order to analyze and characterize different multicore platforms in terms of energy consumption. The models abstract possible algorithms by their operational intensity, the ratio of computation to communication (i.e., flop/byte), and characterize a platform's properties by running a set of micro-benchmarks on the platform. Our new energy model, which abstracts away possible multicore platform and characterize the energy complexity of algorithms based on their {\em work, span} and {\em I/O} complexity, complements the energy roofline models.  

Validated energy models for {\em specific} algorithms have been reported recently \cite{Alonso2014, Malossi2015}. Alonso et al. \cite{Alonso2014} provided an accurate energy model for three key dense matrix factorizations. Malossi et al. \cite{Malossi2015} focused on basic linear-algebra kernels and characterized the kernels by the number of arithmetic operations, memory accesses, reduction and barrier steps. Although the energy models for specific algorithms are accurate for the target algorithms, they are not applicable for other algorithms and therefore cannot be used as general energy complexity models for parallel algorithms.

The {\em energy scalability} of a parallel algorithm has been investigated by Korthikanti et al. \cite{Korthikanti2009, Korthikanti2010}. The energy scalability studies are to find the optimal number of cores for a given algorithm with a real-time constraint  which minimizes energy consumption. Our studies complement the energy scalability studies by addressing the following {\em energy complexity} question: {\em Given two parallel algorithms A and B for a given problem, which algorithm consumes less energy analytically?}. Unlike the energy scalability studies that have not been validated on real platforms, our new energy complexity model is validated on HPC and accelerator platforms, confirming its usability and accuracy.

The energy complexity of {\em sequential} algorithms on a {\em uniprocessor} machine with {\em several memory banks} has been studied by Roy et al. \cite{Roy2013}. Our energy complexity studies complement Roy et al.'s studies by investigating the energy complexity of {\em parallel} algorithms on a {\em multiprocessor} machine with {\em a shared memory bank} and private caches, a machine model that has been widely adopted to study parallel algorithms \cite{Frigo:2006, Arge:2008, Korthikanti2010}.

Our new energy complexity model EPEM for multithreaded algorithms complements the aforementioned seminal studies on energy models. The EPEM model enables algorithm designers to analyze the energy complexity of their multithreaded algorithms without implementing and benchmarking the algorithms on a platform. We prove the model usability and accuracy by demonstrating how to apply the model for different SpMV algorithms and validating the results on HPC and accelerator platforms (i.e., Intel Xeon and Xeon Phi) using different sparse matrix types from Florida Matrix Collection.

%% file: EPEM/EPEM-memory-model.tex
Generally speaking, the energy consumption of a parallel algorithm is the sum of i) static energy (or leakage) $E_{static}$, ii) dynamic energy of computation $E_{comp}$ and iii) dynamic energy of memory accesses $E_{mem}$. The static energy $E_{static}$ is proportional to the execution time of the algorithm while the dynamic energy of computation and the dynamic energy of memory accesses are proportional to the number of computational operations and the number of memory accesses of the algorithm, respectively \cite{Korthikanti2010}. As a result, in the new EPEM energy complexity model the energy complexity of a multithreaded algorithm  is analyzed based on its {\em span complexity} \cite{CormenLRS:2009} (for the static energy), {\em work complexity} \cite{CormenLRS:2009} (for the dynamic energy of computation) and {\em I/O complexity} (for the dynamic energy of memory accesses) (cf. Section \ref{energy-model}).

This section describes shared-memory machine models supporting I/O complexity analysis for parallel algorithms. We first  describe the parallel external memory (PEM) model \cite{Arge:2008} used for  analyzing the energy scalability of parallel algorithms on shared memory multicore platforms \cite{Korthikanti2009} and explain why the PEM model is not appropriate for analyzing the energy complexity of multithreaded algorithms. We then describe the ideal distributed cache (IDC) model \cite{FrigoS:2006} that is used in the EPEM energy complexity model.

\subsubsection{The PEM Model}
\begin{figure}[!t] \centering
\resizebox{0.5\columnwidth}{!}{ \includegraphics{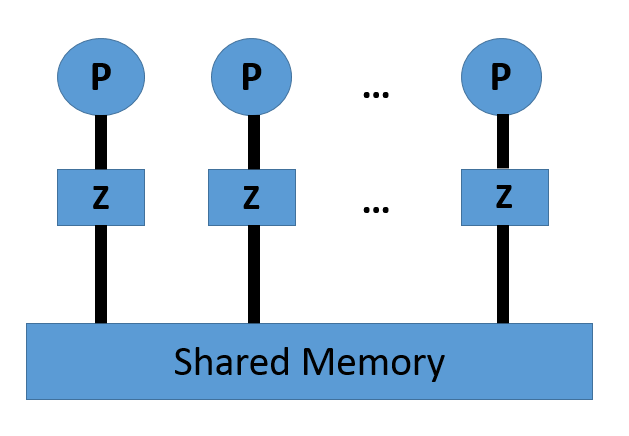}}
\caption{A Shared Memory Machine Model with Private Caches}
\label{fig:machine-model}
\end{figure}

The PEM model \cite{Arge:2008} is an extension of the Parallel Random Access Machine (PRAM) model that includes a two-level memory hierarchy. 
In the PEM model , there are $n$ cores (or processors) each of which has its own {\em private} cache of size $Z$ (in bytes) and shares the main memory with the other cores (cf. Figure \ref{fig:machine-model}). Unlike other I/O models for multicore platforms \cite{BlellochFGS:2011, BlellochGM:1999}, the PEM model enables analyzing the I/O complexity of parallel algorithms without additional assumption on how the cores are connected nor how the algorithm tasks are scheduled. 
In the PEM model, data is transferred between the shared memory and the cache  in the form of blocks of size $B$ (i.e., cache lines). The number of {\em parallel} block transfers between the shared memory and the caches is defined as {\em I/O complexity}. Namely, when $n$ cores access $n$ distinct blocks from the shared memory {\em simultaneously}, the I/O complexity in the PEM model is $O(1)$ instead of $O(n)$.

Like the PRAM shared-memory parallel model, the PEM model has three variations according to how multiple cores access the {\em same} block of shared memory, namely: Concurrent Read, Concurrent Writes (CRCW); Concurrent Read, Exclusive Write (CREW) and Exclusive Read, Exclusive Write (EREW). 
In the cases of exclusive write (i.e., CREW and EREW), there are write conflicts between $n$ simultaneous writes to the same block in the main memory. A solution to the conflict is to serialize the $n$ writes, resulting in $n$ I/Os. The I/O complexity of $n$ conflicting writes can be improved to $O(\log{n})$ by using extra memory to combine the writes in a binary tree fashion \cite{Arge:2008}. 

\subsubsection{The IDC Model}
Although the PEM model is appropriate for analyzing the I/O complexity of parallel algorithms in terms of time performance \cite{Arge:2008}, we have found that the PEM model is not appropriate for analyzing parallel algorithms in terms of the dynamic energy of memory accesses. In fact, even when the $n$ cores can access data from the main memory simultaneously, the {\em dynamic} energy consumption of the access is proportional to the number $n$ of accessing cores (because of the load-store unit activated within each accessing core and the energy compositionality of parallel computations \cite{HaTUTGRWA:2014, chris-eehco14}), rather than a constant as implied by the PEM model.

As a result, we choose the ideal distributed cache (IDC) model \cite{FrigoS:2006} to analyze I/O complexity of multithreaded algorithms in terms of dynamic energy consumption. 
Like the PEM model, the IDC model has $n$ cores and a two-level memory hierarchy as shown in Figure \ref{fig:machine-model}. Each core has its own private cache of size $Z$, which cannot be accessed by the other cores, and shares the main memory with the other cores. 
All the inter-core communication is conducted through writing to and reading from the main memory.
The core must have data in its cache in order to operate on the data and the data is transferred between the main memory and its cache in blocks of size $B$ (i.e., cache line size). 

Unlike the PEM model, the IDC model defines I/O complexity (or cache complexity) of a computation as the number of {\em cache misses} caused by the computation on an {\em ideal cache} starting and ending with an empty cache. An ideal cache is a fully associative cache that uses optimal offline cache replacement policy. If a core does not have the data word it wants to access in its private cache, it incurs a cache miss to bring the data from the main memory to its private cache. The private caches are non-interfering, namely the number of cache misses incurred by a core can be analyzed independently of the other cores' actions. 
Since the cache complexity of $m$ misses is $O(m)$ regardless of whether or not the cache misses are incurred simultaneously by the cores, the IDC model reflects the aforementioned dynamic energy consumption of memory accesses by the cores.  
 
However, the IDC model is mainly designed for analyzing the cache complexity of divide-and-conquer algorithms, making it difficult to apply to general multi-threaded algorithms targeted by our new EPEM energy model. Constraining the new EPEM energy model to the IDC model would limit the applicability of the EPEM model to a wide range of multithreaded algorithms.

In order to make our new EPEM energy model applicable to a wide range of multithreaded algorithms, we show that the cache complexity analysis using the traditional (sequential) ideal cache (IC) model \cite{FrigoLPR:1999} can be used to find an upper bound  on the cache complexity of the same algorithm using the IDC model (cf. Lemma \ref{ePEM}). As the sequential execution of multithreaded algorithms is a valid execution regardless of whether they are divide-or-conquer algorithms, the ability to analyze the cache complexity of multithreaded algorithms via their sequential execution in the EPEM energy model improves the usability of the EPEM model.   
 

Let $Q_1(Alg,B,Z)$ and $Q_P(Alg, B, Z)$ be the cache complexity of a parallel algorithm $Alg$ analyzed in the (uniprocessor) ideal cache (IC) model  \cite{FrigoLPR:1999} with block size $B$ and cache size $Z$ (i.e, running $Alg$ with a single core) and the cache complexity analyzed in the (multicore) IDC model with $P$ cores each of which has a private cache of size $Z$ and block size $B$, respectively. We have the following lemma:

\begin{lemma}
\label{ePEM}
The cache complexity $Q_P(Alg, B, Z)$ of a parallel algorithm $Alg$ analyzed in the ideal distributed cache (IDC) model with $P$ cores is bounded from above by the product of $P$ and the cache complexity $Q_1(Alg, B, Z)$ of the same algorithm analyzed in the ideal cache (IC) model. Namely,
\begin{equation}
Q_P(Alg, B, Z) \leq P*Q_1(Alg,B,Z)
\end{equation}
\end{lemma}

\begin{proof} 
(Sketch)
Let $Q_P^{i}(Alg, B, Z)$ be the number of cache misses incurred by core $i$ during the parallel execution of algorithm $Alg$ in the IDC model. Because caches do not interfere with each other in the IDC model, the number of cache misses incurred by core $i$ when executing algorithm $Alg$ in parallel by $P$ cores is not greater than the number of cache misses incurred by core $i$ when executing the whole algorithm $Alg$ only by core $i$. That is,
\begin{equation}
\label{eq:idc_ic_1}
Q_P^{i}{(Alg, B, Z)} \leq Q_1(Alg,B,Z)
\end{equation}

or

\begin{equation}
\label{eq:idc_ic_2}
\sum_{i=1}^P Q_P^{i}{(Alg, B, Z)} \leq P*Q_1(Alg,B,Z)
\end{equation}

On the other hand, since the number of cache misses incurred by algorithm $Alg$ when it is executed by $P$ cores in the IDC model is the sum of the numbers of cache misses incurred by each core during the $Alg$ execution, we have 

\begin{equation}
\label{eq:idc_ic_3}
Q_P(Alg, B, Z) = \sum_{i=1}^P Q_P^{i}{(Alg, B, Z)}
\end{equation}

From Equations \ref{eq:idc_ic_2} and \ref{eq:idc_ic_3}, we have
\begin{equation}
\label{eq:idc_ic_4}
Q_P(Alg, B, Z) \leq P*Q_1(Alg,B,Z)
\end{equation}
\end{proof}

We also make the following assumptions regarding platforms.
\begin{itemize}
\item Algorithms are executed with the best configuration (e.g., maximum number of cores, maximum frequency) following the race-to-halt strategy.
\item The I/O parallelism is bounded from above by the computation parallelism. Namely, each core can issue a memory request only if its previous memory requests have been served. Therefore, the work and span (i.e., critical path) of an algorithm represent the parallelism for both I/O and computation. 
\end{itemize}

%% file: EPEM/EPEM-energy-complexity-model.tex
This section describes two energy complexity models, a platform-supporting energy complexity model considering both platform and algorithm characteristics and platform-independent energy complexity model considering only algorithm characteristics. The platform-supporting model is used when platform parameters in the model are available while platform-independent model analyses energy complexity of algorithms without considering platform characteristics.
\subsubsection{Platform-supporting Energy Complexity Model} 
This section describes a methodology to find energy complexity of algorithms. The energy complexity model considers three groups of parameters: machine-dependent, algorithm-dependent and input-dependent parameters. The reason to consider all three parameter-categories is that only operational intensity \cite{Williams2009} is insufficient to capture the characteristics of algorithms. Two algorithms with the same values of operational intensity might consume different levels of energy. The reasons are their differences in data accessing patterns leading to performance scalability gap among them. For example, although the sequential version and parallel version of an algorithm may have the same operational intensity, they may have different energy consumption since the parallel version would have less static energy consumption because of shorter execution time.
\begin{table}
\caption{Platform Parameter Description}
\label{table:PowerParameters}
\begin{center}
\begin{tabular}{ll}
\hline\noalign{\smallskip}
Machine & Description \\
\noalign{\smallskip}\hline\noalign{\smallskip}
$P^{sta}$                & Static power of a whole chip         \\ 
$P^{op}$                  & Dynamic power of an operation        \\ 
$P^{I/O}$                  & Power to transfer one cache line \\
$N$                & Maximum number of cores in the platform      \\ 
$M$              & Number of cycles per cache line transfer \\ 
$F$              & Number of cycles per operation      \\
$Freq$                & Platform frequency       \\
$Z$              & Cache size of a single processor         \\
\noalign{\smallskip}\hline\noalign{\smallskip}
\end{tabular}
\end{center}
\end{table}

\begin{table}
\caption{EPEM Model Parameter Description}
\label{table:ModelParameters}
\begin{center}
\begin{tabular}{ll}
\hline\noalign{\smallskip}
Machine & Description \\
\noalign{\smallskip}\hline\noalign{\smallskip}
$\epsilon_{op}$                  & dynamic energy of one operation (1 core) \\ 
$\epsilon_{I/O}$                  & dynamic energy of a random access (1 core)\\ 
$\pi_{op}$                  & static energy when performing one operation  \\ 
$\pi_{I/O}$                  & static energy of a random data access \\ 
$B$              & cache block size         \\ 
\noalign{\smallskip}\hline\noalign{\smallskip}
\hline\noalign{\smallskip}
Algorithm & Description \\
\noalign{\smallskip}\hline\noalign{\smallskip}
$Work$                & Number of work in flops of the algorithm         \\ 
$Span$                 & The critical path of the algorithm        \\ 
$I/O$                  & Number of cache line transfer        \\ 
\noalign{\smallskip}\hline\noalign{\smallskip}
\hline\noalign{\smallskip}
SpMV Input & Description \\
\noalign{\smallskip}\hline\noalign{\smallskip}
$n$                & Number of rows        \\ 
$nz$                 & Number of nonzero elements        \\ 
$nr$                  & Maximum number of nonzero in a row        \\ 
$nc$                  & Maximum number of nonzero in a column        \\ 
$\beta$                  & Size of a block \\ 
\noalign{\smallskip}\hline\noalign{\smallskip}
\end{tabular}
\end{center}
\end{table} 
\begin{table*}
\caption{Platform parameter summary. The parameters of the first nine platforms are derived from \cite{Choi2014} and the parameters of the two new platforms are found in this study.}
\label{table:platform-parameters}
\begin{center}
\begin{tabular}{llllll}
\hline\noalign{\smallskip}
Platform & Processor &  $\epsilon_{op}$(nJ) &$\pi_{op}$(nJ) &$\epsilon_{I/O}$(nJ) & $\pi_{I/O}$(nJ) \\
\noalign{\smallskip}\hline\noalign{\smallskip}
Nehalem i7-950	&Intel i7-950	&0.670 	&2.455	&50.88	&408.80\\
Ivy Bridge i3-3217U	&Intel  i3-3217U	&0.024 	&0.591	&26.75	&58.99\\
Bobcat CPU 	&AMD  E2-1800	&0.199 	&3.980	&27.84	&387.47\\
Fermi GTX 580	&NVIDIA GF100	&0.213 	&0.622	&32.83	&45.66\\
Kepler GTX 680	&NVIDIA GK104	&0.263 	&0.452	&27.97	&26.90\\
Kepler GTX Titan	&NVIDIA GK110	&0.094 	&0.077	&17.09	&32.94\\
XeonPhi KNC	&Intel 5110P	&0.012 	&0.178	&8.70	&63.65\\
Cortex-A9	&TI OMAP 4460	&0.302 &1.152	&51.84	&174.00\\
Arndale Cortex-A15	&Samsung Exynos 5	&0.275 	&1.385	&24.70	&89.34\\
\noalign{\smallskip}\hline\noalign{\smallskip}
Xeon 	&2xIntel E5-2650l v3	&0.263	&0.108	&8.86	&23.29\\
Xeon-Phi	&Intel 31S1P	&0.006	&0.078	&25.02	&64.40\\
\noalign{\smallskip}\hline\noalign{\smallskip}
\end{tabular}
\end{center}
\end{table*}
\begin{equation} \label{eq:BigET}
	E= \epsilon_{op} \times Work + \epsilon_{I/O} \times I/O + P^{sta} \times max(T^{comp} ,T^{mem})
\end{equation}
\begin{equation} \label{eq:BigE0}
	E= \epsilon_{op} \times Work + \epsilon_{I/O} \times I/O + max(\pi_{op} \times Span ,\pi_{I/O} \times \frac{I/O \times Span}{Work})  
\end{equation}

The energy consumption of a parallel algorithm is the sum of i) static energy (or leakage) $E_{static}$, ii) dynamic energy of computation $E_{comp}$ and iii) dynamic energy of memory accesses $E_{mem}$: $E=E_{static}+E_{comp}+E_{mem}$. The static energy $E_{static}$ is the product of the execution time of the algorithm and the static power of the whole processor. The dynamic energy of computation and the dynamic energy of memory accesses are proportional to the number of computational operations $Work$ and the number of memory accesses of the algorithm $I/O$, respectively \cite{Korthikanti2010}. Since computation time and memory-access time can be overlapped, the execution time of the algorithm is the maximum value of computation time and memory-access time. Therefore, the energy consumption of algorithms is computed by Equation \ref{eq:BigET}. 

The computation time of parallel algorithms is proportional to the span complexity of the algorithm, which is $T^{comp}=\frac{Span \times F}{Freq}$ where $Freq$ is the processor frequency, and $F$ is the number of cycles per operation. The memory-access time of parallel algorithms in the EPEM model is is proportional to the I/O complexity of the algorithm divided by its I/O parallelism. As I/O parallelism is bounded by the computation parallelism (cf. Section \ref{EPEM-model}), I/O parallelism is divided by $\frac{Work}{Span}$. The memory-access time $T^{mem}$ becomes: $T^{mem}=\frac{I/O \times Span \times M}{Work \times Freq}$ where $M$ is the number of cycles per cache line transfer. If an algorithm has $T^{comp}$ greater than $T^{mem}$, the algorithm is a CPU-bound algorithm. Otherwise, it is a memory-bound algorithm.

The summary of platform parameters are listed in Table \ref{table:PowerParameters}. The EPEM energy complexity model in Equation \ref{eq:BigET} is simplified to Equation \ref{eq:BigE0}, where the mathematical meaning of $\epsilon_{op}$, $\epsilon_{I/O}$, $\pi_{op}$, and $\pi_{I/O}$ are described in the Equation \ref{eq:epsilon_op}, \ref{eq:epsilon_IO}, \ref{eq:pi_op}, and \ref{eq:pi_IO}. The model considers the parameters listed in Table \ref{table:ModelParameters}. The parameter values of recent computing platforms are summarized in Table \ref{table:platform-parameters}. How to obtain the platform parameters is discussed in Section \ref{experiment-set-up}.

The dynamic energy of one operation by one core $\epsilon_{op}$ is the product of the consumed power of one operation by one active core $P^{op}$ and the time to perform one operation. Equation \ref{eq:epsilon_op} shows how $\epsilon_{op}$ relates to frequency $Freq$ and time per operation $F$. Similarly, the dynamic energy of a random access by one core $\epsilon_{I/O}$ is the product of the consumed power by one active core performing one I/O (i.e., cache-line transfer) $P^{I/O}$ and the time to perform one cache line transfer computed as $M/Freq$ (cf. Equation \ref{eq:epsilon_IO}). The static energy of operations $\pi_{op}$ is the product of the whole chip static power $P^{sta}$ and time per operation. The static energy of one I/O $\pi_{I/O}$ is the product of the whole chip static power and time per I/O, shown by Equation \ref{eq:pi_op} and \ref{eq:pi_IO}. 
\begin{equation} \label{eq:epsilon_op}
	\epsilon_{op}= P^{op} \times \frac{F}{Freq}
\end{equation}
\begin{equation} \label{eq:epsilon_IO}	
	\epsilon_{I/O}= P^{I/O} \times \frac{M}{Freq}
\end{equation}
\begin{equation} \label{eq:pi_op}
	\pi_{op}= P^{sta} \times \frac{F}{Freq}
\end{equation}
\begin{equation} \label{eq:pi_IO}
	\pi_{I/O}= P^{sta} \times \frac{M}{Freq}
\end{equation}
\subsubsection{CPU-bound Algorithms}
If an algorithm has computation time longer than time for accessing data (i.e., CPU-bound algorithms): $T^{comp} \geq T^{mem}$, the EPEM energy complexity model becomes Equation \ref{eq:BigE-cpu-time} and \ref{eq:BigE1}.
\begin{equation} \label{eq:BigE-cpu-time}
	E= \epsilon_{op} \times Work + \epsilon_{I/O} \times I/O +  P^{sta} \times \frac{Span \times F}{Freq}\\
\end{equation}
or
\begin{equation} \label{eq:BigE1}
	E= \epsilon_{op} \times Work + \epsilon_{I/O} \times I/O +  \pi_{op} \times Span\\
\end{equation}

\subsubsection{Memory-bound Algorithms}
If an algorithm has data-accessing time longer than computation time (i.e., memory-bound algorithms): $T^{mem} \geq T^{comp}$, energy complexity becomes Equation \ref{eq:BigE-mem-time} and \ref{eq:BigE2}. 
\begin{equation} \label{eq:BigE-mem-time}
	E= \epsilon_{op} \times Work + \epsilon_{I/O} \times I/O +  P^{sta} \times \frac{I/O \times Span \times M}{Work \times Freq}\\
\end{equation}
or
\begin{equation} \label{eq:BigE2}
	E= \epsilon_{op} \times Work + \epsilon_{I/O} \times I/O +  \pi_{I/O} \times \frac{I/O \times Span}{Work}\\
\end{equation}
\subsubsection{Platform-independent Energy Complexity Model}
This section describes the energy complexity model that is platform-independent and considers only algorithm characteristics. When the platform parameters (i.e., $\epsilon_{op}$, $\epsilon_{I/O}$, $\pi_{op}$, and $\pi_{I/O}$) are unavailable, the energy complexity model is derived from Equation \ref{eq:BigE0}. The platform parameters are constants and can be removed from the Equation \ref{eq:BigE0}. Assuming $\pi_{max} = max(\pi_{op}, \pi_{I/O})$, after removing platform parameters, the platform-independent energy complexity model are shown in Equation \eqref{eq:BigE-non-flatform}.
\begin{equation} \label{eq:BigE-non-flatform}
	E= O(Work+I/O+max(Span, \frac{I/O \times Span}{Work}))  
\end{equation}

%% file: EPEM/SpMV-energy-complexity.tex
SpMV is one of the most common application kernels in Berkeley dwarf list\cite{Asa06}. It computes a vector result $y$ by multiplying a sparse matrix $A$ with a dense vector $x$: $y = A \times x$. SpMV is a data-intensive kernel and has irregular memory-access patterns. The data access patterns for SpMV is defined by its sparse matrix format and matrix input types. 
There are several sparse matrix formats and SpMV algorithms in literature. To name a few, they are Coordinate Format (COO), Compressed Sparse Column (CSC), Compressed Sparse Row (CSR), Compressed Sparse Block (CSB), Recursive Sparse Block (RSB), Block Compressed Sparse Row (BCSR) and so on.
Three popular SpMV algorithms, namely CSC, CSB and CSR are chosen to validate the proposed energy complexity model. They have different data-accessing patterns leading to different values of I/O, work and span complexity. Since SpMV is a memory-bound application kernel, Equation \ref{eq:BigE1} is applied.
\subsubsection{Compressed Sparse Row}
CSR is a standard storage format for sparse matrices which reduces the storage of matrix compared to the tuple representation \cite{Kotlyar:1997}. This format enables row-wise compression of $A$ with size $n \times n$ (or  $n \times m$) to store only the non-zero $nz$ elements. Let {\em nz} be the number of non-zero elements in matrix A. 
The work complexity of CSR SpMV is $\Theta(nz)$ where $nz>=n$ and span complexity is $O(nr + \log{n})$ \cite{Buluc:2009}, where $nr$ is the maximum number of non-zero elements in a row. The I/O complexity of CSR in the sequential I/O model of row-major layout is $O(nz)$ \cite{Bender2010} namely, scanning all non-zero elements of matrix $A$ costs $O(\frac{nz}{B})$ I/Os with B is the cache block size. However, randomly accessing vector $x$ causes the total of $O(nz)$ I/Os.
Applying the proposed model on CSR SpMV, their total energy complexity are computed as Equation \ref{eq:BigE-CSR}.
\begin{equation} \label{eq:BigE-CSR}
	E_{CSR}= O(\epsilon_{op} \times nz + \epsilon_{I/O} \times nz + \pi_{I/O} \times (nr+ \log{}n)) 
\end{equation}
\subsubsection{Compressed Sparse Column}
CSC is the similar storage format for sparse matrices as CSR. However, it compresses the sparse matrix in column-wise manner to store the non-zero elements. The work complexity of CSC SpMV is $\Theta(nz)$ where $nz>=n$ and span complexity is $O(nc + \log{n})$, where $nc$ is the maximum number of non-zero elements in a column. The I/O complexity of CSC in the sequential I/O model of column-major layout is $O(nz)$ \cite{Bender2010}. Similar to CSR, scanning all non-zero elements of matrix $A$ in CSC format costs $O(\frac{nz}{B})$ I/Os. However, randomly updating vector $y$ causing the bottle neck with total of $O(nz)$ I/Os.
Applying the proposed model on CSC SpMV, their total energy complexity are computed as Equation \ref{eq:BigE-CSC}.
\begin{equation} \label{eq:BigE-CSC}
	E_{CSC}= O(\epsilon_{op} \times nz + \epsilon_{I/O} \times nz + \pi_{I/O} \times (nc+ \log{}n)) 
\end{equation}
\subsubsection{Compressed Sparse Block}
Given a sparse matrix $A$, while CSR has good performance on SpMV $y=A\times x$, CSC has good performance on transpose sparse matrix vector multiplication $y=A^{T}\times x$, Compressed sparse blocks (CSB) format is efficient for computing either $Ax$ or $A^{T}x$. CSB is another storage format for representing sparse matrices by dividing the matrix $A$ and vector $x, y$  to blocks. A block-row contains multiple chunks, each chunks contains consecutive blocks and non-zero elements of each block are stored in Z-Morton-ordered \cite{Buluc:2009}.
From Beluc et al. \cite{Buluc:2009}, CSB SpMV computing a matrix with $nz$ non-zero elements, size $n\times n$ and divided by block size $\beta \times \beta$ has span complexity $O(\beta \times \log{\frac{n}{\beta}}+ \frac{n}{\beta})$ and work complexity as $\Theta(\frac{n^2}{b^2}+nz)$.

I/O complexity for CSB SpMV is not available in the literature. We do the analysis of CSB manually by following the master method \cite{CormenLRS:2009}. The I/O complexity is analyzed for the algorithm CSB\_SpMV(A,x,y) from Beluc et al. \cite{Buluc:2009}. The I/O complexity of CSB is similar to work complexity of CSB $O(\frac{n^2}{\beta^2} + nz)$, only that non-zero accesses in a block is divided by B: $O(\frac{n^2}{\beta^2} + {\frac{nz}{B}})$. The reason is that non-zero elements in a block are stored in Z-Morton order which only requires $\frac{nz}{B}$ I/Os. The energy complexity of CSB SPMV is shown in Equation \ref{eq:BigE-CSB}.

From the complexity analysis of SpMV algorithms using different layouts, the complexity of CSR, CSC and CSB are summarized in Table \ref{table:SpMV-complexity}.
\begin{floatEq}
\begin{equation}
\label{eq:BigE-CSB}
	E_{CSB}= O(\epsilon_{op} \times (\frac{n^2}{\beta^2} + nz) + \epsilon_{I/O} \times (\frac{n^2}{\beta^2} + \frac{nz}{B}) + \pi_{I/O} \times \frac{(\frac{n^2}{\beta^2} + \frac{nz}{B})\times (\beta \times \log{\frac{n}{\beta}}+ \frac{n}{\beta})}{(\frac{n^2}{\beta^2} + nz)} )
\end{equation}
\end{floatEq}
\begin{table*}
\caption{SpMV Complexity Analysis}
\label{table:SpMV-complexity}
\begin{center}
\begin{tabular}{llll}
\hline\noalign{\smallskip}
Complexity & CSC & CSB & CSR \\
\noalign{\smallskip}\hline\noalign{\smallskip}
Work  & $\Theta(nz)$ \cite{Buluc:2009} & $\Theta(\frac{n^2}{\beta^2} + nz)$ \cite{Buluc:2009} & $\Theta(nz)$ \cite{Buluc:2009}\\ 
I/O                 & $O(nz)$ \cite{Bender2010} & $O(\frac{n^2}{\beta^2} + {\frac{nz}{B}})$ [this report] & $O(nz)$ \cite{Bender2010} \\ 
Span                  & $O(nc+ \log{}n)$ \cite{Buluc:2009} & $O(\beta \times \log{\frac{n}{\beta}}+ \frac{n}{\beta})$ \cite{Buluc:2009} & $O(nr+ \log{}n)$  \cite{Buluc:2009}\\ 

\noalign{\smallskip}\hline\noalign{\smallskip}
\end{tabular}
\end{center}
\end{table*}

%% file: EPEM/SpMV-implementation.tex
This section describes the experimental study to validate the EPEM model, including: describing SpMV implementation and sparse matrix types used in this validation (cf. Section \ref{SpMV-Implementation}), introducing the two experimental platforms and how to obtain their parameters for the EPEM model(cf. Section \ref{experiment-set-up}) and discussing the validation results.
\subsubsection{SpMV Implementation}
\label{SpMV-Implementation}
\begin{figure}[!t] \centering
\resizebox{0.5\columnwidth}{!}{ \includegraphics{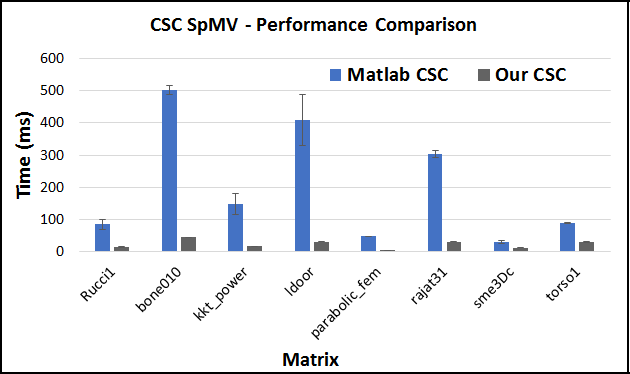}}
\caption{Performance (time) comparison of two parallel CSC SpMV implementations. For a set of different input matrices, the parallel CSC SpMV using Cilk out-performs Matlab parallel CSC.}
\label{fig:Matlab-comparison}
\end{figure}

We want to conduct complexity analysis and experimental study with two SpMV algorithms, namely CSB and CSC. Parallel CSB and sequential CSC implementations are available thanks to the study from Bulu\c{c} et al. \cite{Buluc:2009}. Since the optimization steps of available parallel SpMV kernels (e.g., pOSKI \cite{pOSKI}, LAMA\cite{Forster2011}) might affect the work complexity of the algorithms, we decided to implement a pure parallel CSC using Cilk and $Pthread$. To validate the correctness of our parallel CSC implementation, we compare the vector result $y$ from $y=A*x$ of CSC and CSB implementation. The comparison shows the equality of the two vector results $y$. Moreover, we compare the performance of the our parallel CSC implementation with Matlab parallel CSC-SpMV implementation. Matlab also uses CSC layout as the format for their sparse matrix \cite{Gilbert:1992}. Our CSC implementation has out-performed Matlab parallel CSC kernel when computing the same targeted input matrices. Figure \ref{fig:Matlab-comparison} shows the performance comparison of our CSC SpMV implementation and Matlab CSC SpMV kernel. The experimental study of SpMV energy consumption is then conducted with CSB SpMV implementation from Bulu\c{c} et al. \cite{Buluc:2009} and our CSC SpMV parallel implementation.  
\subsubsection{SpMV Matrix Input Types}
We conducted the experiments with nine different matrix-input types from Florida sparse matrix collection \cite{Davis:2011}. Each matrix input has different properties, including size of the matrix $n\times m$, the maximum number of non-zero of the sparse matrix $nz$, the maximum number of non-zero elements in one column $nc$. Table \ref{table:matrix-type} lists the matrix types in this experimental validation. 
\begin{table}
\caption{Sparse matrix input types. The maximum number of non-zero elements in a column $nc$ is derived from \cite{Buluc:2009}.}
\label{table:matrix-type}
\begin{center}
\begin{tabular}{lllll}
\hline\noalign{\smallskip}
\textbf{Matrix type}	&\textbf{n}	&\textbf{m}	&\textbf{nz}	&\textbf{nc}	\\
\noalign{\smallskip}\hline\noalign{\smallskip}
bone010	&986703	&986703	&47851783	&63	\\
kkt\_power	&2063494	&2063494	&12771361	&90	\\
ldoor	&952203	&952203	&42493817	&77	\\
parabolic\_fem	&525825	&525825	&3674625	&7	\\
pds-100	&156243	&517577	&1096002	&7	\\
rajat31	&4690002	&4690002	&20316253	&1200	\\
Rucci1	&1977885	&109900	&7791168	&108	\\
sme3Dc	&42930	&42930	&3148656	&405	\\
torso1	&116158	&116158	&8516500	&1200	\\
\noalign{\smallskip}\hline\noalign{\smallskip}
\end{tabular}
\end{center}
\end{table}
\subsubsection{Experiment Set-up}
\label{experiment-set-up}
For the validation of the EPEM model, we conduct the experiments on two HPC platforms: one platform with two Intel Xeon E5-2650l v3 processors and one platform with Xeon Phi 31S1P processor. The Intel Xeon platform has two processors Xeon E5-2650l v3 with $2\times12$ cores, each processor has the frequency 1.8 GHz. The Intel Xeon Phi platform has one processor Xeon Phi 31S1P with $57$ cores and its frequency is 1.1 GHz. To measure energy consumption of the platforms, we read the PCM MSR counters for Intel Xeon and MIC power reader for Xeon Phi.
\subsubsection{Identifying Platform Parameters}
We apply the energy roofline approach \cite{Choi2013, Choi2014} to find the platform parameters for the two new experimental platforms, namely Intel Xeon E5-2650l v3 and Xeon Phi 31S1P. The energy roofline study \cite{Choi2014} has also provided a list of other platforms including CPU, GPU, embedded platforms with their parameters considered in the Roofline model. Thanks to authors Choi et al. \cite{Choi2014}, we extract the parameters required for the EPEM energy complexity model from their platform data. Along with the two HPC platforms used in this validation, we provide parameters required in the energy complexity model for a list of available platforms. The parameter values of recent computing platforms for further uses are listed in Table \ref{table:platform-parameters}.


%% file: EPEM/validation.tex
\subsubsection{Validating EPEM Using Different SpMV Algorithms}
\begin{figure}[!t] \centering
\resizebox{0.5\columnwidth}{!}{ \includegraphics{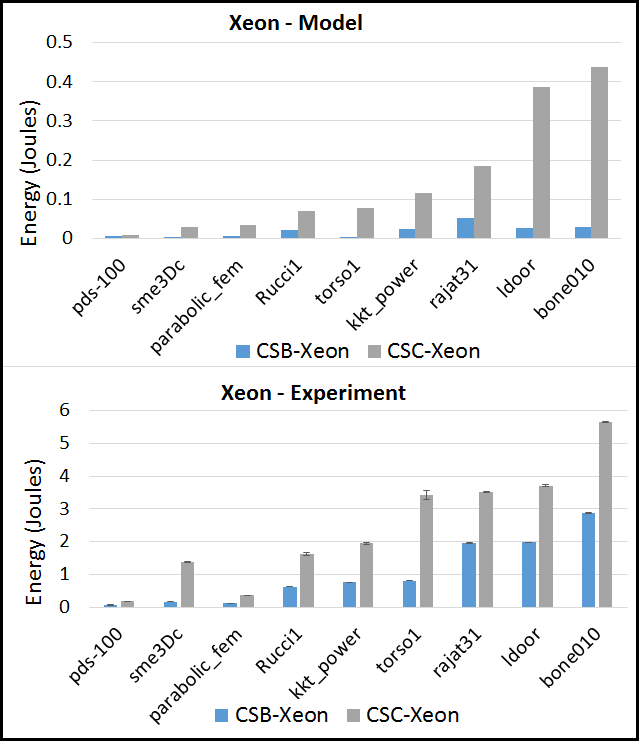}}
\caption{Algorithm comparison of CSB and CSC SpMV energy consumption from the EPEM model estimation and experimental measurement on Intel Xeon platform. The EPEM model is able to predict that the CSC SpMV algorithm consumes more energy than the CSB SpMV algorithm, on different matrix input types.}
\label{fig:CSBvsCSC-Xeon}
\end{figure}

\begin{figure}[!t] \centering
\resizebox{0.5\columnwidth}{!}{ \includegraphics{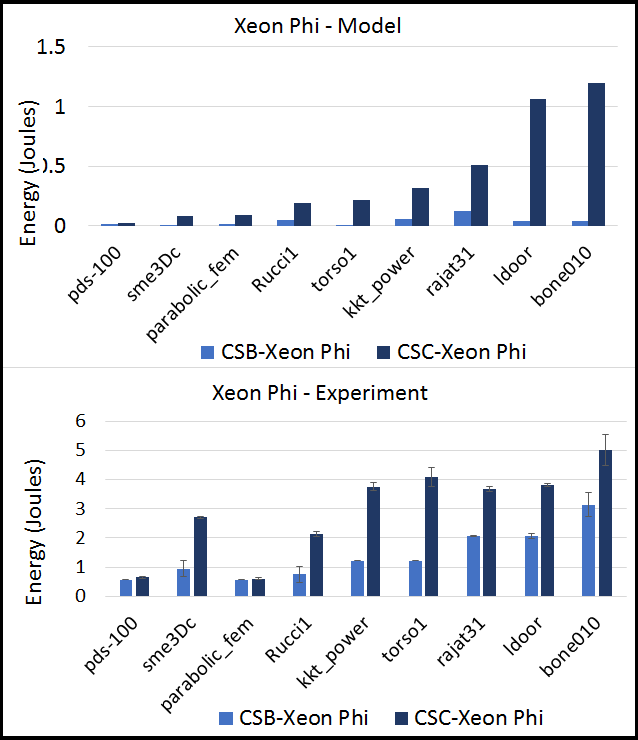}}
\caption{Algorithm comparison of CSB and CSC SpMV energy consumption from the EPEM model estimation and experimental measurement on Intel Xeon Phi platform. The EPEM model is able to predict that the CSC SpMV algorithm consumes more energy than the CSB SpMV algorithm, on different matrix input types.}
\label{fig:CSBvsCSC-XeonPhi}
\end{figure}

Figure \ref{fig:CSBvsCSC-Xeon} and \ref{fig:CSBvsCSC-XeonPhi} show the energy prediction and measurement of CSB SpMV and CSC SpMV algorithms on two platforms Xeon and Xeon Phi. From the model-estimated data, CSB SpMV consumes less energy than CSC SpMV on both platforms. Even though CSB has higher work complexity than CSC, CSB SpMV has less I/O complexity than CSC SpMV. Firstly, the dynamic energy cost of one I/O is much greater than the energy cost of one operation (i.e., $\epsilon_{I/O}>>\epsilon_{op}$) on both platforms.  Secondly, CSB has better parallelism than CSC, computed by $\frac{Work}{Span}$, which results in shorter execution time. Both reasons contribute to the less energy consumption of CSB SpMV. 

The measurement data confirms that CSB SpMV algorithm consumes less energy than CSC SpMV algorithm, shown in the Figure \ref{fig:CSBvsCSC-Xeon} and \ref{fig:CSBvsCSC-XeonPhi}. For all input matrices, the model has predicted precisely that CSB SpMV consumes less energy than CSC SpMV algorithm. 
\subsubsection{Validating EPEM Using Different Input Types}
To validate the EPEM model regarding input types, the experiments have been conducted with nine matrix types listed in Table \ref{table:matrix-type}. The model can capture the energy-consumption relation among different inputs. The increasing order of energy consumption of different matrix-input types are shown in Table \ref{table:Input-Energy-Comparison}, from both model estimation and experimental study.
For instance, in order to validate the comparison of energy consumption for different input types, a validated table as Table \ref{table:Input-Comparison-CSC} is created for CSC SpMV on Xeon to compare model prediction and experimental measurement. For nine input types, there are $\frac{9\times9}{2}-9=36$ input relations. If the relation is correct, meaning both experimental data and model data are the same, the relation value in the table of two inputs is 1. Otherwise, the relation value is 0. From Table \ref{table:Input-Comparison-CSC}, there are 34 out of 36 relations are the same for both model and experiment, which gives 94\% accuracy on the relation of the energy consumption of different inputs. Similarly, the input validation for CSC and CSB on both Xeon and Xeon Phi platforms is provided in Table \ref{table:Input-Comparison-Accuracy}.
\begin{table*}
\caption{Comparison of Energy Consumption of Different Matrix Input Types.}
\label{table:Input-Energy-Comparison}
\centering
\resizebox{\columnwidth}{!}{%
\begin{tabular}{lllllllll}
\hline\noalign{\smallskip}
Algorithm &CSB &CSB &CSC  &CSC  &CSB  &CSB &CSC  &CSC  \\
\noalign{\smallskip}\hline\noalign{\smallskip}
Platform & model-X &  exprmt-X &model-X &exprmt-X  &model-XP &exprmt-XP &model-XP &exprmt-XP  \\
\noalign{\smallskip}\hline\noalign{\smallskip}
Increasing &sme3Dc	&pds-100	&pds-100	&pds-100	&sme3Dc	&pds-100	&pds-100	&parabolic\\

Energy&torso1	&parabolic	&sme3Dc	&parabolic	&torso1	&parabolic	&sme3Dc	&pds-100\\

Consumption&pds-100	&sme3Dc	&parabolic	&sme3Dc	&pds-100	&Rucci1	&parabolic	&Rucci1\\

Order&parabolic	&Rucci1	&Rucci1	&Rucci1	&parabolic	&sme3Dc	&Rucci1	&sme3Dc\\

&Rucci1	&kkt &torso1	&kkt	&ldoor	&kktr	&torso1	&rajat31\\

&kkt	&torso1	&kkt	&torso1	&bone010	&torso1	&kkt	&kkt\\

&ldoor	&rajat31	&rajat31	&rajat31	&Rucci1	&rajat31	&rajat31	&ldoor\\

&bone010	&ldoor	&ldoor	&ldoor	&kkt	&ldoor	&ldoor	&torso1\\

&rajat31	&bone010	&bone010	&bone010	&rajat31	&bone010	&bone010	&bone010\\
\noalign{\smallskip}\hline\noalign{\smallskip}
\end{tabular}
}
\end{table*}
\begin{table}
\caption{Comparison accuracy of SpMV energy consumption computing different input matrix types}
\label{table:Input-Comparison-Accuracy}
\begin{center}
\begin{tabular}{llll}
\hline\noalign{\smallskip}
Algorithm &CSB  &CSC\\
\noalign{\smallskip}\hline\noalign{\smallskip}
Xeon &75\%  &94\%\\
Xeon Phi &63.8\%  &80.5\%\\
\noalign{\smallskip}\hline\noalign{\smallskip}
\end{tabular}
\end{center}
\end{table}
\begin{table*}
\caption{CSC Energy Comparison of Different Input Matrix Types on Xeon}
\label{table:Input-Comparison-CSC}
\begin{center}
\begin{tabular}{llllllllll}
\hline\noalign{\smallskip}
Correctness	&pds-100	&parabolic	&sme3Dc	&Rucci1	&kkt	&torso1	&rajat31	&ldoor	&bone010\\
pds-100	&x	&1	&1	&1	&1	&1	&1	&1	&1\\
parabolic	&	&x	&0	&1	&1	&1	&1	&1	&1\\
sme3Dc		&&	&x	&1	&1	&1	&1	&1	&1\\
Rucci1		&&&		&x	&1	&1	&1	&1	&1\\
kkt			&&&&		&x	&0	&1	&1	&1\\
torso1		&&&&&				&x	&1	&1	&1\\
rajat31		&&&&&&					&x	&1	&1\\
ldoor		&&&&&&&						&x	&1\\
bone010		&&&&&&&&							&x\\
\noalign{\smallskip}\hline\noalign{\smallskip}
\end{tabular}
\end{center}
\end{table*}
\subsubsection{Validating The Applicability of EPEM on Different Platforms}
\begin{figure*}[!t] \centering
\resizebox{1\textwidth}{!}{\includegraphics{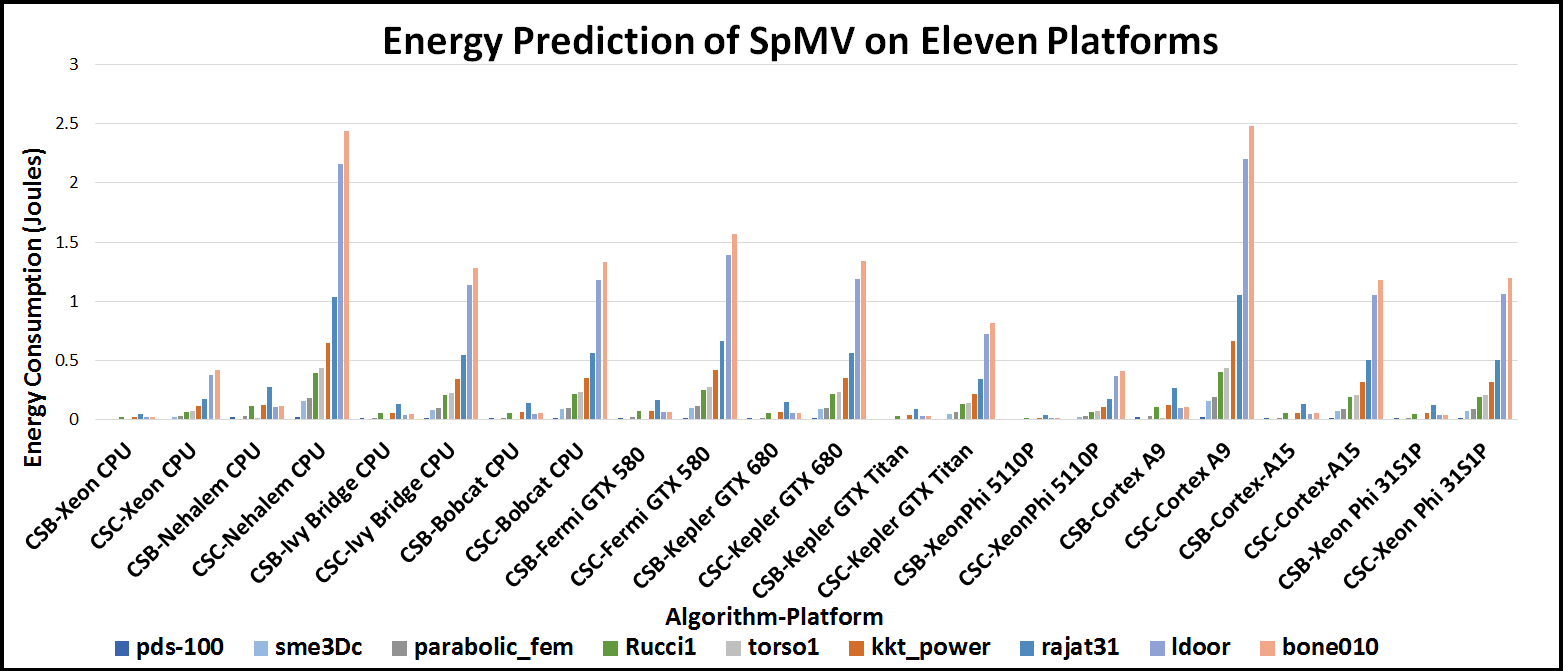}}
\caption{Energy Comparison of CSB and CSC SpMV on eleven different platforms.}
\label{fig:ModelPrediction_Platforms}
\end{figure*}

The energy comparison of CSB and CSC SpMV is predicted for eleven platforms listed in Table \ref{table:platform-parameters}. Like two Xeon and Xeon Phi 31S1P platforms used in experiments, Figure \ref{fig:ModelPrediction_Platforms} shows the prediction that CSB SpMV consumes less energy than CSC SpMV, on all platforms listed in Table \ref{table:platform-parameters}. This confirms the applicability of EPEM model to compare energy consumption of algorithms for different input types on different platforms.
\subsubsection{Validating the Platform-independent Energy Complexity Model}
\begin{figure}[!t] \centering
\resizebox{0.5\columnwidth}{!}{\includegraphics{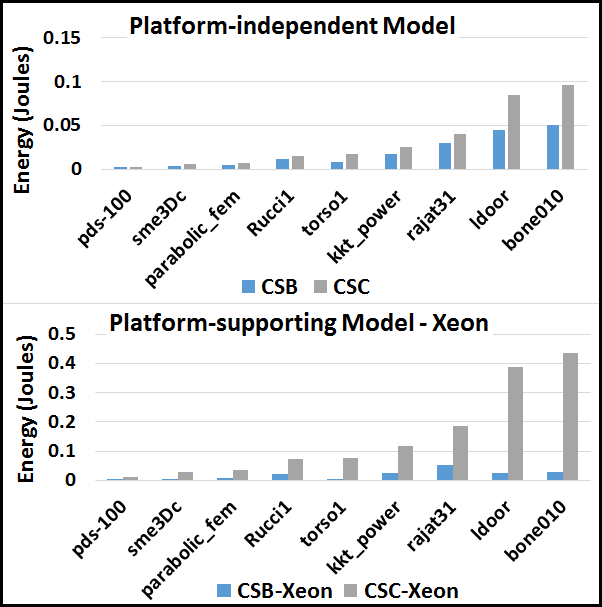}}
\caption{Comparison of platform-dependent and platform-supporting energy complexity model. Both models can predict that CSC SpMV consumes more energy than CSB SpMV.}
\label{fig:model-comparison}
\end{figure}

From Equation \ref{eq:BigE-CSC} and \ref{eq:BigE-CSB}, the platform-independent energy complexity for CSC and CSB SpMV are derived as Equation \ref{eq:BigE-CSC-plat-ind} and \ref{eq:BigE-CSB-plat-ind}, respectively.
\begin{equation} \label{eq:BigE-CSC-plat-ind}
	E_{CSC}= O(2 \times nz + (nc+ \log{}n)) 
\end{equation}
\begin{equation}
\label{eq:BigE-CSB-plat-ind}
	E_{CSB}= O(2 \times \frac{n^2}{\beta^2} + nz \times (1+\frac{1}{B}) + \beta \times \log{\frac{n}{\beta}}+ \frac{n}{\beta}) 
\end{equation}
We validate the platform-independent energy complexity of CSC and CSB SpMV with experimental results. The platform-independent energy complexity also shows the accurate comparison of CSC and CSB SpMV computing different matrix types shown in Figure \ref{fig:model-comparison}. Both platform-independent and platform-supporting models can predict which SpMV algorithm consumes more energy. 
The difference between the energy complexity of CSC and CSB using the platform-independent model is not very clear for all the input types except "ldoor" while in the platform-supporting model, the difference is clear for each input types and consistent with the experiment results in terms of which algorithm consumes less energy for the input types. Comparing energy consumption of different input types requires more detailed information of the platforms. Therefore, the platform-independent model is only applicable to predict which algorithm consumes more energy.

%% file: EPEM/conclusion.tex
In this study, energy models for algorithms to predict energy consumption of an application have been devised. Based on the analysis of application complexity such as \textit{work} complexity, \textit{I/O} complexity and \textit{span} complexity, the energy complexity of applications is predicted.
Moreover, a case study is conducted to demonstrate how to use the model and predict energy consumption of sparse matrix vector multiplication (SpMV) on three different layouts (i.e., CSC, CSR and CSB). This prediction is validated for the two SpMV algorithms on two HPC platform with nine different input matrix types from Florida matrix collection. The results show the precise prediction on which algorithm and which platform consumes more energy. The EPEM energy complexity model gives the algorithm designers the insight to choose which design of algorithm to use for their application to minimize energy consumption.

In the future, we would extend our work in two directions:
\begin{itemize}
\item We want to develop a run-time framework which can choose the least energy-consumption implementations among the available kernels at run-time using the proposed energy model. Selecting the most suitable implementations helps to minimize energy consumption.
\item Nowadays, there are executable frameworks that connect different platforms to one task scheduler. Selecting the most energy-efficient platforms or system configurations to run applications is one of the techniques to achieve energy optimization. 
In order to do so, the energy models need to be able to model the details of each platform. We want to improve and use the energy models to compare the energy consumption of applications on different platforms. 
\end{itemize}

%% file: chalmers-disc.tex
\hyphenation{Abstract-Algorithm}


\subsection{Introduction}
\label{cha-sec:intro}

Lock-free programming provides highly concurrent access to data and
has been increasing its footprint in industrial settings, \eg Intel's
Threading Building Blocks Framework~\cite{itbbf}, the Java concurrency
package~\cite{jav-conc} and the Microsoft .NET
Framework~\cite{mic-net-f}.  Providing a modeling and an analysis
framework capable of describing the practical performance of lock-free
algorithms is an essential, missing resource necessary to the parallel
programming and algorithmic research communities in their effort to
build on previous intellectual efforts.
The definition of lock-freedom mainly guarantees that at least one
concurrent operation on the \ds finishes in a finite number of its own
steps, regardless of the state of the operations. On the
individual operation level, lock-freedom cannot guarantee that an
operation will not starve.
\posrem{The analysis frameworks that currently
exist in the literature focus on such worst-case behavior and are far
from capturing the behavior observed in practice.}{}

The goal of this section is to provide a way to model and analyze the
practically observed performance of lock-free \dss.
In the literature, the common performance measure of a lock-free \ds is
the throughput, \ie the number of successful operations per unit of time.
It is obtained while threads are accessing the \ds according to an
access pattern that interleaves local work between calls to consecutive
operations on the \ds.
Although this access pattern to the data structure is significant,
there is no consensus in the literature on what access to be used when
comparing two data structures.
So, the amount of local work could be
constant (\cite{lf-queue-michael, pq-skiplist-constant}), uniformly
distributed (\cite{scalable-stack-uniform, count-moir}), exponentially
distributed (\cite{Val94, pq-survey-exponential}), null
(\cite{future-ds-nonexist, upp-prio-que}), \etc, and more
questionably, the average amount is rarely scanned, which leads to a
partial covering of the contention domain.

We propose here a common framework enabling a fair comparison between
lock-free \dss, while exhibiting the main phenomena that drive
performance, and particularly the contention, which leads to different
kinds of conflicts. As this is the first step in this direction, we
want to deeply analyze the core of the problem, without impacting
factors being diluted within a probabilistic smoothing. Therefore, we
choose a constant local work, hence implying a constant access rate to
the \dss.
In addition to the prediction of the \ds performance, our model
provides a good back-off strategy, that achieves the peak performance
of a lock-free algorithm.

Two kinds of conflict appear during the execution of a lock-free
algorithm, both of them leading to additional work. \Hardcons occur
when concurrent operations call atomic primitives on the same data:
these calls collide and conduct to stall time, that we name here {\it
  expansion}. \Logcons take place if concurrent operations overlap:
because of the lock-free nature of the algorithm, several concurrent
operations can run simultaneously, but only one can logically
succeed. We show that the additional work produced by the failures is
not necessary harmful for the system-wise performance.

We then show how throughput, that we consider here as the performance
criterion, of a general class of lock-free algorithms can be computed
by connecting these two key factors in an iterative way. We start by
estimating the expansion probabilistically, and emulate the effect of
stall time introduced by the \hardcons as extra work added to each
thread. Then we estimate the number of failed operations, that in
turn lead to additional extra work, produced by the failed retries. We
continue our computation by computing again the expansion on
a system setting where those two new amounts of work have been
incorporated, and reiterate the process; the convergence is ensured by
a fixed-point search.

We consider the class of lock-free algorithms that can be modeled as a
linear composition of fixed size \rls.  This class covers numerous
extensively used lock-free designs such as stacks~\cite{lf-stack}
(\popop, \pushop), queues~\cite{lf-queue-michael} (\enqop, \deqop),
counters~\cite{count-moir} (\incop, \decop) and priority
queues~\cite{upp-prio-que} (\delmin).

To evaluate the accuracy of our model and analysis framework, we
performed experiments both on synthetic tests, that capture a wide
range of possible abstract algorithmic designs, and on several
reference implementations of extensively studied lock-free \dss,
namely stacks, queues, counters and priority queues.
Our evaluation results reveal that our model is able to capture the
behavior of all the synthetic and real designs for all different
numbers of threads and sizes of parallel work (consequently also
contention). \posrem{Our model follows the performance behavior of the \dss
exactly in low contention, when our lower and upper bounds meet in one
line with the observed behavior; and follows closely also the
performance in high contention. }{}We also evaluate the use of our
analysis as a tool for tuning the performance of lock-free code by
selecting the appropriate back-off strategy that will maximize
throughput by comparing our method with against widely known back-off
policies, namely linear and exponential.


The rest of this section is organized as follows.  We discuss related
work in Section~\ref{cha-secrel}, then the problem is formally
described in Section~\ref{cha-secps}. We consider the \logcons in the
absence of \hardcons in Section~\ref{cha-secwt}, while in
Section~\ref{cha-secexp-glue}, we firstly show how to compute the
expansion, then combine hardware and \logcons to obtain the final
throughput estimate. We describe the experimental results in
Section~\ref{cha-secxp}.


\subsection{Related Work}
\label{cha-secrel}

Many studies have been conducted to estimate \rl interferences and
shared memory contention, which are two main components of our
model. Anderson \etal~\cite{anders-rt} evaluate the performance of
lock-free objects in real-time system by emphasizing the impact
of \rl interferences. Tasks can be preempted during the \rl
execution which can lead to interference, thus to an inflation
in \rl execution due to \res. They obtain upper bounds for the
number of interferences under various priority based scheduling
schemes in the uniprocessor setting for periodic real-time tasks.
Also, conflicts in transactional memory and critical section
contention reveal significant conceptual similarities with the \rl
interferences.  Eyerman \etal~\cite{eyer-prob} provide a
probabilistic model which estimates critical section
contention. Assuming total randomness in the critical section entry
times, they formulate the contention in terms of the parallel,
critical section granularities and probability contention for any
two critical sections.  In their model, Yu \etal~\cite{yu-markov}
represent execution as a Markov chain and formulate state
transition probabilities by considering arrival rate and service
time for transactions together with other parameters to come up
with the conflict rate.


Furthermore, performance implications of shared memory contention
have been explored in various
studies. Dwork \etal~\cite{dwork-lcon} provide lower bounds for the
contention on shared resources for well-known
problems. Intel~\cite{intel-emp} conduct an empirical study to
illustrate performance and scalability of locks. It is shown that
the critical section size, the time interval between releasing and
re-acquiring the lock and number of threads contending the lock are
vital parameters. Experiments reveal the increasing significance of
contention with decreasing critical section size.  Empirical
analysis of atomic instructions, such as \cas, is done by
David \etal~\cite{david-emp-atom}, where latencies, which depend on
the cache line state, are illustrated. Also, they experimentally
compare various locks and atomic instructions under different
levels of contention to highlight the impact of memory contention.

Failed \res do not only lead to useless effort but also degrade the
performance of successful ones by contending the shared
resources. By pointing out this fact,
Alemany \etal~\cite{alemany-os} design non-blocking algorithms with
operating system support.

Alistarh \etal~\cite{ali-same} introduce a model for a class of
lock-free structures, which is same as the structure targeted in
our study. Their work is more oriented towards introducing a new
methodology to analyze lock-free structures; in contrast our model
targeting to predict throughput. They model execution as a Markov
chain by considering per-process states under a stochastic
scheduler.  In order to relate per-process performance with system
wide performance, they lift the Markov chain which composes
per-process states using the assumption of process uniformity.
Under stochastic scheduler assumption, they reveal the unfairness
among processes in terms of number of steps taken in short
intervals, which in turn leads to increase in success rate, as one
process takes enough steps to complete its operation. Based on
this, they provide upper bounds on number of steps, system-wise and
thread-wise, to complete an operation. The difference between
scheduling assumptions makes the comparison between their and our
bounds not trivial, not to say incongruous.




\subsection{Problem Statement}
\label{cha-secps}

\subsubsection{Running Program and Targeted Platform}
\label{cha-secprog-plat}

\makeatletter
\newcommand{\removelatexerror}{\let\@latex@error\@gobble}
\makeatother

\begin{figure}[t!]
\removelatexerror
\begin{procedure}[H]
\SetKwData{pet}{execution\_time}
\SetKwData{pdo}{done}
\SetKwData{psucc}{success}
\SetKwData{pcur}{current}
\SetKwData{pnew}{new}
\SetKwData{pacp}{AP}
\SetKwData{pttot}{t}
\SetKwFunction{pinit}{Initialization}
\SetKwFunction{ppw}{Parallel\_Work}
\SetKwFunction{pcw}{Critical\_Work}
\SetKwFunction{pread}{Read}
\SetKwFunction{pcas}{CAS}

\SetAlgoLined
\pinit{}\;\nllabel{alg:li-in}
\While{! \pdo}{
\ppw{}\;\nllabel{alg:li-ps}
\While{! \psucc}{\nllabel{alg:li-bcs}
\pcur $\leftarrow$ \pread{\pacp}\;\nllabel{alg:li-bbcs}
\pnew $\leftarrow$ \pcw{\pcur}\;
\psucc $\leftarrow$ \pcas{\pacp, \pcur, \pnew}\;
}\nllabel{alg:li-ecs}
}
\caption{AbstractAlgorithm()\label{alg:gen-name}}
\end{procedure}
\caption{Thread procedure\label{alg:gen-nb}}
\end{figure}


We aim at evaluating the throughput of a multi-threaded algorithm that
is based on the utilization of a shared lock-free \ds. Such a program
can be abstracted by the Procedure~\ref{alg:gen-name} (see
Figure~\ref{alg:gen-nb}) that represents the skeleton of the function
which is called by each spawned thread. It is decomposed in two main
phases: the {\it \ps}, represented on line~\ref{alg:li-ps}, and the
{\it \rl}, from line~\ref{alg:li-bcs} to line~\ref{alg:li-ecs}. A
{\it \re} starts at line~\ref{alg:li-bbcs} and ends at
line~\ref{alg:li-ecs}.

As for line~\ref{alg:li-in}, the function \pinit shall be seen as an
abstraction of the delay between the spawns of the threads, that is
expected not to be null, even when a barrier is used. We then consider
that the threads begin at the exact same time, but have different
initialization times.

The \ps is the part of the code where the thread does not access the shared
\ds; the work that is performed inside this \ps can possibly depend on the
value that has been read from the \ds, \eg in the case of processing
an element that has been dequeued from a FIFO (First-In-First-Out)
queue.

In each \re, a thread tries to modify the \ds, and does not exit
the \rl until it has successfully modified the \ds. It does that by
firstly reading the access point \DataSty{AP} of the
\ds, then according to the value that has been read, and possibly to other previous computations that occurred in the past, the thread prepares the new desired value as an access point of the \ds. Finally, it atomically tries to perform the change through a call to the \casexp (\cas) primitive. If it
succeeds, \ie if the access point has not been changed by another
thread between the first \rf and the \cas, then it goes to the
next \ps, otherwise it repeats the process. The \rl is composed of at
least one \re, and we number the \res starting from $0$, since the
first iteration of the \rl is actually not a \re, but a try.

\pgfdeclaredecoration{complete sines}{initial}
{
    \state{initial}[
        width=+0pt,
        next state=sine,
        persistent precomputation={\pgfmathsetmacro\matchinglength{
            \pgfdecoratedinputsegmentlength / int(\pgfdecoratedinputsegmentlength/\pgfdecorationsegmentlength)}
            \setlength{\pgfdecorationsegmentlength}{\matchinglength pt}
        }] {}
    \state{sine}[width=\pgfdecorationsegmentlength]{
        \pgfpathsine{\pgfpoint{0.25\pgfdecorationsegmentlength}{0.5\pgfdecorationsegmentamplitude}}
        \pgfpathcosine{\pgfpoint{0.25\pgfdecorationsegmentlength}{-0.5\pgfdecorationsegmentamplitude}}
        \pgfpathsine{\pgfpoint{0.25\pgfdecorationsegmentlength}{-0.5\pgfdecorationsegmentamplitude}}
        \pgfpathcosine{\pgfpoint{0.25\pgfdecorationsegmentlength}{0.5\pgfdecorationsegmentamplitude}}
}
    \state{final}{}
}
\pgfdeclarelayer{background}
\pgfdeclarelayer{foreground}
\pgfsetlayers{background,main,foreground}

\def\amp{5}

\newcounter{wit}
\setcounter{wit}{1}

\renewcommand\rr[1]{#1}
\renewcommand\pp[1]{}

\newcommand{\extth}[3]{%
\checkendxt(#1)
\edef\prev{\cachedata}
\pgfmathparse{\prev+#2}
\endxt(#1)={\pgfmathresult}
\checkendxt(#1)
\edef\nene{\cachedata}
\draw[-,very thick,#3] (\prev,\yt[#1]) -- (\nene,\yt[#1]);
\begin{pgfonlayer}{foreground}
\draw[orange] (\prev,\yt[#1]-.25) -- ++(0,.5);
\end{pgfonlayer}
}
\rr{\def\yt{{-5.5,0,-1.5,-3,-4.5}}}
\pp{\def\yt{{-3,0,-.8,-1.6,-2.4}}}

\begin{figure}
\begin{center}
\begin{tikzpicture} [very thick,scale=1]
\coordinate (Lps) at (0,0);
\coordinate (Lscs) at (0,-.5);
\coordinate (Lfcs) at (0,-1);
\coordinate (Lis) at (0,-1.5);

\node[anchor=west] (Lpst) at (Lps) {\ps};
\node[anchor=west] (Lscst) at (Lscs) {successful \re};
\node[anchor=west] (Lfcst) at (Lfcs) {failed \re};
\node[anchor=west] (List) at (Lis) {initialization};

\draw[blue] (Lps) -- ++(-.5,0);
\draw[green] (Lscs) -- ++(-.5,0);
\draw[red] (Lfcs) -- ++(-.5,0);
\draw[black] (Lis) -- ++(-.5,0);

\draw[very thin] ($(Lpst.north west)+(-1,0.2)$) rectangle ($(List.south east)+(1,-0.2)$);
\end{tikzpicture}
\end{center}
\caption{\label{fig:legend}Legend of Figures~\protect\ref{fig:ex-f2}, \protect\ref{fig:ex-f1}, \protect\ref{fig:ex-lem.1}, \protect\ref{fig:ex-lem.3}, \protect\ref{fig:ex-lem.2}.}
\end{figure}

\rr{\begin{figure}}
\pp{ \begin{figure*}}
\hspace*{.3cm}\begin{tikzpicture} [scale=0.6, font=\small, thdec/.style={ thick, decoration={complete sines, segment length=.1cm,amplitude=\amp},decorate},
par/.style={blue}, csu/.style={green}, cfa/.style={red}, ini/.style={black}, ]
\def\cle{1.8}
\def\ple{3.75}
\def\prevend{0}
\newarray\endxt
\expandarrayelementtrue
\readarray{endxt}{0&0&0&0}
\newarray\wpl

\draw[<->] (10.1,\yt[0]) -- ++ (3*\cle+\ple,0)  node[midway,fill=white] {Cycle};
\node [left] at (0,\yt[1]) {\thr{0}};
\node [left] at (0,\yt[2]) {\thr{1}};
\node [left] at (0,\yt[3]) {\thr{2}};
\node [left] at (0,\yt[4]) {\thr{3}};




\extth{1}{\cle}{csu}

\extth{2}{1.05}{ini} 
\extth{2}{\cle}{cfa}
\extth{2}{\cle}{csu}

\extth{3}{1.2}{ini} 
\extth{3}{\cle}{cfa}
\extth{3}{\cle}{cfa}
\extth{3}{\cle}{csu}

\extth{4}{6.75}{ini} 
\extth{4}{\cle}{csu}

\extth{1}{\ple}{par}
\extth{1}{\cle}{cfa}
\extth{1}{\cle}{cfa}
\extth{1}{\cle}{csu}

\extth{2}{\ple}{par}
\extth{2}{\cle}{cfa}
\extth{2}{\cle}{cfa}
\extth{2}{\cle}{csu}

\extth{3}{\ple}{par}
\extth{3}{\cle}{cfa}
\extth{3}{\cle}{cfa}
\extth{3}{\cle}{csu}

\extth{4}{\ple}{par}
\extth{4}{\cle}{cfa}
\extth{4}{\cle}{cfa}
\extth{4}{\cle}{csu}

\extth{1}{\ple}{par}
\extth{1}{\cle}{cfa}
\extth{1}{\cle}{cfa}
\extth{1}{\cle}{csu}

\extth{2}{\ple}{par}
\extth{2}{\cle}{cfa}
\extth{2}{\cle}{cfa}
\extth{2}{\cle}{csu}

\extth{3}{\ple}{par}
\extth{3}{\cle}{cfa}
\extth{3}{\cle}{cfa}
\extth{3}{\cle}{csu}

\extth{4}{\ple}{par}
\extth{4}{\cle}{cfa}
\extth{4}{\cle}{cfa}
\extth{4}{\cle}{csu}

\extth{1}{\ple}{par}
\extth{1}{\cle}{cfa}

\extth{2}{\ple}{par}

\extth{3}{2}{par}

\draw[draw=none,fill=white]  (25.45,\yt[0]) rectangle (28,\yt[1]+.25);

\end{tikzpicture}
\caption{Execution with one \caca, and one inevitable failure\label{fig:ex-f2}}
\rr{\end{figure}
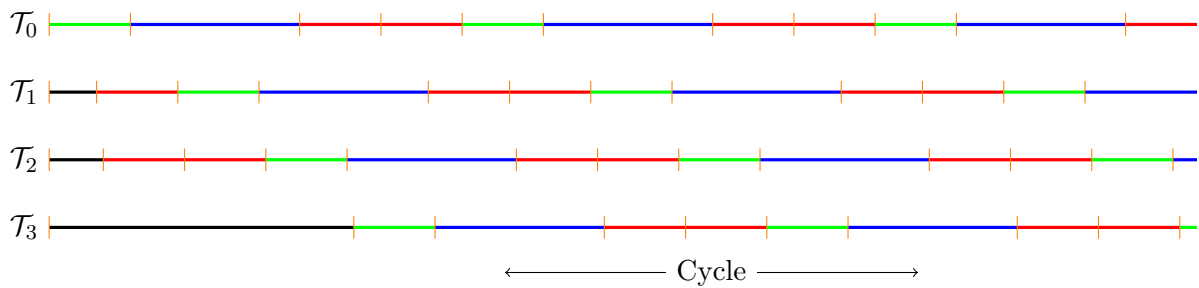}
\pp{\end{figure*}}


\setcounter{wit}{1}

\rr{\begin{figure}\begin{center}
\hspace*{1cm}\begin{tikzpicture} [scale=0.6, font=\small, thdec/.style={thick, decoration={complete sines, segment length=.1cm,amplitude=\amp},decorate}, par/.style={blue}, csu/.style={green}, cfa/.style={red}, ini/.style={black}, ]
\def\cle{2.4}
\def\ple{5}
\def\prevend{0}
\newarray\endxt
\expandarrayelementtrue
\readarray{endxt}{0&0&0&0}
\newarray\wpl



\draw[<->] (.6+3*\cle,\yt[0]) -- ++ (2*\cle+\ple,0)  node[midway,fill=white] {Cycle};
\node [left] at (0,\yt[1]) {\thr{0}};
\node [left] at (0,\yt[2]) {\thr{1}};
\node [left] at (0,\yt[3]) {\thr{2}};
\node [left] at (0,\yt[4]) {\thr{3}};

\extth{1}{\cle}{csu}

\extth{2}{.2}{ini}
\extth{2}{\cle}{cfa}
\extth{2}{\cle}{csu}

\extth{3}{.4}{ini}
\extth{3}{\cle}{cfa}
\extth{3}{\cle}{cfa}
\extth{3}{\cle}{csu}

\extth{4}{.6}{ini}
\extth{4}{\cle}{cfa}
\extth{4}{\cle}{cfa}
\extth{4}{\cle}{cfa}
\extth{4}{\cle}{csu}

\extth{1}{\ple}{par}
\extth{1}{\cle}{cfa}
\extth{1}{\cle}{csu}

\extth{2}{\ple}{par}
\extth{2}{\cle}{cfa}
\extth{2}{\cle}{csu}

\extth{3}{\ple}{par}
\extth{3}{\cle}{cfa}
\extth{3}{\cle}{csu}

\extth{4}{\ple}{par}
\extth{4}{\cle}{cfa}
\extth{4}{\cle}{csu}

\extth{1}{\ple}{par}
\extth{1}{\cle}{cfa}
\extth{1}{\cle}{csu}

\extth{2}{\ple}{par}
\extth{2}{\cle}{cfa}

\extth{3}{\ple}{par}

\extth{4}{\ple}{par}

\draw[draw=none,fill=white]  (22,\yt[4]-.25) rectangle (25.5,\yt[1]+.25);

\end{tikzpicture}
\caption{Execution with minimum number of failures\label{fig:ex-f1}}
\end{center}\end{figure}
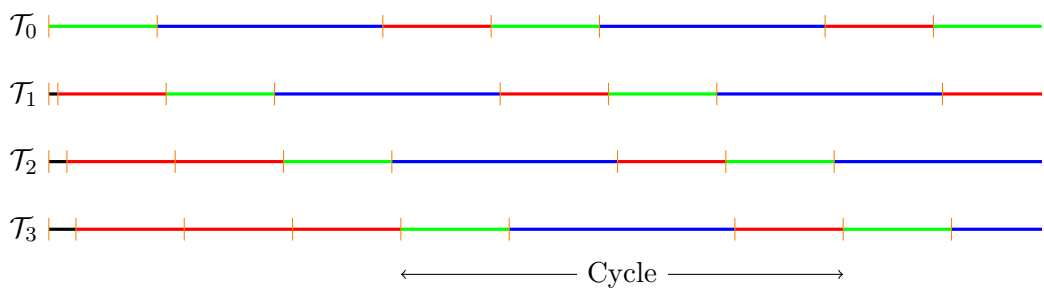}


\rr{
We analyze the behavior of \ref{alg:gen-name} from a
throughput
perspective, which
is defined as the number of successful \ds operations per unit of
time. In the context of Procedure~\ref{alg:gen-name}, it is equivalent to the
number of successful \cas{}s.

\medskip}

The throughput of the lock-free algorithm\pp{, \ie the number of successful \ds operations per unit of
time}, that we denote by \thru, is impacted by several
parameters.
\begin{itemize}
\item {\it Algorithm parameters}: the amount of work inside a call to
  \FuncSty{Parallel\_Work} (resp. \FuncSty{Critical\_Work}) denoted by \pw
  (resp. \cw).
\item {\it Platform parameters}: \rf and \cas latencies (\rc and \cc
  respectively), and the number \ct of processing units (cores). We assume
  homogeneity for the latencies, \ie every thread experiences the same
  latency when accessing an uncontended shared data, which is achieved
  in practice by pinning threads to the same
  socket.
\end{itemize}

\medskip

\subsubsection{Examples and Issues}

We first present two straightforward upper bounds on the throughput,
and describe the two kinds of conflict that keep the actual throughput
away from those upper bounds.

\paragraph{Immediate Upper Bounds}
\label{cha-secimm-bou}

Trivially, the minimum amount of work \rlwp in a given \re is $\rlwp
= \rc+\cw+\cc$, as we should pay at least the memory accesses
(hence \rf latency \rc and \cas latency \cc) and the critical work \cw
in between.

\falseparagraph{Thread-wise} A given thread can at most perform one
successful \re every $\pw+\rlwp$ units of time. In the best case,
\ctot threads can then lead to a throughput of $\ctot/(\pw+\rlwp)$.

\falseparagraph{System-wise} By definition, two successful \res cannot
overlap, hence we have at most $1$ successful \re every \rlwp units of
time.

\medskip

Altogether, the throughput \thru is bounded by
\rr{\[ \thru \leq \min \left( \frac{1}{\rc+\cw+\cc} , \frac{\ctot}{\pw+\rc+\cw+\cc}\right),\]
which can be rewritten as
\hspace*{-1cm}\begin{equation}
\label{eq:imm-bou}
 \thru \leq
\left\{ \begin{array}{ll}
\frac{1}{\rc+\cw+\cc} & \rr{\quad} \text{ if } \pw \leq (\ctot-1)(\rc+\cw+\cc)\\
\frac{\ctot}{\pw+\rc+\cw+\cc} & \rr{\quad} \text{ otherwise.}
\end{array} \right.
\end{equation}}
\pp{\begin{equation}
\label{eq:imm-bou}
\thru \leq \min \left( \frac{1}{\rc+\cw+\cc} , \frac{\ctot}{\pw+\rc+\cw+\cc}\right),
\text{ \ie, }
 \thru \leq
\left\{ \begin{array}{ll}
\frac{1}{\rc+\cw+\cc} & \rr{\quad} \text{ if } \pw \leq (\ctot-1)(\rc+\cw+\cc)\\
\frac{\ctot}{\pw+\rc+\cw+\cc} & \rr{\quad} \text{ otherwise.}
\end{array} \right.
\end{equation}}

\paragraph{Conflicts}

\subparagraph{\Logcons}

Equation~\ref{eq:imm-bou} expresses the fact that when \pw is small
enough, \ie when $\pw \leq (\ctot-1) \rlwp$, we cannot expect that
every thread performs a successful \re every $\pw+\rlwp$ units of
time, since it is more than what the \rl can afford.
As a result, some \logcons, hence unsuccessful \res, will be
inevitable, while the others, if any, are called {\it wasted}.

\rr{However, different executions can lead to different numbers of failures, which end up with
different throughput values. Figures~\ref{fig:ex-f2} and~\ref{fig:ex-f1} depict two
executions,}
\pp{Figure~\ref{fig:ex-f2} depicts an execution,} where the black parts are the calls to \pinit, the blue parts are the \pss,
and the \res can be either unsuccessful --- in red --- or successful
--- in green (the legend is displayed in Figure~\ref{fig:legend}).
\rr{We experiment different initialization times, and
observe different synchronizations, hence different numbers
of \cacas.} After the initial transient state, \rr{the execution
depicted in Figure~\ref{fig:ex-f1} comprises only the inevitable
unsuccessful \res, while the execution of Figure~\ref{fig:ex-f2}
contains one \caca.}\pp{the execution contains actually, for each
thread, one inevitable unsuccessful \re, and one \caca, because there
exists a set of initialization times that lead to a cyclic execution
with a single failure per thread and per period.}

We can see on \rr{those two examples}\pp{this example} that a cyclic
execution is reached after the transient behavior; actually, we show
in Section~\ref{cha-secwt} that, in the absence of \hardcons, every
execution will become periodic, if the initialization times are spaced
enough. In addition, we prove that the shortest period is such that,
during this period, every thread succeeds exactly once. This finally
leads us to define the additional failures as wasted, since we can
directly link the throughput with this number of \cacas: a higher
number of \cacas implying a lower throughput.

\subparagraph{\Hardcons}

\pp{\begin{wrapfigure}{r}{.5\textwidth}\begin{center}
\begin{tikzpicture} [scale=0.7, font=\scriptsize]}
\rr{\begin{figure}[h!]\begin{center}\begin{tikzpicture}}

\def\minh{1.5}
\def\maxh{2.5}
\def\midh{2}

\draw (-1,\minh) rectangle (3.5,\maxh) node[pos=.5] {\rf \& \calrl};
\draw (6.5,\minh) rectangle (11,\maxh) node[midway, align=center] {Previously expanded\\CAS};
\draw [dashed, red](3.5,\minh) rectangle (6.5,\maxh) node[pos=.5] {Expansion};

\def\minh{3}
\def\maxh{3.5}
\def\midh{3.25}

\draw (3,\minh) rectangle (6.5,\maxh) node[pos=.5] {CAS};
\draw[->] (3,\minh) -- ++(0,-.5);
\draw[dotted] (6.5,\minh) -- ++(0,-.5);
\end{tikzpicture}
\caption{Expansion\label{fig:expansion}}
\end{center}
\pp{\end{wrapfigure}
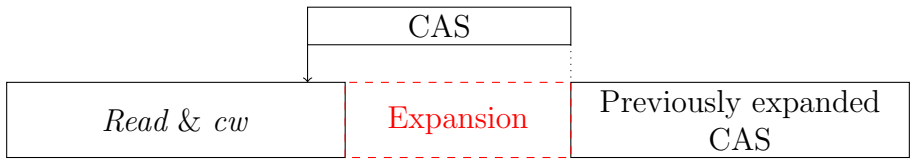}
\rr{\end{figure}}

The requirement of atomicity compels the ownership of the data in an
exclusive manner by the executing core.\rr{ This fact prohibits
concurrent execution of atomic instructions if they are operating on
the same data.} Therefore, overlapping parts of atomic instructions
are serialized by the hardware, leading to stalls in subsequently
issued ones. For our target lock-free algorithm, these stalls that we
refer to as expansion become an important slowdown factor in case
threads interfere in the \rl. As illustrated in
Figure~\ref{fig:expansion}, the latency for \cas can expand and cause
remarkable decreases in throughput since the
\cas of a successful thread is then expanded by others; for this reason, the amount of work inside
a \re is not constant, but is, generally speaking, a function depending on the number of threads
that are inside the \rl.

\paragraph{Process}

We deal with the two kinds of conflicts separately and connect them
together through the fixed-point iterative convergence.

In Section~\ref{cha-secexp}, we compute the expansion in execution
time of a \re, noted \expa, by following a probabilistic approach. The
estimation takes as input the expected number of threads inside
the \rl at any time, and returns the expected increase in the
execution time of a \re due to the serialization of atomic primitives.

In Section~\ref{cha-secwt}, we are given a program without \hardcon
described by the size of the \ps \psiz and the size of
a \re \rlsiz. We compute upper and lower bounds on the
throughput \thru, the number of wasted \res $w$, and the average
number of threads inside the \rl \trl. Without loss of generality, we
can normalize those execution times by the execution time of a \re,
and define the \ps size as $\psiz = q + r$, where $q$ is a
non-negative integer and $r$ is such that $0 \leq r < 1$. This pair
(together with the number of threads \ct) constitutes the actual input
of the estimation.

Finally, we combine those two outcomes in Section~\ref{cha-secthr} by
emulating expansion through work not prone to \hardcons and obtain the
full estimation of the throughput\posrem{ according to the model
parameters that have been described in
Section~\ref{cha-secprog-plat}}{}.

\subsection{Execution without \hardcon}
\label{cha-secwt}


\newcommand{\suc}[1]{\ema{S_{#1}}}
\newcommand{\att}[2]{\ema{R_{#1}^{#2}}}

\newcommand{\fis}[1]{\suc{#1}}

\newcommand{\gap}[2]{\ema{G_{#1}^{(#2)}}}
\newcommand{\gapt}[2]{\ema{\widetilde{G}_{#1}^{(#2)}}}
\newcommand{\tinit}[1]{\ema{t_{#1}}}

\newcommand{\wfs}{well-formed seed\xspace}
\newcommand{\wfss}{well-formed seeds\xspace}

\newcommand{\wifs}{weakly-formed seed\xspace}
\newcommand{\cyex}[2]{$(#1,#2)$-cyclic execution\xspace}
\newcommand{\cyexs}[2]{$(#1,#2)$-cyclic executions\xspace}

\newcommand{\fa}[1]{\ema{f\left( #1\right)}}
\newcommand{\soe}{\ema{\mathcal{S}}}
\newcommand{\wt}{\ema{\mathit{WT}}}
\newcommand{\wati}{\textit{wasted time}\xspace}

\newcommand{\pl}{\ema{n_0}}
\newcommand{\fl}{\ema{f_0}}

\renewcommand\rr[1]{}
\renewcommand\pp[1]{#1}

\begin{figure}
\begin{center}
\hspace*{.1cm}\begin{tikzpicture} [scale=0.6, font=\small, thdec/.style={ thick, %
decoration={complete sines, segment length=.06cm,amplitude=\amp},decorate},%
par/.style={blue}, csu/.style={green}, cfa/.style={red}, ini/.style={black}, ]
\draw[draw=none]  (0,\yt[4]-.25) rectangle (27.1,\yt[1]+.25);
\def\prevend{0}
\newarray\endxt
\expandarrayelementtrue
\readarray{endxt}{0&0&0&0}
\def\cle{1}
\input{inp-chalmers/ex-lemma-4.tex}
\draw[draw=none,fill=white]  (27.1,\yt[4]-.25) rectangle (29.6,\yt[1]+.25);
\end{tikzpicture}
\caption{Reordering and transient state\label{fig:ex-lem.2}}
\end{center}
\end{figure}
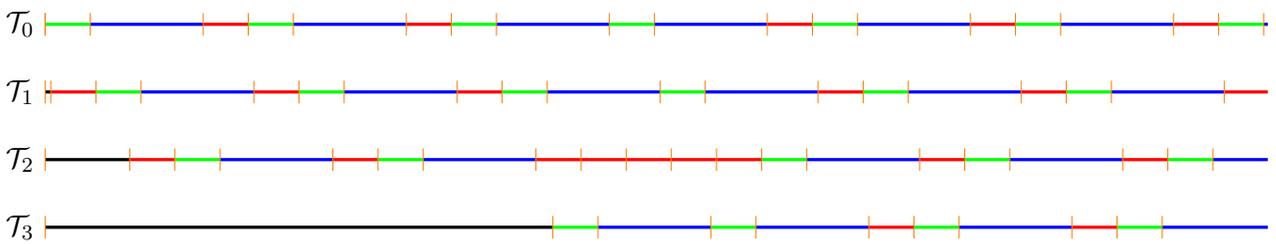

We show in this section that, in the absence of \hardcons, the
execution becomes periodic, which eases the calculation of the
throughput.  We start by defining some concepts: \cyexs{f}{\ct} are a
special kind of periodic executions such that within the shortest
period, each thread performs exactly $f$ unsuccessful \res and $1$
successful \re. The {\it \wfs} is a set of events that allows us to
detect an {\it \cyex{f}{\ct}} early, and the {\it gaps} are a measure
of the quality of the synchronization between threads. The idea is to
iteratively add threads into the execution and show that the
periodicity is maintained.
Theorem~\ref{th.wfs.gap}, on page~\pageref{th.wfs.gap}, establishes a
fundamental relation between gaps and \wfss, while
Theorem~\ref{th.cyc.exec}, on page~\pageref{th.cyc.exec}, proves the
periodicity, relying on the disjoint cases \pp{depicted on
Figures~\ref{fig:ex-lem.1},~\ref{fig:ex-lem.3} and~\ref{fig:ex-lem.2}.
We recall that the complete version of the proofs can be found
in~\cite{our-long}, together with additional Lemmas.}
\rr{of Lemma~\ref{lem.cyc.exec.1},~\ref{lem.cyc.exec.3}, and~\ref{lem.cyc.exec.2}.} Finally, we exhibit upper and lower bounds on throughput and number of failures, along with the average number of threads inside the \rl.

\begin{figure}
\begin{center}
\hspace*{.2cm}\begin{tikzpicture} [scale=0.6, font=\small, thdec/.style={ thick, %
decoration={complete sines, segment
length=.06cm,amplitude=\amp},decorate}, par/.style={blue},
csu/.style={green}, cfa/.style={red}, ini/.style={black}, ]
\draw[draw=none, use as bounding box]  (0,\yt[3]-.25) rectangle (18.9,\yt[1]+.25);
\def\prevend{0}
\newarray\endxt
\expandarrayelementtrue
\readarray{endxt}{0&0&0&0}
\def\cle{1}
\input{inp-chalmers/ex-lemma-1.tex}
\draw[draw=none,fill=white]  (18.9,\yt[3]-.25) rectangle (21,\yt[1]+.25);
\end{tikzpicture}
\caption{New thread does not lead to a reordering \label{fig:ex-lem.1}}
\end{center}
\end{figure}
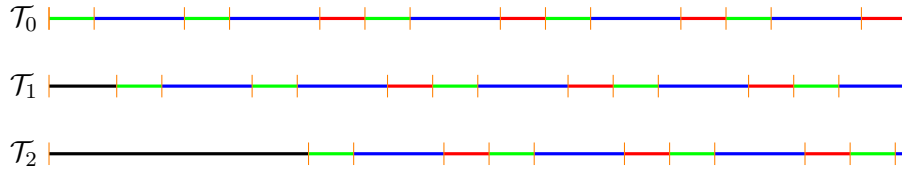

\begin{figure}
\begin{center}
\hspace*{.3cm}\begin{tikzpicture} [scale=0.6, font=\small, thdec/.style={ thick, %
decoration={complete sines, segment length=.06cm,amplitude=\amp},decorate},%
par/.style={blue}, csu/.style={green}, cfa/.style={red}, ini/.style={black}, ]
\draw[draw=none, use as bounding box]  (0,\yt[4]-.25) rectangle (26.6,\yt[1]+.25);
\def\prevend{0}
\newarray\endxt
\expandarrayelementtrue
\readarray{endxt}{0&0&0&0}
\def\cle{1}
\input{inp-chalmers/ex-lemma-3.tex}
\draw[draw=none,fill=white]  (26.6,\yt[4]-.25) rectangle (29.5,\yt[1]+.25);
\end{tikzpicture}
\caption{Reordering and immediate new seed \label{fig:ex-lem.3}}
\end{center}
\end{figure}
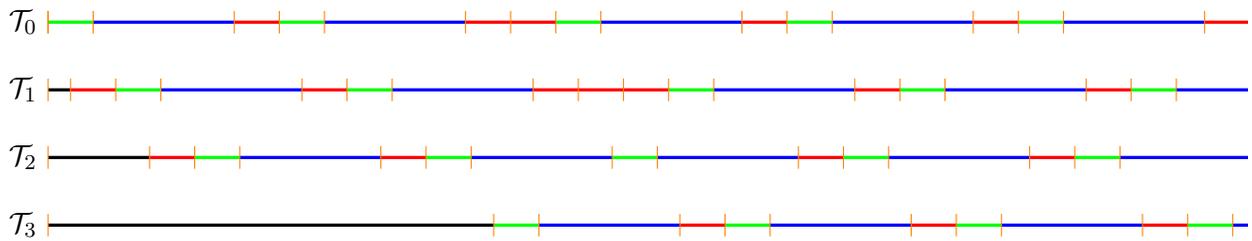

\subsubsection{Setting}
\label{cha-secwt-set}

\rr{

\paragraph{Initial Restrictions}
\begin{remark}
Concerning correctness, we assume that the reference point of the \rf
and the \cas occurs when the thread enters and exits any \re,
respectively.
\end{remark}

\begin{remark}
We do not consider simultaneous events, so all inequalities that refer
to time comparison are strict, and can be viewed as follows: time
instants are real numbers, and can be equal, but every event is
associated with a thread; also, in order to obtain a strict order
relation, we break ties according to the thread numbers (for instance
with the relation $<$).
\end{remark}
}

\pp{ In preamble, note that the events are strictly ordered (according to their instant of occurrence, with the thread id as a tie-breaker). As for correctness, \ie to decide for
  the success or the failure of a \re, we need instants of occurrence
  for \rf and \cas; we consider that the entrance (resp. exit) time of
  a \re is the instant of occurrence of the \rf (resp. \cas).}

\paragraph{Notations and Definitions}

We recall that \ct threads are executing the pseudo-code described in
Procedure~\ref{alg:gen-name}, one \re is of unit-size, and the \ps is of size $\psiz = q
+r$, where $q$ is a non-negative integer and $r$ is such that  $0 \leq r < 1$.
Considering a thread \thr{n} which succeeds at time \suc{n}; this thread completes a
whole \re in $1$ unit of time, then executes the \ps of size \psiz, and attempts to perform
again the operation every
unit of time, until one of the attempt is successful. We note \att{n}{k} the \kth{k} \re
so that $\att{n}{k} = \suc{n} + 1 + q +r + k$.
Also, at a given time $t$ where not any thread is currently succeeding, the next
successful attempt will be at time
\[\min_{n \in \inte{\ctot-1}} \{ \att{n}{k} = \suc{n}+1+q+r+k > t \; ; \;
\text{\suc{n} is the last success of \thr{n}} \},\]
 and $n$ gives the thread number of the
successful thread.

\begin{definition}
An execution with \ct threads is called \cyex{C}{\ct} if and only if
(i) the execution is periodic, \ie at every time, every thread is in
the same state as one period before, (ii) the shortest period contains
exactly one successful attempt per thread, (iii) the shortest period
is $1+q+r+C$.
\end{definition}

\begin{definition}
Let $\soe = \left( \thr{i}, \fis{i} \right)_{i \in \inte{\ct-1}}$,
where \thr{i} are threads and \fis{i} ordered times, \ie such that
$\fis{0} < \fis{1} < \dots < \fis{\ct-1}$.
\soe is a {\it seed} if and
only if for all $i \in \inte{\ct -1}$, \thr{i} does not succeed
between \fis{0} and \fis{i}, and starts a \re at \fis{i}.

We define \fa{\soe} as the smallest non-negative integer such that
$\fis{0} + 1 + q + r + \fa{\soe} > \fis{\ct-1} + 1$, \ie $ \fa{\soe} =
\max\left( 0 , \left\lceil \fis{\ct-1} - \fis{0} - q
-r\right\rceil\right)$. When \soe is clear from the context, we denote
\fa{\soe} by $f$.
\end{definition}

\begin{definition}
\soe is a {\it \wfs} if and only if for each $i \in \inte{\ct-1}$, the execution of thread
\thr{i} contains the following sequence: firstly a success beginning at \fis{i}, the
parallel section,  $f$ unsuccessful \res, and finally a successful \re.
\end{definition}


Those definitions are coupled through the two natural following properties:
\begin{property}
\label{pr.wfs.soe1}
Given a \cyex{C}{\ct}, any seed \soe including \ct consecutive successes is a \wfs,
with $\fa{\soe} = C$.
\end{property}
\begin{proof}
Choosing any set of \ct consecutive successes, we are ensured, by the
definition of a \cyex{f}{\ct}, that for each thread, after the first success,
the next success will be obtained after $f$ failures. The order will be
preserved, and this shows that a seed including our set of successes is
actually a \wfs.
\end{proof}

\begin{property}
\label{pr.wfs.soe2}
If there exists a \wfs in an execution, then after every thread has succeeded once, the
execution coincides with an \cyex{f}{\ct}.
\end{property}
\begin{proof}
By the definition of a \wfs, we know that the threads will first
succeed in order, fails $f$ times, and succeed again in the same
order. Considering the second set of successes in a new \wfs, we
observe that the threads will succeed a third time in the same order,
after failing $f$ times. By induction, the execution coincides with an
\cyex{f}{\ct}.
\end{proof}

\newcommand{\lag}{lagging time\xspace}
\newcommand{\lagt}{\ema{\ell}}

Together with the seed concept, we define the notion of \textit{gap}
that we will use extensively in the next subsection. The general idea
of those gaps is that within an \cyex{f}{\ct}, the period is higher
than $\ct \times 1$, which is the total execution time of all the
successful \res within the period. The difference between the period
(that lasts $1+q+r+f$) and \ct, reduced by $r$ (so that we obtain an
integer), is referred as \textit{\lag} in the following. If the
threads are numbered according to their order of success (modulo \ct),
as the time elapsed between the successes of two given consecutive
threads is constant (during the next period, this time will remain the
same), this \lag can be seen in a circular manner\rr{ (see
Figure~\ref{fig:gaps})}: the threads are represented on a circle whose
length is the \lag increased by $r$,
and the length between two consecutive threads is the time between the
end of the successful \re of the first thread and the begin of the
successful \re of the second one. More formally, for all
$(n,k) \in \inte{\ct-1}^2$, we define the gap
\gap{n}{k} between \thr{n} and its \kth{k} predecessor based on the gap with the first
predecessor:
\rr{\begin{figure}
\begin{center}
\begin{tikzpicture}[arrows = {<->}]
\def\cim{2}
\def\cii{1.4}
\def\cio{2.5}
\draw[thick] (0,0) circle (\cim);
\draw[red]
 (90:\cii) arc (91:449:\cii);
\node[red] at (-60:\cii-.8) {$\displaystyle \sum_{n=0}^{\ct-1} \gap{n}{1}$};
\draw[fill=white,thick] (90:\cim) circle (.25) node {\thr{0}};
\draw[fill=white,thick] (45:\cim) circle (.25) node {\thr{1}};
\draw[fill=white,thick] (-45:\cim) circle (.25) node {\thr{2}};
\draw[fill=white,thick] (180:\cim) ellipse (.4 and .3) node {\thr{\ctot-1}};
\draw[blue] (89.5:\cio) arc (89.5:45.5:\cio);
\node[blue] at (66:1.2*\cio) {\gap{1}{1}};
\draw[blue] (44.5:\cio) arc (44.5:-44.5:\cio);
\node[blue] at (0:1.2*\cio) {\gap{2}{1}};
\draw[blue] (90.5:\cio) arc (90.5:314.5:\cio);
\node[blue] at (200:1.2*\cio) {\gap{0}{2}};
\end{tikzpicture}
\end{center}
\caption{Gaps\label{fig:gaps}}
\end{figure}}%

\[
\left\{ \begin{array}{l}
\forall n \in \inte[1]{\ct-1} \quad ; \quad \gap{n}{1} = \suc{n} - \suc{n-1} - 1\\
\gap{0}{1} = \suc{0} + q + r + f - \suc{\ct-1}
\end{array} \right.,
\]
which leads to the definition of higher order gaps:
\[ \forall n \in \inte{\ct-1} \rr{\quad} ; \rr{\quad} \forall k > 0 \rr{\quad} ; \rr{\quad} \gap{n}{k} = \sum_{j=n-k+1}^n \gap{j \bmod \ct}{1}.\]

For consistency, for all $n \in \inte{\ct-1}$, $\gap{n}{0} = 0$.

Equally, the gaps can be obtained directly from the successes: for all $k \in \inte[1]{\ct-1}$,
\begin{equation}
\gap{n}{k} = \left\{ \begin{array}{ll}
\suc{n} - \suc{n-k} - k & \mathrm{~if~} n>k\\
\suc{n} - \suc{\ct + n-k} +1+q+r+f-k & \mathrm{ otherwise}\\
\end{array} \right.
\label{eq.gats}
\end{equation}

Note that, in an \cyex{f}{\ct}, the \lag is the sum of all first order gaps, reduced by $r$.

\rr{

Now we extend the concept of \wfs to \wifs.

\begin{definition}
Let $\soe = \left( \thr{i}, \fis{i} \right)_{i \in \inte{\ct-1}}$ be a seed.

\soe is a \wifs for \ct threads if and only if: $\left( \thr{i}, \fis{i} \right)_{i \in \inte{\ct-2}}$
is a \wfs for $\ct-1$ threads, and the first thread succeeding after
\thr{\ct-2} is \thr{\ct-1}.
\end{definition}

\begin{property}
Let $\soe = \left( \thr{i}, \fis{i} \right)_{i \in \inte{\ct-1}}$ be a \wifs.

Denoting $f = \fa{\left( \thr{i}, \fis{i} \right)_{i \in \inte{\ct-2}}}$, for each $n \in
\inte{\ct-1}$, $\gap{n}{f} < 1$.
\end{property}
\rr{\begin{proof}
We have $\suc{\ct-2} + 1 < \suc{\ct-1} < \att{0}{f}$,
and if we note indeed \gapt{n}{k} the gaps within $\left( \thr{i}, \fis{i} \right)_{i \in \inte{\ct-2}}$,
the previous \wfs with $\ct-1$ threads, we know that for
all $n \in \inte[1]{\ct-2}$, $\gapt{n}{1} = \gap{n}{1}$, and $\gap{\ct-1}{1} +
\gap{0}{1} = \gapt{0}{1}$, which leads to $\gap{n}{k} \leq \gapt{n}{k}$, for all
$n \in \inte{\ct-1}$ and $k$; hence the weaker property.
\end{proof}}

\begin{lemma}
\label{lem.cyc.exec.1}
Let \soe be a \wifs, and $f=\fa{\left( \thr{i}, \fis{i} \right)_{i \in \inte{\ct-2}}}$. If, for all
$n \in \inte{\ct-1}$, $\gap{n}{f+1} < 1$, then there exists later in the execution a
\wfs $\soe'$ for \ct threads such that $\fa{\soe'} = f +1$.
\end{lemma}
\begin{proof}
The proof is straightforward; \soe is actually a \wfs such that $\fa{\soe} = f +
1$. \rr{Since $\att{0}{f} - \suc{\ct-1} < \gap{0}{1} <1 $, the first success of
\thr{0} after the success of \thr{\ct-1} is its \kth{f+1} \re.}
\end{proof}

} 

\subsubsection{Cyclic Executions}
\label{cha-secwt-cyc}

\begin{theorem}
\label{th.wfs.gap}
Given a seed $\soe = \left( \thr{i}, \fis{i} \right)_{i \in \inte{\ct-1}}$,
\soe is a \wfs if and only if for all $n \in \inte{\ct-1}$,
$0 \leq \gap{n}{f}<1$.
\end{theorem}
\begin{proof}
Let $\soe = \left( \thr{i}, \fis{i} \right)_{i \in \inte{\ct-1}}$ be a seed.

\noindent$(\Leftarrow)$
We assume that for all $n \in \inte{\ct-1}$, $0<\gap{n}{f}<1$, and we
first show that the first successes occur in the following
order: \thr{0} at \suc{0}, \thr{1} at \suc{1}, $\dots$, \thr{\ct-1}
at \suc{\ct-1}, \thr{0} again at $\att{0}{f}$. The first threads that
are successful executes their \ps after their success, then enters
their second \rl: from this moment, they can make the first attempt of
the threads, that has not been successful yet, fail. Therefore, we
will look at which \re of which already successful threads could have
an impact on which other threads.

We can notice that for all $n \in \inte{\ct -1}$, if the first success of
\thr{n} occurs at \suc{n}, then its next attempts will potentially occur at
$\att{n}{k} = \suc{n} + 1 + q + r + k$, where $k\geq 0$. More
specifically, thanks to Equation~\ref{eq.gats}, for all $n \leq f$,
$\att{n}{k} = \suc{\ct+n-f} + \gap{n}{f} + k$. Also, for all $k \leq f
-n$,
\begin{align}
\att{n}{k}  - \suc{\ct+n-f+k} &= -\left(\suc{\ct+n-f+k} - \suc{\ct+n-f} -k \right) +\gap{n}{f}\nonumber\\
&= \gap{n}{f} - \gap{\ct+n-f+k}{k}\nonumber\\
\att{n}{k}  - \suc{\ct+n-f+k} &= \gap{n}{f-k}, \label{eq.att.suc}
\end{align}
and this implies that if $k>0$,
\begin{equation}
\suc{\ct+n-f+k} - \att{n}{k-1} = 1- \gap{n}{f-k}. \label{eq.att.suc2}
\end{equation}

We know, by hypothesis, that $0<\gap{n}{f-k}<1$, equivalently
$0<1-\gap{n}{f-k}<1$. Therefore Equation~\ref{eq.att.suc} states that
if a thread \thr{n'} starts a successful attempt at \suc{\ct+n-f+k},
then this thread will make the \kth{k} \re of \thr{n} fail,
since \thr{n} enters a \re while \thr{n'} is in a successful \re. And
Equation~\ref{eq.att.suc2} shows that, given a thread \thr{n'}
starting a new \re at \suc{\ct+n-f+k}, the only \re of \thr{n} that
can make \thr{n'} fail on its attempt is the \kth{(k-1)} one. There is
indeed only one \re of \thr{n} that can enter a \re before the
entrance of \thr{n'}, and exit the \re after it.

\thr{0} is the first thread to succeed at \suc{0}, because no other thread is in the \rl at this time. Its next attempt will occur at \att{0}{0}, and
all thread attempts that start before \suc{\ct-f} (included) cannot
fail because of \thr{0}, since it runs then the \ps. Also, since all
gaps are positive, the threads \thr{1} to \thr{\ct-f} will succeed in
this order, respectively starting at times \suc{1} to
\suc{\ct-f}.

Then, using induction, we can show that \thr{\ct-f+1},
$\dots$, \thr{\ct-1} succeed in this order, respectively starting at
times \suc{\ct-f+1}, $\dots$, \suc{\ct-1}. For $j \in \inte{f-1}$,
let \pro{j} be the following property: for all
$n \in \inte{\ct-f+j}$, \thr{n} starts a successful \re at \suc{n}.
We assume that for a given $j$, \pro{j} is true, and we show that it
implies that \thr{\ct-f+j+1} will succeed at \suc{\ct-f+j+1}. The
successful attempt of \thr{\ct-f+j} at \suc{\ct-f+j} leads, for all
$j' \in \inte{j}$, to the failure of the \kth{j'} \re of
\thr{j-j'} (explanation of Equation~\ref{eq.att.suc}). But for each \thr{j'}, this attempt was precisely the one that could have made
\thr{\ct-f+j+1} fail on its attempt at \suc{\ct-f+j+1} (explanation of Equation~\ref{eq.att.suc}).
Given that all threads \thr{n}, where $n>\ct-f+j+1$, do not start
any \rl before \suc{\ct-f+j+1}, \thr{\ct-f+j+1} will succeed
at \suc{\ct-f+j+1}.  By induction, \pro{j} is true for all
$j \in \inte{f-1}$.

Finally, when \thr{\ct-1} succeeds, it makes the
\kth{(f-1-n)} \re of \thr{n} fail, for all $n \in \inte{f-1}$;
also the next potentially successful
attempt for \thr{n} is at \att{n}{f-n}. (Naturally, for all $n \in
\inte[f]{\ct-1}$, the next potentially successful attempt for \thr{n} is at
\att{n}{0}.)

We can observe that for all $n < \ct$,  $j \in \inte{\ct-1-n}$, and all $k \geq j$,
\begin{align}
\att{n+j}{k-j} - \att{n}{k} &= \suc{n+j} + k - j - \left( \suc{n}  +k\right)\nonumber\\
\att{n+j}{k-j} - \att{n}{k} &= \gap{n+j}{j}, \label{eq.att.att}
\end{align}
hence for all $n \in \inte[1]{f}$, $\att{n}{f-n} - \att{0}{f} = \gap{n}{n} > 0$.
\rr{\[ \att{n}{f-n} - \att{0}{f} = \gap{n}{n} > 0. \]}

As we have as well, for all $n \in\inte[f+1]{\ct-1}$,
$\att{n}{0} > \att{f}{0}$, we obtain that among all the threads, the
earliest possibly successful attempt is \att{0}{f}. Following
\thr{\ct-1}, \thr{0} is consequently the next successful thread in its
\kth{f} \re.

To conclude this part, we can renumber the threads (\thr{n+1} becoming
now \thr{n} if $n>0$, and \thr{0} becoming \thr{\ct-1}), and follow
the same line of reasoning. The only difference is the fact
that \thr{\ct-1} (according to the new numbering) enters the \rl $f$
units of time before \suc{\ct-1}, but it does not interfere with the
other threads, since we know that those attempts will fail.

There remains the case where there exists $n \in \inte{\ct-1}$ such
that $\gap{n}{f}=0$.  This implies that $f=0$, thus we have a \wfs.

\bigskip

\noindent$(\Rightarrow)$ We prove now the implication by contraposition; we assume that there exists $n
\in \inte{\ct-1}$ such that $\gap{n}{f}>1$ or $\gap{n}{f}<0$, and show that \soe
is not a \wfs.

We assume first that an \kth{f} order gap is negative. As it is a sum of \kst{1} order
gaps, then there exists $n'$ such that \gap{n'}{1} is negative; let $n''$ be the highest
one.

If $n''>0$, then either the threads $\thr{0}, \dots, \thr{n''-1}$ succeeded in order at
their \kth{0} \re, and then \thr{n''-1} makes \thr{n''} fail at its \kth{0} \re (we have
a seed, hence by definition, $\suc{n''-1}<\suc{n''}$, and $\gap{n''}{1}<0$, thus $\suc{n''-1}<\suc{n''}<\suc{n''-1}+1$ ),
or they did not succeed in order
at their first try. In both cases, \soe is not a \wfs.

If $n''=0$, let us assume that \soe is a \wfs. Let also a new seed be $\soe' =
\left( \thr{i}, \suc{i}' \right)_{i \in \inte{\ct-1}}$, where for all $n \in \inte{\ct-2}$,
$\suc{n+1}'=\suc{n}$, and $\suc{0}' = \suc{\ct-1} - (q+1+f+r) $. Like \soe, $\soe'$ is a
\wfs; however, \gap{1}{1} is negative, and we fall back into the previous case, which
shows that $\soe'$ is not a \wfs. This is absurd, hence \soe is not a \wfs.

\medskip

We assume now that every gap is positive and choose \pl defined by: $\pl = \min \{ n \; ;
\; \exists k \in \inte{\ct-1} / \gap{n+k}{k} >1 \}$, and $\fl = \min \{ k \; ; \; \gap{\pl
  + k}{k} >1 \}$: among the gaps that exceed $1$, we pick those that concern the earliest
thread, and among them the one with the lowest order.

Let us assume that threads \thr{0}, $\dots$, \thr{\ct-1} succeed at their \kth{0} \re in
this order, then \thr{0}, $\dots$, \thr{\pl} complete their second successful \rl at their \kth{f}
\re, in this order. If this is not the case, then \soe is not a \wfs, and the proof is
completed. According to Equation~\ref{eq.att.att}, we have, on the one hand,
$\att{\pl+1}{\fl-1} - \att{\pl}{\fl} = \gap{\pl+1}{1}$, which implies $\att{\pl+1}{\fl} -1
- \att{\pl}{\fl} = \gap{\pl+1}{1}$, thus $\att{\pl+1}{f} - (\att{\pl}{f}+1) =
\gap{\pl+1}{1}$; and on the other hand, $\att{\pl+\fl}{0} - \att{\pl}{\fl} =
\gap{\pl+\fl}{\fl}$ implying $\att{\pl+\fl}{f-\fl} - \left(\att{\pl}{f}+1\right) =
\gap{\pl+\fl}{\fl} -1$. As we know that $\gap{\pl+\fl}{\fl} - \gap{\pl+1}{1} =
\gap{\pl+\fl}{\fl-1} < 1$ by definition of \fl (and \pl), we can derive that
$\att{\pl+1}{f} - (\att{\pl}{f}+1) > \att{\pl+\fl}{f-\fl} - (\att{\pl}{f}+1)$. We have
assumed that \thr{\pl} succeeds at its \kth{f} \re, which will end at
$\att{\pl}{f}+1$. The previous inequality states then that \thr{\pl+1} cannot be
successful at its \kth{f} \re, since either a thread succeeds before \thr{\pl+\fl} and
makes both \thr{\pl+\fl} and \thr{\pl+1} fail, or \thr{\pl+\fl} succeeds and makes
\thr{\pl+1} fail. We have shown that \soe is not a \wfs.

\end{proof}

\rr{
\begin{lemma}
\label{lem.event.succ}
Assuming $r \neq 0$, if a new thread is added to an \cyex{f}{\ct}, it will
eventually succeed.
\end{lemma}
\begin{proof}


Let \att{\ct}{0} be the time of the \kth{0} \re of the new thread, that we
number \thr{\ct}. If this \re is successful, we are done; let us assume now
that this \re is a failure, and let us shift the thread numbers (for the
threads \thr{0}, $\dots$, \thr{\ct-1}) so that \thr{0} makes \thr{\ct} fail on
its first attempt. We distinguish two cases, depending on whether $\gap{0}{\ct}
> \att{\ct}{0} - \suc{0}$ or not.

We assume that $\gap{0}{\ct} > \att{\ct}{0} - \suc{0}$. We know that $n \mapsto
\gap{n}{n}$ is increasing on \inte{\ct-1} and that $\gap{0}{0} = 0$, hence let
$\pl = \min \{ n \in \inte{\ct-1} \; ; \; \gap{n}{n} < \att{\ct}{0} - \suc{0} \}$.
For all $k \in \inte{\pl}$, we have
$\att{\ct}{k} - \suc{k} = k + \att{\ct}{0} - (\gap{k}{k} + \suc{0} + k) =
\att{\ct}{0} - \suc{0} - \gap{k}{k}$ hence $\att{\ct}{k} - \suc{k}>0$ and
$\att{\ct}{k} - \suc{k} < \att{\ct}{0} - \suc{0} < 1$.  This shows that \thr{0},
$\dots$, \thr{\pl}, because of their successes at \suc{0}, $\dots$, \suc{\pl},
successively make \kth{0}, $\dots$, \kth{\pl} \res (respectively) of \thr{\ct} fail.
The next attempt for \thr{\ct} is at \att{\ct}{\pl+1}, which fulfills the
following inequality: $\att{\ct}{\pl+1} - (\suc{\pl}+1) < \suc{\pl+1} - (\suc{\pl}+1)$
since
\begin{align*}
\att{\ct}{\pl+1} - \suc{\pl+1} &=  (\pl+1 + \att{\ct}{0}) -(\gap{\pl+1}{\pl+1} + \suc{0} + \pl+1)\\
\att{\ct}{\pl+1} - \suc{\pl+1} &> 0.
\end{align*}
\thr{\pl+1} should have been the successful thread, but \thr{\ct} starts a
\re before \suc{\pl+1}, and is therefore succeeding.

We consider now the reverse case by assuming that $\gap{0}{\ct} < \att{\ct}{0} -
\suc{0}$. With the previous line of reasoning, we can show that \thr{0}, $\dots$,
\thr{\ct-1}, because of their successes at \suc{0}, $\dots$, \suc{\ct-1}, successively
make \kth{0}, $\dots$, \kth{(\ct-1)} \res (respectively) of \thr{\ct} fail.  Then we are
back in the same situation when \thr{0} made \thr{\ct} fail for the first time (\thr{0}
makes \thr{\ct} fail), except that the success of \thr{0} starts at $\suc{0}' = \suc{0} +
\gap{0}{\ct}$. As $\gap{0}{\ct} = q+r+f-\ct >0$ and $q$, f and \ct are integers, we have
that $\gap{0}{\ct} \geq r$. By the way, if we had $\gap{0}{\ct} > r$, we would have
$\gap{0}{\ct} \geq 1+r > \att{\ct}{0} - \suc{0}$, which is absurd. \suc{0} makes indeed
\att{\ct}{0} fail, therefore \gap{0}{\ct} should be less than $1$. Consequently, we are
ensured that $\gap{0}{\ct} = r$. We define
\[ k_0 = \left\lfloor \frac{\att{\ct}{0} - \suc{0}}{r} \right\rfloor;  \]
also, for every $k \in \inte[1]{k_0}$, $r < \att{\ct}{0} - (\suc{0} +
k\times r)$ and $r > \att{\ct}{0} - (\suc{0} + (k_0+1)\times r)$: the cycle of
successes of \thr{0}, $\dots$, \thr{\ct-1} is executed $k_0$ times. Then the
situation is similar to the first case, and \thr{\ct} will succeed.

\end{proof}
} 

\newcommand{\perfun}{\ema{\sigma}}
\newcommand{\perb}[1]{\ema{\perfun \left( #1 \right)}}
\newcommand{\iperfun}{\ema{\perfun^{-1}}}
\newcommand{\iper}[1]{\ema{\iperfun \left( #1 \right)}}
\newcommand{\enfun}{\ema{m_2}}
\newcommand{\en}[1]{\ema{\enfun \left( #1 \right)}}

\newcommand{\enofun}{\ema{m_1}}
\newcommand{\eno}[1]{\ema{\enofun \left( #1 \right)}}

\newcommand{\evafun}{\ema{\mathit{rank}}}
\newcommand{\eva}[1]{\ema{ \evafun \left( #1 \right)}}
\newcommand{\card}[1]{\ema{\# #1}}
\newcommand\restr[2]{{
    \left.\kern-\nulldelimiterspace 
  #1 
  \vphantom{\big|} 
  \right|_{#2} 
}}

\rr{

\begin{lemma}
\label{lem.cyc.exec.3}
Let \soe be a \wifs, and $f=\fa{\left( \thr{i}, \fis{i} \right)_{i \in \inte{\ct-2}}}$. If
$\gap{f}{f+1} > 1$,  and if the second success of
\thr{\ct-1} does not occur before the second success of \thr{f-1}, then we can find in
the execution a \wfs $\soe'$ for \ct threads such that $\fa{\soe'} = f$.
\end{lemma}
\begin{proof}


\newcommand{\seth}[1]{\ema{\mathcal{S}_{#1}}}

\rr{Let us first remark that, by}\pp{By} the definition of a \wifs, all threads will succeed once, in
order.  Then two ordered groups of threads will compete for each of the next successes,
until \thr{f-1} succeeds for the second time.
\pp{The rationale is that the gaps will shrink successively until the \kth{f} order
gaps become smaller than one, leading to the appearance of a \wfs.}

\rr{Let $e$ be the smallest integer of \inte[f]{\ct-1} such that the second success of \thr{e}
occurs after the second success of \thr{f-1}. Let then \seth{1} and \seth{2} be
the two groups of threads that are in competition, defined by
\begin{align*}
\seth{1} = \{ \thr{n} \; ; \; n \in \inte{f-1} \}\\
\seth{2} = \{ \thr{n} \; ; \; n \in \inte[f]{e-1} \}\\
\end{align*}

For all $n \in \inte{e-1}$, we note
\[ \eva{n}  = \left\{ \begin{array}{ll}
\gap{n}{n+1} & \text{ if } \thr{n} \in \seth{1}\\
\gap{n}{n+1} -1 \quad & \text{ if } \thr{n} \in \seth{2}
\end{array} \right. . \]
We define \perfun, a permutation of \inte{e-1} that describes the reordering of
the threads during the round of the second successes, such that, for all $(i,j)
\in \inte{e-1}^2$, $\perb{i} < \perb{j}$ if and only if $\eva{i} < \eva{j}$.

We also define a function that will help in expressing the \iper{k}'s:
\[
\begin{array}{cccc}
\enfun:& \inte{e-1} & \longrightarrow & \inte[f]{e-1}\\
 & k & \longmapsto & \max {\{ \ell \in \inte[f]{e-1} \; ; \; \thr{\ell} \in \seth{2} \; ; \; \perb{\ell} \leq k  \} }
\end{array}.
\]

We note that $\restr{\evafun}{\inte{f-1}}$ is increasing, as well as
$\restr{\evafun}{\inte[f]{e-1}}$. This shows that
$\card \{ \thr{\ell} \in \seth{2} \; ; \; \perb{\ell} \leq k  \} = \en{k} -(f -1)$.
Consequently, if $\thr{\iper{k}} \in \seth{2}$, then
\begin{align*}
\en{k} &= \card \{ \thr{\ell} \in \seth{2} \; ; \; \perb{\ell} \leq k  \} +f -1\\
&= \card \{ \thr{\ell} \in \seth{2} \; ; \; \ell \leq \iper{k}  \} +f -1\\
&= \iper{k}-f+1+f-1\\
\en{k} &= \iper{k}.
\end{align*}

Conversely, if $\thr{\iper{k}} \in \seth{1}$, among $\{ \thr{\perb{n}} \; ; \; n
\in \inte{k} \}$, there are exactly $\en{k}-f+1$ threads in \seth{2}, hence
\[ \iper{k} = k+1-(\en{k}-f+1)-1 = f + k -\en{k} -1.\]

In both cases, among $\{ \thr{\perb{n}} \; ; \; n \in \inte{k} \}$, there are
exactly $\en{k}-f+1$ threads in \seth{2}, and $\eno{k} = k - (\en{k}-f)$ threads
in \seth{1}.

We prove by induction that after this first round, the next successes will be,
respectively, achieved by \thr{\iper{0}}, \thr{\iper{1}}, $\dots$,
\thr{\iper{e-1}}. In the following, by ``\kth{k} success'', we mean \kth{k}
success after the first success of \thr{\ct-1}, starting from $0$, and the
\att{i}{j}'s denote the attempts of the second round.

Let \pro{K} be the following property: for all $k \leq K$, the \kth{k} success
is achieved by \thr{\iper{k}} at \att{\iper{k}}{f+k-\iper{k}}. We assume \pro{K}
true, and we show that the \kth{(K+1)} success is achieved by \thr{\iper{K+1}} at
\att{\iper{K+1}}{f+K+1-\iper{K+1}}.

We first show that if $\thr{\iper{K}} \in \seth{1}$, then
\begin{equation}
 \att{\en{K}+1}{\eno{K}-1} > \att{\iper{K}}{f+K-\iper{K}} > \att{\en{K}}{\eno{K}}.
\label{eq.inv.fm1}
\end{equation}
On the one hand,
\begin{align*}
\att{\iper{K}}{f+K-\iper{K}} &= K - \iper{K} + \att{\iper{K}}{f}\\
&= K - \iper{K} + \att{0}{f} + \iper{K} + \gap{\iper{K}}{\iper{K}}\\
&= K + \suc{\ct-1} +1 + \gap{0}{1} + \gap{\iper{K}}{\iper{K}}\\
\att{\iper{K}}{f+K-\iper{K}}
&= K + \suc{\ct-1} +1 + \gap{\iper{K}}{\iper{K}+1}.\\
\end{align*}
On the other hand,
\begin{align*}
\att{\en{K}}{f+K-\en{K}} &= (\en{K}-f) + \att{f}{K-(\en{K}-f)}  + \gap{\en{K}}{\en{K}-f}\\
&= (\en{K}-f)  + K - (\en{K}-f) + \att{f}{0} + \gap{\en{K}}{\en{K}-f}\\
&= (\en{K}-f)  + K - (\en{K}-f) + \suc{\ct-1} + 1 + (\gap{f}{f+1}-1) + \gap{\en{K}}{\en{K}-f}\\
\att{\en{K}}{f+K-\en{K}}
&= K + \suc{\ct-1} + 1 + \gap{\en{K}}{\en{K}+1} -1.\\
\end{align*}
Therefore,
\begin{align*}
\att{\iper{K}}{f+K-\iper{K}} - \att{\en{K}}{\eno{K}}
&= \att{\iper{K}}{f+K-\iper{K}} - \att{\en{K}}{f+K-\en{K}}\\
&= \gap{\iper{K}}{\iper{K}+1} - \left( \gap{\en{K}}{\en{K}+1} -1  \right)\\
\att{\iper{K}}{f+K-\iper{K}} - \att{\en{K}}{\eno{K}}&= \eva{\iper{K}} - \eva{\en{K}}.
\end{align*}

In a similar way, we can obtain that if $\thr{\iper{K}} \in \seth{2}$, then
\begin{equation}
\att{\eno{K}}{\en{K}} > \att{\iper{K}}{f+K-\iper{K}} > \att{\eno{K}-1}{\en{K}+1}.
\label{eq.inv.em1}
\end{equation}

In addition, we recall that if $\thr{\iper{K}} \in \seth{2}$, $\iper{K} = \en{K}$,
thus the second inequality of Equation~\ref{eq.inv.fm1} becomes an equality, and
if $\thr{\iper{K}} \in \seth{1}$, $\iper{K} = f + K -\en{K} -1$, hence the second
inequality of Equation~\ref{eq.inv.em1} becomes an equality.

Now let us look at which attempt of other threads \thr{\iper{K}} made fail. From
now on, and until explicitly said otherwise, we assume that $\thr{\iper{K}} \in
\seth{1}$. According to Equation~\ref{eq.inv.fm1}, we have
\begin{align*}
\att{\en{K}+1}{\eno{K}-1} &>& \att{\iper{K}}{f+K-\iper{K}} &>& \att{\en{K}}{\eno{K}}\\
\att{\en{K}+j}{\eno{K}-j} - \att{\en{K}+1}{\eno{K}-1} &<& \att{\en{K}+j}{\eno{K}-j} - \att{\iper{K}}{f+K-\iper{K}}
&<& \att{\en{K}+j}{\eno{K}-j} - \att{\en{K}}{\eno{K}}\\
\gap{\en{K}+j}{j-1} &<& \att{\en{K}+j}{\eno{K}-j} - \att{\iper{K}}{f+K-\iper{K}} &<& \gap{\en{K}+j}{j}
\end{align*}
This holds for every $j \in \inte[1]{\eno{K}}$, implying $j \leq f$, since there
could not be more than $f$ threads in \seth{1}.
Therefore, as by assumptions gaps of at most \kth{f} order are between $0$ and $1$,
\[ 0 < \att{\en{K}+j}{\eno{K}-j} - \att{\iper{K}}{f+K-\iper{K}} < 1; \]
showing that the success of \thr{\iper{K}} makes thread \thr{\en{K}+j} fail on
its attempt at \att{\en{K}+j}{\eno{K}-j}, for all $j \in \inte[1]{\eno{K}}$.

Since $\thr{\iper{K}} \in \seth{1}$, $\iper{K} = \eno{K}-1$. Also, for all $j
\in \inte{f-1-\eno{K}}$,
\begin{align*}
\att{\eno{K} + j}{\en{K} - j} - \att{\iper{K}}{f+K-\iper{K}}
&= \att{\eno{K} + j}{\en{K} - j} - \att{\eno{K} - 1}{\en{K} + 1} \\
&= \left( \att{\eno{K} -1}{\en{K} - j} + (j+1) + \gap{\eno{K}+j}{j+1} \right)
 - \left( \att{\eno{K} - 1}{\en{K} -j} + (j+1)  \right) \\
\att{\eno{K} + j}{\en{K} - j} - \att{\iper{K}}{f+K-\iper{K}}
&= \gap{\eno{K}+j}{j+1}
\end{align*}
As a result, \thr{\iper{K}} makes \thr{\eno{K} + j} fail on its attempt at
\att{\eno{K}+j}{\en{K} - j}, for all $j\in \inte{f-1-\eno{K}}$, and the next
attempt will occur at \att{\eno{K}+j}{\en{K} - j+1}.

Altogether, the next attempt after the end of the success of \thr{\iper{K}} for
\thr{\eno{K}+j} is \att{\eno{K}+j}{\en{K}-j+1}, for $j\in \inte{f-1-\eno{K}}$,
and for \thr{\en{K}+j} is \att{\en{K}+j}{\eno{K}-j+1}, for all $j \in
\inte[1]{\eno{K}}$.

Additionally, a thread will begin a new \rl, the \kth{0} \re being
at $\att{\en{K}+\eno{K}+1}{0} = \att{f+K+1}{0}$. We note that $f+K+1$ could be
higher than $\ct-1$, referring to a thread whose number is more than $\ct-1$.
Actually, if $n>\ct-1$, \att{n}{j} refers to the \kth{j} \re of
\thr{\eva{n-\ct+1}}, after its first two successes.

The two heads, \ie the two smallest indices, of $\seth{1} \cap \iper{\inte[K+1]{e-1}}$ and
$\seth{2} \cap \iper{\inte[K+1]{e-1}}$ will then compete for being successful. Indeed,
within \seth{1}, for $j\in \inte{f-1-\eno{K}}$,
\[ \att{\eno{K}+j}{\en{K}-j+1} - \att{\eno{K}}{\en{K}+1} = \gap{\eno{K}+j}{j} > 0, \]
thus if someone succeeds in \seth{1}, it will be \thr{\eno{K}}.
In the same way, for all $j \in \inte[1]{\eno{K}+1}$,
\[ \att{\en{K}+j}{\eno{K}-j+1}  - \att{\en{K}+1}{\eno{K}} = \gap{\en{K}+j}{j-1} > 0, \]
meaning that if someone succeeds in \seth{2}, it will be \thr{\en{K}+1}.

Let us compare now those two candidates:
\begin{align*}
\att{\eno{K}}{\en{K}+1} - \att{\en{K}+1}{\eno{K}}
&= \en{K}+1-f+\suc{\ct-1} + \eno{K} + \gap{\eno{K}}{\eno{K}+1}\\
& \quad \quad - \left( \eno{K} + \att{f}{0} + \en{K}+1 -f  + \gap{\en{K}+1}{\en{K}+1-f} \right)\\
&= \suc{\ct-1} -1 + \gap{\eno{K}}{\eno{K}+1}\\
& \quad \quad - \left(  \suc{\ct-1} + \gap{f}{f+1} -1  + \gap{\en{K}+1}{\en{K}+1-f} \right)\\
&= \gap{\eno{K}}{\eno{K}+1} - \left( \gap{\en{K}+1}{\en{K}+2} -1 \right)\\
\att{\eno{K}}{\en{K}+1} - \att{\en{K}+1}{\eno{K}}&= \eva{\eno{K}} - \eva{\en{K}+1}.\\
\end{align*}

By definition, \iper{K+1} is either \eno{K} or $\en{K}+1$ and corresponds to the
next successful thread.
We can follow the same line of reasoning in the case where $\thr{\iper{K}} \in \seth{2}$
and prove in this way that \pro{K+1} is true.

\pro{0} is true, and the property spreads until \pro{e-1}, where all threads of
\seth{1} and \seth{2} have been successful, in the order ruled by \iperfun, \ie
\thr{\iper{0}}, $\dots$, \thr{\iper{e-1}}. And before those successes the threads
\thr{e-1}=\thr{\iper{e-1}}, $\dots$, \thr{\ct-1} have been successful as
well. The seed composed of those successes is a \wfs. Given a thread, the gap
between this thread and the next one in the new order could indeed not be higher
than the gap in the previous order with its next thread. Also the \kth{f} order
gaps remain smaller than $1$. And as \thr{e-1} succeeds the second time after
$f$ failures, it means that the new seed $\soe''$ is such that $\fa{\soe''} =
f$.}


\end{proof}

\begin{lemma}
\label{lem.cyc.exec.2}
Let \soe be a \wifs, and $f=\fa{\left( \thr{i}, \fis{i} \right)_{i \in \inte{\ct-2}}}$. If
$\gap{f}{f+1} > 1$ and if the second success of \thr{\ct-1} occurs before the
second success of \thr{f-1}, then we can find in the execution a \wfs $\soe'$ for \ct
threads such that $\fa{\soe'} = f$.
\end{lemma}
\begin{proof}
\pp{In~\cite{our-long}, we show that in this case, the gap between two successes of
\thr{\ct-1} is $r$, hence all gaps at any order less than \ct will eventually become less
than $r$, which builds the \wfs.
}
\rr{Until the second success of \thr{\ct-1}, the execution follows the same pattern as in
  Lemma~\ref{lem.cyc.exec.3}. Actually, the case invoked in the current lemma could have
  been handled in the previous lemma, but it would have implied tricky notations, when we
  referred to \thr{\eva{n-\ct+1}}. Let us deal with this case independently then, and come
  back to the instant where \thr{\ct-1} succeeds for the second time.

We had $0 < \att{f-1}{0} - \suc{\ct-1} = \gap{f-1}{f} < 1$. For the thread \thr{\perb{j}}
to succeed at its \kth{k} \re after the first success of \thr{\ct-1} and before \thr{f-1},
it should necessary fill the following condition: $j+1 < \att{\perb{j}}{k} - \suc{\ct-1} <
j+1 + \gap{f-1}{f}$. This holds also for the second success of \thr{\ct-1}, which implies
that $P' < \suc{\ct-1} + 1 + q + r + h - \suc{\ct-1} < P' + \gap{f-1}{f}$, where $h$ is
the number of failures of \thr{\ct-1} before its second success and $P'$ is the number of
successes between the two successes of \thr{\ct-1}. As $\gap{f-1}{f}<1$, and $q$, $P'$ and
$h$ are non-negative integers, we have $r< \gap{f-1}{f}$ and $h = P' - 1 -q$.

To conclude, as any gap at any order is less than the gap between the two
successes of \thr{\ct-1}, which is $r < 1$, we found a \wfs for $P'$ threads.

Finally any other thread will eventually succeed (see
Lemma~\ref{lem.event.succ}). We can renumber the threads such that
\thr{P'} is the first thread that is not in the \wfs to succeed, and
the threads of the \wfs succeeded previously as \thr{0}, $\dots$,
\thr{P'-1}. As explained before, for all $(k,n) \in \inte{P'-1}^2$,
$\gap{n}{k} < \gap{n}{n} = r$. With the new thread, the first order
gaps are changed by decomposing \gap{0}{1} into \gap{P'}{1} and the
new \gap{0}{1}. All gaps can only be decreased, hence we have a new
\wfs for $P'+1$ threads. We repeat the process until all threads have
been encountered, and obtain in the end $\soe'$, a \wfs with \ct
threads such that $\fa{\soe'} = \ct -1 -q$, which is an optimal cyclic
execution.

Still, as \thr{f} succeeds between two successes of \thr{\ct-1} that are separated by $r$,
we had, in the initial configuration: $\gap{\ct-1}{\ct-1-f} < r$. As, in addition, we have
both $\gap{f-1}{f} <1 $ and $\gap{f}{1} <1 $, we conclude that the \lag was
initially less than $2+r$. By hypothesis, we know that $\gap{f}{f+1} >1 $, which implies
that, before the entry of the new thread, the \lag was $1+r$.
In the final execution with one more thread, the \lag is $r$ and we have one more
success in the cycle, thus $\fa{\soe'} = f$.}
\end{proof}

} 

\begin{theorem}
\label{th.cyc.exec}
Assuming $r \neq 0$, if a new thread is added to an \cyex{f}{\ct-1}, then all the
threads will eventually form either an \cyex{f}{\ct}, or an \cyex{f+1}{\ct}.
\end{theorem}
\rr{ 
\begin{proof}
According to Lemma~\ref{lem.event.succ}, the new thread will eventually
succeed.  In addition, we recall that Properties~\ref{pr.wfs.soe1} and~\ref{pr.wfs.soe2} ensure that
before the first success of the new thread, any set of $\ct -1$ consecutive
successes is a \wfs with $\ct-1$ threads. We then consider a seed (we number the
threads accordingly, and number the new thread as \thr{\ct-1}) such that the
success of the new thread occurs between the success of \thr{\ct-2} and \thr{0};
we obtain in this way a \wifs $\soe = \left( \thr{n}, \fis{n} \right)_{n \in \inte{\ct-1} \& }$. We differentiate between two cases.

Firstly, if for all $n \in \inte{\ct-1}$, $\gap{n}{f+1} < 1$, according to
Lemma~\ref{lem.cyc.exec.1}, we can find later in the execution a \wfs $\soe'$
for \ct threads such that $\fa{\soe'} = f +1$, hence we reach eventually an
\cyex{f+1}{\ct}.

Let us assume now that this condition is not fulfilled. There exists $n_0 \in
\inte{\ct-1}$ such that $\gap{n_0}{f+1} > 1$. We shift the thread numbers, such
that $n_0$ is now $f$, and we have then $\gap{f}{f+1} > 1$. Then two cases are
feasible. If the second success of \thr{\ct-1} occurs before the second success
of \thr{f-1}, then Lemma~\ref{lem.cyc.exec.3} shows that we will reach an
\cyex{f}{\ct}. Otherwise, from Lemma~\ref{lem.cyc.exec.3}, we conclude that an
\cyex{f}{\ct} will still occur.
} 
\pp{
\begin{proof}
We only give the sketch of the proof, that decomposes the
  Theorem into three Lemmas which we describe here graphically:
\begin{itemize}
\item If all gaps of \kth{(f+1)} order are less than 1, then every existing thread will
  fail once more, and the new steady-state is reached immediately. See
  Figure~\ref{fig:ex-lem.1}.
\item Otherwise
\begin{itemize}
\item Either: everyone succeeds once, whereupon a new \cyex{f}{\ct} is formed. See
  Figure~\ref{fig:ex-lem.3}.
\item Or: before everyone succeeds again, a new \cyex{f}{\ct'}, where $\ct' \leq \ct$,
  is formed, which finally leads to an \cyex{f}{\ct}. See Figure~\ref{fig:ex-lem.2}.\qedhere
\end{itemize}
\end{itemize}
}
\end{proof}

\subsubsection{Throughput Bounds}
\label{cha-secwt-thr}

\renewcommand\rr[1]{#1}
\renewcommand\pp[1]{}

\pp{
The periodicity offers an easy way to compute the expected number of
threads inside the \rl, and to bound the number of failures and the
throughput.

\begin{lemma}
\label{lem:av-thr}
In an \cyex{f}{\ct}, the throughput is $\thru = \frac{\ct}{q+r+1+f}$, and the average number of threads in the \rl
$\trl = \ct \times \frac{f+1}{q+r+f+1}$.
\end{lemma}

\begin{lemma}
\label{lem:wt-bounds}
The number of failures is bounded by $\fup\leq f\leq\flo$, and throughput
by $\tup\leq\thru\leq\tlo$, where
\begin{align*}
\fup = \left\{ \begin{array}{ll}
\ct-q-1 \quad& \text{if } q \leq \ct-1\\
0 \quad& \text{otherwise}
\end{array}\right., & \qquad
 \tup =
\left\{ \begin{array}{ll}
\frac{\ct}{\ct+r} \quad& \text{if } q \leq \ct-1\\
\frac{\ct}{q+r+1} \quad& \text{otherwise.}
\end{array}\right.\\
\flo = \left\lfloor \frac{1}{2} \left( (\ct-1-q-r) +
\sqrt{(\ct-1-q-r)^2 + 4\ct} \right) \right\rfloor, & \qquad
\tlo = \frac{\ct}{q+r+1+\flo},
\end{align*}
and those bounds are tight.
\end{lemma}





\newcommand{\wupu}{\ema{\tilde{w}}}
\newcommand{\wupf}[1]{\ema{\wupu(#1)}}
\newcommand{\wup}{\ema{\wupf{\ct}}}
\newcommand{\wupp}{\ema{\wupu'(\ct)}}

\newcommand{\hfu}{\ema{a}}
\newcommand{\hf}{\ema{\hfu(\ct)}}
\newcommand{\hfp}{\ema{\hfu'(\ct)}}

\newcommand{\hhfu}{\ema{h}}
\newcommand{\hhff}[1]{\ema{\hhfu(#1)}}
\newcommand{\hhf}{\ema{\hhff{\ct}}}
\newcommand{\hhfp}{\ema{\hhfu'(\ct)}}

\newcommand{\flofun}[1]{\ema{\left\lfloor #1 \right\rfloor}}
\newcommand{\ceifun}[1]{\ema{\left\lceil #1 \right\rceil}}

\smallskip
} 

\rr{
Firstly we calculate the expression of throughput and the expected
number of threads inside the \rl (that is needed when we gather
expansion and \cacas). Then we exhibit upper and lower bounds on both
throughput and the number of failures, and show that those bounds are
reached. Finally, we give the worst case on the number of \cacas.

\begin{lemma}
\mbox{In an \cyex{f}{\ct}, the throughput is}
\begin{equation}
\thru = \frac{\ct}{q+r+1+f}.\label{eq:cye-thr}
\end{equation}
\end{lemma}
\begin{proof}
By definition, the execution is periodic, and the period lasts $q+r+1+f$ units of time. As
\ct successes occur during this period, we end up with the claimed expression.
\end{proof}

\begin{lemma}
\label{lem:av-thr}
In an \cyex{f}{\ct}, the average number of threads \trl in the \rl
is given by
\[ \trl = \ct \times \frac{f+1}{q+r+f+1}.  \]
\end{lemma}
\begin{proof}
Within a period, each thread spends $f+1$ units of time in the \rl, among the $q+r+f+1$
units of time of the period, hence the Lemma.
\end{proof}

\bigskip

\begin{lemma}
\label{lem:wt-upp}
\mbox{The number of failures is not less than \fup, where}
\begin{equation}
\fup = \left\{ \begin{array}{ll}
\ct-q-1 \quad& \text{if } q \leq \ct-1\\
0 \quad& \text{otherwise}
\end{array}\right., \text{ and accordingly, }\qquad
 \thru \leq
\left\{ \begin{array}{ll}
\frac{\ct}{\ct+r} \quad& \text{if } q \leq \ct-1\\
\frac{\ct}{q+r+1} \quad& \text{otherwise.}
\end{array}\right. \label{eq:wt-upp}
 \end{equation}
\end{lemma}
\begin{proof}
According to Equation~\ref{eq:cye-thr}, the throughput is maximized when the number of
failures is minimized. In addition, we have two lower bounds on the number of failures:
(i) $f \geq 0$, and (ii) \ct successes should fit within a period, hence $q+1+f \geq
\ct$. Therefore, if $\ct-1-q < 0$, $\thru \leq \ct/(q+r+1+0)$, otherwise,
\[ \thru \leq \frac{\ct}{q+r+1+\ct-1-q} = \frac{\ct}{\ct+r}.\]
\end{proof}

\begin{remark}
We notice that if $q > \ct-1$, the upper bound in
Equation~\ref{eq:wt-upp} is actually the same as the immediate upper
bound described in Section~\ref{cha-secimm-bou}. However, if $q \leq
\ct-1$, Equation~\ref{eq:wt-upp} refines the immediate upper bound.
\end{remark}




\begin{lemma}
\label{lem:wt-low}
The number of failures is bounded by
\[ f \leq \flo = \left\lfloor \frac{1}{2} \left( (\ct-1-q-r) +
\sqrt{(\ct-1-q-r)^2 + 4\ct} \right) \right\rfloor, \text{ and accordingly,}\]
the throughput is bounded by
\[\thru \geq \frac{\ct}{q+r+1+\flo}. \]
\end{lemma}
\begin{proof}
We show that a necessary condition so that an \cyex{f}{\ct},
whose \lag is \lagt, exists, is $f \times (\lagt+r)
< \ct$. \rr{According to Properties~\ref{pr.wfs.soe1}
and~\ref{pr.wfs.soe2}, any set of \ct consecutive successes is a \wfs
with $\ct$ threads. Let \soe be any of them. As we have $f$ failures
before success, Theorem~\ref{th.wfs.gap} ensures that for all $n \in
\inte{\ct-1}$, $\gap{n}{f} <1$. We recall that for all $n \in \inte{\ct-1}$, we also have
$\gap{n}{\ct} = \lagt+r$.

On the one hand, we have
\begin{align*}
\sum_{n=0}^{\ct-1} \gap{n}{f} &= \sum_{n=0}^{\ct-1} \sum_{j=n-f+1}^{n} \gap{j\bmod \ct}{1}   \\
&= f \times \sum_{n=0}^{\ct-1}  \gap{n}{1} \\
\sum_{n=0}^{\ct-1} \gap{n}{f} &= f \times (\lagt+r).
\end{align*}
On the other hand, $\sum_{n=0}^{\ct-1} \gap{n}{f} < \sum_{n=0}^{\ct-1} 1 = \ct$.

Altogether, the necessary condition states that $f \times (\lagt+r) < \ct$, which can be
rewritten as $f \times (q+1+f-\ct+r) < \ct$. The proof is complete since minimizing the
throughput is equivalent to maximizing the number of failures.
\end{proof}

\begin{lemma}
\label{lem:wt-reach}
For each of the bounds defined in Lemmas~\ref{lem:wt-upp}
and~\ref{lem:wt-low}, there exists an \cyex{f}{\ct} that reaches the
bound.
\end{lemma}
\begin{proof}
According to Lemmas~\ref{lem:wt-upp} and~\ref{lem:wt-low}, if an \cyex{f}{\ct} exists, then
the number of failures is such that $\fup \leq f \leq \flo$.
\\We show now that this double
necessary condition is also sufficient. We consider $f$ such that $\fup \leq f \leq
\flo$, and build a \wfs $\soe = \left( \thr{i}, \fis{i} \right)_{i \in \inte{\ct-1}}$.

For all $n \in \inte{\ct-1}$, we define \fis{i} as}%

\[ \fis{n} = n \times \left( \frac{q+1+f-\ct+r}{\ct} + 1 \right). \]

We first show that $\fa{\soe} = f$.
By definition, $\fa{\soe} = \max \left(0, \left\lceil \fis{\ct-1}-\fis{0}-q-r \right\rceil \right)$;
we have then
\begin{align*}
\fa{\soe} &=
\max \left(0, \left\lceil (\ct -1) \times \left( \frac{q+1+f-\ct+r}{\ct} + 1 \right) -q-r \right\rceil \right)\\
&= \max \left(0, \left\lceil (\ct -1 - q -r) + (q+1+f-\ct+r) - \frac{q+1+f-\ct+r}{\ct} \right\rceil \right)\\
\fa{\soe}&= \max \left(0, \left\lceil f - \frac{q+1+f-\ct+r}{\ct} \right\rceil \right).\\
\end{align*}
Firstly, we know that $q+1+f-\ct \geq 0$, thus if $f=0$, then the second term of the
maximum is not positive, and $\fa{\soe}=0=f$. Secondly, if $f>0$, then according to
Lemma~\ref{lem:wt-upp}, $(q+1+f-\ct+r)/\ct<1/f\leq 1$. As we also have
$(q+1+f-\ct+r)/\ct\geq 0$, we conclude that $\fa{\soe} = \left\lceil f -
\frac{q+1+f-\ct+r}{\ct} \right\rceil = f$.



Additionally, for all $n \in \inte{\ct-1}$,
\begin{align*}
\gap{n}{f}  &=
\left\{ \begin{array}{ll}
\fis{n} - \fis{n-f} - f \quad& \text{if } n>f \\
\fis{n} - \fis{\ct+n-f} +1+q+r \quad& \text{otherwise} \\
\end{array}\right.\\
&=
\left\{ \begin{array}{l}
n \times \left( \frac{q+1+f-\ct+r}{\ct} + 1 \right) - (n-f) \times \left( \frac{q+1+f-\ct+r}{\ct} + 1 \right) - f\\
n \times \left( \frac{q+1+f-\ct+r}{\ct} + 1 \right) - (\ct+n-f)\times \left( \frac{q+1+f-\ct+r}{\ct} + 1 \right) +1+q+r \\
\end{array}\right.\\
&=
\left\{ \begin{array}{l}
f \times \frac{q+1+f-\ct+r}{\ct}\\
- (\ct-f) - (q+1+f-\ct+r) + f \times \frac{q+1+f-\ct+r}{\ct} +1+q+r\\
\end{array}\right.\\
\gap{n}{f} &=
f \times \frac{w+r}{\ct}\\
\end{align*}
As $w \leq 0$ and $f\leq 0$, $\gap{n}{f} > 0$. Since $f \leq \flo$, $\gap{n}{f} < 1$.
Theorem~\ref{th.wfs.gap} implies that \soe is a \wfs that leads to an \cyex{f}{\ct}.

We have shown that for all $f$ such that $\fup \leq f \leq \flo$ there exists an
\cyex{f}{\ct}; in particular there exist an \cyex{\flo}{\ct} and an \cyex{\fup}{\ct}.
\end{proof}


\newcommand{\wupu}{\ema{\tilde{w}}}
\newcommand{\wupf}[1]{\ema{\wupu(#1)}}
\newcommand{\wup}{\ema{\wupf{\ct}}}
\newcommand{\wupp}{\ema{\wupu'(\ct)}}

\newcommand{\hfu}{\ema{a}}
\newcommand{\hf}{\ema{\hfu(\ct)}}
\newcommand{\hfp}{\ema{\hfu'(\ct)}}

\newcommand{\hhfu}{\ema{h}}
\newcommand{\hhff}[1]{\ema{\hhfu(#1)}}
\newcommand{\hhf}{\ema{\hhff{\ct}}}
\newcommand{\hhfp}{\ema{\hhfu'(\ct)}}

\newcommand{\flofun}[1]{\ema{\left\lfloor #1 \right\rfloor}}
\newcommand{\ceifun}[1]{\ema{\left\lceil #1 \right\rceil}}

\smallskip

\begin{corollary}
\label{th:compar}
The highest possible number of wasted repetitions is $\left\lceil\sqrt{\ct}-1\right\rceil$
and is achieved when $\ct=q+1$.
\end{corollary}
\begin{proof}


The highest possible number of wasted repetitions \wup with \ct threads is given by
\[ \wup = \flo-\fup = \flofun{\frac{1}{2} \left( - \hf + \sqrt{\hf^2 + 4\ct} \right) -\fup} .\]

Let \hfu and \hhfu be the functions respectively defined as $\hf = q+1-\ct+r$, which
implies $\hfp=-1$, and $\hhf = (- \hf + \sqrt{\hf^2 + 4\ct})/2 - \fup$, so that
$\wup = \flofun{\hhf}$.

Let us first assume that $\hf>0$. In this case, $q\leq \ct-1$, hence $\fup = 0$.
We have
\begin{align*}
2\hhfp &= 1 + \frac{-2\hf + 4}{2 \sqrt{\hf^2 + 4\ct}}\\
2\hhfp&= 2 \times \frac{2 - \hf + \sqrt{\hf^2 + 4\ct}}{2 \sqrt{\hf^2 + 4\ct}}
\end{align*}
Therefore, \hhfp is negative if and only if $\sqrt{\hf^2 + 4\ct} < \hf -2$. It cannot be
true if $\hf<2$. If $\hf\geq 2$, then the previous inequality is equivalent to $\hf^2 +
4\ct < \hf^2 - 4\hf +4$, which can be rewritten in $q+1+r<1$, which is absurd. We have
shown that \hhfu is increasing in $]0,q+1]$.

Let us now assume that $\hf\leq0$. In this case, $q > \ct-1$, hence $\fup = \ct -q -1$,
and $\hhfp = \left( \hf + \sqrt{\hf^2 + 4\ct} \right) /2 - r$. Assuming \hhfp to be
positive leads to the same absurd inequality $q+1+r<1$, which proves that \hhfu is
decreasing on $[q+2,+\infty[$.

Also, the maximum number of wasted repetitions is achieved as $\ct=q+1$ or $\ct=q+2$.
Since
\[ \hhff{q+1} = \frac{1}{2} \left( -r + \sqrt{r^2 + 4\ct} \right) >
\frac{1}{2} \left(-(r+1) + \sqrt{r^2 + 4\ct}\right) = \hhff{q+2}, \]
the maximum number of wasted repetitions is \wupf{q+1}.
In addition,
\begin{equation*}
\bgroup
\def\arraystretch{1.7}%
\begin{array}{rcccl}
\dfrac{1}{2} \left( -r + \sqrt{4\ct} \right) &<& \hhff{q+1} &<&
\dfrac{1}{2} \left( -r + \sqrt{r^2} + \sqrt{4\ct} \right)\\
\sqrt{\ct}-\dfrac{r}{2} &<& \hhff{q+1} &<&
 \sqrt{\ct}\\
\sqrt{\ct}-1 &\leq& \hhff{q+1} &<&
 \sqrt{\ct}\\
\end{array}
\egroup
\end{equation*}
We conclude that the maximum number of wasted repetitions is \ceifun{\sqrt{\ct}-1}.
\end{proof}
}

\renewcommand\rr[1]{}
\renewcommand\pp[1]{#1}

\subsection{Expansion and Complete Throughput Estimation}
\label{cha-secexp-glue}


\subsubsection{Expansion}
\label{cha-secexp}

Interference of threads does not only lead to \logcons but also to
\hardcons which impact the performance significantly.

We model the behavior of the cache coherency protocols which determine
the interaction of overlapping \rf{}s and \cas{}s.  By taking
MESIF~\cite{mesif} as basis, we come up with the following
assumptions.  When executing an atomic \cas, the core gets the cache
line in exclusive state and does not forward it to any other
requesting core until the instruction is retired. Therefore, requests
stall for the release of the cache line which implies
serialization. On the other hand, ongoing \rf{}s can overlap with
other operations. As a result, a \cas introduces expansion only to
overlapping \rf and \cas operations that start after it, as
illustrated in Figure~\ref{fig:expansion}.  As a remark, we ignore
memory bandwidth issues which are negligible for our study.

Furthermore, we assume that \rf{}s that are executed just after a \cas
do not lead to expansion (as the thread already owns of the data),
which takes effect at the beginning of a \re following a failing
attempt.  Thus, read expansions need only to be considered before
the \kth{0} \re. In this sense, read expansion can be moved to
parallel section and calculated in the same way as \cas expansion is
calculated.

To estimate expansion, we consider the delay that a thread can
introduce, provided that there is already a given number of threads in
the \rl.  The starting point of each \cas is a random variable which
is distributed uniformly within an expanded \re. The cost
function \shiftf provides the amount of delay that the additional
thread introduces, depending on the point where the starting point of
its \cas hits. By using this cost function we can formulate the
expansion increase that each new thread introduces and derive the
differential equation below to calculate the expansion of a \cas.


\begin{lemma}
\label{lem.1}
The expansion of a \cas operation is the solution of the following system of equations:
\[ \left\{
\begin{array}{lcl}
\expansionp{\trl} &=& \fcas \times \dfrac{\frac{\fcas}{2} + \expansion{\trl}}{ \mem +  \calrl + \scas + \expansion{\trl}}\\
\expansion{\trlo} &=& 0
\end{array} \right., \text{ where \trlo is the point where expansion begins.}
\]
\end{lemma}
\begin{proof}

We compute $\expansion{\trl + h}$, where $h\leq1$, by assuming that there are
already \trl threads in the \rl, and that a new thread attempts
to \cas during the \re, within a probability $h$.
\pp{\begin{multline*}
\expansion{\trl + h}= \expansion{\trl} + h\times
\rint{0}{\rlsiz}{\frac {\shift{t}}{\rlsiz}}{t}\\
  = \expansion{\trl}
     + \Big( \rint{0}{\mem+\calrl-\fcas}{\frac{\shift{t}}{\rlsiz}}{t}
     + \rint{\mem+\calrl-\fcas}{\mem+\calrl}{\frac{\shift{t}}{\rlsiz}}{t}
    + \rint{\mem+\calrl}{\mem+\calrl+\expansion{\trl}}{\frac{\shift{t}}{\rlsiz}}{t}
  + \rint{\mem+\calrl+\expansion{\trl}}{\rlsiz}{\frac{\shift{t}}{\rlsiz}}{t}\Big) h\\
 = \expansion{\trl} + \Big(
    \rint{\mem+\calrl-\fcas}{\mem+\calrl}{\frac{t}{\rlsiz}}{t}
      + \rint{\mem+\calrl}{\mem+\calrl+\expansion{\trl}}{\frac{\fcas}{\rlsiz}}{t} \Big) h
 = \expansion{\trl} + h \times \frac{ \frac{\fcas^2}{2} + \expansion{\trl}\times\fcas}{\rlsiz} .
\end{multline*}}
\rr{\begin{flalign*}
&\expansion{\trl + h}\\
=& \expansion{\trl} + h\times
\rint{0}{\rlsiz}{\frac {\shift{t}}{\rlsiz}}{t} \\
  =& \expansion{\trl}
     + h \times \Big( \rint{0}{\mem+\calrl-\fcas}{\frac{\shift{t}}{\rlsiz}}{t}\\
   & \quad  +
      \rint{\mem+\calrl-\fcas}{\mem+\calrl}{\frac{\shift{t}}{\rlsiz}}{t}\\
    & \quad  + \rint{\mem+\calrl}{\mem+\calrl+\expansion{\trl}}{\frac{\shift{t}}{\rlsiz}}{t}\\
 & \quad + \rint{\mem+\calrl+\expansion{\trl}}{\rlsiz}{\frac{\shift{t}}{\rlsiz}}{t}\Big)\\
  =& \expansion{\trl} + h \times \Big(
    \rint{\mem+\calrl-\fcas}{\mem+\calrl}{\frac{t}{\rlsiz}}{t}\\
    & \quad  + \rint{\mem+\calrl}{\mem+\calrl+\expansion{\trl}}{\frac{\fcas}{\rlsiz}}{t} \Big)\\
 =& \expansion{\trl} + h \times \frac{ \frac{\fcas^2}{2} + \expansion{\trl}\times\fcas}{\rlsiz}
\end{flalign*}}
This leads to
$\displaystyle \quad\frac{\expansion{\trl + h}- \expansion{\trl}}{ h} = \frac{ \frac{\fcas^2}{2} + \expansion{\trl}\times\fcas}{\rlsiz}$.
When making $h$ tend to $0$, we finally obtain
\[ \expansionp{\trl} = \fcas \times \frac{\frac{\fcas}{2} + \expansion{\trl}}{ \mem +  \calrl + \scas + \expansion{\trl}}. \pp{\qedhere}\]
\end{proof}

\subsubsection{Throughput Estimate}
\label{cha-secthr}

\newcommand{\antm}[1]{\ema{h_{#1}(\trl)}}

\newcommand{\antuf}{\ema{h^{\exppl}}}
\newcommand{\antlf}{\ema{h^{\expmi}}}

\newcommand{\antuo}[1]{\ema{h^{\exppl}(#1)}}
\newcommand{\antlo}[1]{\ema{h^{\expmi}(#1)}}

\newcommand{\antu}{\ema{\antuf(\trl)}}
\newcommand{\antl}{\ema{\antlf(\trl)}}

\newcommand{\trlu}{\ema{\trl^{\exppl}}}
\newcommand{\trll}{\ema{\trl^{\expmi}}}

There remains to combine hardware and \logcons in order to obtain the
final upper and lower bounds on throughput.
We are given as an input an expected number of threads \trl inside
the \rl. We firstly compute the expansion accordingly, by solving
numerically the differential equation of Lemma~\ref{lem.1}.
As explained in the previous subsection, we have $\psiz = \pw
+ \expa$, and $\rlsiz = \rc + \cw + \expa + \cc$.
We can then compute $q$ and $r$, that are the inputs (together with
the total number of threads \ct) of the method described in
Section~\ref{cha-secwt}. Assuming that the initialization times of the
threads are spaced enough, the execution will superimpose
an \cyex{f}{\ct}. Thanks to Lemma~\ref{lem:av-thr}, we can compute the
average number of threads inside the \rl, that we note by \antm{f}.
A posteriori, the solution is consistent if this average number of
threads inside the \rl
\antm{f} is equal to the expected number of threads \trl that has been given as an input.

Several \cyexs{f}{\ct} belong to the domain of the possible outcomes,
but we are interested in upper and lower bounds on the number of
failures $f$. We can compute them through Lemmas~\ref{lem:wt-upp}
and~\ref{lem:wt-low}, along with their corresponding throughput and
average number of threads inside the \rl. We note by \antu and \antl
the average number of threads for the lowest number of failures and
highest one, respectively. Our aim is finally to find \trll and \trlu,
such that $\antuo{\trlu} = \trlu$ and $\antlo{\trll} = \trll$. If
several solutions exist, then we want to keep the smallest, since
the \rl stops to expand when a stable state is reached.

Note that we also need to provide the point where the expansion
begins. It begins when we start to have failures, while reducing
the \ps. Thus this point is $(2 \ct -1 ) \rlwp$ (resp. $(\ct
-1) \rlwp$) for the lower (resp. upper) bound on the throughput.

\begin{theorem}
Let $(x_n)$ be the sequence defined recursively by $x_0=0$ and $x_{n+1} = \antuo{x_n}$. If
$\pw\geq\rc+\cw+\cc$, then
\rr{\[ \trlu = \lim_{n \rightarrow +\infty} x_n. \]}
\pp{$ \trlu = \lim_{n \rightarrow +\infty} x_n$. }
\end{theorem}
\begin{proof}
First of all, the average number of threads belongs to $]0,\ct[$, thus for all $x \in
  [0,\ct]$, $0<\antuo{x}<\ct$. In particular, we have $\antuo{0}>0$, and
  $\antuo{\ct}<\ct$, which proves that there exist one fixed point for \antuf.

In addition, we show that \antuf is a non-decreasing function. According to Lemma~\ref{lem:av-thr},
\[ \antu = \ct \times \frac{1+\fup}{q+r+\fup+1}, \]
where all variables except \ct depend actually on \trl. We have
\[ q=\flofun{\frac{\pw + \expa}{\rlwp + \expa}} \text{  and  }
r = \frac{\pw + \expa}{\rlwp + \expa} - q,\]
hence, if $\pw\geq\rlwp$, $q$ and $r$ are non-increasing as \expa is non-decreasing,
which is non-decreasing with \trl. Since \fup is non-decreasing as a function of $q$, we
have shown that if $\pw\geq\rlwp$, \antuf is a non-decreasing function.

Finally, the proof is completed by the theorem of Knaster-Tarski.
\end{proof}
The same line of reasoning holds for \antlf as well.
As a remark, we point out that when $\pw<\rlwp$, we scan the interval
of solution, and have no guarantees about the fact that the solution
is the smallest one; still it corresponds to very extreme cases.

\renewcommand\rr[1]{#1}
\renewcommand\pp[1]{}

\rr{
\subsubsection{Several \RLs}
\label{cha-secsev-rl}

We consider here a lock-free algorithm that, instead of being a loop
over one \ps and one \rl, is composed of a loop over a sequence of
alternating \pss and \rls.  We show that this algorithm is equivalent
to an algorithm with only one \ps and one \rl, by proving the
intuition that the longest \rl is the only one that fails and hence
expands.

\paragraph{Problem Formulation}

\newcommand{\maxs}{\ema{S}}

\begin{figure}[b!]
\removelatexerror
\begin{procedure}[H]
\SetKwData{pet}{execution\_time}
\SetKwData{pdo}{done}
\SetKwData{psucc}{success}
\SetKwData{pcur}{current}
\SetKwData{pnew}{new}
\SetKwData{pacp}{AP}
\SetKwData{pttot}{t}
\SetKwFunction{pinit}{Initialization}
\SetKwFunction{ppw}{Parallel\_Work}
\SetKwFunction{pcw}{Critical\_Work}
\SetKwFunction{pread}{Read}
\SetKwFunction{pcas}{CAS}
\SetKwData{pmaxs}{S}

\SetAlgoLined
\pinit{}\;
\While{$\mathit{not}(\pdo)$}{
\For{i $\leftarrow$ 1 \KwTo \pmaxs}{
\ppw{i}\;\nllabel{alg:lis-ps}
\While{$\mathit{not}(\psucc)$}{\nllabel{alg:lis-bcs}
\pcur $\leftarrow$ \pread{$\pacp[i]$}\;
\pnew $\leftarrow$ \pcw{i,\pcur}\;
\psucc $\leftarrow$ \pcas{\pacp, \pcur, \pnew}\;
}
}\nllabel{alg:lis-ecs}
}
\caption{Combined()\label{alg:gen-name-s}}
\end{procedure}
\caption{Thread procedure with several \rls\label{alg:gen-nb-s}}
\end{figure}

In this subsection, we consider an execution such that each spawned thread runs
Procedure~\ref{alg:gen-name-s} in Figure~\ref{alg:gen-nb-s}. Each thread executes a linear
combination of \maxs independent \rls, \ie operating on separate variables, interleaved
with \pss. We note now as $\rlsiz_i$ and $\psiz_i$ the size of a \re of the \kth{i} \rl and the size of
the \kth{i} \ps, respectively, for each $i \in \inte[1]{\maxs}$. As previously, $q_i$ and
$r_i$ are defined such that $\psiz_i = (q_i + r_i) \times \rlsiz_i$, where $q_i$ is a
non-negative integer and $r_i$ is smaller than $1$.

The Procedure~\ref{alg:gen-name-s} executes the \rls and \pss in a cyclic fashion, so we
can normalize the writing of this procedure by assuming that a \re of the \kst{1} \rl is the
longest one. More precisely, we consider the initial algorithm, and we define $i_0$ as
\[ i_0 = \min \operatorname{argmax}_{i \in \inte[1]{\maxs}}  \rlsiz_i. \]
We then renumber the \rls such that the new ordering is $i_0, \dots, \maxs, 1, \dots,
i_0-1$, and we add in \pinit the first \pss and \rls on access points from $1$ to $i_0$
--- according to the initial ordering.


One success at the system level is defined as one success of the last \cas, and the
throughput is defined accordingly. We note that in steady-state, all \rls have the same
throughput, so the throughput can be computed from the throughput of the \kst{1} \rl
instead.

\paragraph{Wasted \REs}

\newcommand{\seqt}[1]{\ema{t_{#1}}}
\newcommand{\seqs}[1]{\ema{s_{#1}}}

\begin{lemma}
\label{lem:no-fail}
Unsuccessful \rls can only occur in the \kst{1} \rl.
\end{lemma}
\begin{proof}

We note $(\seqt{n})_{n \in [1,+\infty[}$ the sequence of the thread numbers that succeeds
    in the \kst{1} \rl, and $(\seqs{n})_{n \in [1,+\infty[}$ the sequence of the
        corresponding time where they exit the \rl. We notice that by construction, for
        all $n \in [1,+\infty[$, $\seqs{n}<\seqs{n+1}$. Let, for $i \in \inte[2]{\maxs}$
            and $n \in [1,+\infty[$, \pro{i,n} be the following property: for all $i' \in
                \inte[2]{i}$, and for all $n' \in \inte[1]{n}$, the thread \thr{\seqt{n'}}
                succeeds in the \kth{i} \rl at its first attempt.

We assume that for a given $(i,n)$, \pro{i+1,n} and \pro{i,n+1} is true, and show that
\pro{i+1,n+1} is true. As the threads \thr{\seqt{n}} and \thr{\seqt{n+1}} do not have any
failure in the first $i$ \rls, their entrance time in the \kth{i+1} \rl is given by
\[ \seqs{n} + \sum_{i'=1}^{i} ( \rlsiz_{i'} + \psiz_{i'} ) + \psiz_{i+1} = X_1 \text{  and  }
\seqs{n+1} + \sum_{i'=1}^{i} ( \rlsiz_{i'} + \psiz_{i'} ) + \psiz_{i+1} = X_2, \]
respectively. Thread \thr{\seqt{n}} does not fail in the \kth{i+1} \rl, hence exits at
\[X_1 + \rlsiz_{i+1} < X_1 + \rlsiz_1 = \seqs{n} + X_2 - \seqs{n+1} < X_2. \]
As the previous threads $\thr{n-1}, \dots, \thr{1}$ exits the \kth{i} \rl before \thr{n},
and next threads \thr{n'}, where $n'>n+1$, enters this \rl after \thr{n+1}, this implies
that the thread \thr{\seqt{n+1}} succeeds in the \kth{i+1} \rl at its first attempt, and
\pro{i+1,n+1} is true.

Regarding the first thread that succeeds in the first \rl, we know that he successes in
any \rl since there is no other thread to compete with. Therefore, for all $i \in
\inte[2]{\maxs}$, \pro{i,1} is true. Then we show by induction that all \pro{2,n} is true,
then all \pro{3,n}, \etc, until all \pro{\maxs,n}, which concludes the proof.
\end{proof}

\begin{theorem}
\label{th:mult-eq}
The multi-\rl Procedure~\ref{alg:gen-name-s} is equivalent to the
Procedure~\ref{alg:gen-name}, where
\[ \psiz = \psiz_1 + \sum_{i=2}^\maxs \left( \psiz_i + \rlsiz_i \right) \quad \text{and}\quad
\rlsiz = \rlsiz_1.\]
\end{theorem}
\begin{proof}
According to Lemma~\ref{lem:no-fail} there is no failure in other \rl than the first one;
therefore, all \rls have a constant duration, and can thus be considered as \pss.
\end{proof}

\paragraph{Expansion}
\label{cha-secsev-exp}

The expansion in the \rl starts as threads fail inside this \rl. When threads are launched,
there is no expansion, and Lemma~\ref{lem:no-fail} implies that if threads fail, it
should be inside the first \rl, because it is the longest one. As a result, there will be
some stall time in the memory accesses of this first \rl, \ie expansion, and it will get
even longer. Failures will thus still occur in the first \rl: there is a positive feedback
on the expansion of the first \rl that keeps this first \rl as the longest one among all
\rls. Therefore, in accordance to Theorem~\ref{th:mult-eq}, we can compute the expansion
by considering the equivalent single-\rl procedure described in the theorem.

} 

\subsection{Experimental Evaluation}
\label{cha-secxp}


We validate our model and analysis framework through a set of successive steps, from
synthetic tests, capturing a wide range of possible
abstract algorithmic designs,
to several reference implementations of
extensively studied lock-free \ds
designs that include cases with non constant \ps and \rl.
\pp{The complete results can be found in the appendix.}

\pp{\vspace*{-.2cm}}\subsubsection{Setting}

\rr{
\paragraph{Single \rl}
\label{cha-secsynt-rls}
}
\label{cha-secsynt-rl}

\rr{\begin{figure}[t!]}
\pp{\begin{wrapfigure}{l}{.5\textwidth}}
\begin{center}
\rr{\includegraphics[width=.8\textwidth]{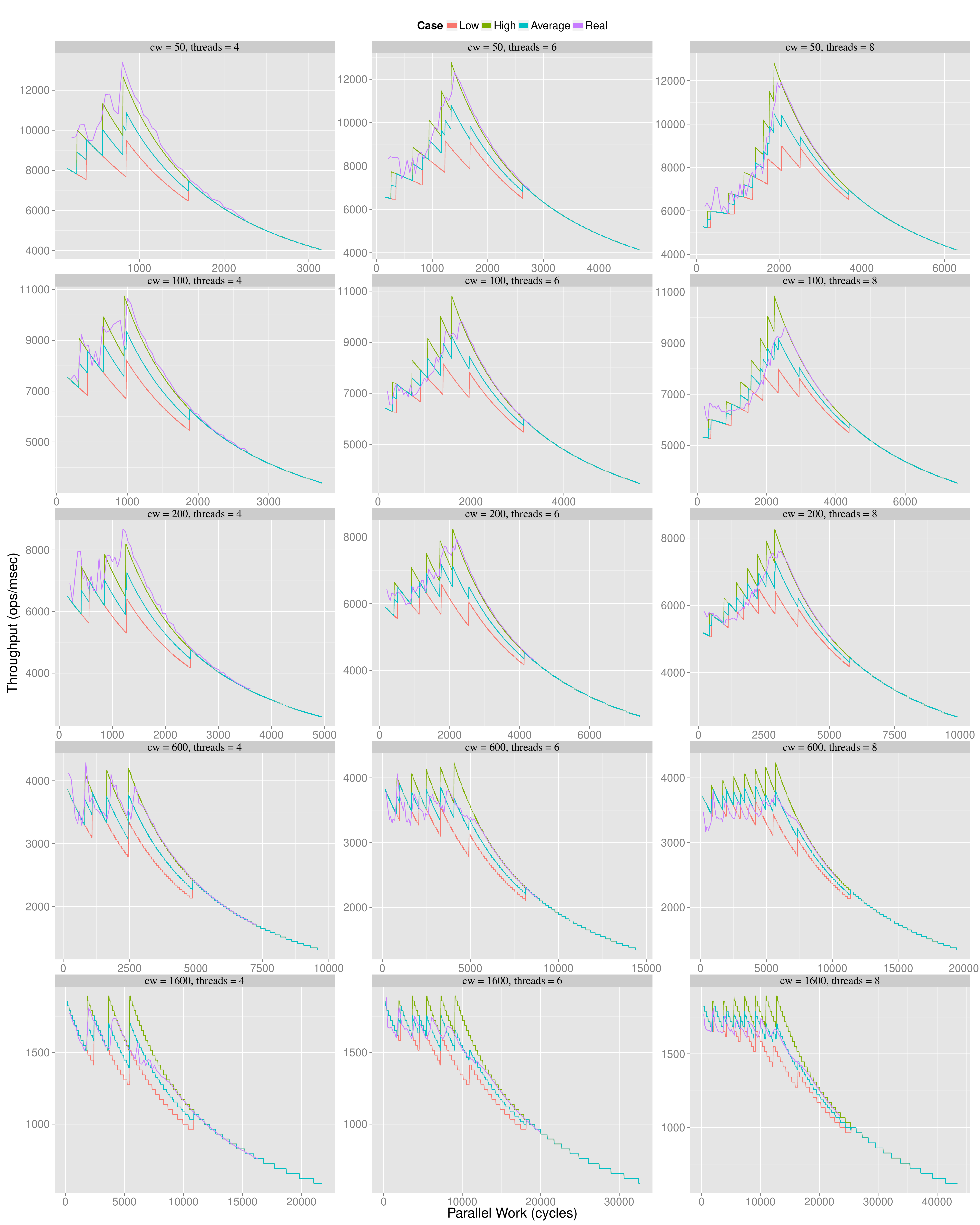}}
\pp{\vspace*{.05cm}\fuckspaa\includegraphics[width=.5\textwidth]{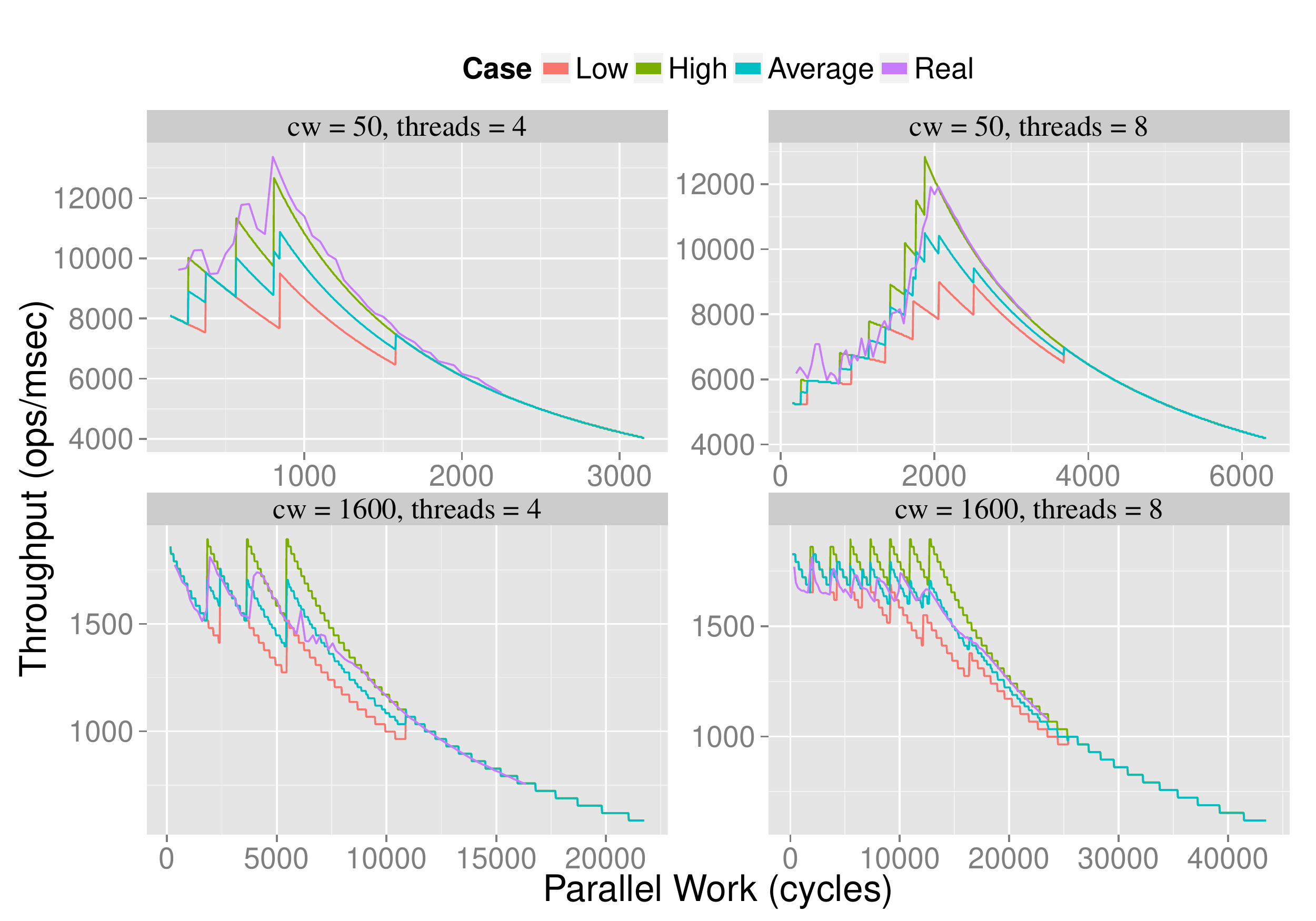}\vspace*{-.27cm}}
\end{center}
\caption{Synthetic program\label{fig:synt_smart}}
\pp{\end{wrapfigure}}
\rr{\end{figure}}


We have conducted experiments on an Intel ccNUMA workstation system. The system
is composed of two sockets, that is equipped with Intel Xeon E5-2687W v2
CPUs\pp{.}\rr{ with frequency band \ghz{1.2\text{-}3.4}. The physical cores
have private L1, L2 caches and they share an L3 cache, which is \megb{25}.}
In a socket, the ring interconnect provides L3 cache accesses and core-to-core
communication. \rr{Due to the bi-directionality of the ring
interconnect, uncontended latencies for intra-socket communication between cores
do not show significant variability.}

\pp{ Threads are pinned to a single socket to minimize non-uniformity in \rf and \cas latencies.
  Due to the bi-directionality of the ring that interconnects L3
 caches, uncontended latencies for intra-socket communication between cores do not
 show significant variability.}
\pp{The methodology in~\cite{david-emp-atom} is used to
  measure the \cas and \rf latencies, while the work inside the \ps is
  implemented by a for-loop of {\it Pause} instructions.}
\rr{Our model assumes uniformity in the \cas and \rf latencies on the shared
cache line. Thus, threads
are pinned to a single socket to minimize non-uniformity in \rf and \cas latencies.
In the experiments, we vary the number of threads between 4 and 8
since the maximum number of threads that can be used in the experiments are
bounded by the number of physical cores that reside in one socket.}

In all figures, y-axis provides the throughput,\rr{ which is the
  number of successful operations completed per millisecond. Parallel
  work}\pp{ while the parallel work} is represented in x-axis in
cycles. As mentioned in Section~\ref{cha-secwt}, the graphs contain the
high and low estimates, corresponding to the lower and upper bound on the \cacas,
respectively,
and an additional curve that shows the average of them.


\rr{
As mentioned before, the latencies of \cas and \rf are parameters of our model. We used
the methodology described in~\cite{david-emp-atom} to measure
latencies of these operations in a benchmark program by using two threads that
are pinned to the same socket. The aim is to bring the cache line into the
state used in our model. Our assumption is that the \rf is
conducted on an invalid line. For \cas, the state of the cache line could be
exclusive, forward, shared or invalid. Regardless of the state of the cache line,
\cas requests it for ownership, that compels invalidation
in other cores, which in turn incurs a two-way communication and a memory fence afterwards
to assure atomicity. Thus, the latency of \cas does not show negligible variability
with respect to the state of the cache line, as also revealed in our latency benchmarks.

As for the computation cost, the work inside the \ps is implemented by
a dummy for-loop of {\it Pause} instructions.

}

\newcommand\fuckspaa{}

\pp{\vspace{-.3cm}}\subsubsection{Synthetic Tests}
\label{cha-secsynt}

For the evaluation of our model, we first create synthetic tests that emulate different
design patterns of lock-free data structures (value of \cw) and different application
contexts (value of \pw).
\rr{As described in the previous subsection, in the
  Procedure~\ref{alg:gen-name}, the amount of work in both the \ps and
  the \rl are implemented as dummy loops, whose costs are adjusted
  through the number of iterations in the loop.}

Generally speaking, in Figure~\ref{fig:synt_smart}, we observe two main behaviors: when \pw is high,
the \ds is not contended, and threads can operate without
failure. When \pw is low, the \ds is contended, and depending on the
size of \cw (that drives the expansion)  a steep decrease
in throughput or just a roughly constant bound on the performance is observed.

The position of the experimental curve between the high and low
estimates, depends on \calrl. It can be observed that the experimental
curve mostly tends upwards as \calrl gets smaller, possibly because
the serialization of the \cas{}s helps the synchronization of the
threads.

\posrem{For the cases with considerable expansion, it is expected to have
unfairness among threads.
This fact loosens the validity of our deterministic model that assumes
uniformity and presumably leads to underestimation of throughput.}{}

Another interesting fact is the waves appearing on the experimental
curve, especially when the number of threads is low or the critical work
big. This behavior is originating because of the variation of $r$
with the change of parallel work, a fact that is captured by our analysis.

\rr{
\paragraph{Several \rls}
\label{cha-secsynt-rls}

\begin{figure}[h!]
\begin{center}
\includegraphics[width=.95\textwidth]{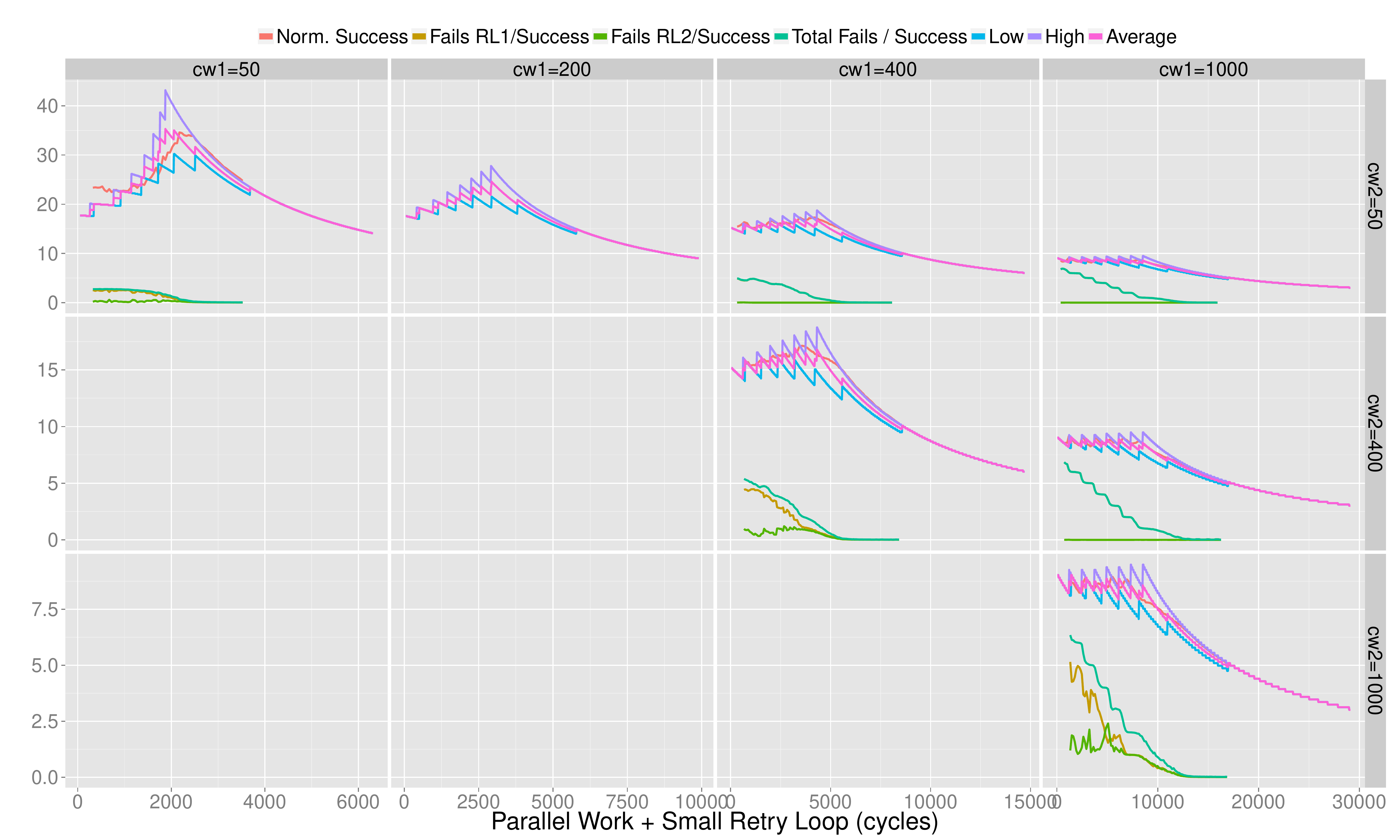}
\end{center}
\caption{Multiple \rls with 8 threads}
\label{fig:multiple_retry}
\end{figure}

We have created experiments by combining several \rls, each operating on an independent variable
which is aligned to a cache line. In Figure~\ref{fig:multiple_retry}, results are compared
with the model for single \rl case where the single \rl is equal to the longest \rl, while
the other \rls are part of the \ps.
The distribution of fails in the \rls are illustrated and all throughput curves are normalized
with a factor of 175 (to be easily seen in the same graph). Fails per success values are
not normalized and a success is obtained after completing all \rls.
}

\subsubsection{Treiber's Stack}
\label{cha-sectrei}

\rr{\begin{figure}[t!]}
\pp{\begin{wrapfigure}{l}{.5\textwidth}}
\begin{center}
\rr{\includegraphics[width=.85\textwidth]{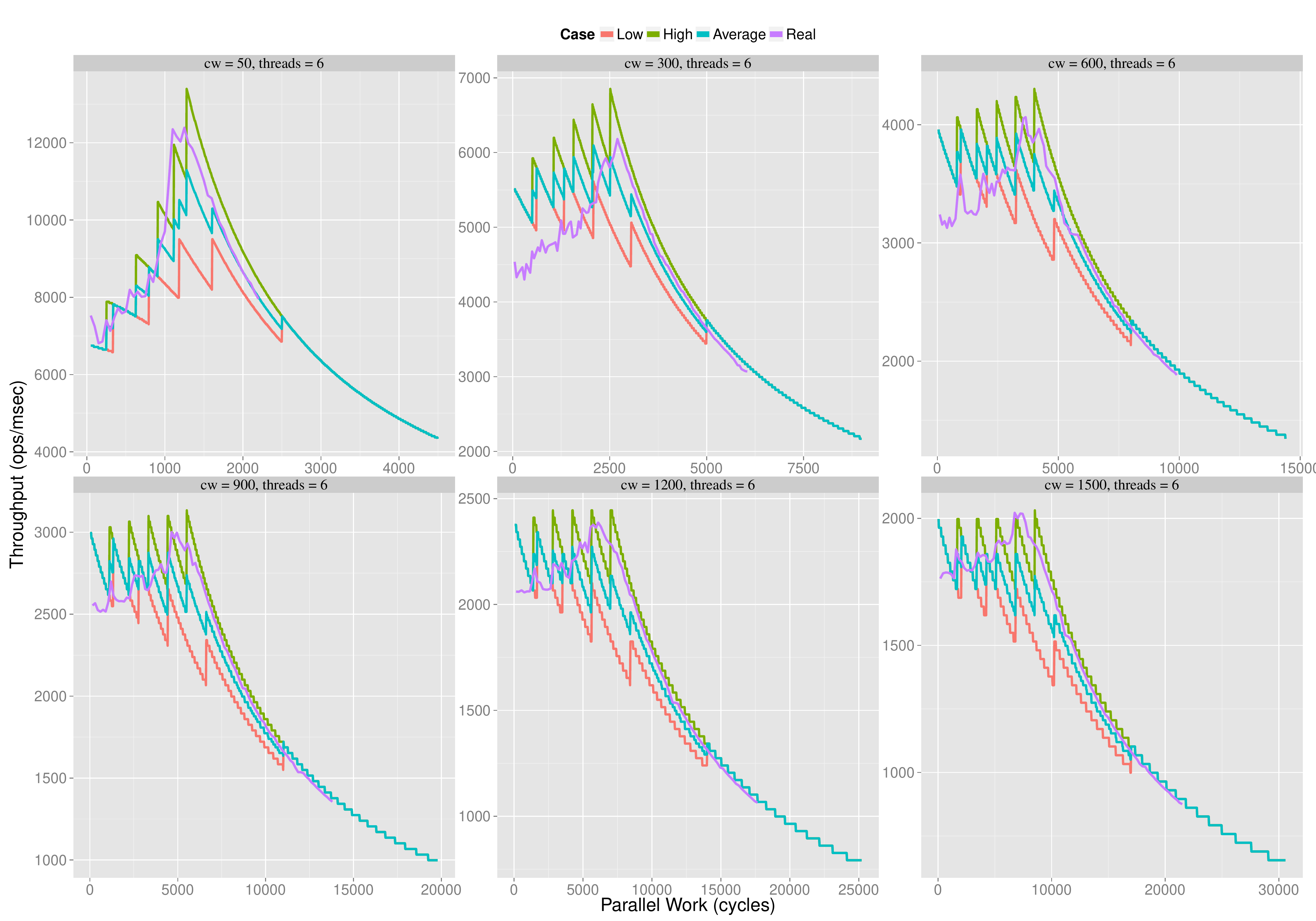}}
\pp{\fuckspaa\includegraphics[width=.5\textwidth]{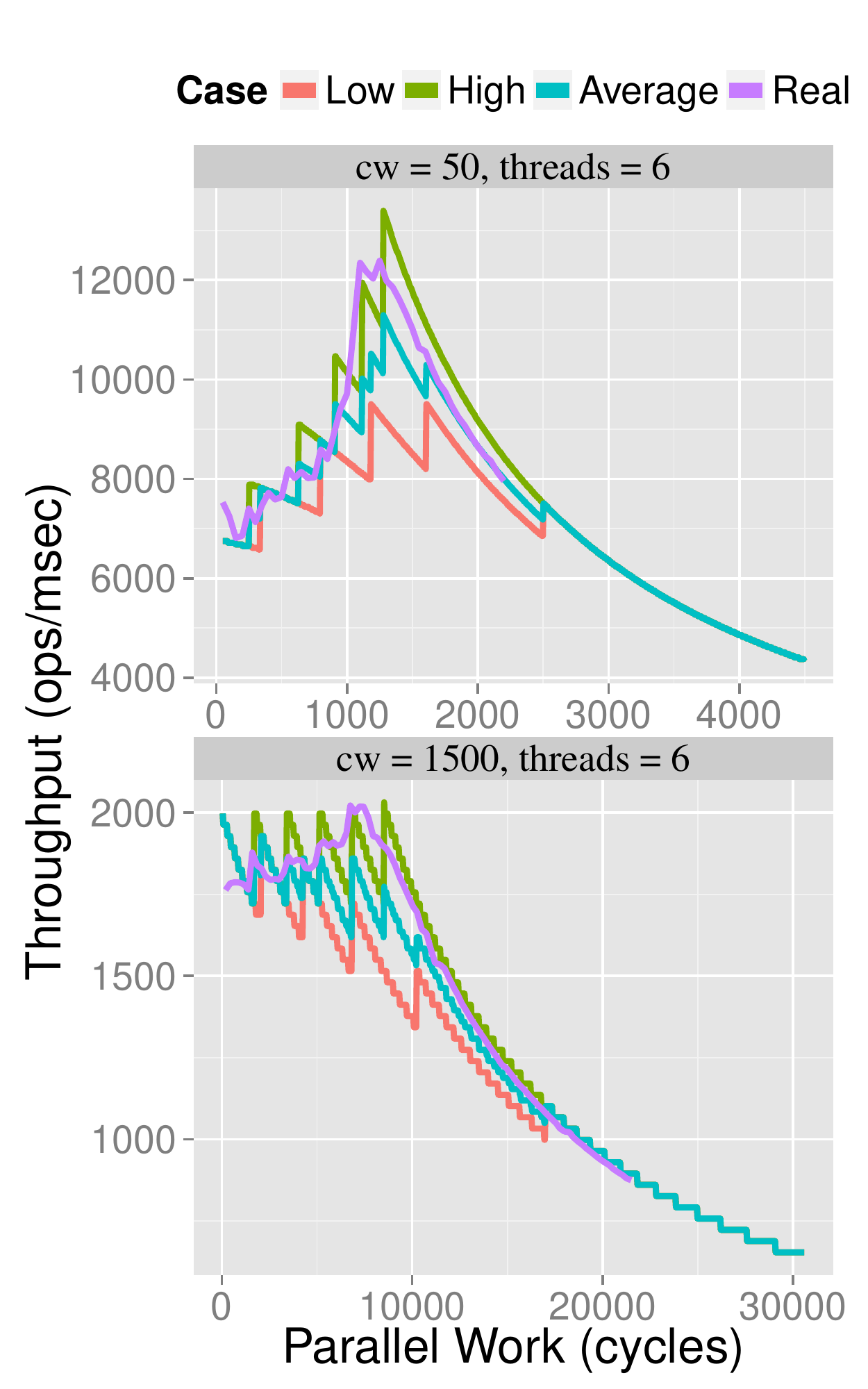}\vspace*{-.27cm}}
\end{center}
\caption{\popop on Treiber's stack}
\label{fig:treiber}
\pp{\end{wrapfigure}}
\rr{\end{figure}}

\pp{ The lock-free stack by Treiber~\cite{lf-stack} is typically the
  first example that is used to validate a freshly-built model on
  lock-free \dss. A \popop contains a \rl that first reads the top
  pointer and gets the next pointer of the element to obtain the
  address of the second element in the stack, before attempting to
  \cas with the address of the second element. The access to the next
  pointer of the first element occurs between the \rf and the
  \cas. Thus, it represents the work in \calrl. By varying the
  number of elements that are popped at the same time, and the
  cache misses implied by the reads, 
  we vary the  work in \calrl and obtain the results depicted in
  Figure~\ref{fig:treiber}.}

\rr{The lock-free stack by Treiber~\cite{lf-stack} is one of the most
  studied efficient \dss.
\popop and \pushop both contain a \rl, such that each \re starts with
a \rf and ends with \cas on the shared top pointer. In order to
validate our model, we start by using \popop{}s. From a stack
which is initiated with 50 million elements, threads continuously pop
elements for a given amount of time. We count the total number of pop
operations per millisecond. Each \popop first reads the top
pointer and gets the next pointer of the element to obtain the address
of the second element in the stack, before attempting to \cas with the
address of the second element. The access to the next pointer of the
first element occurs in between the \rf and the \cas. Thus, it
represents the work in \calrl. This memory access can possibly
introduce a costly cache miss depending on the locality of the popped
element.

To validate our model with different \calrl values, we make use of
this costly cache miss possibility. We allocate a contiguous chunk of
memory and align each element to a cache line. Then, we initialize the
stack by pushing elements from contiguous memory either with a single
or large stride to disable the prefetcher.  When we measure the
latency of \calrl in \popop for single and large stride cases, we
obtain the values that are approximately 50 and 300 cycles,
respectively.  As a remark, 300 cycles is the cost of an L3 miss in
our system when it is serviced from the local main memory module. To
create more test cases with larger \calrl, we extended the stack
implementation to pop multiple elements with a single operation.
Thus, each access to the next element could introduce an additional L3
cache miss while popping multiple elements. By doing so, we created
cases in which each thread pops 2, 3, \etc elements, and \calrl goes to
600, 900, \etc cycles, respectively. In Figure~\ref{fig:treiber},
comparison of the experimental results from Treiber's stack and our
model is provided.

As a remark, we did not implemented memory reclamation for our
experiments but one can implement a stack that allows pop and push of
multiple elements with small modifications using hazard
pointers~\cite{hazard}.  Pushing can be implemented in the same way as
single element case.  A \popop requires some modifications for memory
reclamation. It can be implemented by making use of hazard pointers
just by adding the address of the next element to the hazard list
before jumping to it. Also, the validity of top pointer should be
checked after adding the pointer to the hazard list to make sure that
other threads are aware of the newly added hazard pointer. By
repeating this process, a thread can jump through multiple elements
and pop all of them with a \cas at the end.

\begin{algorithm}
\SetAlgoLined
Pop (multiple)

\While{true} {

    t = Read(top)\;
    \For { multiple } {
      \If {t = NULL} {return EMPTY\;}
      hp* = t\;
      \If {top != t} {break\;}
      hp++\;
      next = t.next\;
    }
    \If {CAS(\&top, t, next)} {break\;}
}
RetireNodes (t, multiple)\;
\caption{Multiple Pop\label{fig:stack}}
\end{algorithm}
}

\rr{
\subsubsection{Shared Counter}

\def\shishi{.4}

\begin{figure}[b!]
\begin{center}
\begin{minipage}{\shishi\textwidth}\begin{center}
\includegraphics[width=\textwidth]{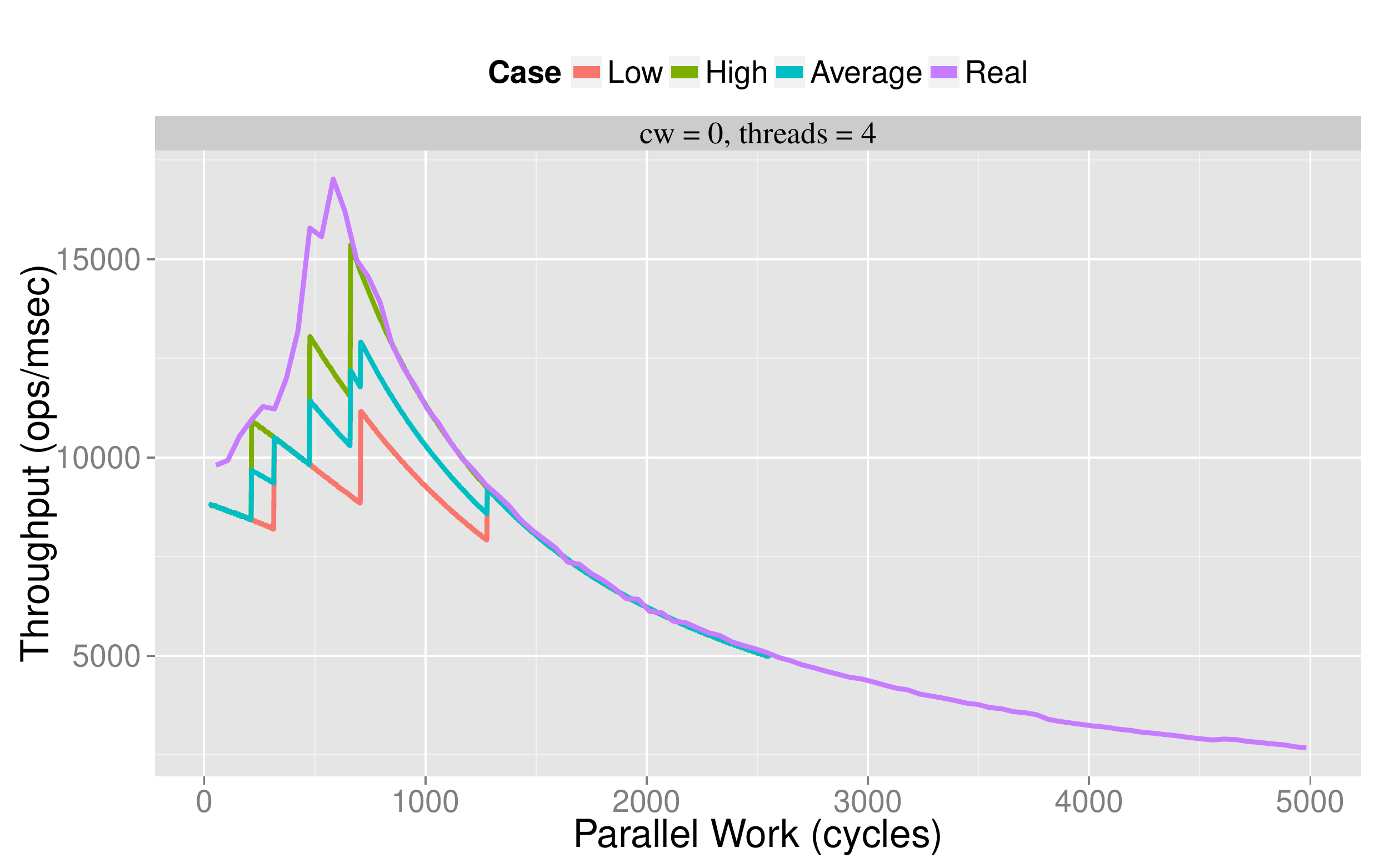}

(a) $4$ threads
\end{center}\end{minipage}\hfill%
\begin{minipage}{\shishi\textwidth}\begin{center}
\includegraphics[width=\textwidth]{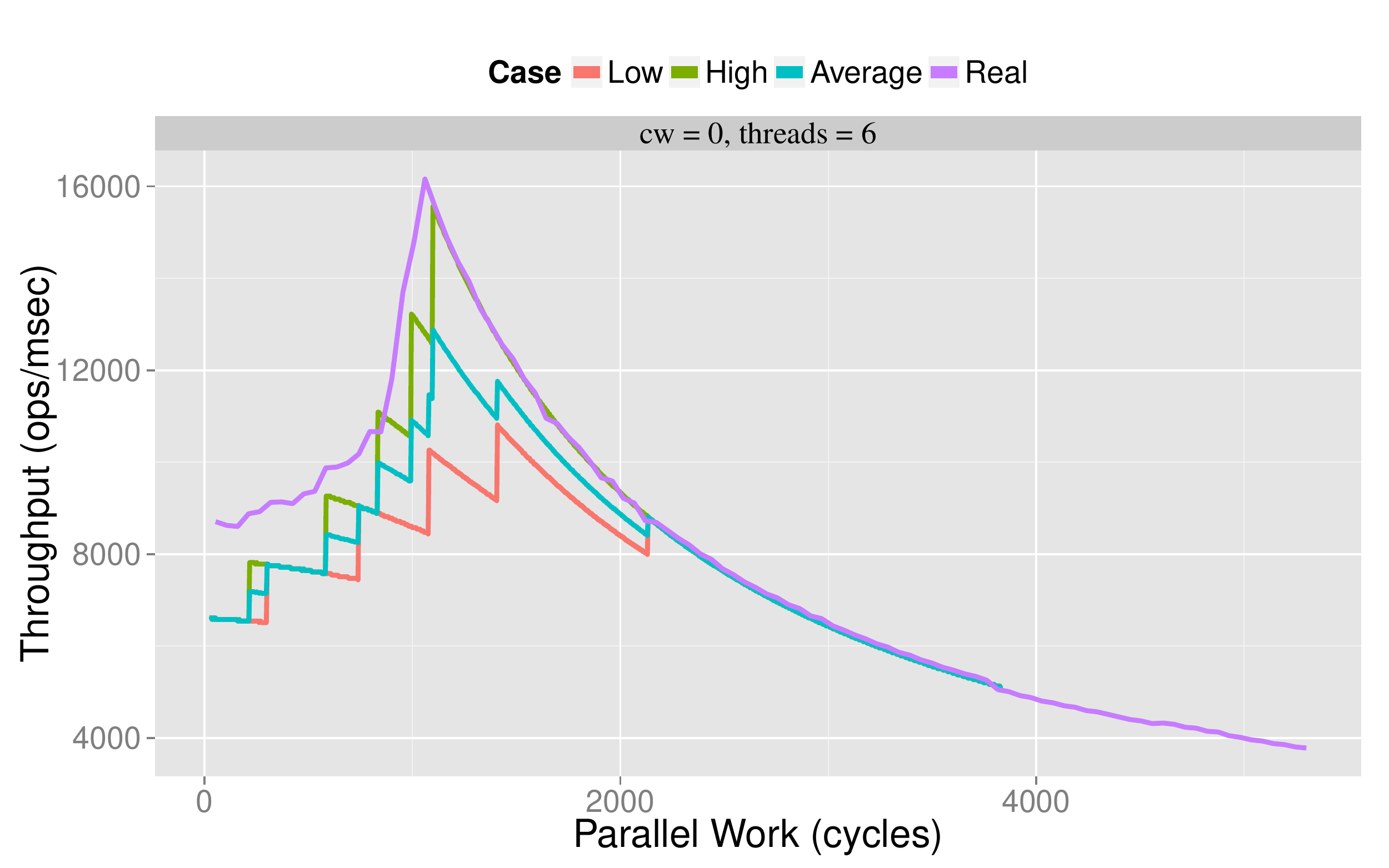}

(b) $6$ threads
\end{center}\end{minipage}

\vspace*{-.1cm}\includegraphics[width=\shishi\textwidth]{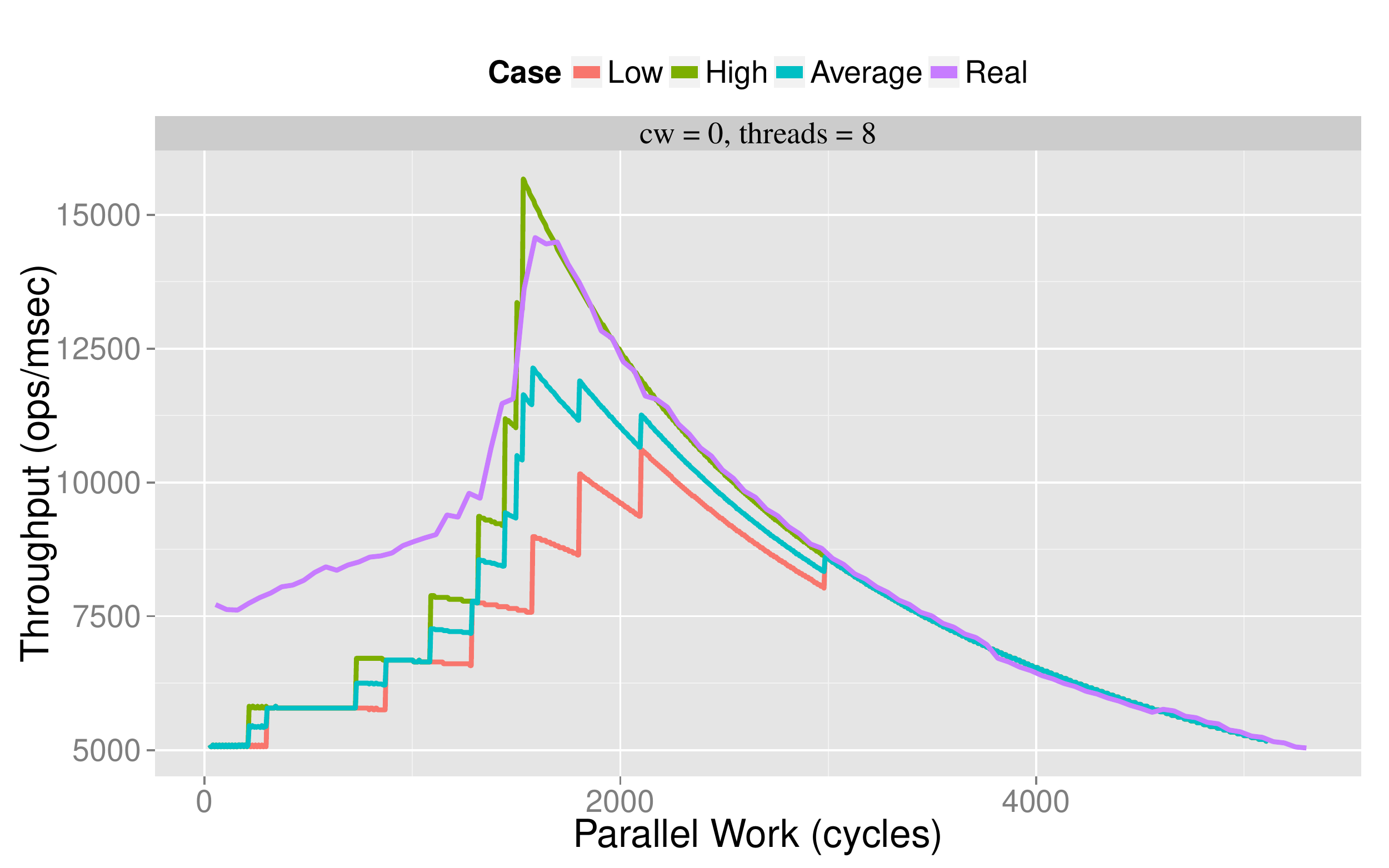}

(c) $8$ threads
\end{center}
\caption{\incop on a shared counter\label{fig:sc}}
\end{figure}

In~\cite{count-moir}, the authors have implemented a ``scalable
statistics counters'' relying on the following idea: when contention
is low, the implementation is a regular concurrent counter with a
\cas; when the counter starts to be contended, it switches to a
statistical implementation, where the counter is actually incremented
less frequently, but by a higher value. One key point of this
algorithm is the switch point, which is decided thanks to the number
of failed increments; our model can be used by providing the peak
point of performance of the regular counter implementation as the
switch point.
We then have implemented a shared counter which is basically a \faa
using a \cas, and compared it with our analysis. The result is
illustrated in Figure~\ref{fig:sc}, and shows that the \ps size corresponding to the peak point is correctly
estimated using our analysis.



\subsubsection{\delmin in Priority List}

\begin{figure}[h!]
\begin{center}
\begin{minipage}{.47\textwidth}\begin{center}
\includegraphics[width=\textwidth]{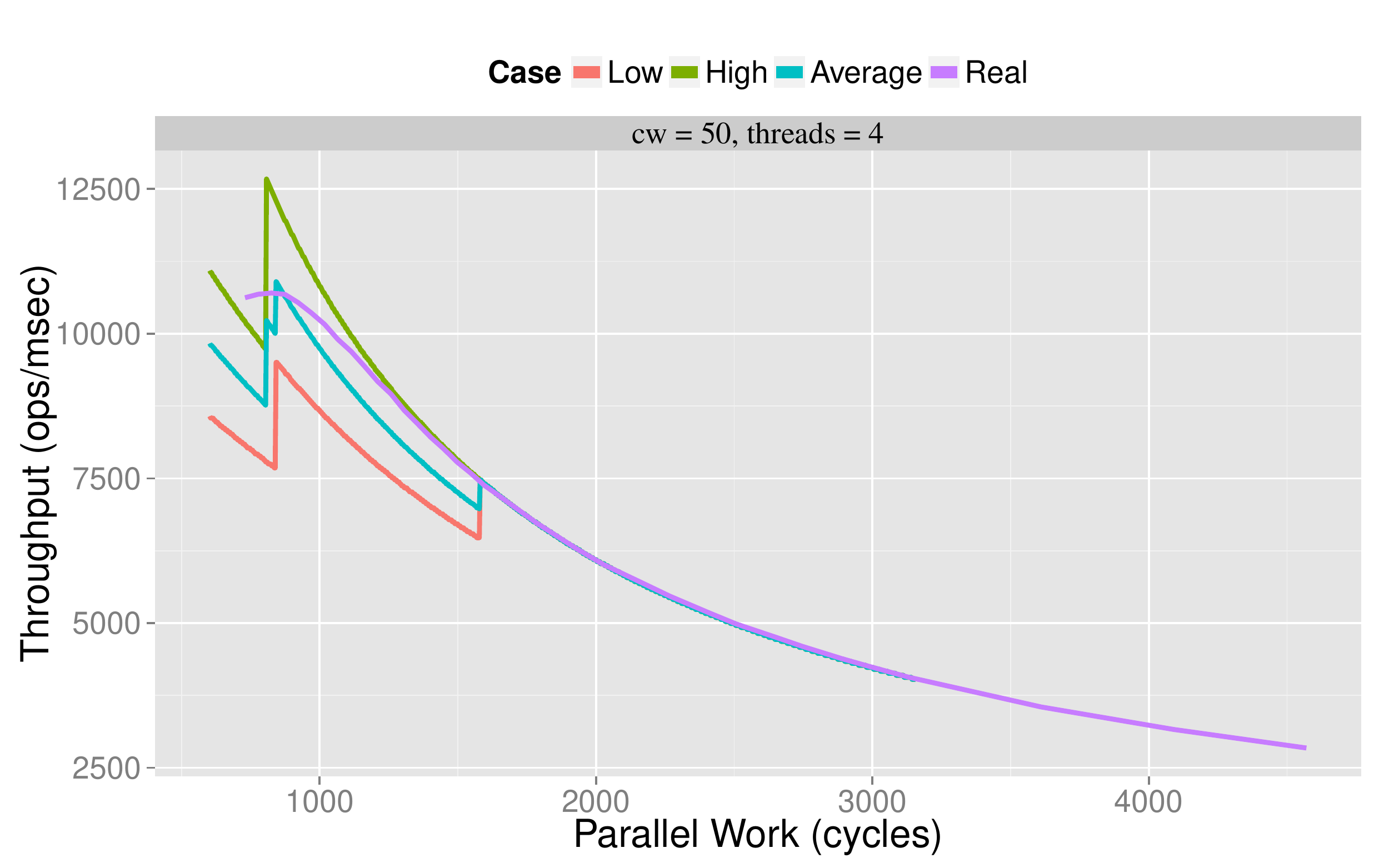}

(a) $4$ threads
\end{center}\end{minipage}\hfill%
\begin{minipage}{.47\textwidth}\begin{center}
\includegraphics[width=\textwidth]{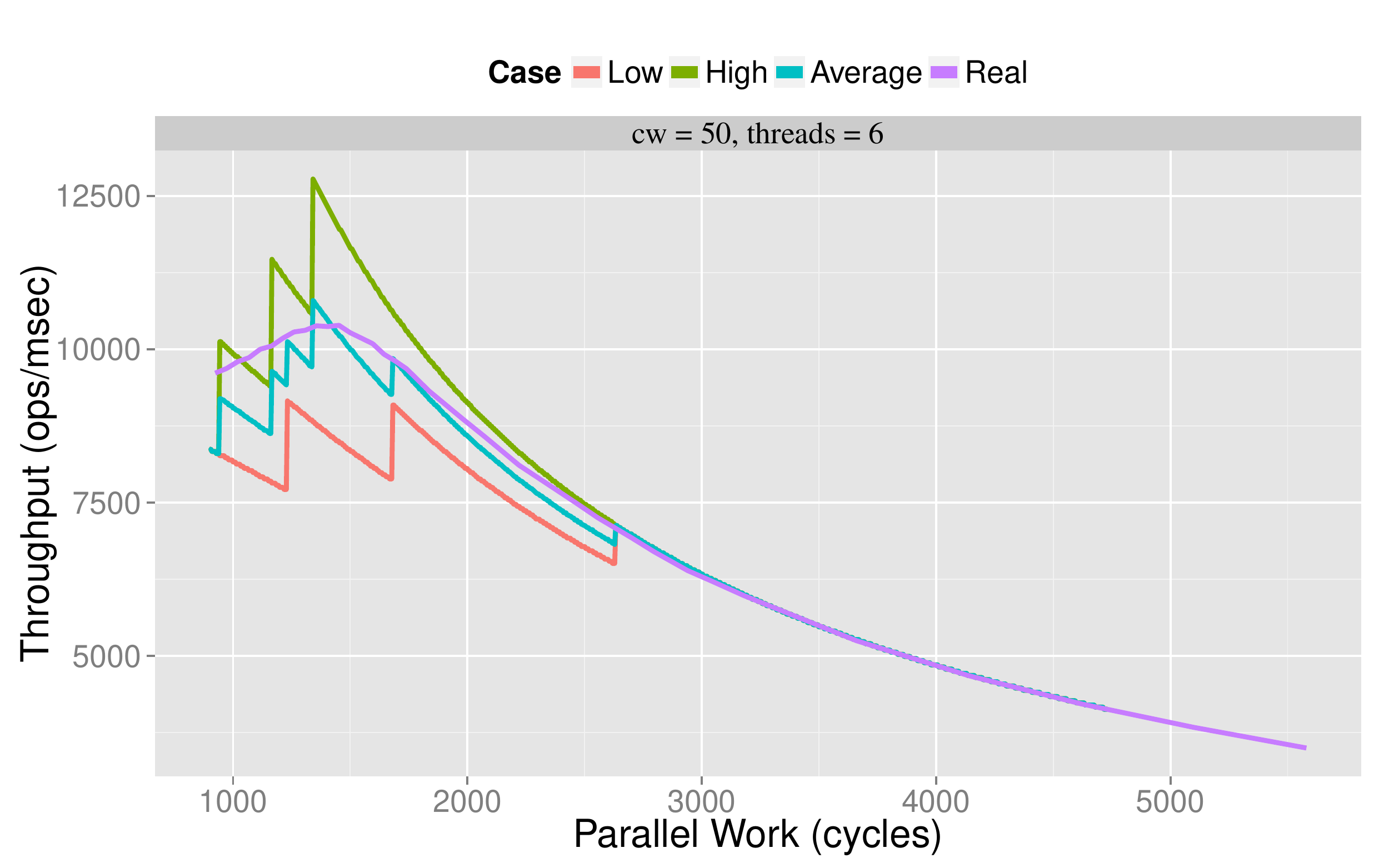}

(b) $6$ threads
\end{center}\end{minipage}

\includegraphics[width=.47\textwidth]{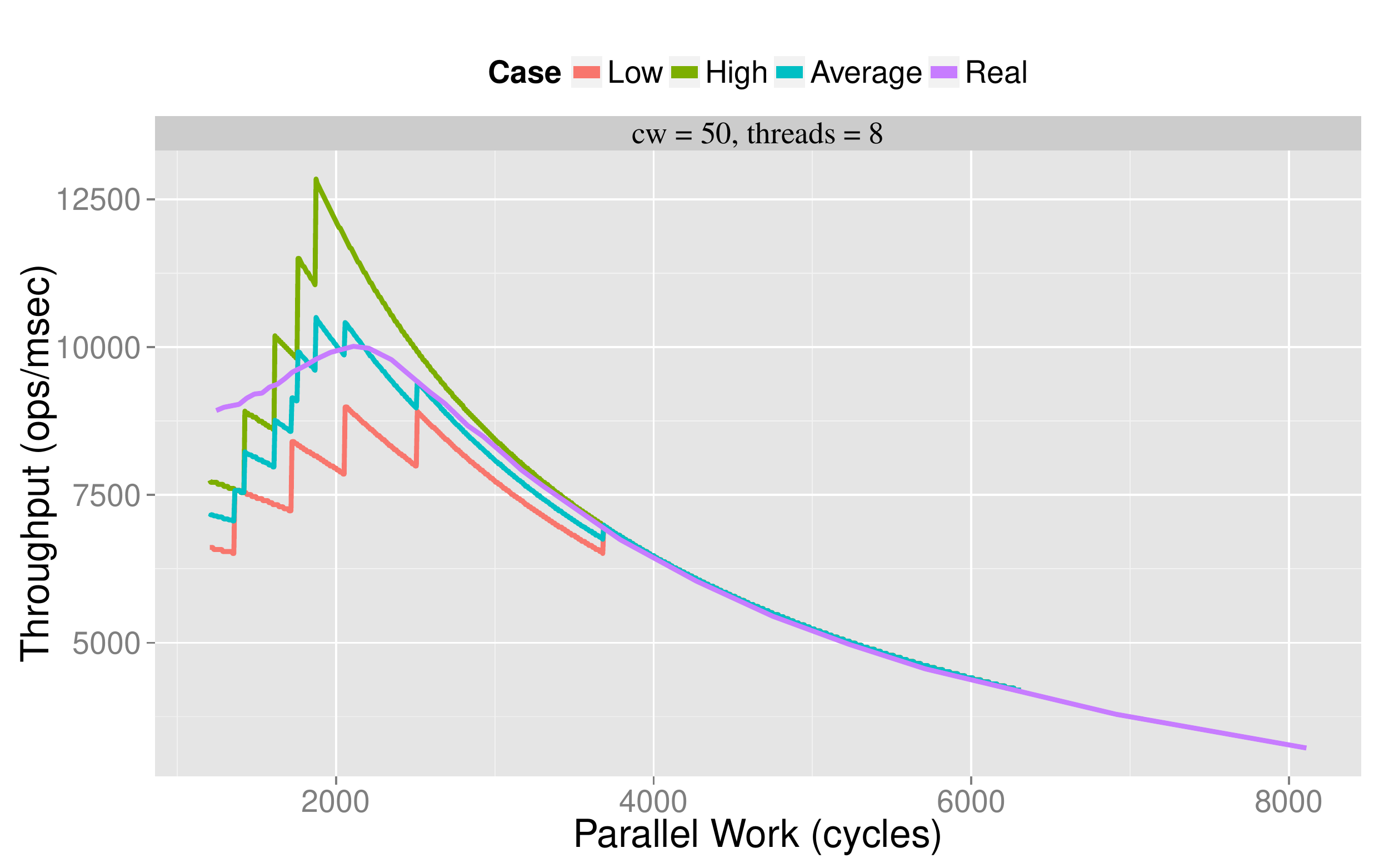}

(c) $8$ threads
\end{center}
\caption{\delmin on a priority list \label{fig:pq}}
\end{figure}

We have applied our model to \delmin of the skiplist based priority queue
designed in~\cite{upp-prio-que}. \delmin traverses the list from the beginning
of the lowest level, finds the first node that is not logically deleted, and
tries to delete it by marking. If the operation does not succeed, it continues
with the next node. Physical removal is done in batches when reaching a
threshold on the number of deleted prefixes, and is followed by a restructuring
of the list by updating the higher level pointers, which is conducted by the
thread that is successful in redirecting the head to the node deleted by itself.

We consider the last link traversal before the logical deletion as critical
work, as it continues with the next node in case of failure. The rest of
the traversal is attributed to the \ps as the threads can proceed concurrently without
interference. We measured the average cost of a traversal under low contention
  for each number of threads, since traversal becomes expensive with more
  threads. In addition, average cost of restructuring is also included in the
\ps since it is executed infrequently by a single thread.

We initialize the priority queue with a large set of elements. As illustrated in
Figure~\ref{fig:pq}, the smallest \pw value is not zero as the average cost of traversal
and restructuring is intrinsically included. The peak point is in the estimated
place but the curve does not go down sharply under high contention. This
presumably occurs as the traversal might require more than one steps (link
access) after a failed attempt, which creates a back-off effect.

} 

\subsubsection{\enqop-\deqop on a Queue}

\pp{\begin{wrapfigure}{l}{.5\textwidth}
\begin{center}
\fuckspaa\includegraphics[width=.4\textwidth]{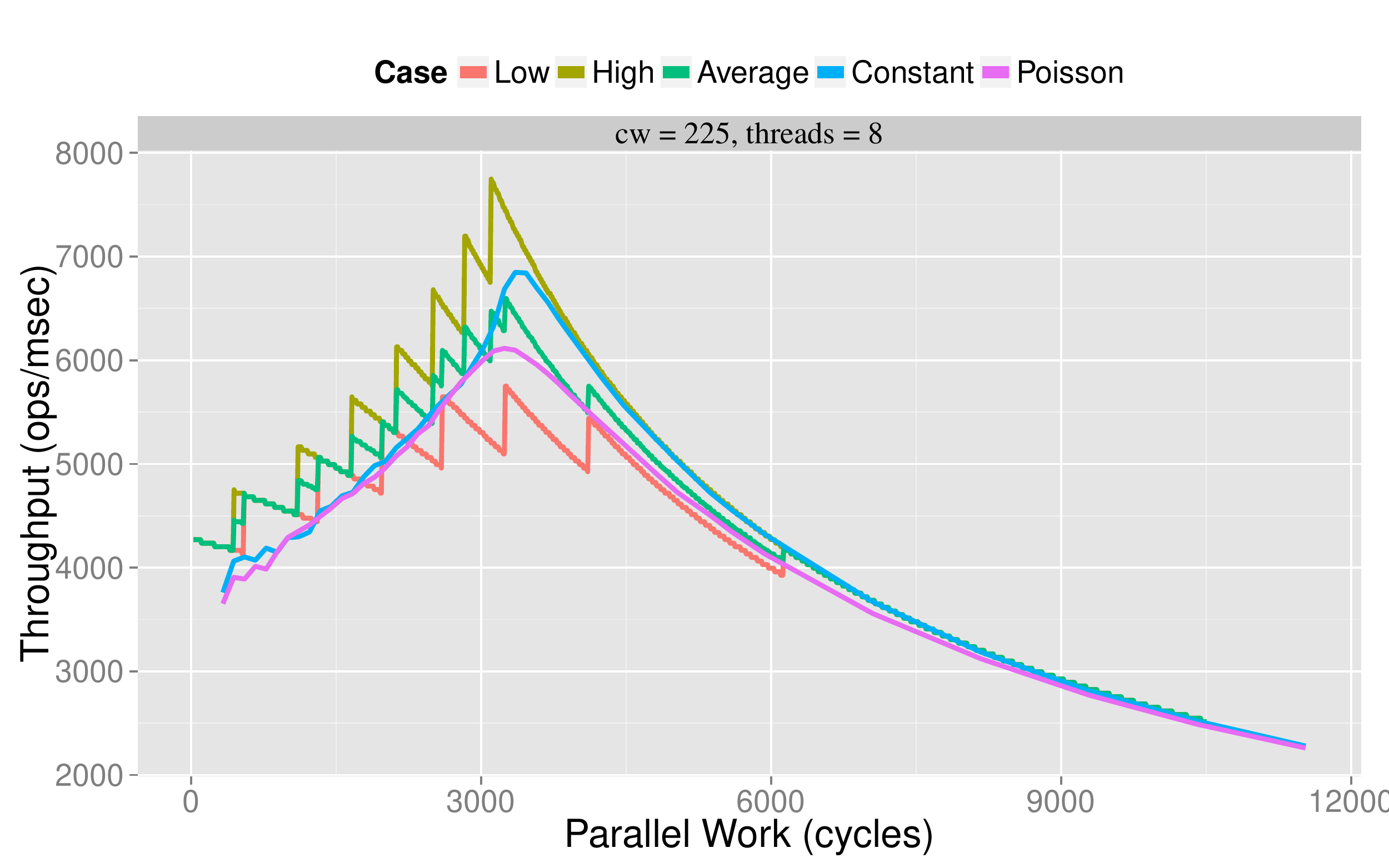}\vspace*{-.27cm}
\end{center}
\caption{\enqop-\deqop on Michael and Scott queue\label{fig:ms}}
\end{wrapfigure}}

\rr{\begin{figure}[h!]
\begin{center}
\begin{minipage}{.47\textwidth}\begin{center}
\includegraphics[width=\textwidth]{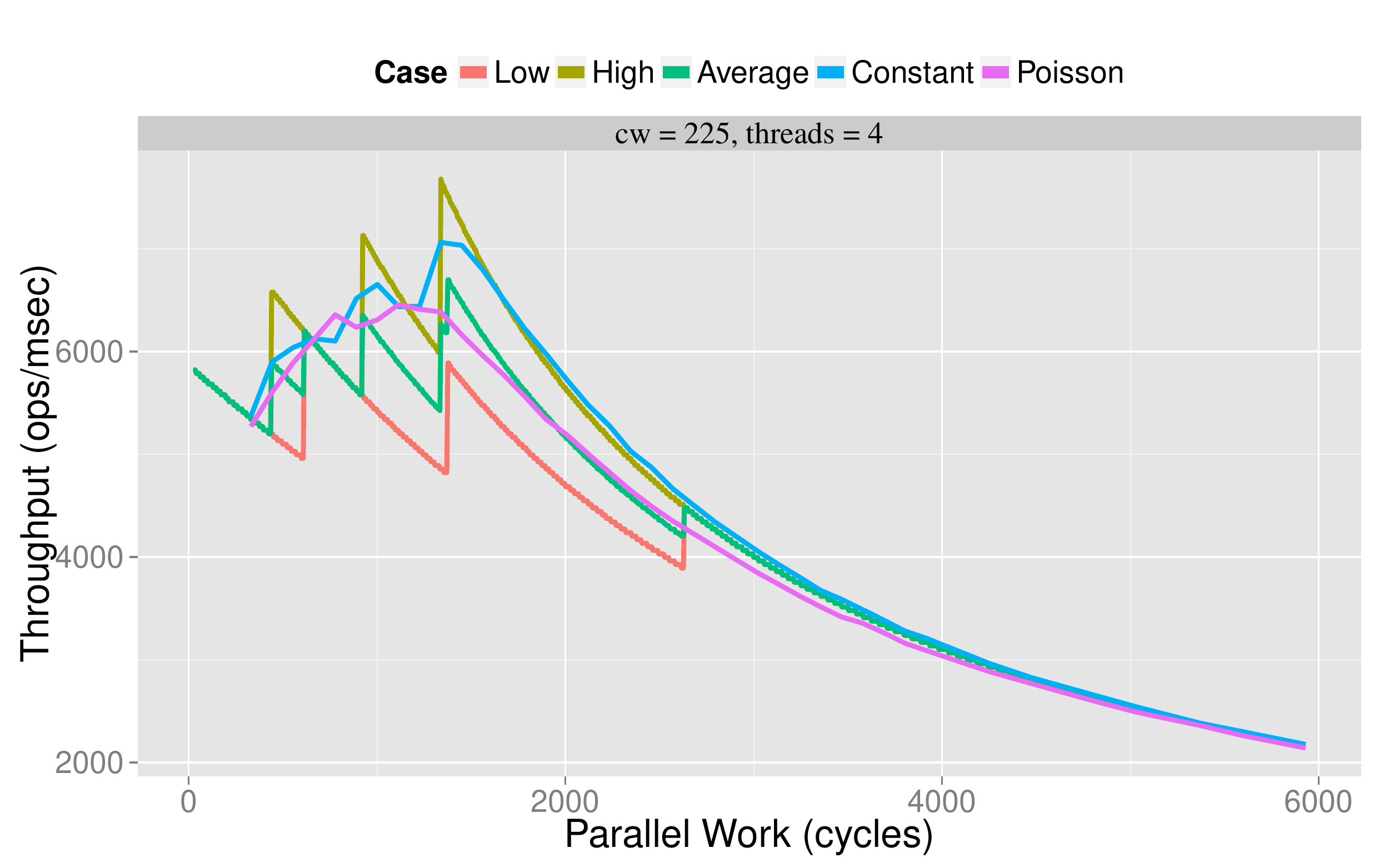}

(a) $4$ threads
\end{center}\end{minipage}\hfill%
\begin{minipage}{.47\textwidth}\begin{center}
\includegraphics[width=\textwidth]{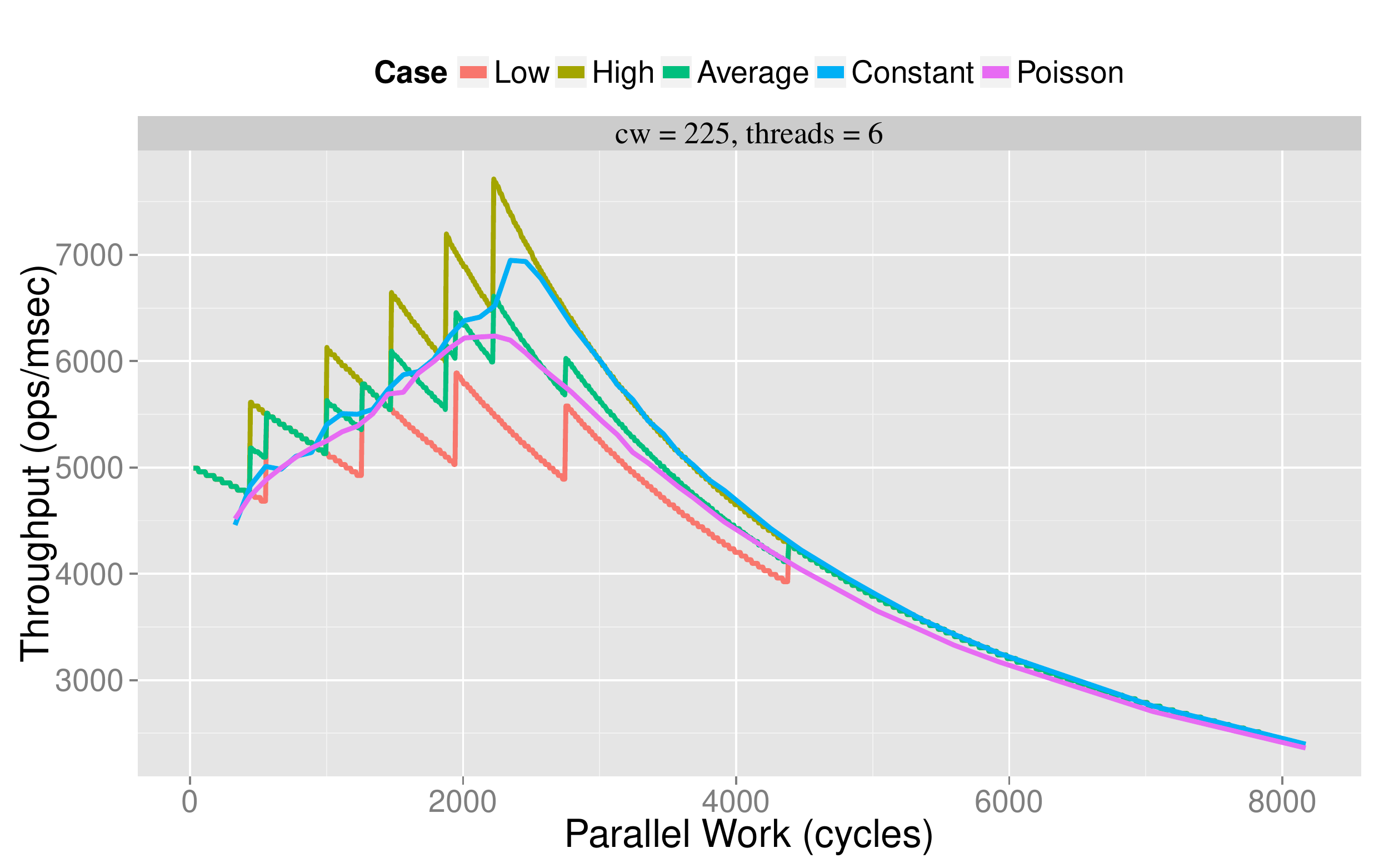}

(b) $6$ threads
\end{center}\end{minipage}

\includegraphics[width=.47\textwidth]{MS_8.pdf}

(c) $8$ threads
\end{center}
\caption{\enqop-\deqop on Michael and Scott queues\label{fig:ms}}
\end{figure}}

In order to demonstrate the validity of the model with several \rls
(see Section~\ref{cha-secsev-rl}), and that the results covers a wider
spectrum of application and designs from the ones we focused in our
model, we studied the following setting: the threads share a queue,
and each thread enqueues an element, executes the parallel section,
dequeues an element, and reiterates.
We consider the queue implementation by Michael and
Scott~\cite{lf-queue-michael}, that is usually viewed as the reference
queue while looking at lock-free queue implementations.

\pp{\vspace{.15cm}}
\deqop operations fit immediately into our model but \enqop operations need an adjustment due to
the helping mechanism. Note that without this helping mechanism, a simple queue
implementation would fit directly, but we also want to show that the model is
malleable, \ie the fundamental behavior remains unchanged even if we divert
slightly from the initial assumptions. We consider an equivalent execution that
catches up with the model, and use it to approximate the performance of the
actual execution of \enqop.\pp{ Due to lack of space, the full process is explained
in~\cite{our-long}.}

\rr{
\bigskip
\enqop is composed of two steps. Firstly, the new node is attached to the last
node of the queue via a \cas, that we denote by \casop{A}, leading to a
transient state. Secondly, the tail is redirected to point to the new node via
another \cas, that we denote by \casop{B}, which brings back the queue into a
steady state.

A new \enqop can not proceed before the two steps of previous success are
completed. The first step is the linearization point of operation and the second
step could be conducted by a different thread through the helping mechanism. In
order to start a new \enqop, concurrent \enqop{}s help the completion of the
second step of the last success if they find the queue in the transient
state. Alternatively, they try to attach their node to the queue if the queue is
in the steady state at the instant of check. This process continues until they
manage to attach their node to the queue via a retry loop in which state is checked
and corresponding \casop{} is executed.

The flow of an \enqop is determined by this state checks. Thus, an \enqop could
execute multiple \casop{B} (successful or failing) and multiple \casop{A}
(failing) in an interleaved manner, before succeeding in
\casop{A} at the end of the last \re. If we assume that both states are
equally probable for a check instant which will then end up with a retry, the
number of \casop{}s that ends up with a retry are expected to be
distributed equally among \casop{A} and \casop{B} for each thread. In addition,
each thread has a successful \casop{A} (which linearizes the \enqop)
and a \casop{B} at the end of the operation which could either be successful or
failed by a concurrent helper thread.

We imitate such an execution with an equivalent execution in which threads keep
the same relative ordering of the invocation, return from \enqop
together with same result. In equivalent execution, threads alternate between
\casop{A} and \casop{B} in their \res, and both steps of successful operation is
conducted by the same thread. The equivalent execution can be obtained by
thread-wise reordering of \casop{}s that leads to a \re and exchanging successful
\casop{B}s with the failed counterparts at the end of an \enqop, as the latter ones
indeed fail because of this success of helper threads. The model can be applied to
this equivalent execution by attributing each \casop{A}-\casop{B} couple to a single
iteration and represent it as a larger retry loop since the successful couple
can not overlap with another successful one and all overlapping ones fail. With
a straightforward extension of the expansion formula, we accomodate the \casop{A} in
the critical work which can also expand, and use \casop{B} as the \cas of our model.
\bigskip}

In addition, we take one step further outside the analysis by including a new
case, where the \ps follows a Poisson distribution, instead of being
constant. \rr{\pw is chosen as the mean to generate Poisson distribution instead of
taking it constant. }The results are illustrated in Figure~\ref{fig:ms}. Our
model provides good estimates for the constant \pw and also reasonable results
for the Poisson distribution case, although this case deviates from (/extends) our
model assumptions. The advantage of regularity, which brings synchronization to
threads, can be observed when the constant and Poisson distributions are compared. In the Poisson
distribution, the threads start to fail with larger \pw, which smoothes the curve
around the peak of the throughput curve.

\subsubsection{Discussion}
\label{cha-secdisc}

\pp{\begin{wrapfigure}{l}{.5\textwidth}}
\rr{\begin{figure}[b!]}
\begin{center}
\pp{\fuckspaa\includegraphics[width=.45\textwidth]{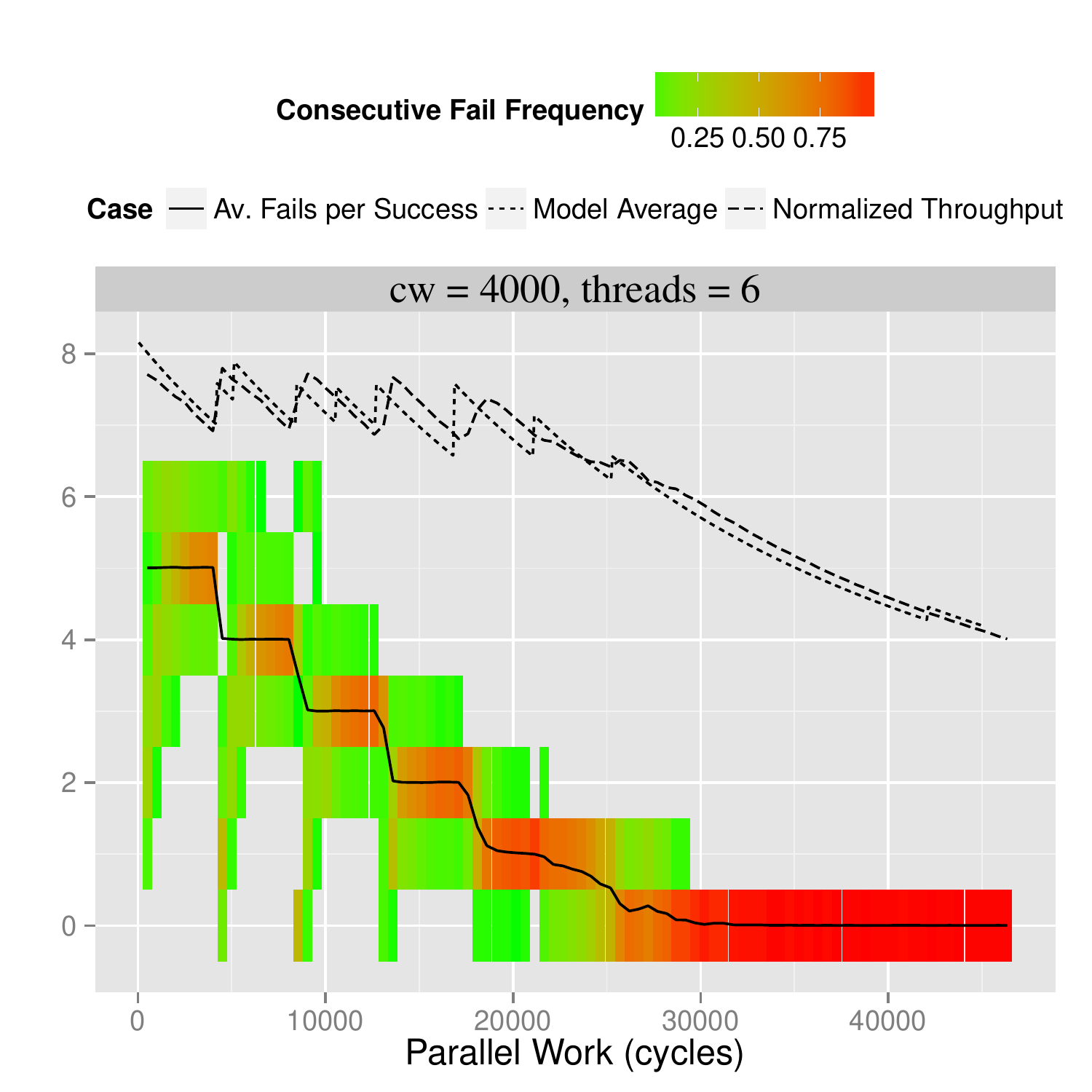}\vspace*{-.27cm}}
\rr{\fuckspaa\includegraphics[width=\textwidth]{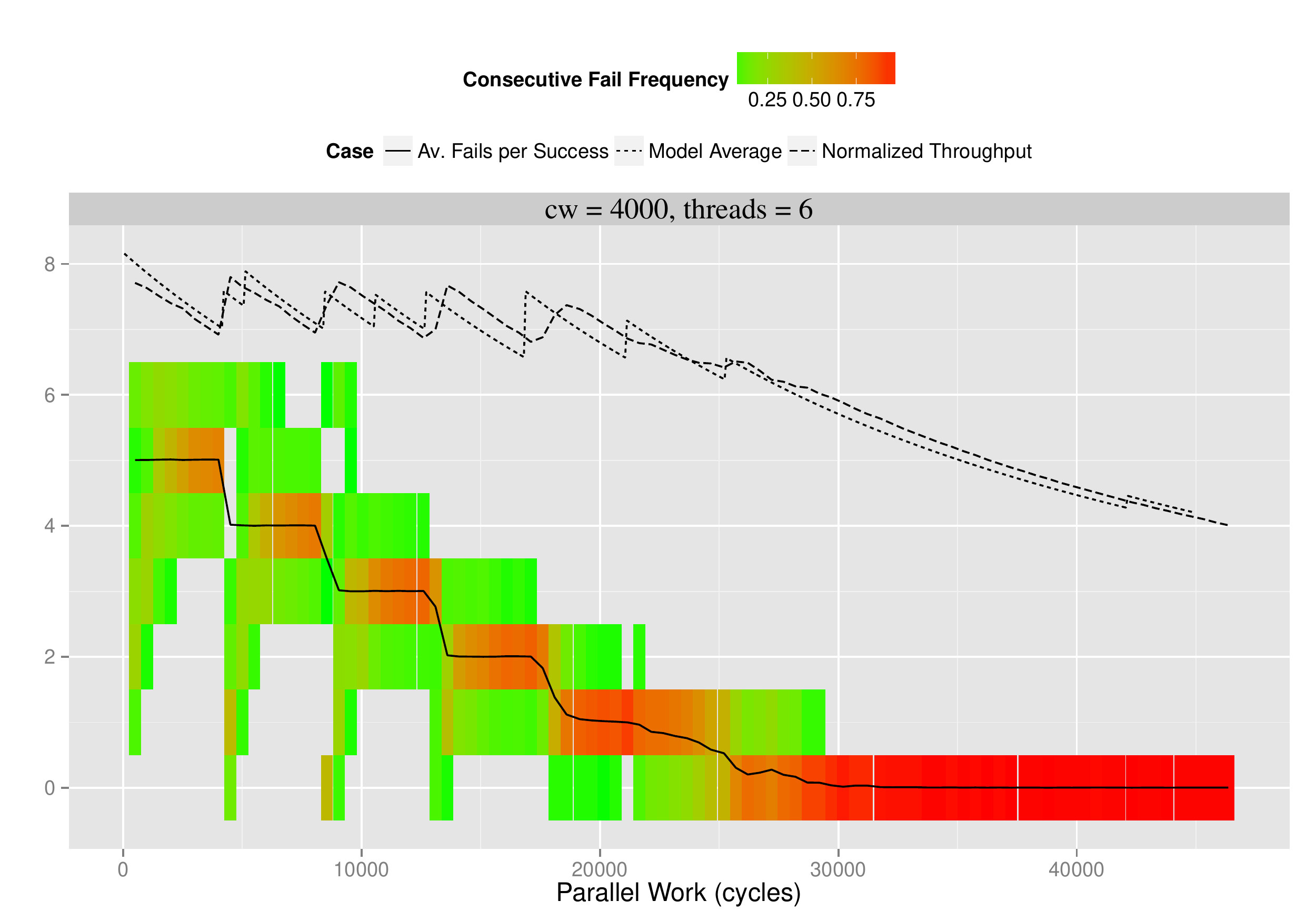}}
\end{center}
\caption{Consecutive Fails Frequency\label{fig:failFreq}}
\pp{\end{wrapfigure}}
\rr{\end{figure}}

In this subsection we discuss the adequacy of our model, specifically
the cyclic argument, to capture the behavior that we observe in practice.
Figure~\ref{fig:failFreq} illustrates the frequency of occurrence of a given number of consecutive
fails, together with average fails per success values and the throughput
values, normalized by a constant factor so that they can be seen on the graph. In the background, the
frequency of occurrence of a given number of consecutive fails before success 
is presented. As a remark, the frequency of 6+ fails is gathered with
6. We expect to see a frequency distribution concentrated around the
average fails per success value, within the bounds computed by our
model.

While comparing the distribution of failures with the throughput, we
could conjecture that the bumps come from the fact that the failures
spread out. However, our model captures correctly the throughput
variations and thus strips down the right impacting factor.
The spread of the distribution of failures indicates the violation of
a stable cyclic execution (that takes place in our model), but in
these regions, $r$ actually gets close to $0$, as well as the minimum
of all gaps. The scattering in failures shows that, during the
execution, a thread is overtaken by another one. Still, as gaps are
close to $0$, the imaginary execution, in which we switch the two
thread IDs, would create almost the same performance effect. This
reasoning is strengthened by the fact that the actual average number
of failures follows the step behavior, predicted by our model.
This shows that even when the real execution is not cyclic and the
distribution of failures is not concentrated, our model that results
in a cyclic execution remains a close approximation of the actual
execution.

\subsubsection{Back-Off Tuning}
\label{cha-secbo}

\pp{\begin{wrapfigure}{l}{.5\textwidth}
\begin{center}
\fuckspaa\includegraphics[width=.5\textwidth]{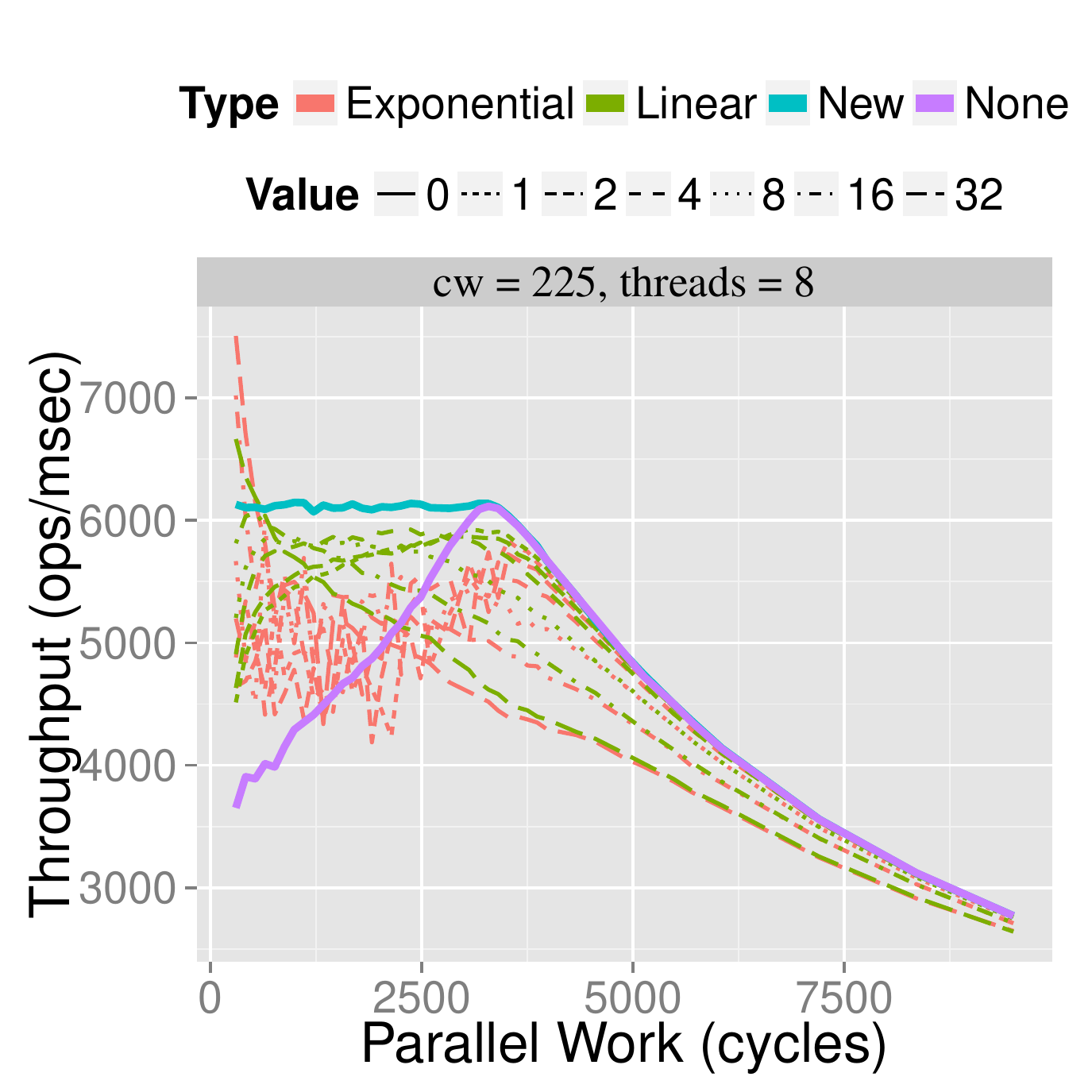}\vspace*{-.27cm}
\end{center}
\caption{Comparison of back-off schemes}
\label{fig:bo}
\end{wrapfigure}}

\rr{
\begin{figure}[h!]
\begin{center}
\begin{minipage}{.47\textwidth}\begin{center}
\includegraphics[width=\textwidth]{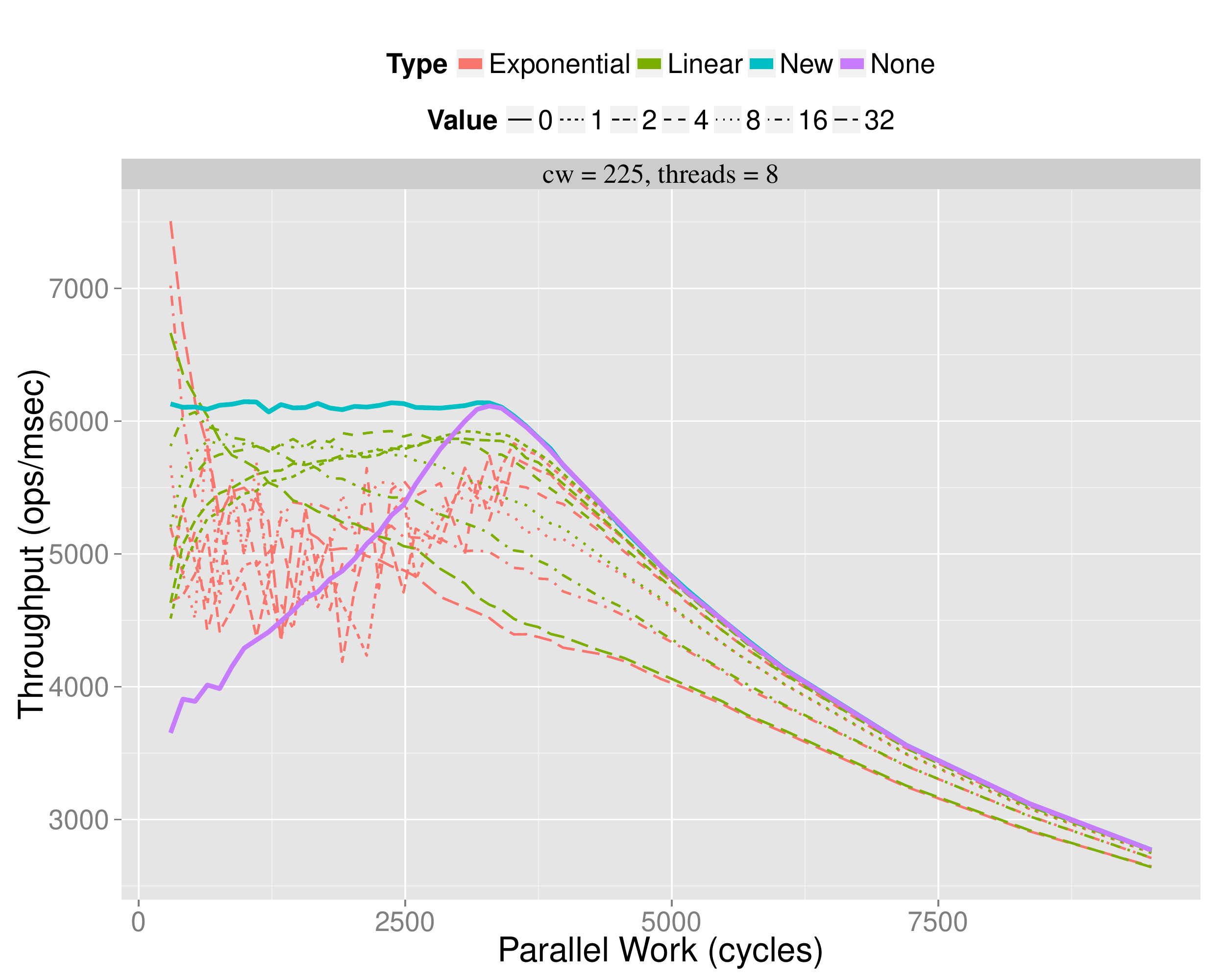}

(a) $8$ threads
\end{center}\end{minipage}\hfill%
\begin{minipage}{.47\textwidth}\begin{center}
\includegraphics[width=\textwidth]{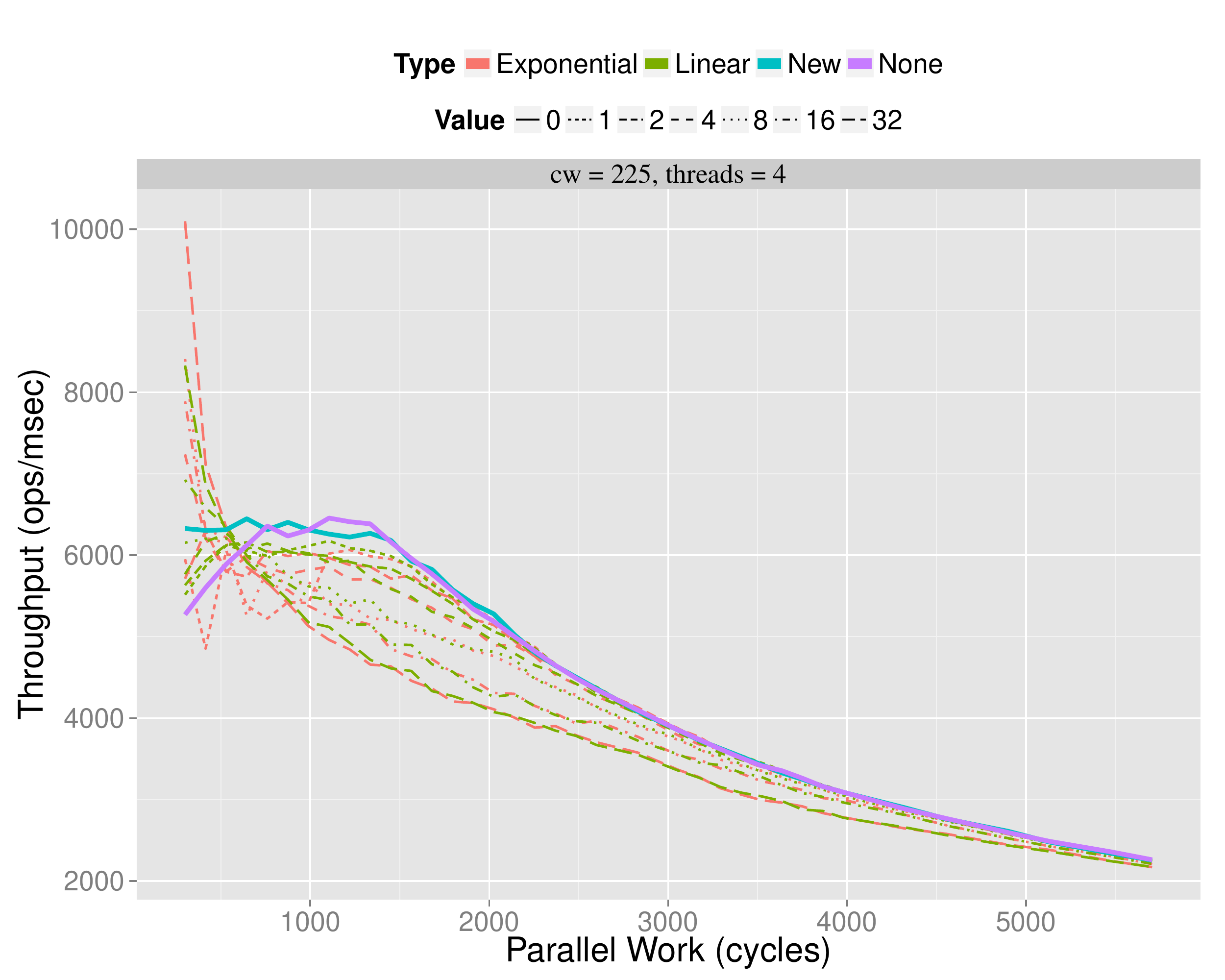}

(b) $4$ threads
\end{center}\end{minipage}
\end{center}
\caption{Comparison of back-off schemes for Poisson Distribution\label{fig:bo-poi}}
\end{figure}
}

Together with the analysis comes a natural back-off strategy: we estimate the
\pw corresponding to the peak point of the average curve, and when the \ps is smaller than the
corresponding \pw, we add a back-off in the \ps, so that the new \ps is at the
peak point.

We have applied exponential, linear and our back-off strategy to the
\enqop/\deqop experiment specified above. Our back-off estimate
provides good results for both types of distribution. In
Figure\pp{~\ref{fig:bo}}\rr{~\ref{fig:bo-poi}} (where the values of
back-off are steps of $115$ cycles), the comparison is plotted for the
Poisson distribution, which is likely to be the worst for our
back-off. Our back-off strategy is better than the other, except for
very small \pss, but other back-off strategies should be tuned for
each value of \pw.

\rr{
We obtained the same shapes while removing the distribution law and considering constant values.
The results are illustrated in Figure~\ref{fig:bo-con}.

\begin{figure}[h!]
\begin{center}
\begin{minipage}{.47\textwidth}\begin{center}
\includegraphics[width=\textwidth]{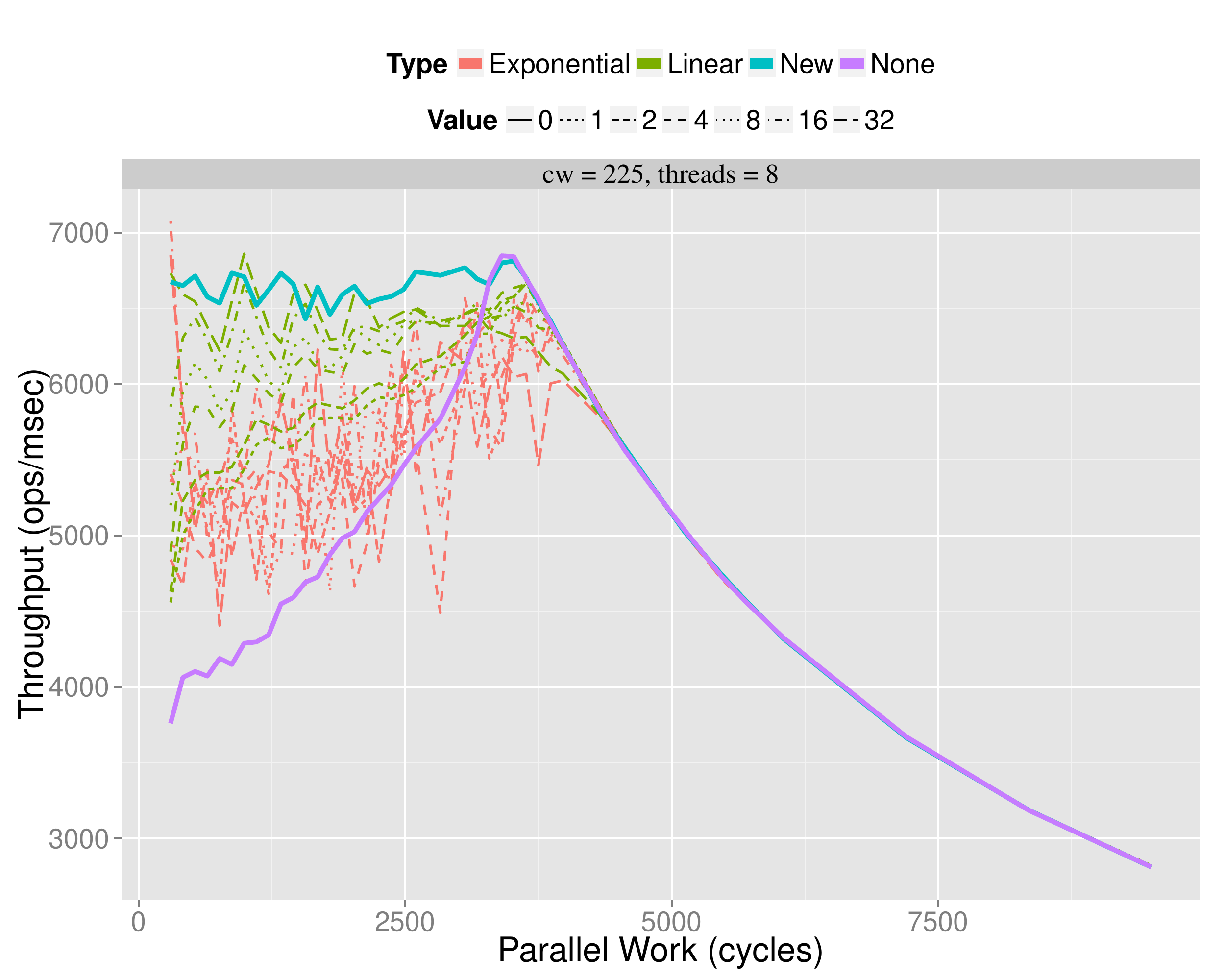}

(a) $8$ threads
\end{center}\end{minipage}\hfill%
\begin{minipage}{.47\textwidth}\begin{center}
\includegraphics[width=\textwidth]{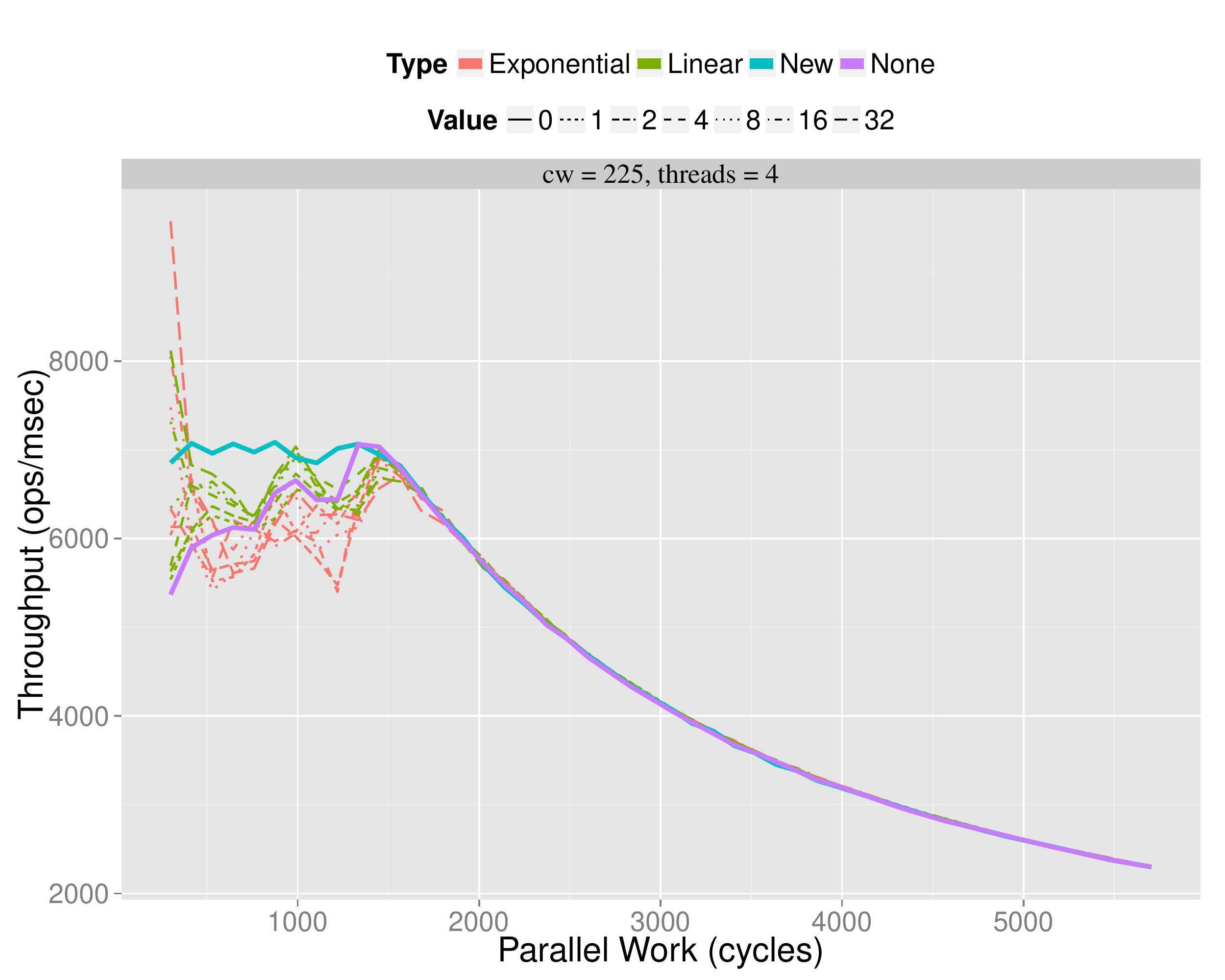}

(b) $4$ threads
\end{center}\end{minipage}
\end{center}
\caption{Comparison of back-off schemes for constant \pw \label{fig:bo-con}}
\end{figure}

}



\subsection{Conclusion on Throughput Modeling}

We have modeled and analyzed the performance of a general class of
lock-free algorithms. Thanks to this analysis, we have been able to
predict the throughput of such algorithms, on actual executions. The
analysis relies on the estimation of two impacting factors that lower
the throughput: on the one hand, the expansion, due to the
serialization of the atomic primitives that take place in the \rls; on
the other hand, the \cacas, due to a non-optimal synchronization
between the running threads. We have derived methods to calculate
those parameters, along with the final throughput estimate, that is
calculated from a combination of these two previous parameters. As a
side result of our work, this accurate prediction enables the design
of a back-off technique that performs better than other well-known
techniques, namely linear and exponential back-offs.

As a future work, we envision to enlarge the domain of validity of the
model, in order to cope with \dss whose operations do not have
constant \rl, as well as the framework, so that it includes more
various access patterns.  The fact that our results extend outside the
model allows us to be optimistic on the identification of the right
impacting factors.  Finally, we also foresee studying back-off
techniques that would combine a back-off in the \ps (for lower
contention) and in the \rls (for higher
robustness).



\subsection{Energy Modelling and Empirical Evaluation}

We introduced our power model and the power impacting factors in
D2.1~\cite{EXCESS:D2.1}. Here, we combine them with our performance
model that is illustrated in Section~\ref{sec:chalmers-disc}. By doing
so, we aim to come up with the average power consumption predictions
for the parallel programs that use fundamental concurrent
lock-free \dss.

\begin{figure}[t!]
\begin{center}
\includegraphics[width=.8\textwidth]{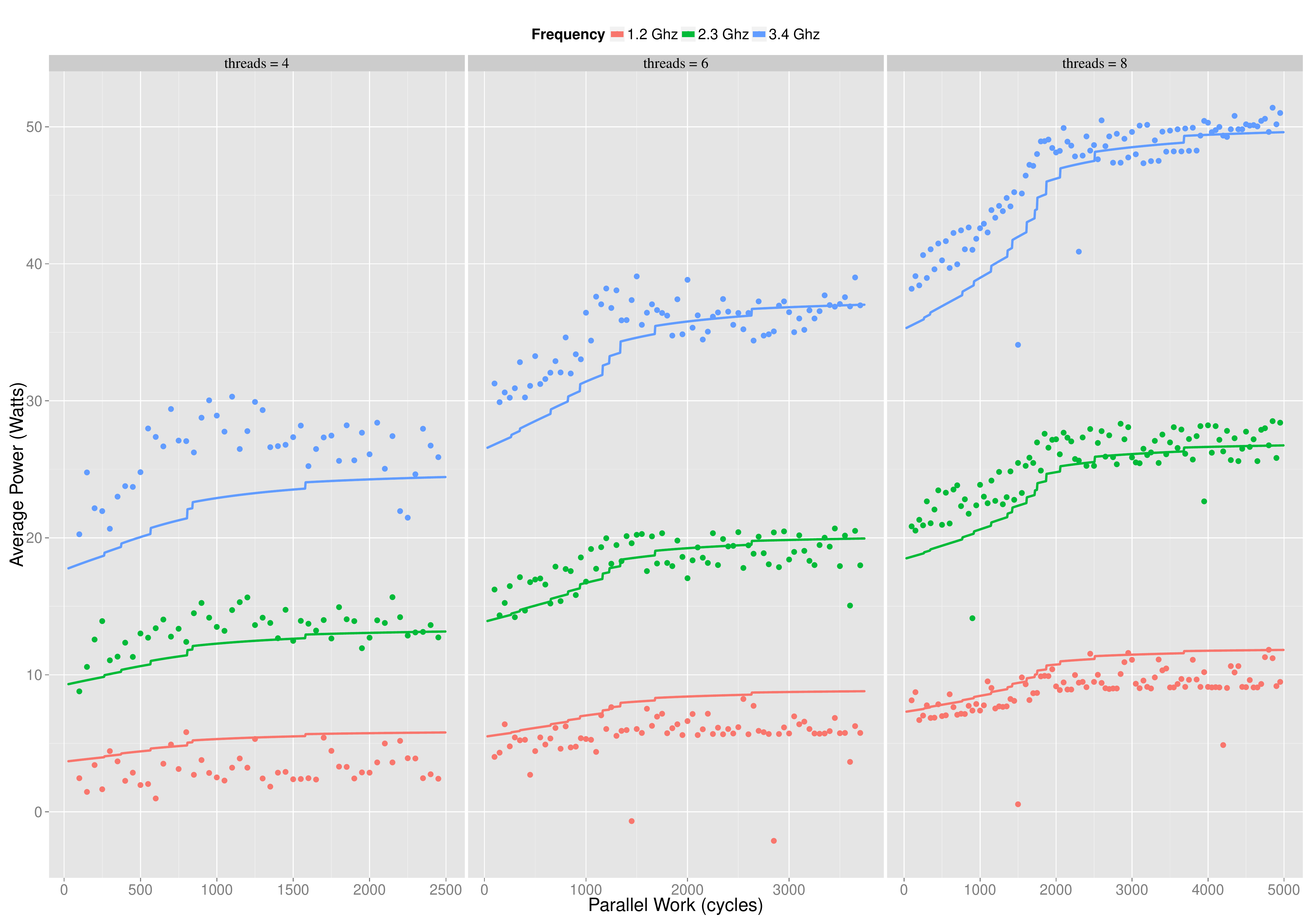}
\end{center}
\caption{Average Power Consumption for Treiber's Stack (Pop operation)\label{fig:en_stack}}
\end{figure}

In D2.1, we decompose the power into two orthogonal bases, each base
having three dimensions. On the one hand, we define the model base by
separating the power into static, active and dynamic power. On the
other hand, the measurement base corresponds to the components that
actually dissipates the power,\ie CPU, memory and uncore, in
accordance with RAPL energy counters. We recall that we are interested
only in the dynamic component of power, since we determine the static
power and the activation power, that do not depend on the \ds
implementation or the application that uses the concurrent \ds. Our
performance model does not cover the cases where the inter-socket
communication takes place. Here, we do not present the dynamic memory
and uncore power evaluations because they are insignificant (\ie close
to 0 for all cases) when there are not memory accesses (parallel work
is composed of multiplication instructions) or inter-socket
communication (threads are pinned to the same socket).

\begin{figure}[h!]
\begin{center}
\includegraphics[width=.8\textwidth]{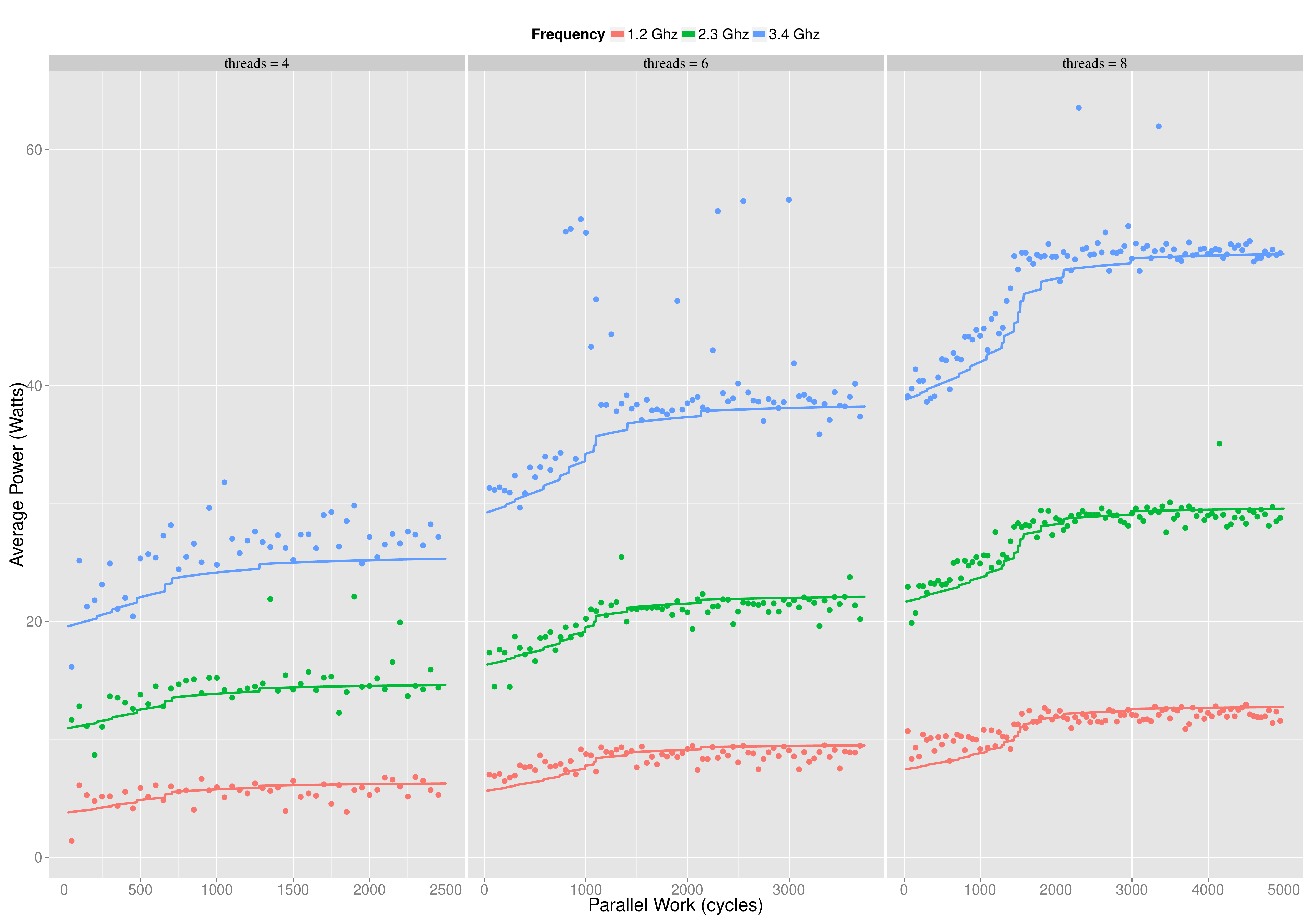}
\end{center}
\caption{Average Power Consumption for Shared Counter (Increment operation)\label{fig:en_sc}}
\end{figure}

In D2.1, we illustrate that CPI (cycles per instruction) is the main
impacting factor for the dynamic CPU power. And, CPI reaches to a high
value during the execution of \ds operations. This is because the
retry loops are composed of read accesses and \cas{s} which typically
lead to costly cache misses in a concurrent environment. In contrast,
one could expect to observe a lower value of CPI in the application
specific parts of the parallel program, \ie parallel work. This is
because computations presumably are executed in the application specific
part and the concurrent \dss are used for the communication. That is
why, we emulate the parallel work with a for-loop of multiplication
instructions. In this region of the parallel program, CPI gets low as
a multiplication instruction can be executed within almost a cycle so
we observe an increase in the dynamic CPU power during the execution
of the parallel work. As we expect two different average power
behaviours in these two regions, we build our reasoning over \ the
ratio of execution time that the threads spend in the retry
loops. Thanks to our performance model, we can predict this ratio for
each value of the parallel work, the number of threads and the \ds
operation. We can also generalize our performance predictions to a whole
clock frequency domain with a straightforward evaluation of the model
parameters for each frequency (\ie \pw, \cw, \cas and \mem).

\begin{figure}[t!]
\begin{center}
\includegraphics[width=.8\textwidth]{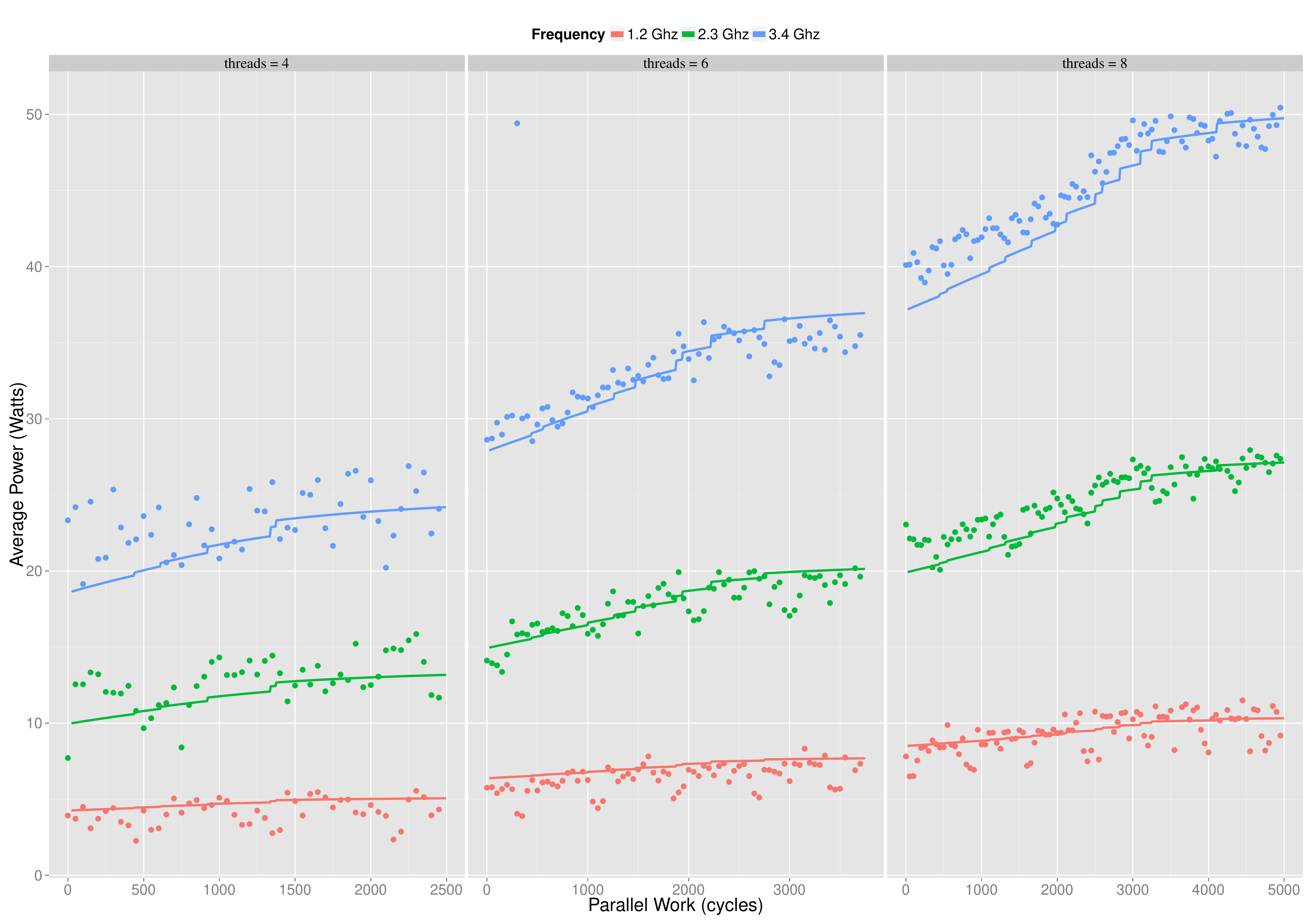}
\end{center}
\caption{Average Power Consumption for MS Queue (Enqueue operation)\label{fig:en_enqueue}}
\end{figure}

Dynamic CPU power does not scale linearly with the frequency, instead
shows a superlinear behaviour (see D2.1). We handle this issue by
taking power measurements for each frequency. For each \ds operation
and frequency, we run the parallel program for two values of \pw that
corresponds to a low and a high contention case. respectively (\ie where the ratio
of time spent in retry loop is $0.05$ and $0.5$) with $8$ threads for
a given duration. Using RAPL energy counters, we measure the energy
consumption for these cases. Then, we use these values to extract the
average power consumption for each phase (retry loop and parallel
work) of the parallel program.  Recall that we know the static and
activation component of power so we can extract the dynamic component for
memory, uncore and CPU power.

The average power consumption (for each of the memory, CPU and uncore
components) for the retry loop ($Pow_{RL}$) and the parallel work
($Pow_{PW}$) can be obtained as follows: (by using the two energy
measurements for \pw values corresponding to two different ratios of
execution time spent in the retry loops):

\[ \mathit{Pow}_{\mathit{Average}} = \mathit{Threads} \times (\mathit{Pow}_{RL} \times \mathit{Ratio}_{RL} + \mathit{Pow}_{PW} \times (1-\mathit{Ratio}_{RL})) \]

Based on the estimation of $Pow_{PW}$ and $Pow_{RL}$, for each
frequency and \ds operation, we provide average power predictions that
span the whole parallel work and the number of threads domain. This is
simply done by plugging the $Ratio_{RL}$, that is provided by our
performance model, to the formula above. In
Figures~\ref{fig:en_stack},~\ref{fig:en_sc},~\ref{fig:en_enqueue},~\ref{fig:en_dequeue},
we present the results for a set of fundamental lock-free \ds operations,
namely for Micheal and Scott Queue (Enqueue and Dequeue operations),
Treiber's Stack (Pop operation) and Shared Counter (Increment
operation). In the figures, lines and points represent predictions and
actual measurements, respectively.

\begin{figure}[h!]
\begin{center}
\includegraphics[width=\textwidth]{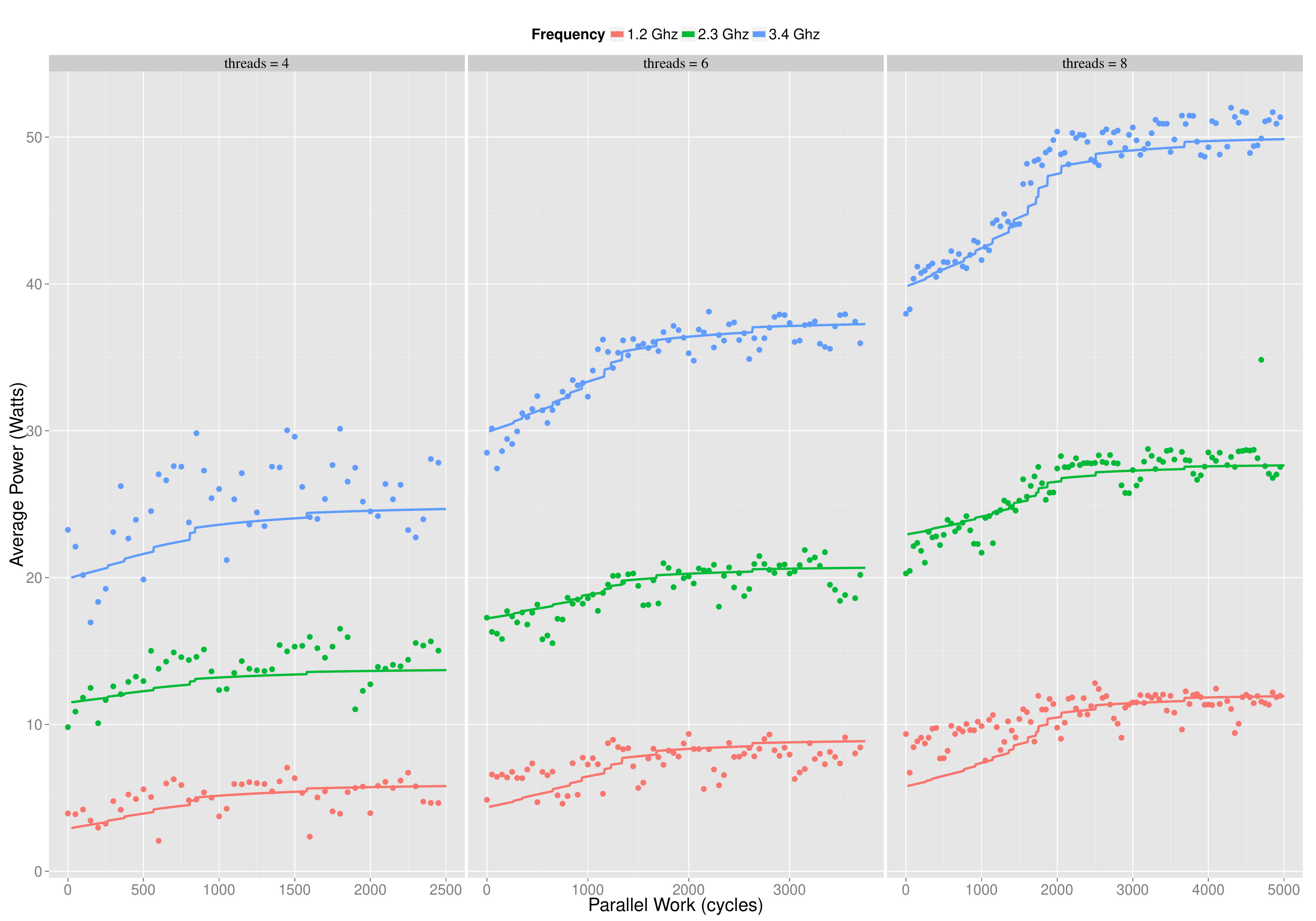}
\end{center}
\caption{Average Power Consumption for MS Queue (Dequeue operation)\label{fig:en_dequeue}}
\end{figure}

In all the figures, we observe a similar behaviour. Dynamic CPU power
decreases when
\pw decreases. We know that \pw is a  key aspect that influences the contention on the \ds, equally with the
the ratio of time that threads spend in the retry loop. With the
decrease of \pw, CPI increases and dynamic CPU power reduces.

\clearpage

%% file: inp-chalmers/ex-lemma-1.tex
\node [left] at (0,\yt[1]) {\thr{0}};
\node [left] at (0,\yt[2]) {\thr{1}};
\node [left] at (0,\yt[3]) {\thr{2}};
\pgfmathparse{2.000000*\cle}
\edef\ple{\pgfmathresult}
\pgfmathparse{0.000000*\cle}
\extth{1}{\pgfmathresult}{ini}
\extth{1}{\cle}{csu}
\extth{1}{\ple}{par}
\extth{1}{\cle}{csu}
\extth{1}{\ple}{par}
\extth{1}{\cle}{cfa}
\extth{1}{\cle}{csu}
\extth{1}{\ple}{par}
\extth{1}{\cle}{cfa}
\extth{1}{\cle}{csu}
\extth{1}{\ple}{par}
\extth{1}{\cle}{cfa}
\extth{1}{\cle}{csu}
\extth{1}{\ple}{par}
\extth{1}{\cle}{cfa}
\pgfmathparse{1.500000*\cle}
\extth{2}{\pgfmathresult}{ini}
\extth{2}{\cle}{csu}
\extth{2}{\ple}{par}
\extth{2}{\cle}{csu}
\extth{2}{\ple}{par}
\extth{2}{\cle}{cfa}
\extth{2}{\cle}{csu}
\extth{2}{\ple}{par}
\extth{2}{\cle}{cfa}
\extth{2}{\cle}{csu}
\extth{2}{\ple}{par}
\extth{2}{\cle}{cfa}
\extth{2}{\cle}{csu}
\extth{2}{\ple}{par}
\pgfmathparse{5.750000*\cle}
\extth{3}{\pgfmathresult}{ini}
\extth{3}{\cle}{csu}
\extth{3}{\ple}{par}
\extth{3}{\cle}{cfa}
\extth{3}{\cle}{csu}
\extth{3}{\ple}{par}
\extth{3}{\cle}{cfa}
\extth{3}{\cle}{csu}
\extth{3}{\ple}{par}
\extth{3}{\cle}{cfa}
\extth{3}{\cle}{csu}
\extth{3}{\ple}{par}

%% file: inp-chalmers/ex-lemma-3.tex
\node [left] at (0,\yt[1]) {\thr{0}};
\node [left] at (0,\yt[2]) {\thr{1}};
\node [left] at (0,\yt[3]) {\thr{2}};
\node [left] at (0,\yt[4]) {\thr{3}};
\pgfmathparse{3.125000*\cle}
\edef\ple{\pgfmathresult}
\pgfmathparse{0.000000*\cle}
\extth{1}{\pgfmathresult}{ini}
\extth{1}{\cle}{csu}
\extth{1}{\ple}{par}
\extth{1}{\cle}{cfa}
\extth{1}{\cle}{csu}
\extth{1}{\ple}{par}
\extth{1}{\cle}{cfa}
\extth{1}{\cle}{cfa}
\extth{1}{\cle}{csu}
\extth{1}{\ple}{par}
\extth{1}{\cle}{cfa}
\extth{1}{\cle}{csu}
\extth{1}{\ple}{par}
\extth{1}{\cle}{cfa}
\extth{1}{\cle}{csu}
\extth{1}{\ple}{par}
\extth{1}{\cle}{cfa}
\pgfmathparse{0.500000*\cle}
\extth{2}{\pgfmathresult}{ini}
\extth{2}{\cle}{cfa}
\extth{2}{\cle}{csu}
\extth{2}{\ple}{par}
\extth{2}{\cle}{cfa}
\extth{2}{\cle}{csu}
\extth{2}{\ple}{par}
\extth{2}{\cle}{cfa}
\extth{2}{\cle}{cfa}
\extth{2}{\cle}{cfa}
\extth{2}{\cle}{csu}
\extth{2}{\ple}{par}
\extth{2}{\cle}{cfa}
\extth{2}{\cle}{csu}
\extth{2}{\ple}{par}
\extth{2}{\cle}{cfa}
\extth{2}{\cle}{csu}
\extth{2}{\ple}{par}
\pgfmathparse{2.250000*\cle}
\extth{3}{\pgfmathresult}{ini}
\extth{3}{\cle}{cfa}
\extth{3}{\cle}{csu}
\extth{3}{\ple}{par}
\extth{3}{\cle}{cfa}
\extth{3}{\cle}{csu}
\extth{3}{\ple}{par}
\extth{3}{\cle}{csu}
\extth{3}{\ple}{par}
\extth{3}{\cle}{cfa}
\extth{3}{\cle}{csu}
\extth{3}{\ple}{par}
\extth{3}{\cle}{cfa}
\extth{3}{\cle}{csu}
\extth{3}{\ple}{par}
\pgfmathparse{9.875000*\cle}
\extth{4}{\pgfmathresult}{ini}
\extth{4}{\cle}{csu}
\extth{4}{\ple}{par}
\extth{4}{\cle}{cfa}
\extth{4}{\cle}{csu}
\extth{4}{\ple}{par}
\extth{4}{\cle}{cfa}
\extth{4}{\cle}{csu}
\extth{4}{\ple}{par}
\extth{4}{\cle}{cfa}
\extth{4}{\cle}{csu}
\extth{4}{\ple}{par}

%% file: inp-chalmers/ex-lemma-4.tex
\node [left] at (0,\yt[1]) {\thr{0}};
\node [left] at (0,\yt[2]) {\thr{1}};
\node [left] at (0,\yt[3]) {\thr{2}};
\node [left] at (0,\yt[4]) {\thr{3}};
\pgfmathparse{2.500000*\cle}
\edef\ple{\pgfmathresult}
\pgfmathparse{0.000000*\cle}
\extth{1}{\pgfmathresult}{ini}
\extth{1}{\cle}{csu}
\extth{1}{\ple}{par}
\extth{1}{\cle}{cfa}
\extth{1}{\cle}{csu}
\extth{1}{\ple}{par}
\extth{1}{\cle}{cfa}
\extth{1}{\cle}{csu}
\extth{1}{\ple}{par}
\extth{1}{\cle}{csu}
\extth{1}{\ple}{par}
\extth{1}{\cle}{cfa}
\extth{1}{\cle}{csu}
\extth{1}{\ple}{par}
\extth{1}{\cle}{cfa}
\extth{1}{\cle}{csu}
\extth{1}{\ple}{par}
\extth{1}{\cle}{cfa}
\extth{1}{\cle}{csu}
\extth{1}{\ple}{par}
\pgfmathparse{0.125000*\cle}
\extth{2}{\pgfmathresult}{ini}
\extth{2}{\cle}{cfa}
\extth{2}{\cle}{csu}
\extth{2}{\ple}{par}
\extth{2}{\cle}{cfa}
\extth{2}{\cle}{csu}
\extth{2}{\ple}{par}
\extth{2}{\cle}{cfa}
\extth{2}{\cle}{csu}
\extth{2}{\ple}{par}
\extth{2}{\cle}{csu}
\extth{2}{\ple}{par}
\extth{2}{\cle}{cfa}
\extth{2}{\cle}{csu}
\extth{2}{\ple}{par}
\extth{2}{\cle}{cfa}
\extth{2}{\cle}{csu}
\extth{2}{\ple}{par}
\extth{2}{\cle}{cfa}
\pgfmathparse{1.875000*\cle}
\extth{3}{\pgfmathresult}{ini}
\extth{3}{\cle}{cfa}
\extth{3}{\cle}{csu}
\extth{3}{\ple}{par}
\extth{3}{\cle}{cfa}
\extth{3}{\cle}{csu}
\extth{3}{\ple}{par}
\extth{3}{\cle}{cfa}
\extth{3}{\cle}{cfa}
\extth{3}{\cle}{cfa}
\extth{3}{\cle}{cfa}
\extth{3}{\cle}{cfa}
\extth{3}{\cle}{csu}
\extth{3}{\ple}{par}
\extth{3}{\cle}{cfa}
\extth{3}{\cle}{csu}
\extth{3}{\ple}{par}
\extth{3}{\cle}{cfa}
\extth{3}{\cle}{csu}
\extth{3}{\ple}{par}
\pgfmathparse{11.250000*\cle}
\extth{4}{\pgfmathresult}{ini}
\extth{4}{\cle}{csu}
\extth{4}{\ple}{par}
\extth{4}{\cle}{csu}
\extth{4}{\ple}{par}
\extth{4}{\cle}{cfa}
\extth{4}{\cle}{csu}
\extth{4}{\ple}{par}
\extth{4}{\cle}{cfa}
\extth{4}{\cle}{csu}
\extth{4}{\ple}{par}

%% file: chalmers-aggregate.tex


\subsection{Introduction}

\subsubsection{Background}

Nowadays due to the fact that multiprocessors can reach high levels of
performance and efficiency, they are employed in almost all areas of
engineering. The increase in number of transistors was until recently
a widely used approach in order to improve performance. This trend
introduced power consumption issues, additional design complexity of
computing systems and wire delays because of complicated designs
adopted. Due to all these aforementioned issues arising, a shift has
been observed towards multicore systems, improving in this way
parallelism and performance \cite{fastforward}.




Having multiple processors working in parallel on the same or
different tasks, translates into multiple processes trying to
communicate and synchronize, a trend that provides an interesting
challenge especially when space availability and energy consumption
are limited resources, as is the case for embedded systems.


The data structures that are used play essential role for
the communication to become more efficient. In the context of
streaming applications, data flows from one stage to the other via data
structures in order to be processed and used further by an
application. The way this communication takes place defines the
effectiveness of the application, the level of difficulty it can deal
with and its applicability to crucial projects with major impact for
the society.

Extensive research on how to increase performance of a computing
system has already been conducted within the High Performance
Computing (HPC) community and the results and solutions obtained match
the needs of the embedded systems
field \cite{message_passing}. Furthermore power consumption becomes a
major issue in HPC as the energy cost of a supercomputer facility's
operation after some years almost equals the cost of the hardware
infrastructure.

Due to the fact that power consumption consists one of the main
constraints the HPC community will have to deal with in future systems
and as this is an important issue that embedded systems try to
confront, it is the turn of HPC to adopt solutions utilized by
embedded systems \cite{performance_and_power_consumption}.

So far embedded platforms utilize mutual exclusion or interrupt
handling in order to achieve synchronization, while lock-based and
lock-free approaches are adopted by the HPC community.  As there is a
convergence between the two fields and one can benefit from the other,
it is interesting to search how solutions provided by the HPC field
can be adjusted to embedded multicore platforms in this context and
how much improvement we can get by such an initiative. Additionally
the improvement in power consumption that we can get out of such an
investigation can be useful to HPC in the future.




The work presented here aims to make a research on data structures
suited for embedded systems and investigate trade-offs between
different implementations in terms of energy consumption, memory
utilization and performance. Through this investigation the focus will
be on data streaming applications implementing multiway aggregation of
the received data.

Although efficient data structures for a concurrent environment have
been studied extensively, the issue of appropriate data structures for
data streaming applications has been
neglected~\cite{Cederman2013}. Concurrent data structures play a major
role between aggregation stages, through the parallelism and the load
balancing role that they can offer in this kind of applications. For
this reason such an application will be developed and based on the
research conducted on concurrent data structures, an efficient
solution providing lower latency, bigger throughput and energy
efficiency at the data aggregation function of the application will
try to be achieved.


We examin already existing algorithms and solutions of concurrent data
structures, with specific interest in energy, space and performance
trade-offs. The shared data structure is the queue as it consists one
of the most widely used data structures in embedded applications. The
algorithms analyzed provide solutions for 
the Single-Producer-Single-Consumer (SPSC) problem.

The data streaming application runs on Myriad 2 platform and
evaluation is realized in terms of how many messages containing
application data can be processed each millisecond for given fixed
workloads. Furthermore power consumption is measured in order to
evaluate how many Joules are needed for the processing of a message.

Next section will introduce latest researches conducted on the
subject. Section~\ref{chag-ch2} explains the type of application this
work takes into consideration and how concurrent data structures are
used within such kind of applications. Following,
Section~\ref{chag-ch3} gives a brief overview of the platform on which
evaluation will be conducted, together with the issues concerning
memory handling when cache memory is used. Subsequently
Section~\ref{chag-ch5} contains theory concerning algorithms for
concurrent data structures, essential to keep up with the rest of the
document.
Section~\ref{chag-ch7} continues with the evaluation
results of the chosen algorithms in order to observe their performance
on the platform before they are used in the final application whose
evaluation comes right after in Section~\ref{chag-ch8}. Finally, a
discussion on the results and some final thoughts are provided in
Section~\ref{chag-ch9}.

\subsubsection{Related Work}

In 2013 Cederman \etal~\cite{Cederman2013} investigate about the
neglected field of concurrent data structures on the context of
efficiency in data streaming aggregation. As it has already been
mentioned, data structures play a major role between aggregation
stages and parallelism in some of the stages is a challenging issue to
address. Through this work it is shown that for this type of
applications lock-based or lock-free approaches do not matter as much
as the data structure itself that is used. New types of data
structures have been implemented giving better throughput and less
latency than already existing queue based approaches.

In 2014 Papadopoulos \etal~\cite{message_passing} try to look
into the future where the number of cores per chip will be
increased. As currently used lock based techniques for synchronization
between contending threads over shared data has some disadvantages and
do not scale well as the number of cores increases (due to increased
contention), lock-free solutions 
introduced by the HPC field are adjusted
for use in a multicore embedded platform in order to find out how much
we can improve in terms of performance depending on the solution
tested. The shared data structure used for this work is the queue. As
a result of the investigation $29.6\%$ increase in the performance of
the platform was obtained.

Again in 2014
Papadopoulos \etal~\cite{performance_and_power_consumption}
conduct an investigation on concurrent queue implementations inspired
by the HPC domain adjusted to an embedded multicore platform. This
time along with performance evaluation, power consumption is taken
into consideration and $6.8\%$ less power dissipation is achieved while
the lock free implementations provide lower execution times by
$28.2\%$ compared to lock based approaches. As mentioned to this work,
as the number of cores per chip will increase in the future, lock free
solutions will be more attractive and there is plenty of room for
improvement.


\subsection{Data streaming aggregation}
\label{chag-ch2}
In this section multiway data streaming aggregation is explained in
order to further understand the type of problem we aim to
investigate and give a solution to.

\subsubsection{Data streaming}

Data streaming constitutes a new paradigm which processes incoming information of a system in real time instead of following the classic way of storing the received data in order to process them later on. This need is imperative nowadays as the amount of data that needs to be processed on a daily basis by a contemporary system can be really big rendering solutions that abide by the store-and-process paradigm non-practical.

\subsubsection{A Stream and Multiway Streams }
 
 A stream is a flow of incoming tuples containing fields related to data that need to be communicated by an application along with a timestamp provided by the producer of a tuple \cite{aggregation}. In the context of a multiway aggregation, multiple input streams are sent to an aggregator which processes them in order to produce a deterministic output depending on what the application wants to extract from the given data.

\subsubsection{Aggregation}

An example of an aggregation application would be smart metering data out of which an aggregator sums up the amount of Watts that have been consumed by a house or a whole district. 
An aggregator may be stateful or stateless depending on the nature of an application and whether it needs an aggregator to hold some kind of state throughout the whole processing phase. 

In case of stateful aggregators, time is divided in windows within
which state is held. Windows are set either according to time or
according to tuples. They have a certain size declaring the amount of
time or incoming tuples for which they are valid and they also have an
advance parameter which sets the boundaries of the next window
depending on the previous one.

As an example, if tuples of last ten minutes are grouped together every 5 minutes this means that we have windows of size ten and advance five. The windows that would be created in this way would be $[0,10)$, $[5,15)$, $[10,20)$ and so on. The same holds for tuple based approaches where for example ten last tuples are grouped together every five incoming tuples.

According to \cite{aggregation} the functionality of an aggregator consists of four main stages :

\begin{enumerate}
  \item \textit{Add stage}: Fetch tuples from each input stream.
  \item \textit{Merge stage} : Merge and sort fetched tuples according to timestamp.
  \item \textit{Update stage} : Update the state of windows a tuple contributes to.
  \item \textit{Output stage} : Forward output tuples to the next aggregation stage.
\end{enumerate}

In this work, the focus is on the third and fourth stages of an aggregator. The first and second stages are out of scope and are not taken into consideration as the application deployed simulates already merged and sorted data.

\subsubsection{Concurrent Data Structures}

Under the context of multiway data streaming aggregation, concurrent data structures are used between the different stages of the aggregation process in order for the communication between different parties to be achieved.

The data structures used need to provide as much parallelism as
possible, ease the communication of tuples between different stages of
the process and load balance the workload so that all processes deal
with similar amount of workload and none is potentially choked
creating a bottleneck. Lock-free approaches 
prove to increase throughput and start arising research interest more
and more, something that led to mainstream programming languages to
incorporate implementations in their standard
libraries \cite{aggregation}.

\begin{figure}[h!]
\centering
\includegraphics[width=140mm]{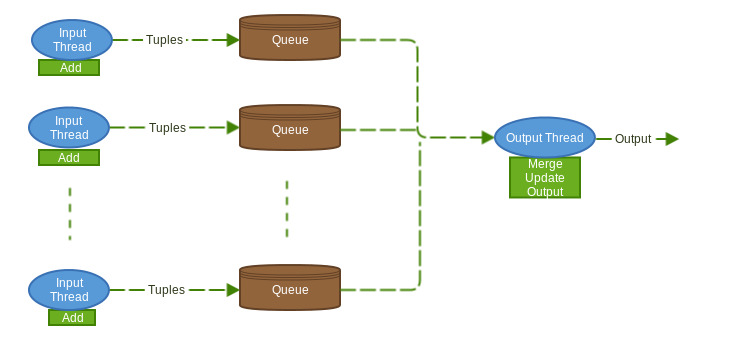}
\caption{Aggregation example \label{chag-sae}}
\end{figure}

Two typical stream processing engines are Borealis \cite{borealis} and StreamCloud \cite{streamcloud}. For multiway aggregation to be achieved in these implementations, as depicted in Figure \ref{chag-sae}, tuples from each input stream are placed in queues by multiple threads. On the other side there is one consumer thread performing the merge, update and output stages of aggregation by dequeuing each time the first tuple of each queue in order to make a decision on which tuple should be processed next. 

\subsection{Myriad2 Hardware Platform}
\label{chag-ch3}

The complete specifications of Myriad2 Movidius processor can be found
in Deliverable D4.2~\cite{EXCESS:D4.2}. We recall here important
information needed in our context.

The SHAVEs are processors designed for handling efficiently VLIW instructions,
containing registers and functional units that enable SIMD operations
and provide high parallelism and throughput. Used for performing
intensive computational tasks. SHAVES also have access to a 2KB L1
instruction cache which enables running code residing in DDR without
much delay and a 1KB L1 data cache. Additionally shaves have access to
a 256KB L2 cache memory following a write-back policy and featuring a
64-byte cache line. Applications on SHAVES can be written in Assembly
or C/C++.

CMX memory allows LEON and SHAVE processors to have low memory access
cost in contrast to DDR memory which depending on whether retrieved
data are already cached or not may incur high cost. For this reason
CMX should be preferred for application data manipulated by SHAVES.

CMX is divided in sixteen 128KB parts. Each SHAVE has a preferential
port attached to one of these parts, thus $128 \times 12 = 1536$KB are
preferentially used by SHAVES. The remaining 512KB can be used for
other general purposes such as LEON\_OS timing critical code.

A SHAVE may have access to any CMX part with the same cost. On the
other hand the resources used for routing between different CMX parts
are finite. Furthermore an access to a local CMX slice happens with
lower energy consumption and in order to achieve optimal performance,
memory should be manipulated in such a way so that a SHAVE mostly
accesses data residing in its local part.

There exists a cache coherency protocol that handles the movement of
data between the DDR memory and the L1 and L2 caches of every SHAVE.
However, due to the fact that using DDR is much slower, CMX should be
preferred. As CMX is not cached, cache coherency is not an issue any
more so the overhead incured by cache coherency protocols is
avoided. Furthermore all SHAVES have very fast access to all CMX
memory.

\begin{figure}[ht!]
\centering
\includegraphics[width=100mm]{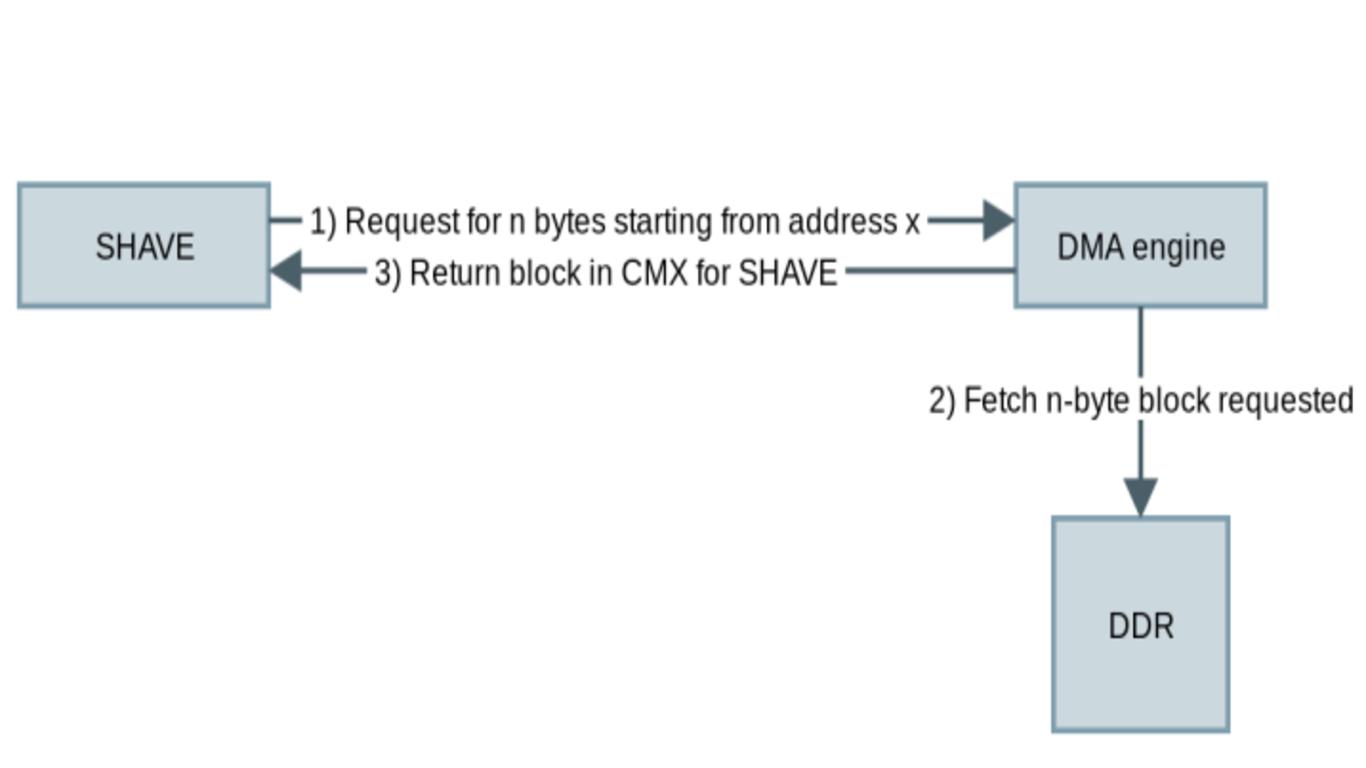}
\caption{DMA usage\label{chag-dma}}
\end{figure}

In case DDR should be used (e.g too much data to be held by CMX) a
better alternative would be to use the DMA engine which allows whole
blocks of memory to be fetched from DDR to CMX much faster than when
DDR is accessed directly and data get cached. The way DMA engine can
be used is depicted in Figure \ref{chag-dma}.

\subsection{Single-Producer-Single-Consumer algorithms}
 \label{chag-ch5}


We present here some widely-used implementations of a FIFO queue,
where a single entity can enqueue elements and another single entity
can dequeue elements.

A bounded buffer with a fixed number of slots is used and the producer
can only insert elements when the buffer is not full (in case the
buffer is static) while the consumer may consume when the buffer is
not empty. Furthermore the FIFO property needs to be guaranteed,
through which the elements removed by the consumer appear in the same
order as inserted by the producer.

In this case synchronization is a less complicated issue which can be overcome through primitive linearizable stores which are faster than hardware synchronization primitives. As CAS is not needed and there is only one producer and one consumer, the ABA problem, which is a factor contributing to the previous algorithms being slower, is not taken into consideration anymore.

\subsubsection{Lamport's algorithm}
\label{chag-lamp}
In 1983 Lamport introduces a lock-free algorithm that solves the SPSC problem \cite{lamport}. The approach uses a cyclic array as the data structure, is pretty straightforward and is depicted below in Listing~\ref{chag-lamport}.

\begin{lstlisting}[caption=Lamport's algorithm, label=chag-lamport]
    int head = 0;
    int tail = 0;
    enqueue (data) {
        while (AFTER(tail) == head); // wait, queue is full
        queue[tail] = data;
        tail = AFTER(tail);
        return 0;
    }
    dequeue (data) {
        while (head == tail); // wait, queue is empty
        data = queue[head];
        head = AFTER(head);
        return 0;
    }
 }
\end{lstlisting}

At the software level the proposed solution avoids explicit
synchronization between the processes, which communicate indirectly
through atomic read and write operations on control variables, namely
head and tail according to Listing~\ref{chag-lamport}. On the other
hand, the algorithm does not take into consideration memory and cache
coherency, a fact which renders the solution the slowest one compared
to the algorithms presented in this section.

Due to the fact that the control variables are shared, there are a lot
of cache line transfers induced as the producer and the consumer need
to manipulate these variables (either read them or write them) at
every operation, generating a lot of invalidations and cache coherency
traffic. To make things worse, in case the control variables are not
placed in separate cache lines, there is a possibility their values
get invalidated even when they are not actually changed because of
changes to a nearby memory location residing in the same cache
line \cite{batchqueue}.

Furthermore, in case the producer and the consumer work at the same
speed to nearby memory locations operations will be happening on the
same cache line, obviously leading to even more invalidations and
cache line thrashing. For this phenomenon to be avoided, the consumer
should avoid manipulating elements in cache lines that will possibly
change (see temporal slip in Section~\ref{chag-ff}) \cite{batchqueue}.
Finally the approach does not work for systems running on weaker
memory consistency models than sequential consistency and additional
fence operations would be needed if this would be the case,
introducing more overhead.

\subsubsection{Fast Forward algorithm}
\label{chag-ff}
In 2008 Giaconomi \etal \cite{fastforward}, design an efficient
algorithm in order to provide better throughput in a pipeline parallel
application where packet processing is realized. In order for the
algorithm to be applicable to some weaker memory consistency models a
memory barrier needs to be placed before an enqueue takes place. A
pseudocode of the algorithm is provided below in
Listing~\ref{chag-fforward}.

\begin{lstlisting}[caption=Fast Forward algorithm, label=chag-fforward]
    int head = 0;
    int tail = 0;
    enqueue (data) {
        while (queue[tail] != NULL); // wait, queue is full
        queue[tail] = data;
        tail = AFTER(tail);
        return 0;
    }
    dequeue (data) {
        do {
            data = queue[head];
           }while (data == NULL); // wait, queue is empty   
        queue[head] == NULL;
        head = AFTER(head);
        return 0;
    }
 }
\end{lstlisting}

The approach is array based again and manages to decrease the amount
of cache thrashing, in comparison with Lamport's
algorithm in Listing~\ref{chag-lamp}. This is achieved through a special element
introduced, which indicates whether an array cell is or is not
available for an operation to occur. By synchronizing the processes in
that way, head and tail are manipulated only by the producer and the
consumer respectively so no invalidations occur because of these
variables.

By introducing this special element though, a constraint is imposed, due to the fact that the datatypes of elements that can be used are limited and the value used as a special element is not possible to be used by the application \cite{mcringbuffer}. A solution for this issue the designers propose, is to have an additional array containing pointers indicating whether a slot is empty or not. In case this solution is adopted, the issue that arises is that one more access is necessary in order for an operation on the data structure to occur (i.e. one access to the array holding the pointers and one access to the array holding actual data) and the memory needs of the algorithm increase.

Cache thrashing may still occur though when the producer and the consumer operate on array elements residing in the same cache line. For this reason temporal slip is suggested by the designers, according to which the consumer is delayed for as much time as needed, in order for the producer to fill one cache line. At this point it has to be noted that as the algorithm is destined for pipeline parallel applications,  in case stages of the pipeline differ significantly in duration, temporal slip needs to be checked more often, increasing the overhead incurred and eventually introducing cache line thrashing which the solution tries to avoid in the first place \cite{batchqueue}. 

Compared to Lamport's algorithm, the approach proves to be approximately $3.7$ times more efficient performance wise and exhibits more stable behaviour. 
The algorithm's performance does not seem to be affected in cases where memory fences and variations to the queue size, the workload, the core allocation or the pipeline stage duration are introduced. As far as temporal slip is concerned, the overhead is minimal (1 ns). Finally the algorithm scales well for processes that reside on the same or different chips. 

\subsection{Batch Queue algorithm}
\label{chag-bq}
In 2010 Preud'Homme \etal \cite{batchqueue}, design an array based
algorithm, which as the name implies, conscripts batch processing. The
algorithm proves to be fast and takes into consideration memory usage
and cache coherency. A pseudocode demonstrating the logic of the
algorithm is given below in Listing~\ref{chag-batch}.

\begin{lstlisting}[caption=Batch queue, label=chag-batch]
    int enq_index = 0;
    int deq_index = 0;
    bool isFull = false;
    enqueue (data) {
        queue[enq_index++] = data;
        enq_index = enq_index mod (2*N) // N equals half size of the array
        if (enq_index mod N == 0) { // half queue filled
            while (isFull);
            isFull = true; // allow consumer to consume
        }
    }
    dequeue (data) {
        int i;
        while(!isFull); // wait for producer to fill half queue 
        for (i = deq_index; i < deq_index + N; i++)  
            copy_buf[i - deq_index] = queue[i];
        deq_index = (deq_index + N) mod (2*N);
        isFull = false; // notify producer half array is emptied
 }
\end{lstlisting}

The algorithm processes a whole batch of elements at a time. The queue is split in two sub-queues where either the 
producer or the consumer works at a time. After the producer has finished enqueuing in one sub-queue it signals this 
to the consumer who can now dequeue the elements of the sub-queue while the producer keeps enqueuing to the other sub-queue.

Synchronization is achieved through one shared flag (the isFull variable in Listing~\ref{chag-batch}) in order for the swapping of the sub-queues to happen. When the producer finishes with filling one sub-queue, it sets the flag in order to notify its part is full. Subsequently it waits for the consumer to signal it emptied the other half of the queue by unsetting the flag, notifying it is finished dequeuing and the producer can start filling the sub-queue again .
 
In order to achieve optimal performance two full cache lines are
needed and the two variables holding the indexes along with the flag,
need to be placed in different cache lines, otherwise a lot of
unnecessary invalidations and cache coherency traffic will take place
decreasing performance at a substantial level. The flag ensures the
two processes work in different cache lines, playing the role temporal
slip does in Listing~\ref{chag-fforward}.

The algorithm provides high throughput and has small memory needs. On the other hand it exhibits increased latency due to batch processing which increases overhead and requires high level of communication through a shared flag in order to work. So the algorithm is suitable for applications with high L1 cache usage and relaxed latency requirements.

Performance wise an improvement of up to 10 times is observed in relation to Lamport's algorithm and a word sized data element can be sent through the queue within $19-125$ cycles and $12.5-40.6$ nanoseconds depending on L1 cache pressure.

\subsection{MCRingBuffer}
Again in 2010, Patrick PC Lee \etal \cite{mcringbuffer}, propose an algorithm, aiming to keep up with the bandwidth of the communication link for packet processing purposes (line rate packet processing).  The algorithmic design is presented below in pseudocode \ref{chag-mcring}.

\begin{lstlisting}[caption=MCRingBuffer, label=chag-mcring]
    /* Variable definitions */
    char cachePad0[CACHE LINE];
    
    /* shared control variables */
    volatile int read;
    volatile int write;
    char cachePad1[CACHE LINE - 2 * sizeof(int)];
 
    /* consumer local variables */
    int localWrite;
    int nextRead;
    int rBatch;
    char cachePad2[CACHE LINE - 3 * sizeof(int)];
    
    /* producer local variables */
    int localRead;
    int nextWrite;
    int wBatch;
    char cachePad3[CACHE LINE - 3 * sizeof(int)];
 
    /* constants */
    int batchSize;
    char cachePad4[CACHE LINE -  sizeof(int)];

    /* Enqueue function */
    enqueue (data) {
        int afterNextWrite = AFTER(nextWrite);
        if (afterNextWrite == localRead) {
            while (afterNextWrite == read); // wait queue is full
            localRead = read;
        }
        queue[nextWrite] = data;
        nextWrite = afterNextWrite;
        wBatch++;
        if (wBatch >= batchSize) {
            write = nextWrite;
            wBatch = 0;
        }
        return SUCCESS;
    }
 
    /* Dequeue function */
    dequeue (data) {
        if (nextRead == localWrite) {
            while (nextRead == write); // wait queue is empty
            localWrite = write;
        }
        data = queue[nextRead];
        nextRead = AFTER(nextRead);
        rBatch++;
        if (rBatch >= batchSize) {
            read = nextRead;
            rBatch = 0;
        }
        return SUCCESS;
    }
\end{lstlisting}

The algorithm is lock-free and tries to take advantage of cache locality in order for access in shared data to become more efficient. As it is shown in pseudocode \ref{chag-mcring}, control variables are carefully laid out in memory, separated by paddings wherever needed in order for cache invalidations to be avoided. The approach of placing variables in memory in such a way is called \textit{cache line protection}. 

By employing this strategy false sharing between the producer and the consumer is avoided. As a consequence, when one of the processes invalidates a whole cache line by changing one variable, the variables local to the other one are not affected as it is ensured that they reside in different cache lines.

Aside from cache line protection which reduces cache coherency traffic, another important design choice with impact on the performance of the algorithm is the \textit{batch updates} of shared control variables. 
Through the local variables to each process, the shared control variables (read and write in the pseudocode provided) are sparsely changed. Each process consults its variable that holds what the process thinks about the progress 
of the other process and only reads the shared variable holding the actual progress of the other process when no 
more progress can be made. The shared variables are updated by each process after a predetermined number of steps 
which separate the buffer in smaller parts. 

As it is obvious this approach of batch updates can lead to wrongly not inserting data to the buffer even if the buffer is not full (and not dequeuing data from the buffer even if data are available) due to the fact that the shared variables are not updated yet. This may happen when the incoming rate of data elements is too small. In order for this scenario to be avoided, the authors suggest that the producer periodically inserts unused elements to the queue, discarded by the consumer, used to make the shared variables be updated and have progress. By making use of batch updates, the same effect as in the temporal slip  
approach (see Section~\ref{chag-ff}) or the flag used in BatchQueue (see Section~\ref{chag-bq}) can be achieved without needing any scheduling of the producer and the consumer.

As control and data elements are not merged like in the fast forward algorithm, the approach supports generic datatypes and does not impose any limitation on the datatypes than an application may use, rendering it completely independent. On the other hand, compared to the other algorithms, the memory needs are increased as six variables are used by the processes in order to synchronise. The approach works with systems supporting the sequential consistency memory model while in order to be adopted to platforms with weaker memory consistency, memory barriers need to be used.

The algorithm outperforms Lamport's approach especially when batch updates are introduced. The same holds for the fast forward algorithm for which though temporal slip has not been used. It does not seem to perform well in case the buffer size is small as it is probable that the producer cannot enqueue elements. In general, through this approach, throughput of conventional lock-free solutions is improved by up to 5 times and cache misses are considerably reduced. Furthermore processing throughput is augmented by up to 5.2 times for the single-threaded case and 1.9 times for the multi-threaded case.

\subsection{Algorithms Evaluation}
\label{chag-ch7}
Due to constraints of the Myriad2 platform (not providing CAS
functionality) and the pipelined nature of the aggregation phase of a
data streaming application, the algorithms this evaluation focuses on
are the four SPSC algorithms that have already been discussed in the
previous section. From now on FastForward is mentioned as FF,
BatchQueue as BQ and MCRingBuffer as MCR. The performance evaluation
of these algorithms along with Lamport's design, is conducted in order
to get an insight on the way they behave on the platform.

In general, in order to take advantage of the platform specifications
and achieve optimal performance, issues already discussed in
Section~\ref{chag-ch3} are taken into consideration. Thus, if
possible, queues are placed in CMX (particularly in the CMX slice of
one of the SHAVES that manipulate the queue) and each shared variable
is placed in the CMX slice of the SHAVE that makes changes to it.  If
the queue size exceeds CMX capacity, the queue is placed into DDR.

Leon OS boots and starts Leon RT from which it awaits a signal in
order to initiate a task that performs the energy measurement of the
platform. Leon RT sends this signal right before it turns on the
SHAVES used for the evaluation. After signaling the beginning of the
power measurement it starts measuring execution time and turns on the
SHAVES. Once the SHAVES finish their executions Leon RT stops
measuring the execution time and signals to Leon OS to stop the power
measurement.

The first two SHAVES (SHAVE0 and SHAVE1) are used for the evaluations
discussed in this section. SHAVE0 plays the role of the producer while
SHAVE1 runs the application of the consumer. In case the queue used
for communication is placed in CMX it is stored in the slice of the
producer (SHAVE0).

For the purposes of the evaluation, by one operation a pair of
enqueue/dequeue is meant and the percentages used for comparison,
provide an approximation of the differences between the algorithms
resulted from the average calculation over all different cases
studied. For each case, an algorithm is evaluated five times and the
average of these evaluations is used as the result in the final
computation of the percentages provided.

\subsubsection{Buffer Size Evaluation}
\label{chag-buffSizeAnalysis}

The behaviour of the algorithms depending on the size of the queue at question is examined in order to investigate how performance is affected.

In the following evaluation in case the buffer size is greater than $2048$ elements the queue is placed in DDR memory as CMX's capacity is not enough. Furthermore DMA engine is not used and data are directly accessed from DDR.

Prior to running the tests the buffers are already filled with data. In case buffer capacity is $128$ elements, the buffer is half filled while for all the other occasions buffers are filled with $150$ elements. In this way temporal slip is achieved (producer begins $128$ or $150$ elements ahead of the consumer) for all the algorithms as the producer and the consumer are working on different cache lines when buffer is placed in DDR. 

Batches are used by BQ (by design batches are equal to half the size of the queue) but not for MCR. By not applying batches to MCR it means that the shared variables are updated at every operation but still they are sparsely consulted by the processes (every $128$ or $150$ operations). Lamport and FF do not incorporate batches in their implementations by design.

Because of the nature of CMX memory, temporal slip does not need to be
taken into consideration when queues are placed in CMX as cache line
thrashing is not an issue anymore. Each element of the queue is $12$
bytes (evaluation on the impact of the size of a data element is
investigated later in Section~\ref{chag-dataElementEval}).

\begin{figure}[ht!]
\centering
\includegraphics[width=140mm]{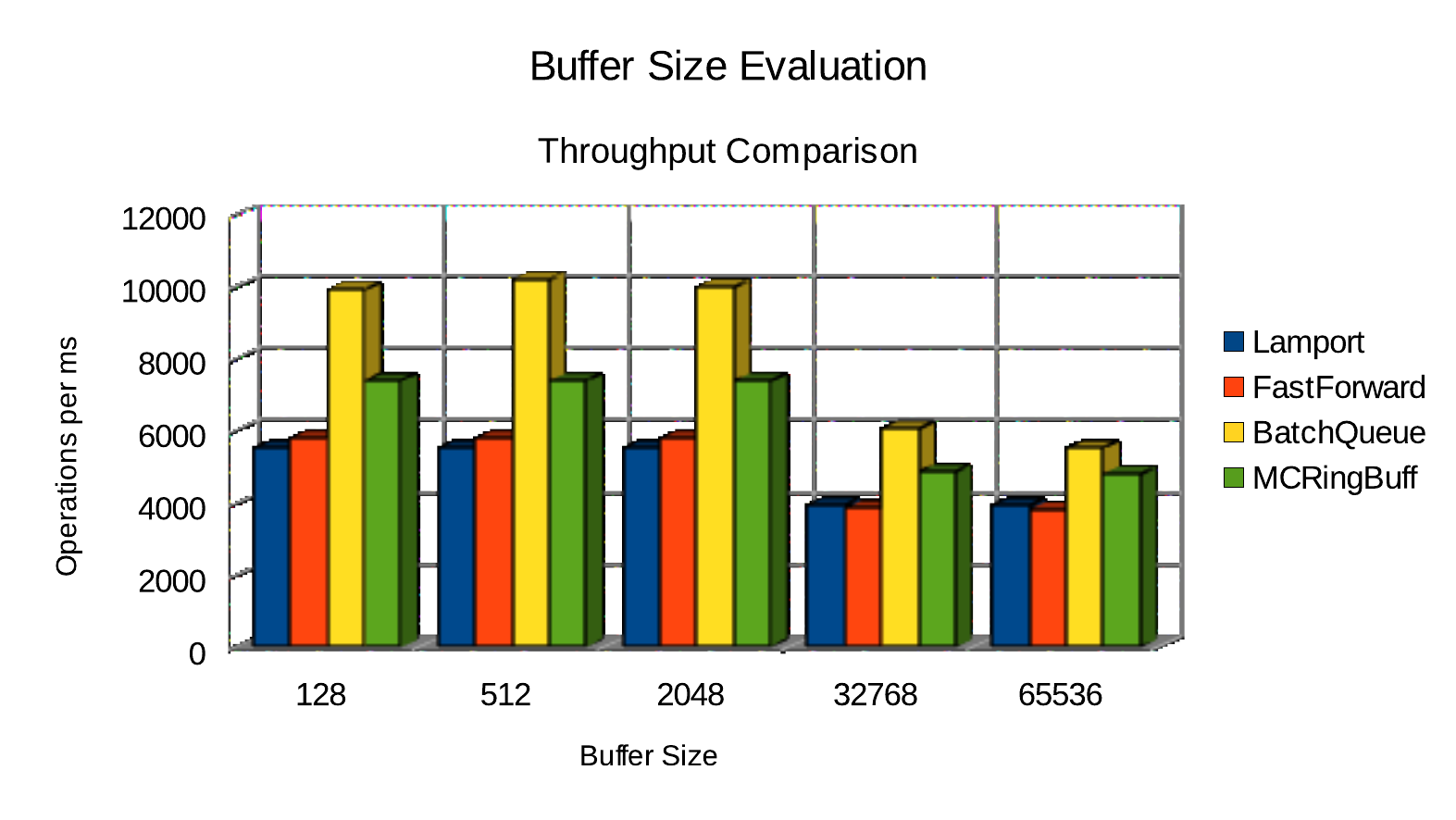}
\caption{Throughput comparison depending on the size of the queue used\label{chag-BuffSizeVelocity}}
\end{figure}

BQ proves to provide better throughput than any other algorithm for every single occasion (Fig. \ref{chag-BuffSizeVelocity}). Specifically, the algorithm achieves better throughput by $24\%$ than MCR, $40\%$ than FF and $42\%$ than Lamport. This leads us to the conclusion that processing data in batches and not immediately when they are available provides much better performance. As the producer and the consumer are constantly working independently, needing to synchronize only when they are done with their batch (a synchronization made through just a single variable), a much more efficient design is achieved compared to all other algorithms which perform checks on synchronization variables much more frequently. 

On the down side of using batches, in case the queue is placed in DDR,
it seems that BQ is the only algorithm whose throughput is affected as
the size increases. This happens because the bigger the queue the more
the latency gets increased. A SHAVE has access to a 1 KB L1 data
cache. As a consequence performance is affected by how long data can
be kept in the L1 cache. As cache misses occur, processing delay is
augmented and due to the fact that long batches are used (half the
size of the queue) the processing of the tuples gets slowed down as
well. As a consequence it can clearly be seen that throughput is
decreased as the size of the queue increases. For all other algorithms
that do not make use of batches, throughput does not seem to be
affected.

It also needs to be noted that the input rate of elements is constant
and the producer runs at the same speed as the consumer. This is not
the case though for the different stages of aggregation that we
investigate and it would not be surprising if the results shown here
would not apply later under the new circumstances.

Additionally it can be seen that FF is less efficient than any other algorithm when the queue is placed in DDR. It performs even worse than Lamport's algorithm even though it surpasses it when the queue is placed in CMX. This happens as FF couples control with data elements (i.e. a data element is checked in order to decide whether an enqueue or a dequeue should happen). While all other algorithms synchronize processes through variables residing in CMX, FF needs to access DDR or a cache in order to synchronize the producer and the consumer, rendering it slower.

Although FF is slower than Lamport when queue is placed in DDR it is more energy efficient by a slight difference although this is not always the case when the queue resides in CMX. For all the test cases, the differences between the two are minor as when CMX is used no caches are involved for cache thrashing to occur from Lamport and when DDR is used and caches are involved, due to the fact that the queues are already filled with data prior to execution, Lamport avoids cache thrashing as well. 

\begin{figure}[ht!]
\centering
\includegraphics[width=140mm]{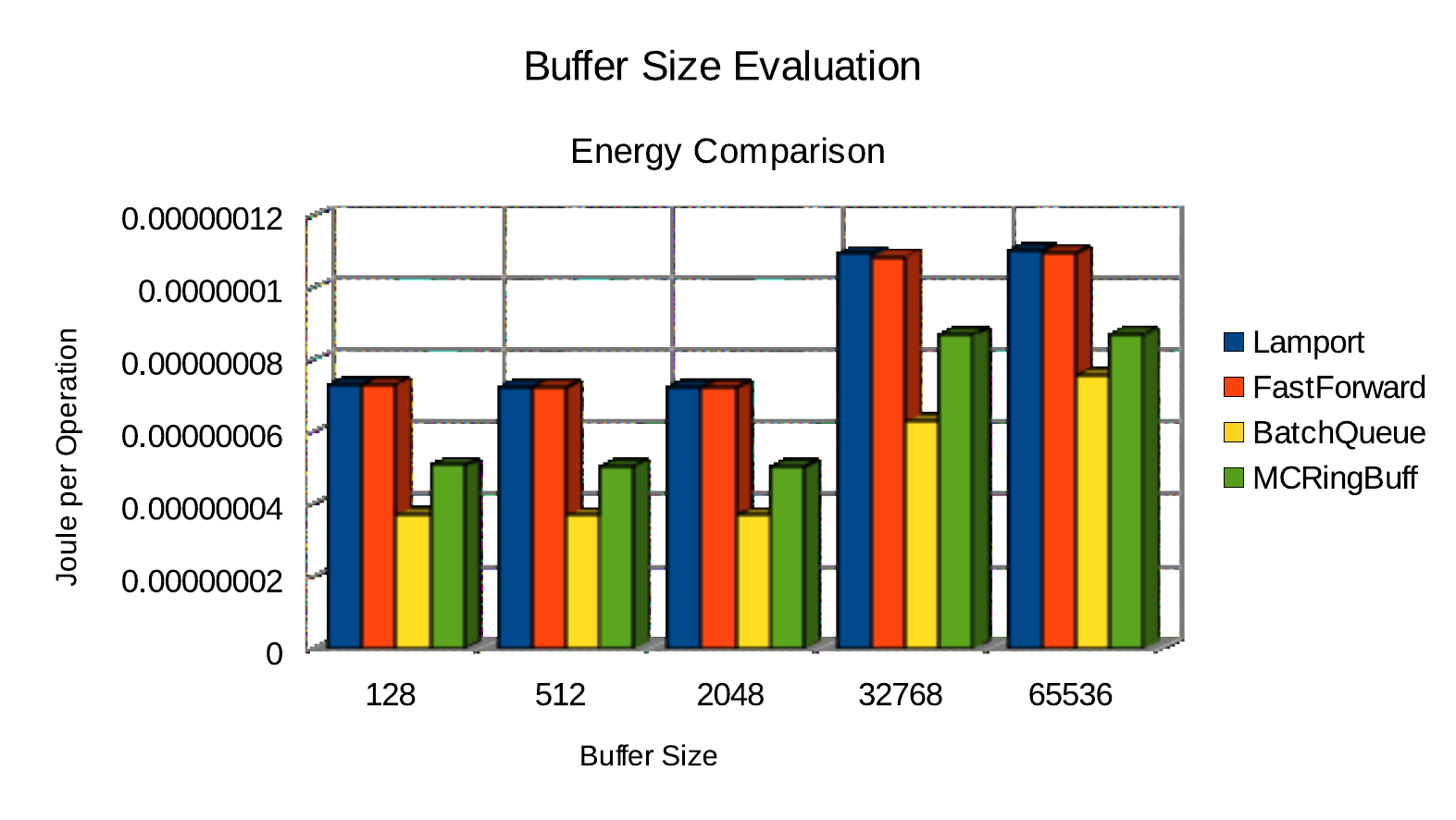}
\caption{Energy comparison depending on the size of the queue used\label{chag-BuffSizeEnergy}}
\end{figure}

BQ also prevails in the energy consumption evaluation, constituting the most energy efficient solution (Fig. \ref{chag-BuffSizeEnergy}). It is more energy efficient by a percentage of $23\%$ from MCR and $43\%$ from Lamport and FF. As for the throughput evaluation earlier, it can be seen that energy efficiency is affected by the size of the queue when it resides in DDR for the same reason that throughput is affected as well. Cache misses mean slower processing of tuples thus more busy waiting by the processes and consequently increased energy needs. The rest of the algorithms that do not make use of batches do not seem to be affected and demonstrate a stable behaviour either when CMX or DDR is used.

Overall, with the exception of BQ when the queue is placed in DDR, the algorithms exhibit a stable behaviour on the platform that is not affected by the size of the buffer used for communication. Furthermore the inferiority of DDR over CMX is clearly shown as throughput decreases when queues are placed in DDR while energy consumption increases. BQ seems to be influenced the most when the queue is placed in DDR as its throughput is decreased by $42\%$ while its energy consumption is increased by $47\%$. On the other hand Lamport is the algorithm affected the least, as its throughput is decreased by $29\%$ while its energy needs are augmented by $34\%$. Throughput for FF and MCR is affected at a similar degree (around $34\%$ slower) while for the former, energy needs increase by $33\%$ and for the latter energy consumption increases by $42\%$. 

\subsubsection{DMA engine test}

Subsequently DMA engine is used when queues are placed in DDR in order to decide whether it is suitable to use it for the algorithms tested, compared to accessing directly data from DDR. The way DMA is designed to be used is already discussed in Figure \ref{chag-dma}.

Tests were run for FF and MCR in order to decide what kind of applications DMA engine favours the most (batch updates or single element update).

\begin{figure}[ht!]
\centering
\includegraphics[width=140mm]{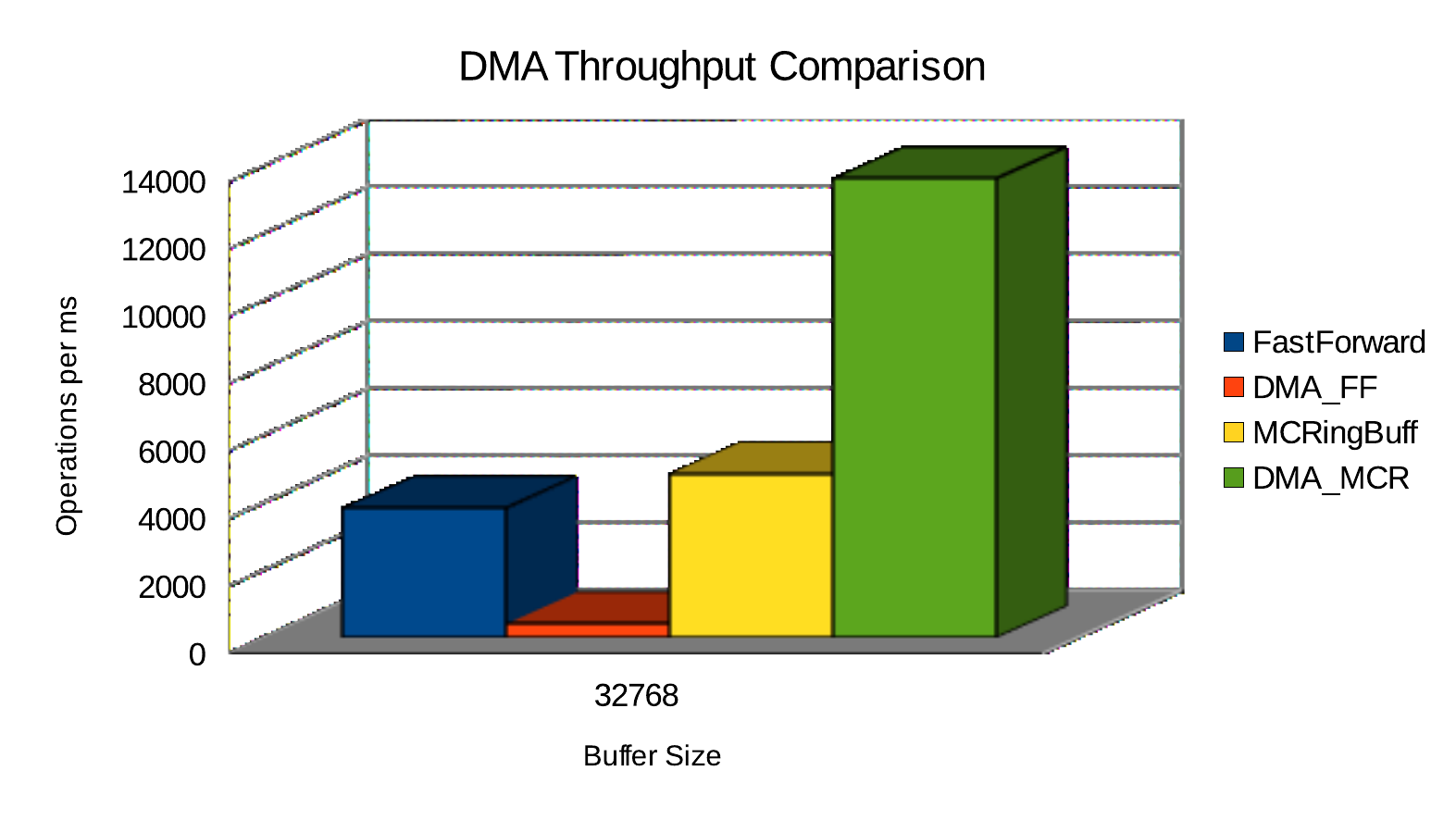}
\caption{Throughput comparison when DMA used\label{chag-DMA_velocity}}
\end{figure}

\begin{figure}[ht!]
\centering
\includegraphics[width=140mm]{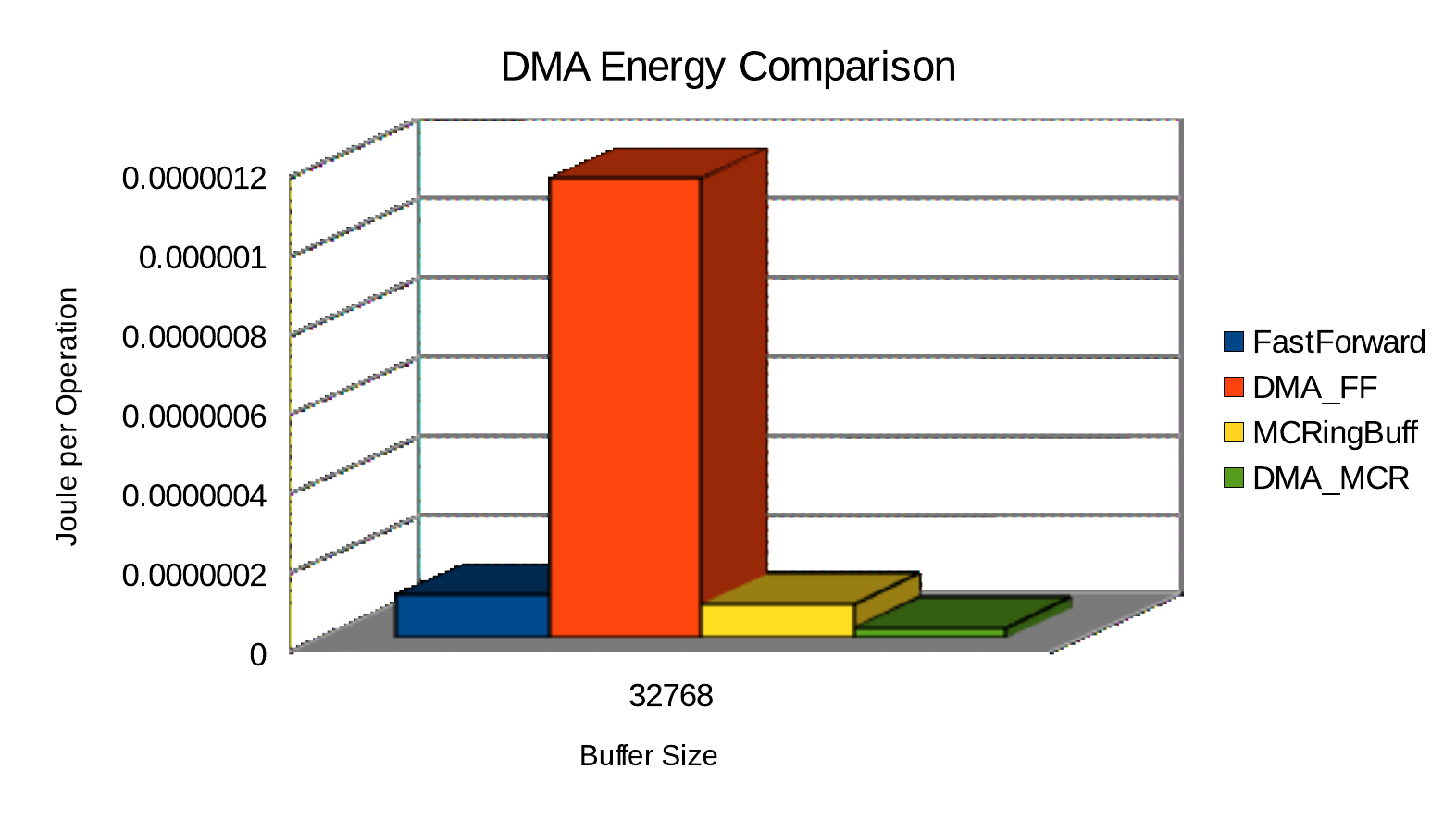}
\caption{Energy comparison when DMA used\label{chag-DMA_energy}}
\end{figure}

DMA is not practical when used for single data element access (i.e Lamport, FastForward), as it is invoked every single time an element needs to be read in order to decide whether an enqueue or a dequeue should happen (FastForward). This is a use case for which DMA is not destined to be used in the first place and the results provided in Figures \ref{chag-DMA_velocity} and \ref{chag-DMA_energy} enforce this fact. FF algorithm's throughput is decreased by $89\%$ when DMA is used while energy consumption is increased by $91\%$.

\begin{figure}[ht!]
\centering
\includegraphics[width=140mm]{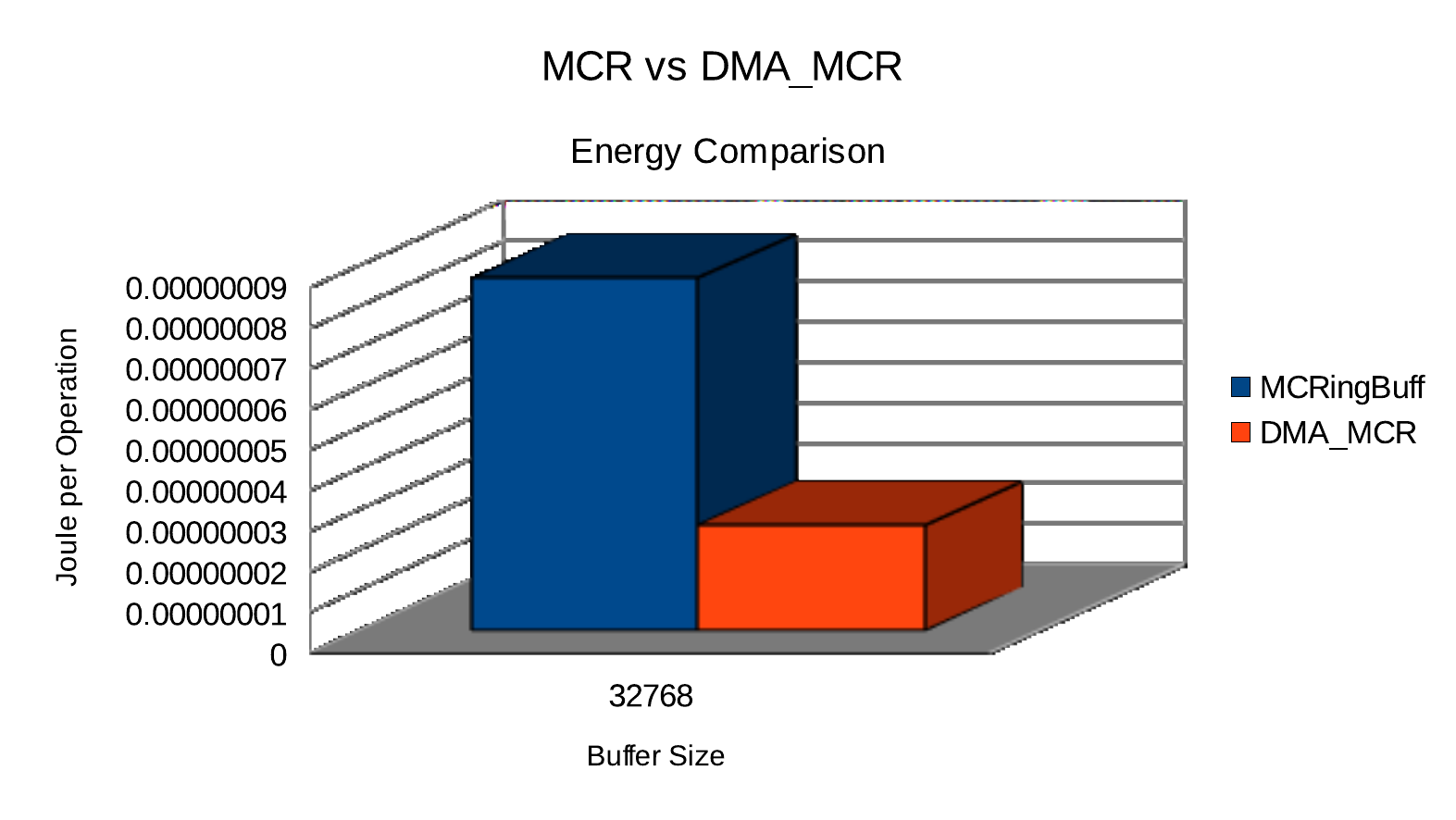}
\caption{Energy comparison between MCR and DMA\_MCR \label{chag-mcr_vs_dma_mcr}}
\end{figure}

On the other hand when batches of elements are fetched (for MCRingBuffer where 128 elements are fetched at a time), MCR design using DMA outperforms the case where data are accessed directly from DDR, performance and energy wise (Fig. \ref{chag-DMA_velocity} and \ref{chag-mcr_vs_dma_mcr}). Throughput is increased by $66\%$ while energy needs are diminished by $70\%$. The larger the blocks requested by DMA are, the more efficiency is increased.

\subsubsection{Data Element Size Evaluation}
\label{chag-dataElementEval}
The performance of the algorithms is evaluated depending on the size of a data element they store. For these tests the buffer size is fixed to $128$ elements and buffers are half filled prior to execution. Buffers always reside in CMX (the producer's slice) so temporal slip is not needed as the memory manipulated is not cached. Batches are only used by BQ as it incorporates them by design (a batch equals half the capacity of the queue thus 64 elements for these experiments). Batches are not used for MCR.

\begin{figure}[ht!]
\centering
\includegraphics[width=140mm,height=60mm,keepaspectratio]{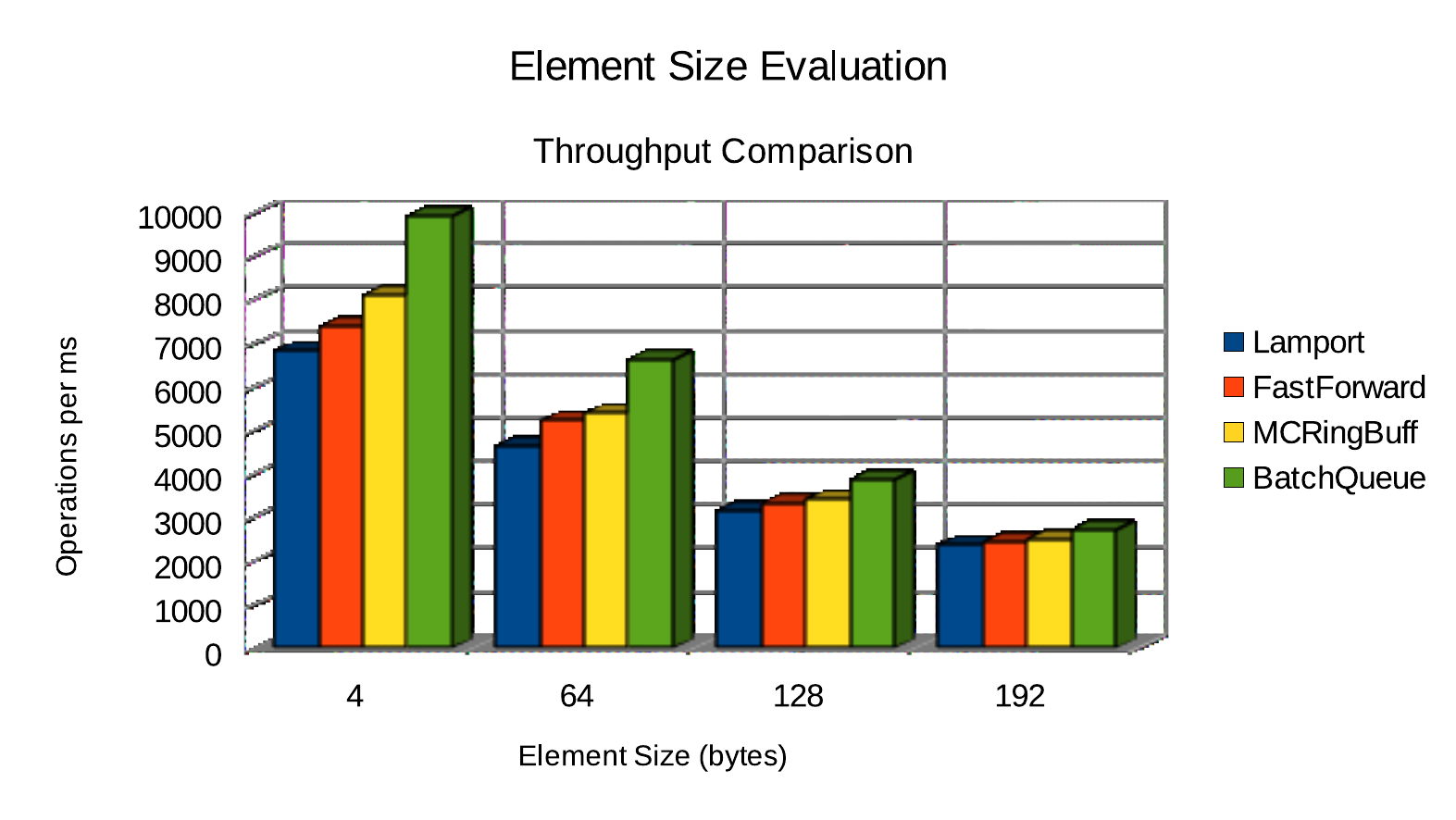}
\caption{Throughput comparison for variable data element size  \label{chag-elementSizeVelocity}}
\end{figure}

Once more BQ is superior than all other algorithms, MCR comes second, with FF and Lamport following (Fig. \ref{chag-elementSizeVelocity}). On average, BQ provides performance enhancement of the size of $27\%$ compared to Lamport, $21\%$ compared to FF and $16\%$ compared to MCR. It is obvious that as the size of a data element increases performance of the algorithms converges as more calculations are performed before operations take place. As a consequence differences in performance are smaller compared to the buffer size evaluation previously.

\begin{figure}[ht!]
\centering
\includegraphics[width=140mm,height=60mm,keepaspectratio]{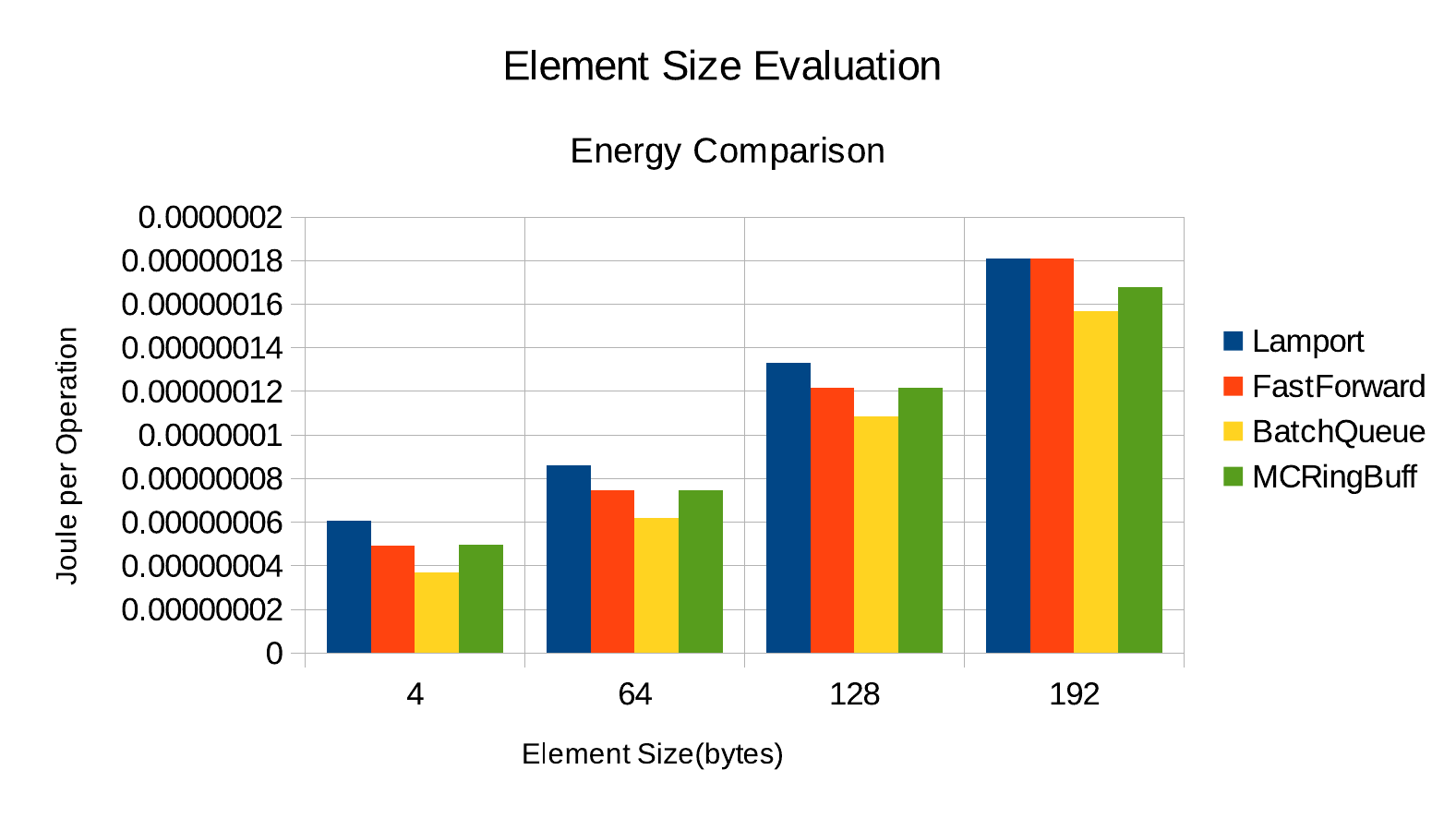}
\caption{Energy comparison for variable data element size \label{chag-elementSizeEnergy}}
\end{figure}

With respect to energy, FF is the only algorithm that seems to be affected as the element size increases. Until the 64-byte element test, it is more energy efficient than MCR something that does not hold for the 128-byte element test with the margin increasing even more for the 192-byte element test. All other algorithms exhibit stable behaviour as the element size increases (Fig. \ref{chag-elementSizeEnergy}). In particular BQ is more energy efficient than Lamport by $21\%$, FF by $15\%$ and MCR by $12\%$. Again the small difference in energy consumption comes from the fact that the algorithms converge as the size of the data elements increases. 

\clearpage

\subsubsection{Algorithms evaluation as the number of tuples to be processed increases}

The processing speed and energy consumption of the algorithms is compared as the number of elements that need to be processed by the system increases. Buffer capacity is fixed to $128$ elements while a data element is $12$ bytes. As for the previous evaluation, batches are only incorporated in the BQ implementation (64 elements batches). Again the queue resides in the CMX slice of the producer, thus temporal slip does not constitute an issue for these tests. 

\begin{figure}[ht!]
\centering
\includegraphics[width=140mm]{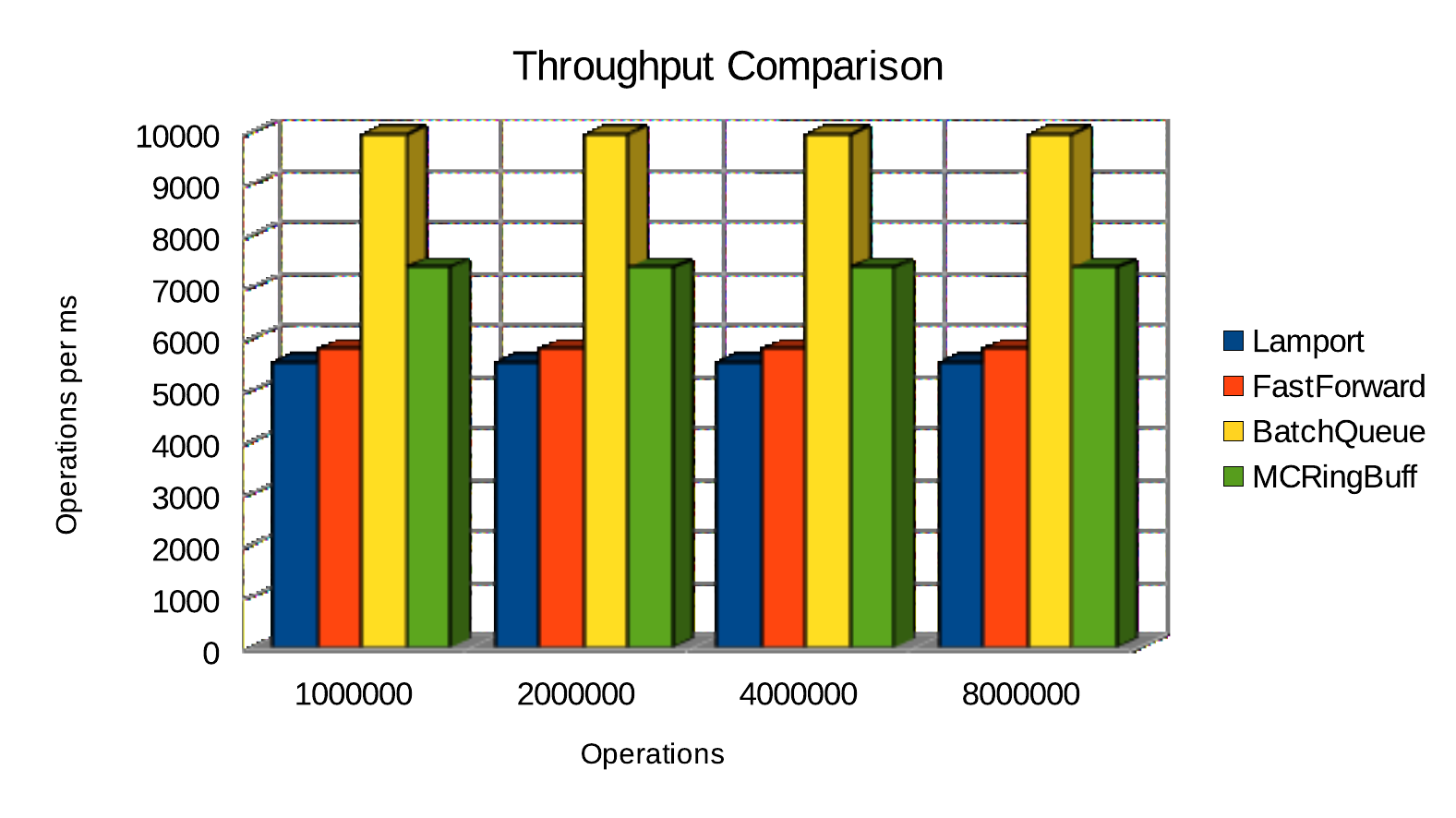}
\caption{Throughput evaluation as the number of tuples in the system increases \label{chag-velocity_Evaluation2}}
\end{figure}

As it is expected from previous evaluations, BQ proves to be the fastest algorithm in this evaluation as well (Fig. \ref{chag-velocity_Evaluation2}). MCR follows next, with FF and Lamport coming third and fourth respectively. BQ exhibits better throughput than MCR by $26\%$, FF by $42\%$ and Lamport by $45\%$.

\begin{figure}[ht!]
\centering
\includegraphics[width=140mm]{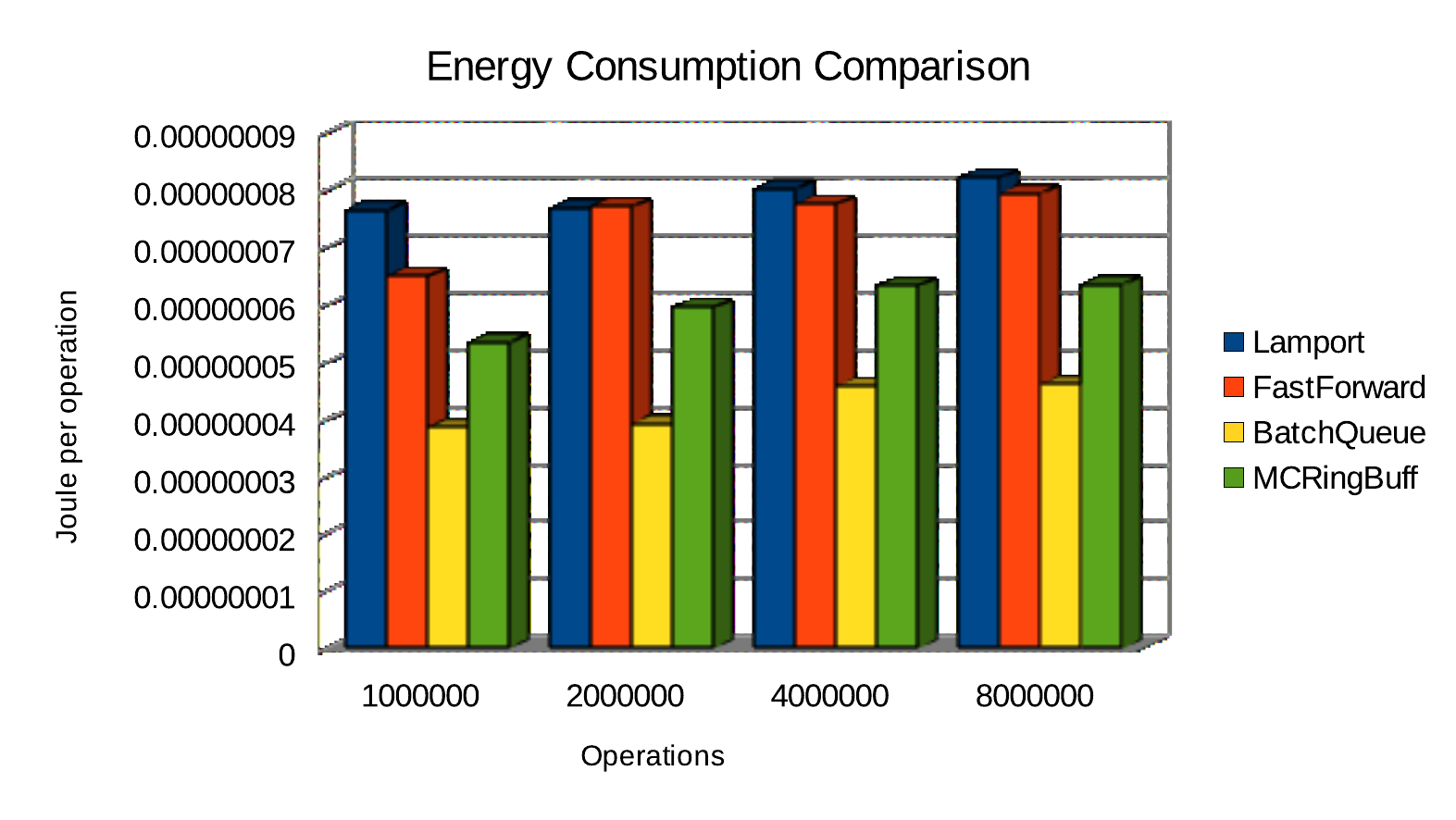}
\caption{Energy evaluation as the number of tuples in the system increases \label{chag-energy_Evaluation2}}
\end{figure}

The same holds for the energy footprint of the algorithms, where BQ is more efficient than MCR by $29\%$, FF by $43\%$ and  Lamport by $46\%$ (Fig. \ref{chag-energy_Evaluation2}). 

\subsection{Data streaming Aggregation Evaluation}
\label{chag-ch8}
The algorithms evaluated in the previous section are now used between the stages of a data streaming aggregation process in order to decide on their suitability and performance under the new circumstances. As previous evaluation has shown that use of data structures is most effective when they reside in CMX, throughout this evaluation the data structures are placed in CMX in order to benefit from a simpler and more efficient model of communication. 

The same process described at Section~\ref{chag-ch7} is used for execution time and energy measurements while all twelve shaves are used for the tests performed. The queues connecting an aggregator with producers and the final aggregator can store 128 elements and they are not filled with elements prior to execution. Batches are used only for the BQ case.

\subsubsection{Approach}

The aggregation process simulates smart metering data that have already been merged and sorted by a node which plays the role of the producer of the tuples that will be emitted to aggregators. The producer constitutes the first stage of the process.

The tuples produced contain a timestamp and a value representing the amount of Watts that have been reported by a smart meter. Once aggregators receive the tuples, they aggregate the values obtained according to timestamp. The aggregators constitute the second stage within the process.

The time is divided in windows of four time units with two time units
of advance. Windows containing the aggregated values are reported to a
final aggregator who reports the total sum of each window and
represents the third and final stage of the process.


The communication between the different stages of the process and their functionality is implemented according to the following description :

\begin{description}
\item[\textbf{Producer-Aggregator Communication}] \hfill \\
A producer communicates with each aggregator through a dedicated SPSC queue. It waits until all the tuples sent have been received and once notified through a shared variable that everything has been consumed, it informs aggregators that production is finished in order for them to report last windows that have not yet expired.
\item[\textbf{Aggregators - Last Aggregator Communication}] \hfill \\
The aggregators while consuming tuples, report for windows expired to the last aggregator through a SPSC queue (same type of queue used by a producer to communicate with an aggregator). They count how many messages are sent over to the last aggregator in order to inform him when he is finished processing everything that has been sent to him. Once this is true they state their inactivity to the last aggregator.
\item[\textbf{Last Aggregator}] \hfill \\
The last aggregator, aggregates values sent to him by every aggregator for each window. Each window in the aggregator's window list encompasses a contribution list stating which second stage aggregator has reported on the specific  window. 

In order for a window to be reported a list holding which aggregator is still active or inactive is checked in order to decide whether there will be somebody else that has not yet reported on the window and will potentially do so.  

The queues are checked in a round-robin fashion and the aggregator finishes only when all second stage aggregators are inactive (which means everything has been processed). Once everything has been processed, final windows that may have not yet been reported are eventually reported.
\end{description}

The algorithm that needs to be treated with special care in this type of application is the BQ. When using this data structure, it is likely that a consumer waiting for a producer to fill a batch, does not eventually consume everything when the producer is finished. Such an event may occur when a producer has not completely filled a batch. Thus a consumer will never be informed for leftovers residing in the last batch which has not yet been processed.

As a consequence a producer cannot wait for consumers to consume
everything in order to inform them to make their final report like in
all other algorithms. For this reason a flag is introduced by a
producer which informs that leftovers exist for a consumer when the
former states completion. Under this occasion, the index of the
producer is consulted by a corresponding aggregator in order to know
which data elements to consume. In case a producer has completely
filled a batch before completion, it informs that no leftovers exist
so an aggregator may proceed straight to the final report.

\begin{figure}[t!]
\centering
\includegraphics[width=140mm]{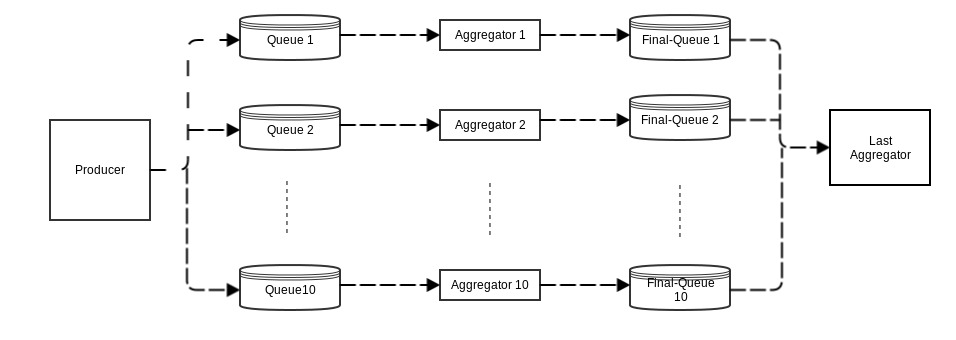}
\caption{Single producer aggregation \label{chag-one_producer_aggregation}}
\end{figure}

\begin{figure}[b!]
\centering
\includegraphics[width=120mm]{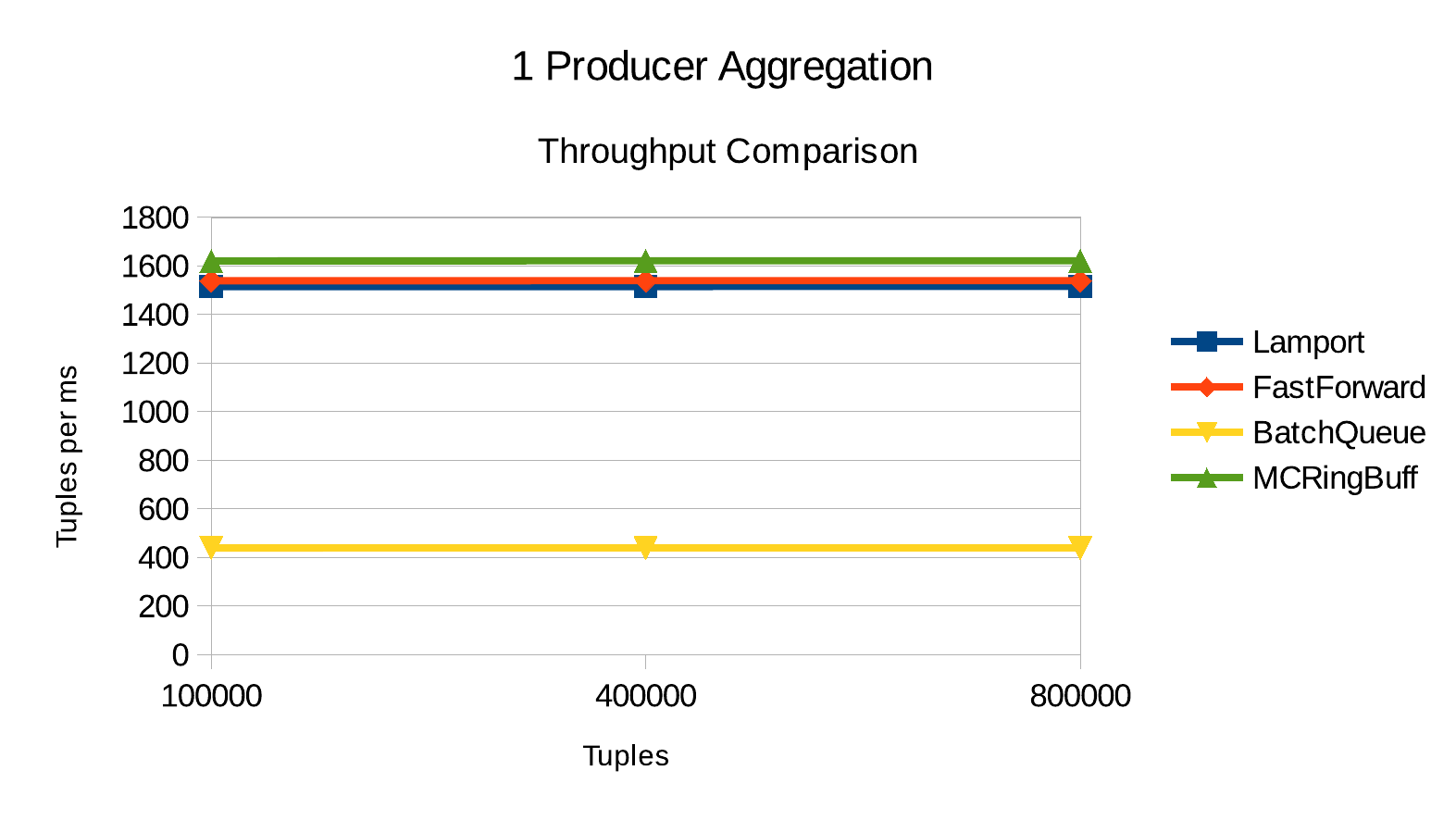}
\caption{Single producer aggregation throughput evaluation \label{chag-1ProducerVelocityComparison}}
\end{figure}

\begin{figure}[t!]
\centering
\includegraphics[width=120mm]{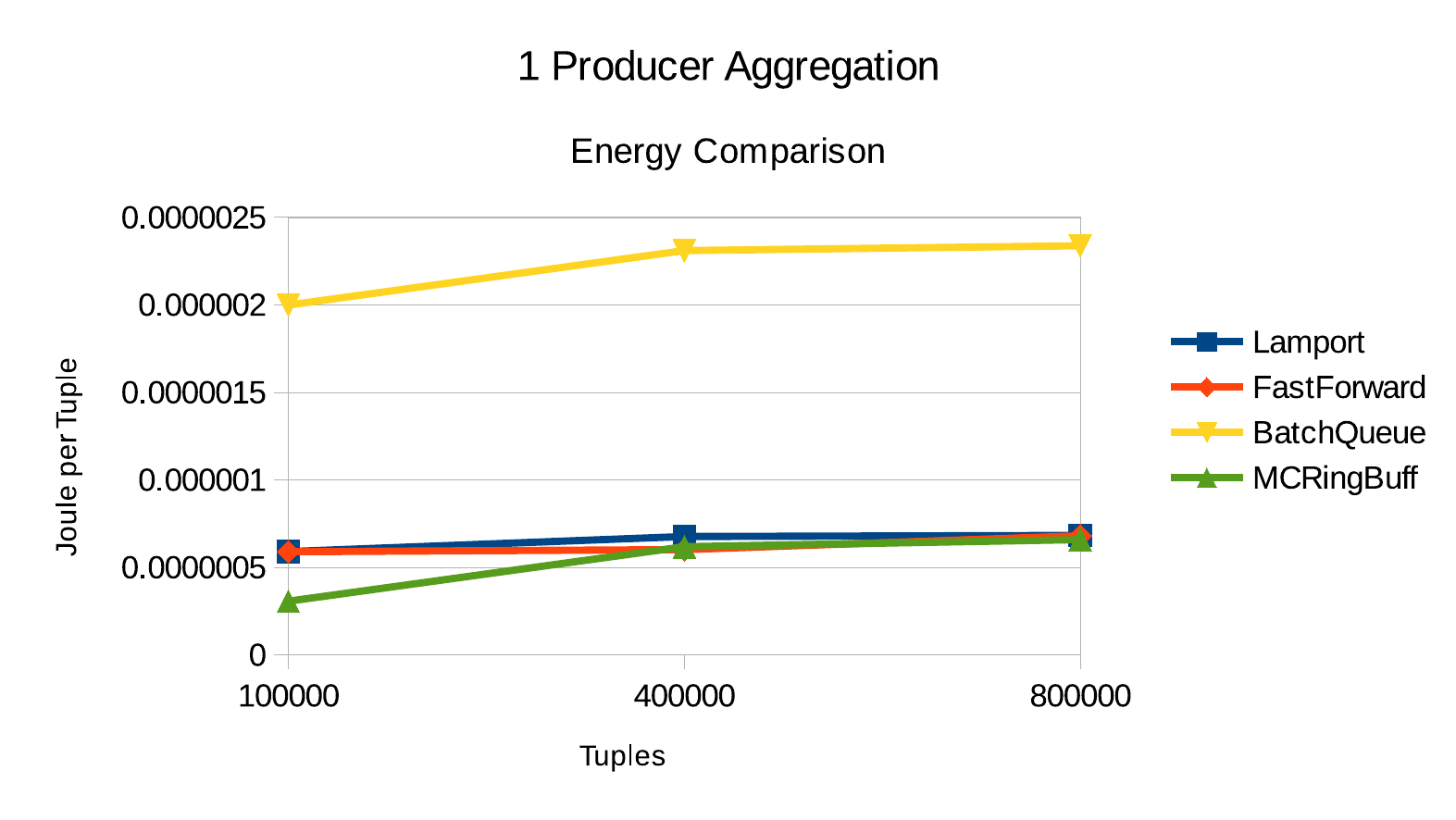}
\caption{Single producer aggregation energy evaluation \label{chag-1ProducerAggregationEnergyComparison}}
\end{figure}

This could have been the case for MCR as well as it separates the buffer in multiple batches. As already discussed in the previous section though, when data structures reside in CMX, it is not so beneficial to incorporate batches in the algorithm. As a consequence shared variables are updated immediately and the above scenario is avoided.

\begin{figure}[b!]
\centering
\includegraphics[width=120mm]{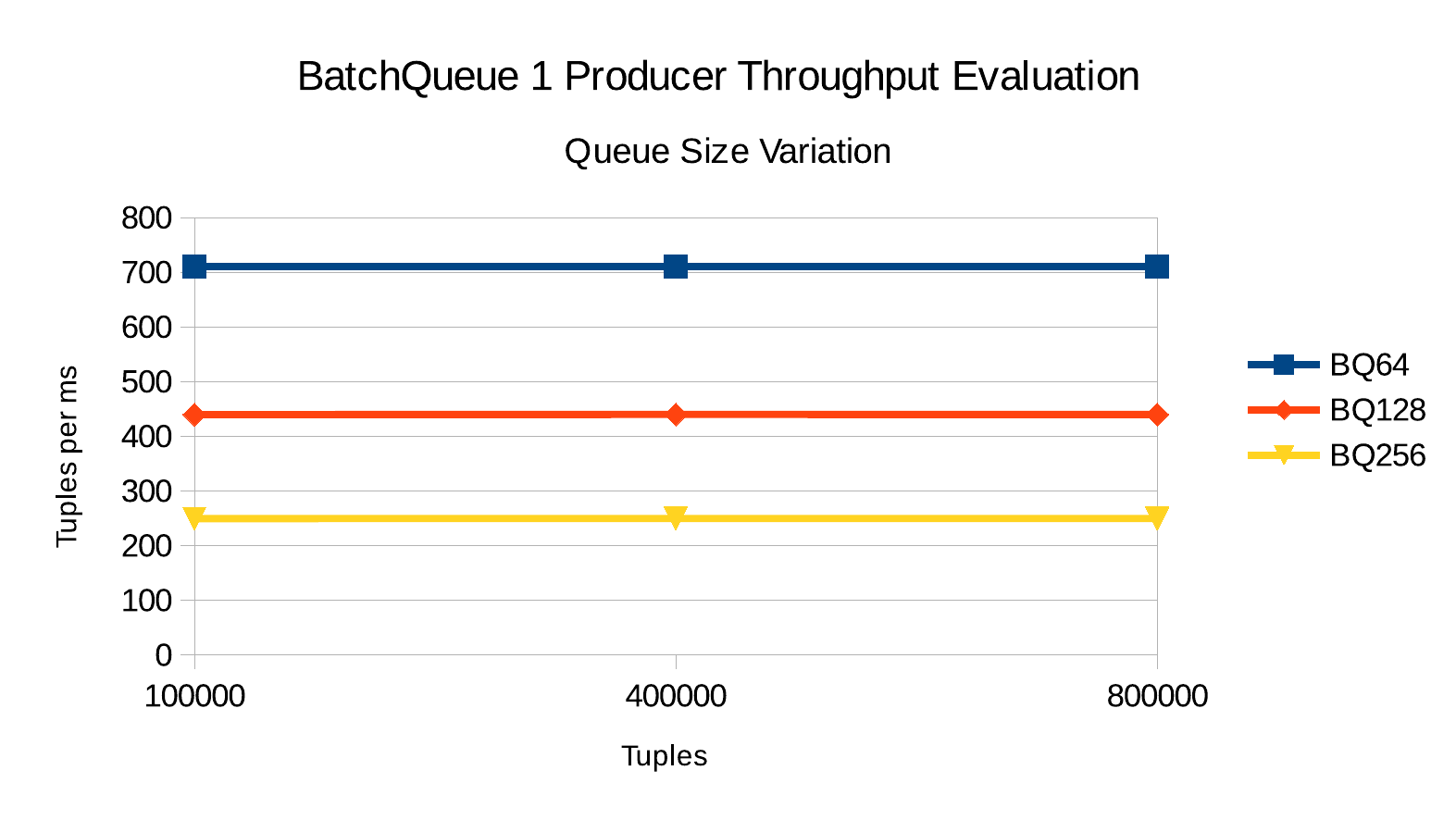}
\caption{BQ buffer variation throughput evaluation \label{chag-BQ_BuffVariation_1Prod_Velocity}}
\end{figure}

\begin{figure}[t!]
\centering
\includegraphics[width=120mm]{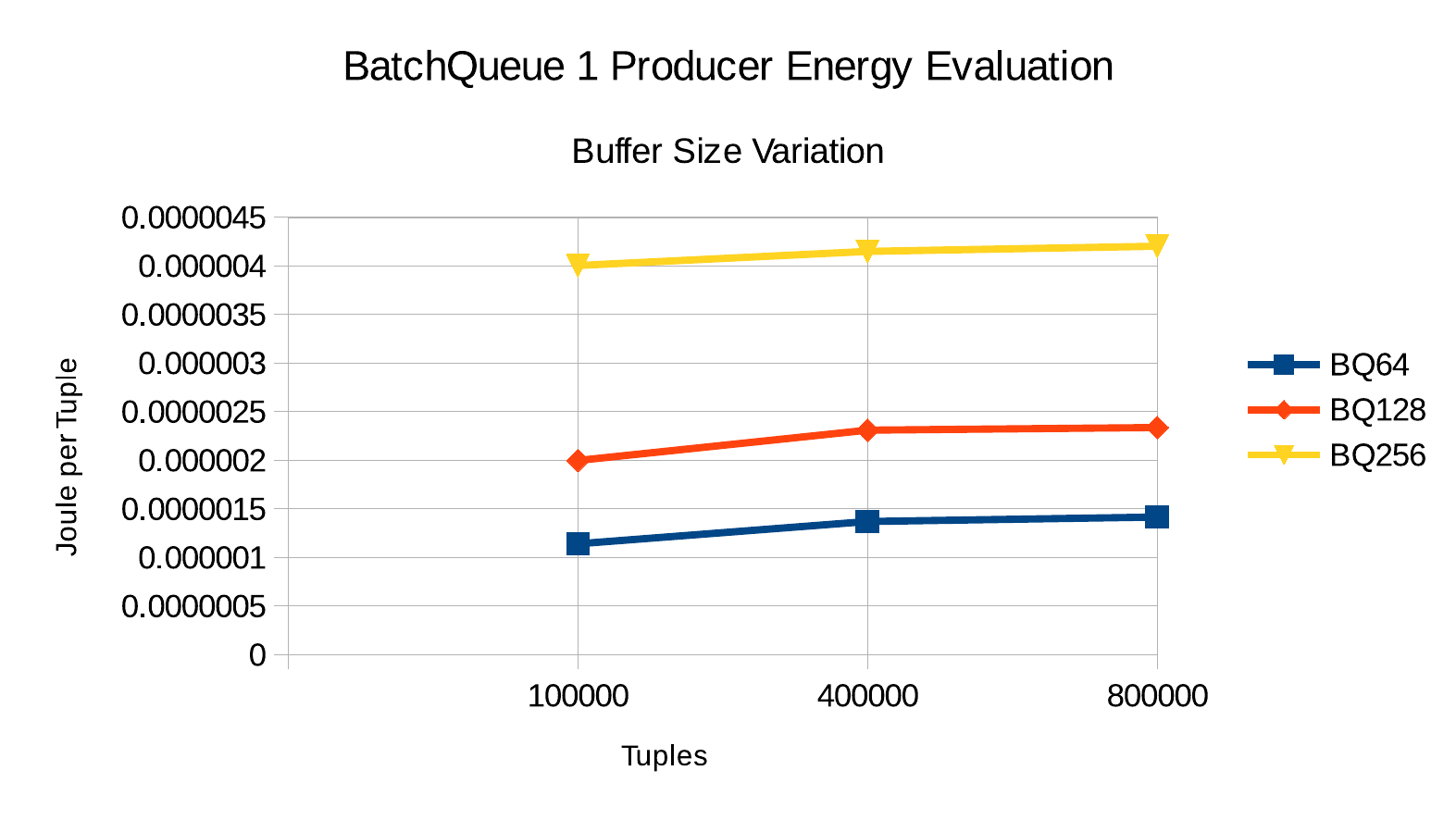}
\caption{BQ buffer variation energy evaluation \label{chag-BQ_BuffVariation_1Prod_Energy}}
\end{figure}

\subsubsection{Single producer variation}
\label{chag-1Prod}
In this variation, one producer feeds tuples to ten aggregators in a round robin fashion and one final aggregator is used (Fig. \ref{chag-one_producer_aggregation}). All processes are run on SHAVES. SHAVE0 acts as the producer while SHAVE1-SHAVE10 run the applications for the aggregators. SHAVE11 runs the application of the final aggregator. The data structures reside on the CMX slices of the second stage aggregators (i.e every aggregator holds the queues that make it communicate with the producer and the final aggregator).

When used within aggregation, BQ performs much worse than every other algorithm despite the fact that it is the most efficient in the evaluation of Section~\ref{chag-ch7} (Fig. \ref{chag-1ProducerVelocityComparison} and \ref{chag-1ProducerAggregationEnergyComparison}). Specifically MCR provides better throughput than all other algorithms, surpassing BQ by $73\%$, Lamport by $7\%$ and FF by $5\%$. Energy wise MCR is the most energy efficient solution as well with BQ exhibiting the highest energy consumption needs. BQ needs $71\%$ more power than Lamport, $72\%$ more power than FF and $76\%$ more power than MCR.

Superiority of MCR holds due to synchroniztion being more efficient as shared variables are consulted more sparsely than in other algorithms. The producer, as both of the processes start from index zero, will consult the shared variable manipulated by the consumer only when he fills the buffer and from then on every time it reaches the index at which it thinks the consumer is currently working. The consumer also consults the shared variable manipulated by the producer more sparsely as the producer is faster and fills some elements before the consumer finishes the aggregation algorithm (producer only executes a for loop while a consumer executes the aggregation algorithm).

Furthermore, the size of the queues used for communication seems to affect BQ's performance and energy consumption, as the bigger the capacity of a queue the longer the consumer should wait to consume and thus more busy waiting is involved. For this reason evaluation of differences between different sizes of queues is conducted in order to visualize its effect in the algorithm's performance in the context of aggregation.

As expected the capacity of the communication queues used have a major impact on the algorithm's performance (Fig. \ref{chag-BQ_BuffVariation_1Prod_Velocity} and \ref{chag-BQ_BuffVariation_1Prod_Energy}). When buffer size is doubled to $256$ elements, a decrease by $43\%$ is observed in throughput while energy consumption is increased by $46\%$. On the contrary when buffer size for communication between the stages of the aggregation is reduced to half ($64$ elements), an increase in throughput by $38\%$ is observed while energy consumption is decreased by $41\%$.

\begin{figure}[b!]
\centering
\includegraphics[width=140mm]{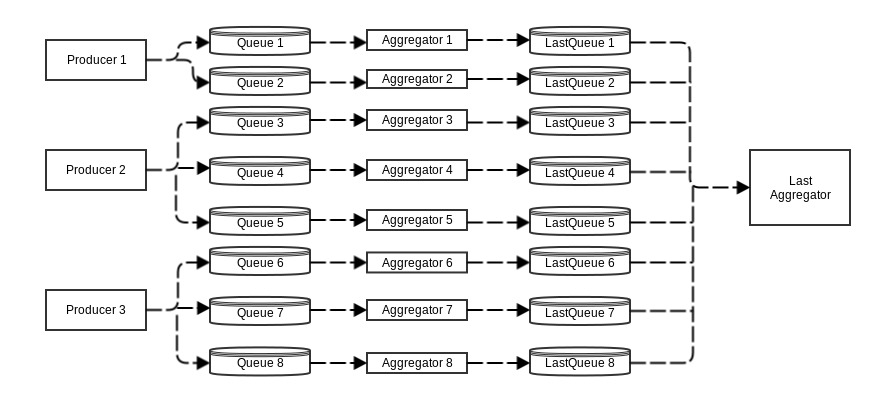}
\caption{Three producers aggregation \label{chag-3_producers_aggregation}}
\end{figure}

\begin{figure}[t!]
\centering
\includegraphics[width=140mm,height=70mm,keepaspectratio]{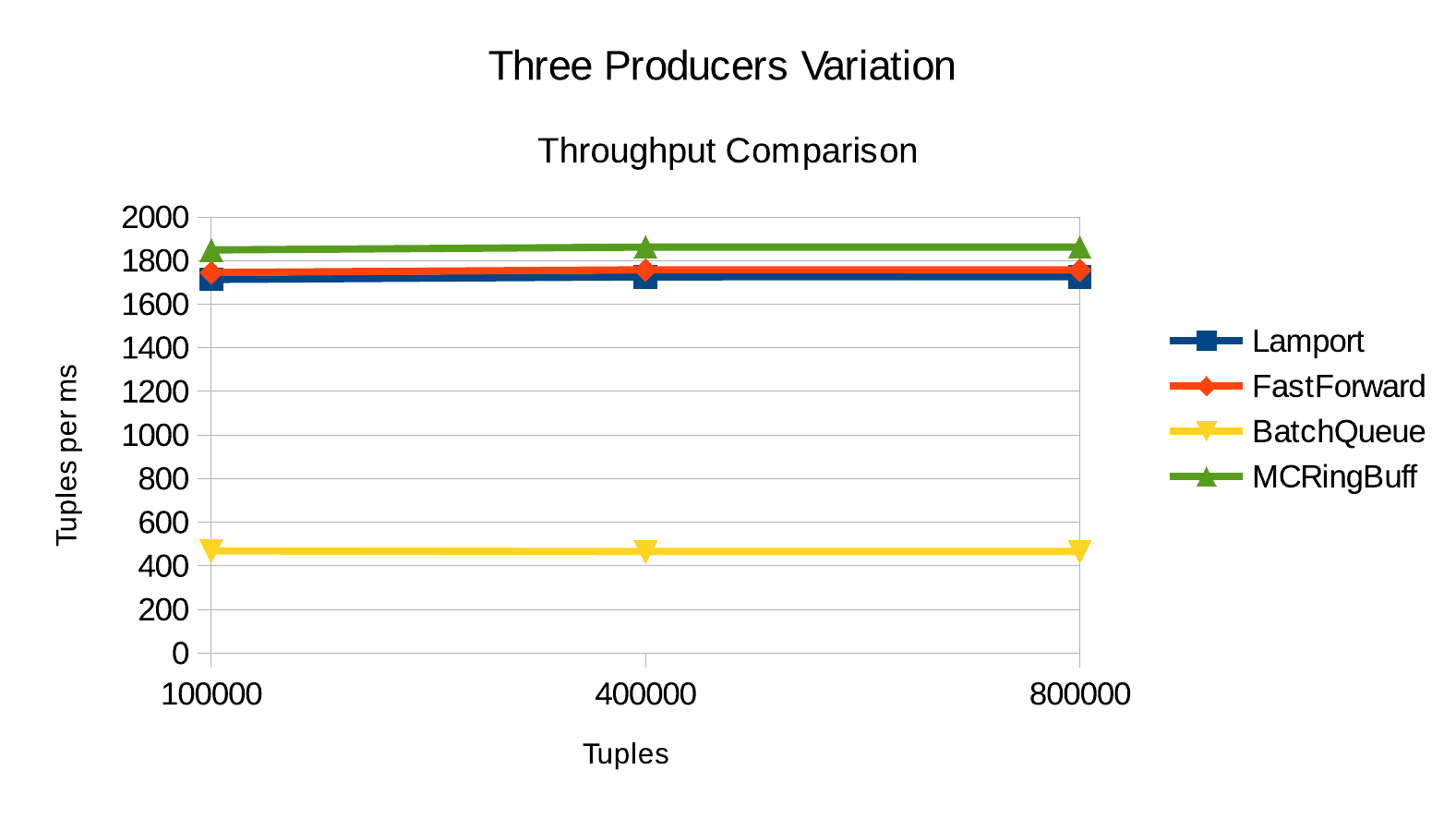}
\caption{Three producers aggregation - throughput comparison  \label{chag-3ProducerAggregationVelocityComparison}}
\end{figure}

\begin{figure}[b!]
\centering
\includegraphics[width=140mm]{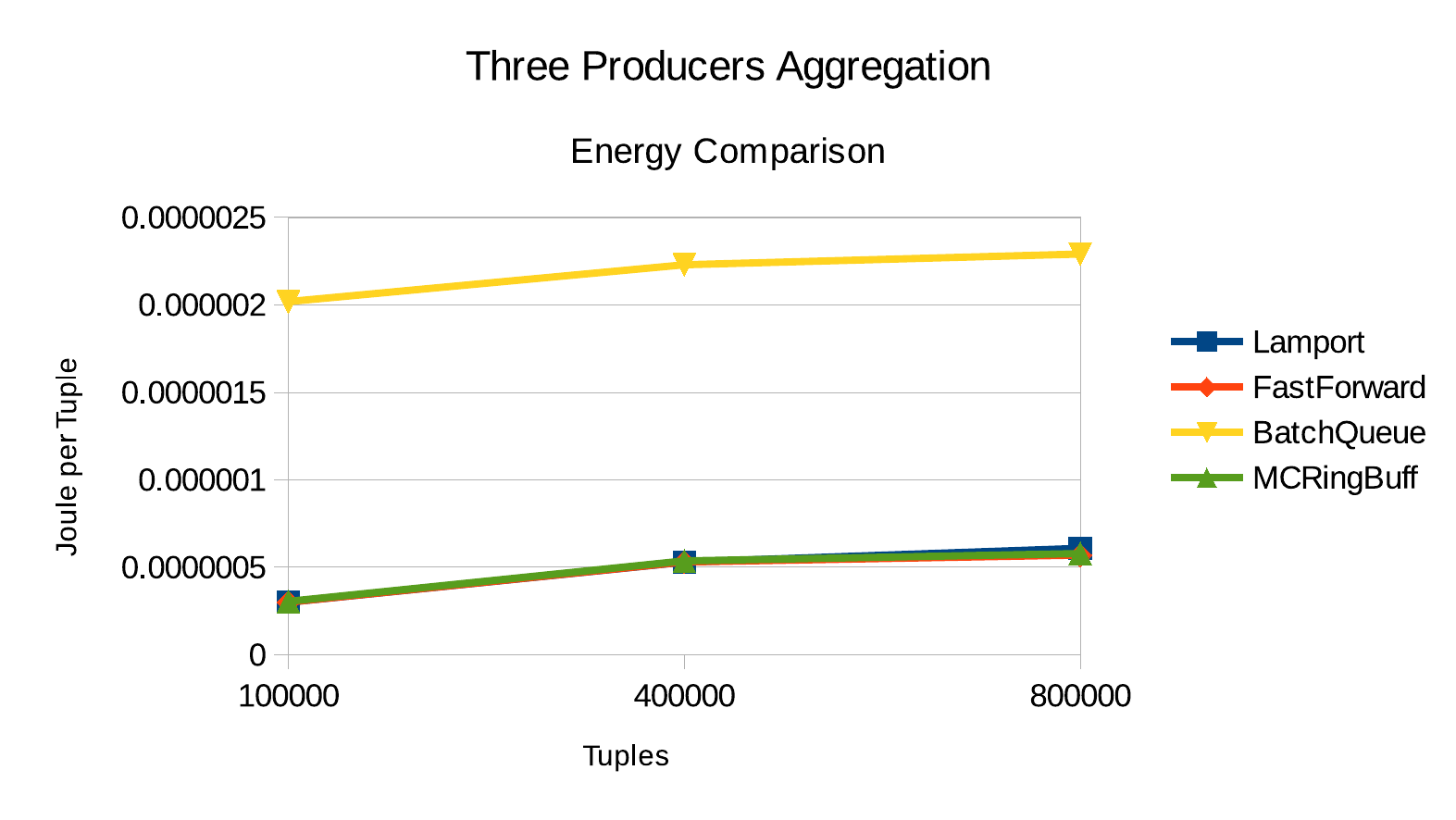}
\caption{Three producers aggregation - energy comparison \label{chag-3ProducerAggregationEnergyComparison}}
\end{figure}
\begin{figure}[t!]
\centering
\includegraphics[width=140mm]{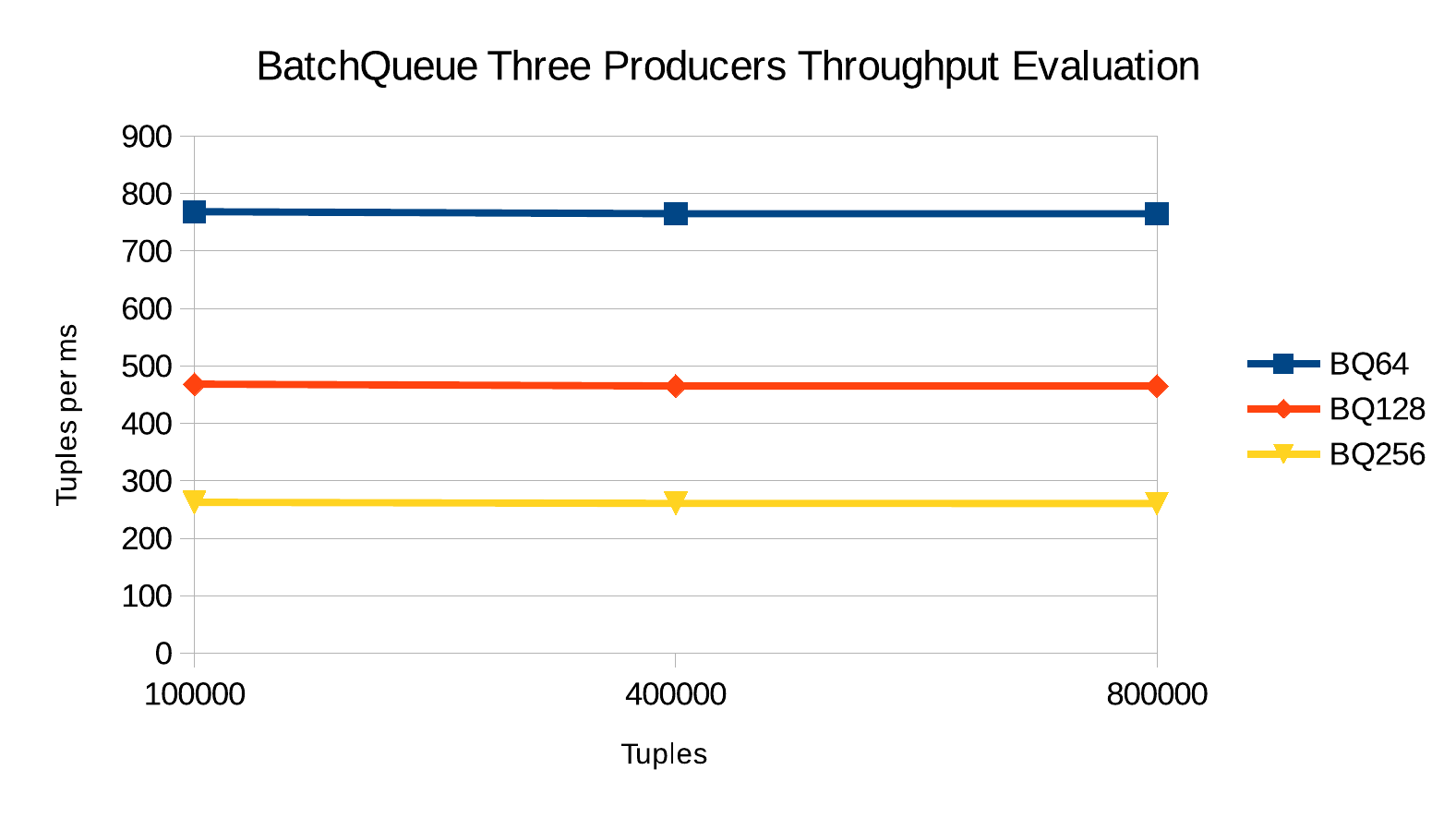}
\caption{BQ buffer variation throughput evaluation \label{chag-BQ_BuffVariation_3Prod_Velocity}}
\end{figure}

\begin{figure}[b!]
\centering
\includegraphics[width=140mm]{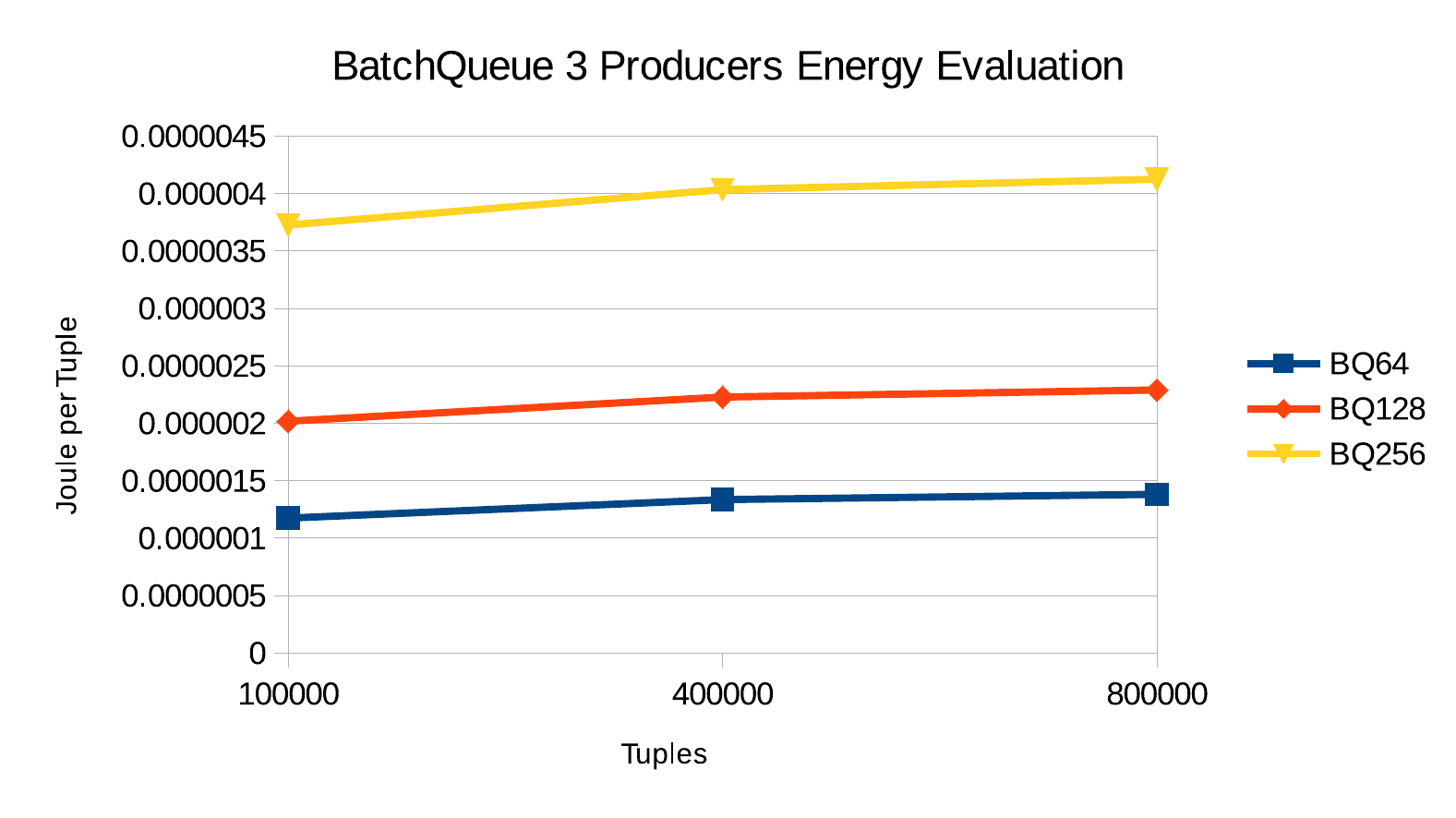}
\caption{BQ buffer variation energy evaluation \label{chag-BQ_BuffVariation_3Prod_Energy}}
\end{figure}

\begin{figure}[t!]
\centering
\includegraphics[width=140mm]{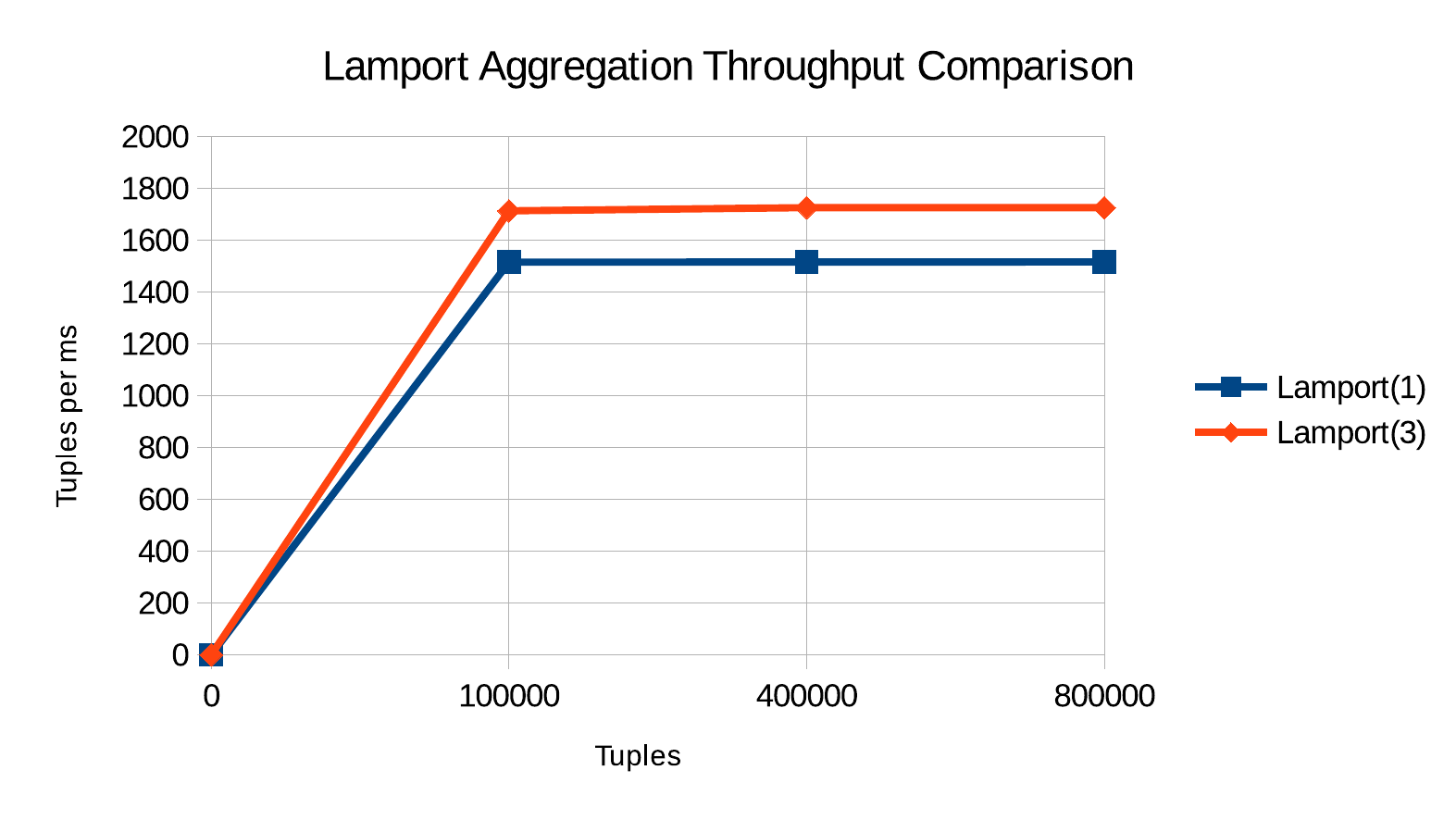}
\caption{Lamport aggregation - throughput comparison \label{chag-LamportAggregationVelocityComparison}}
\end{figure}

\begin{figure}[b!]
\centering
\includegraphics[width=140mm]{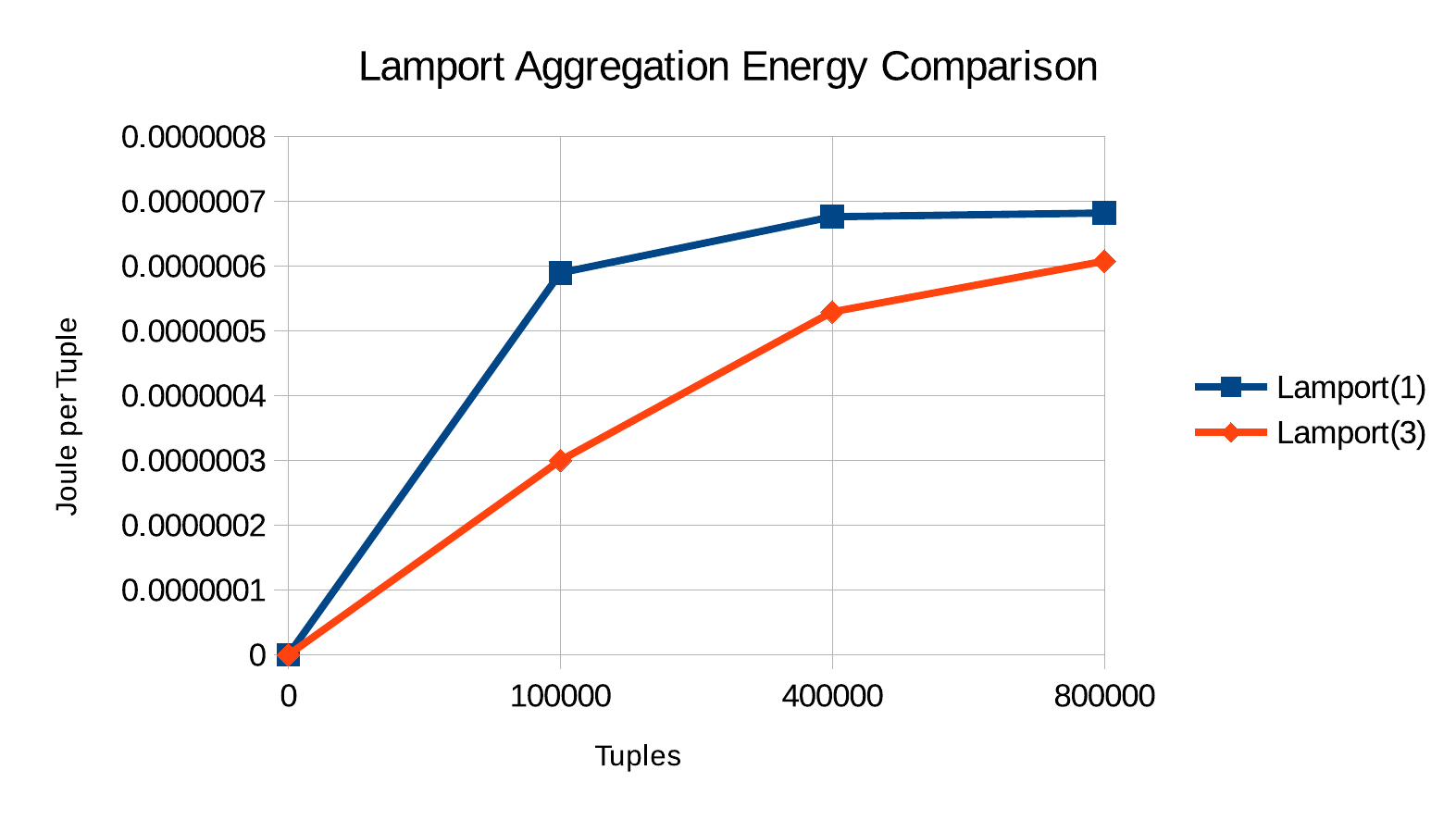}
\caption{Lamport aggregation - energy comparison \label{chag-LamportAggregationEnergyComparison}}
\end{figure}

\begin{figure}[t!]
\centering
\includegraphics[width=140mm]{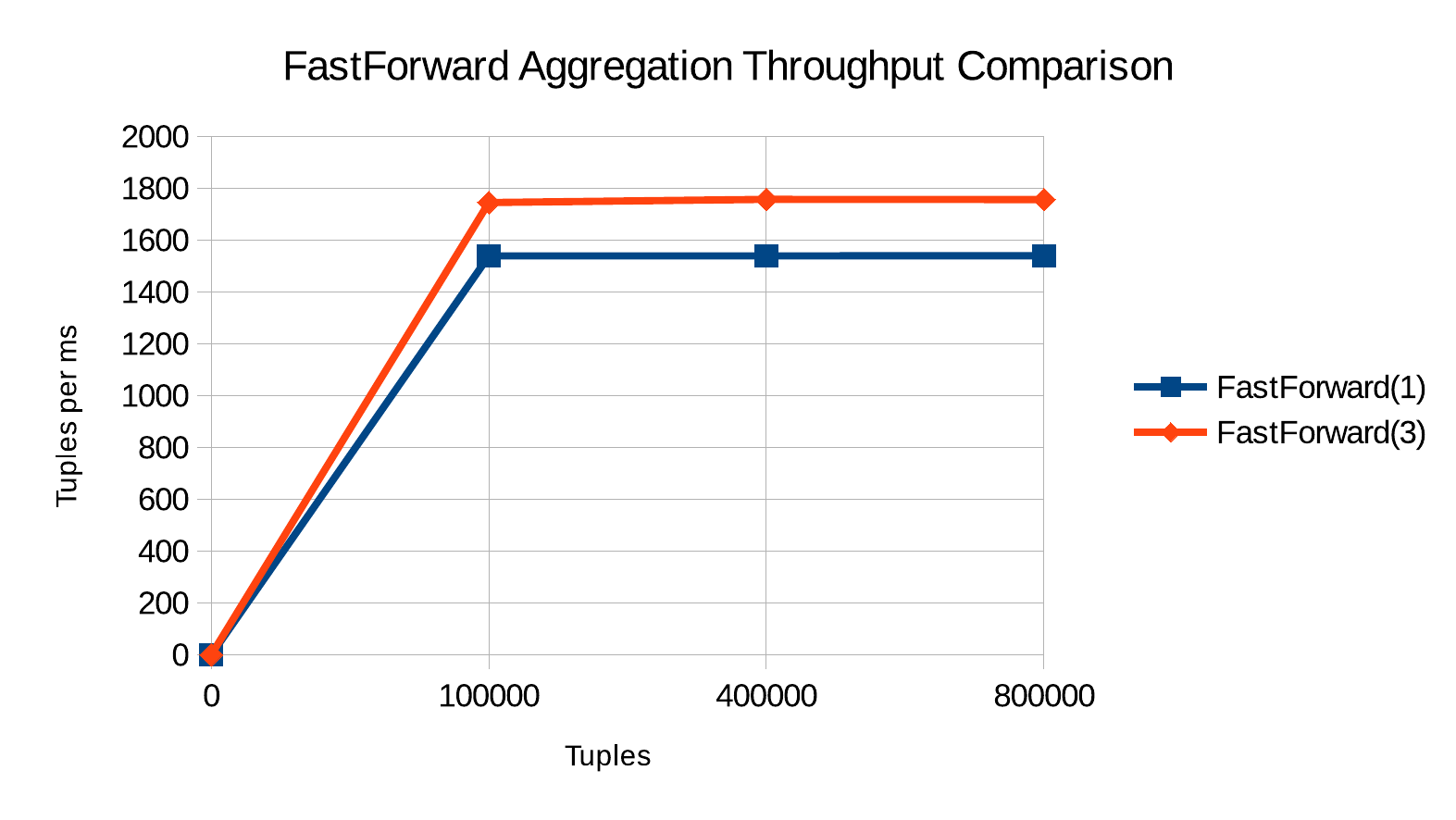}
\caption{FF aggregation - throughput comparison \label{chag-FFAggregationVelocityComparison}}
\end{figure}

\begin{figure}[b!]
\centering
\includegraphics[width=140mm]{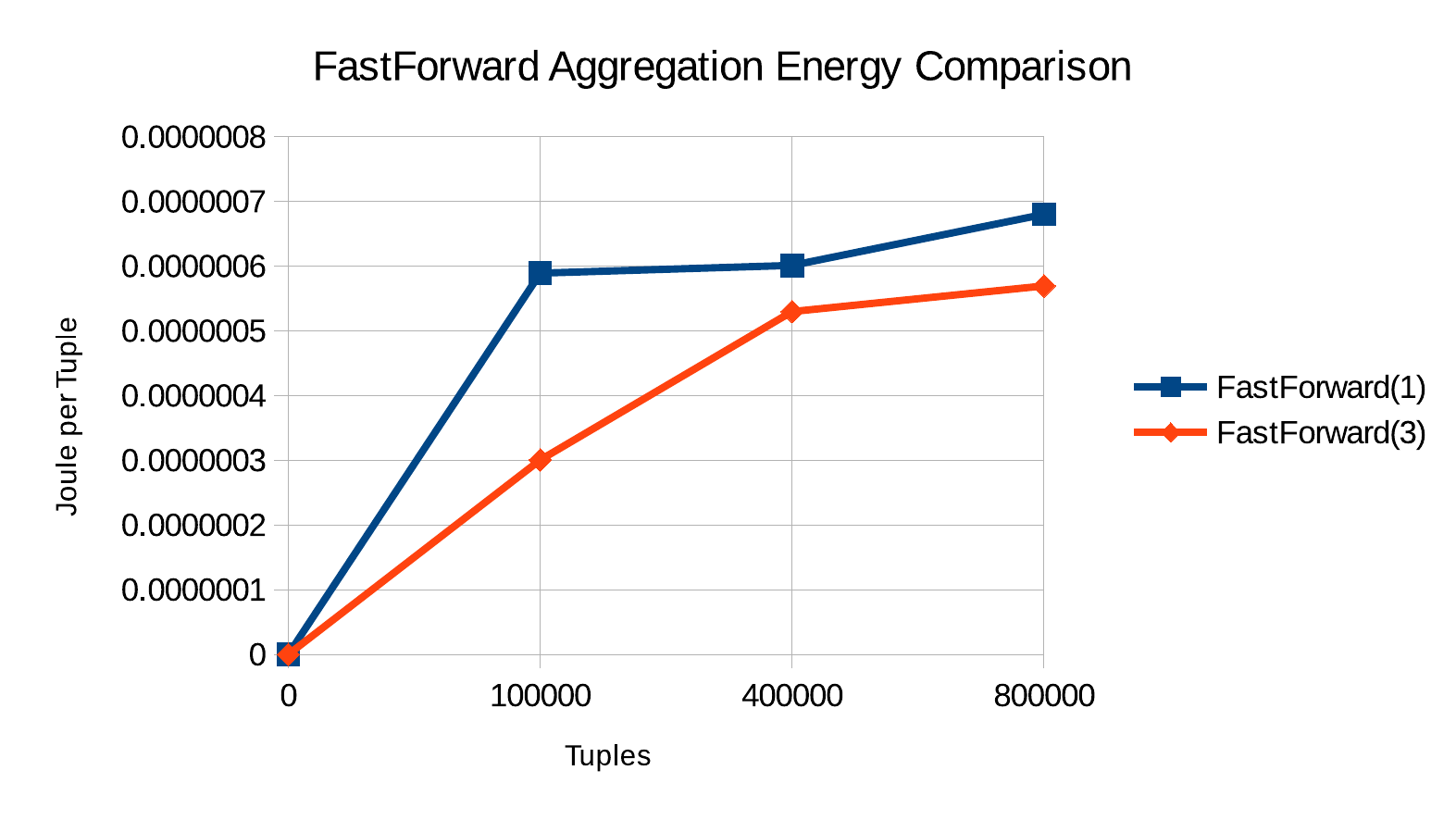}
\caption{FF aggregation - energy comparison \label{chag-FFAggregationEnergyComparison}}
\end{figure}

\begin{figure}[t!]
\centering
\includegraphics[width=140mm]{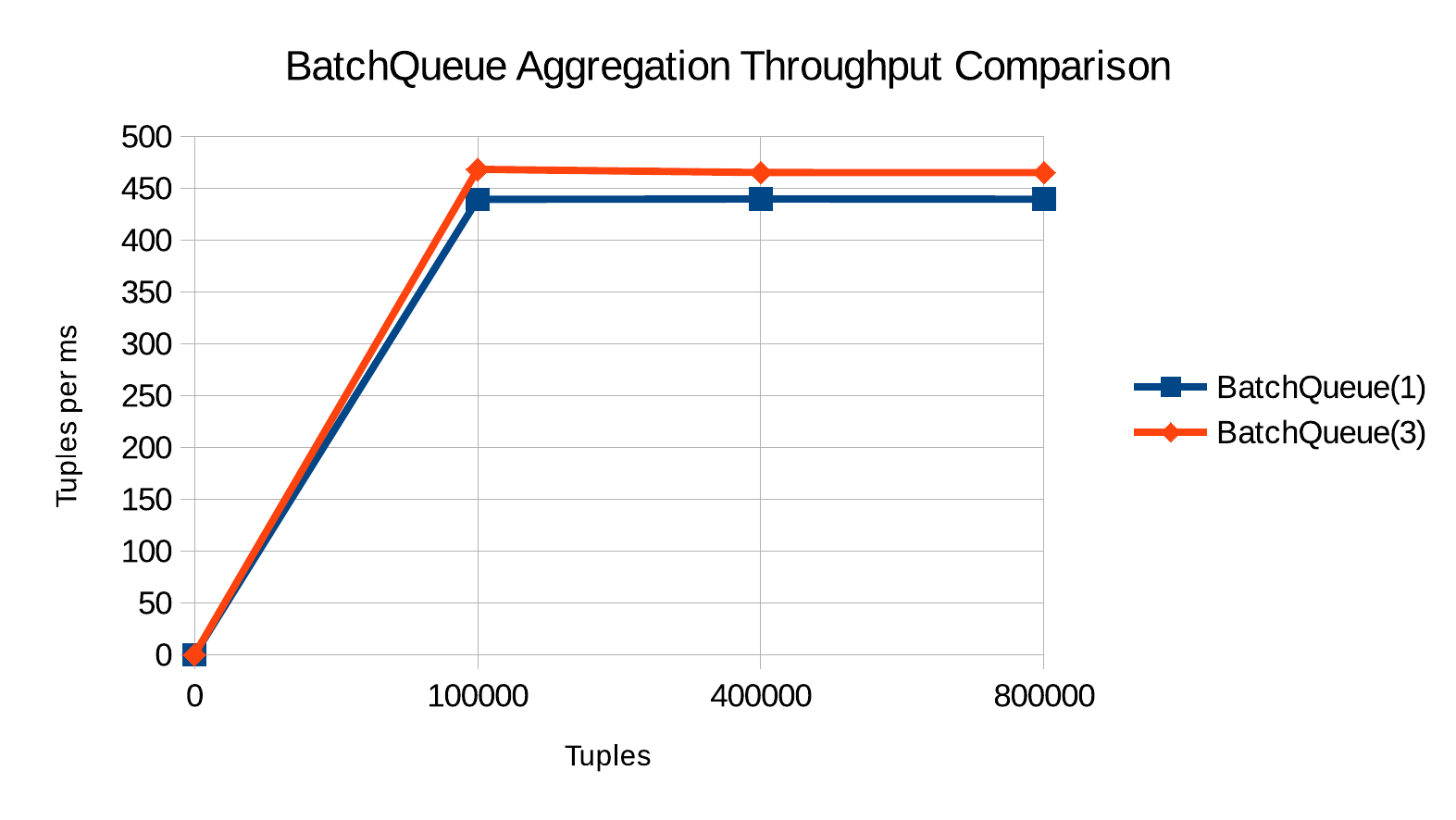}
\caption{BQ aggregation - throughput comparison \label{chag-BQAggregationVelocityComparison}}
\end{figure}

\begin{figure}[b!]
\centering
\includegraphics[width=140mm]{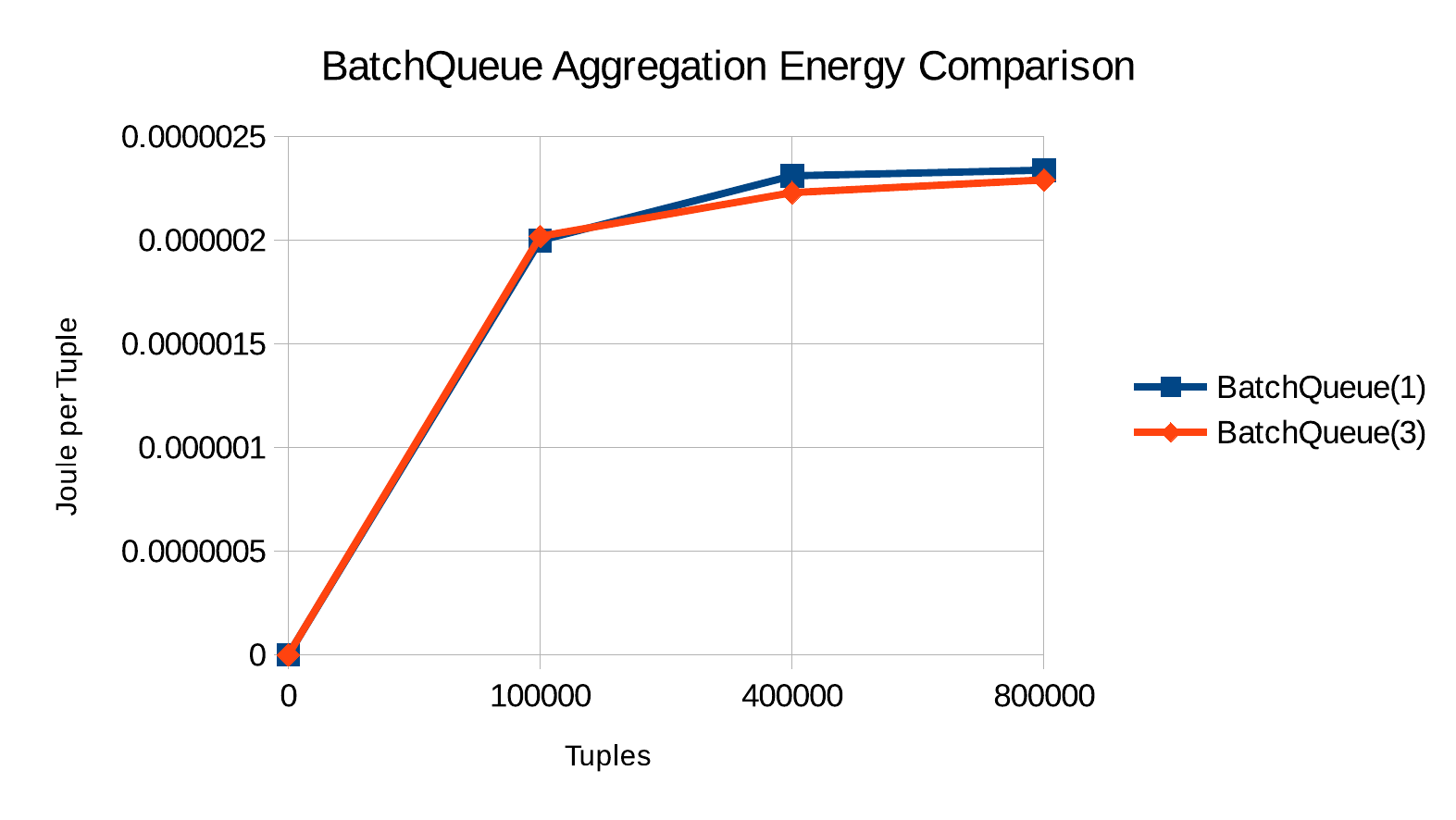}
\caption{BQ aggregation - energy comparison \label{chag-BQAggregationEnergyComparison}}
\end{figure}

\begin{figure}[t!]
\centering
\includegraphics[width=140mm,height=70mm,keepaspectratio]{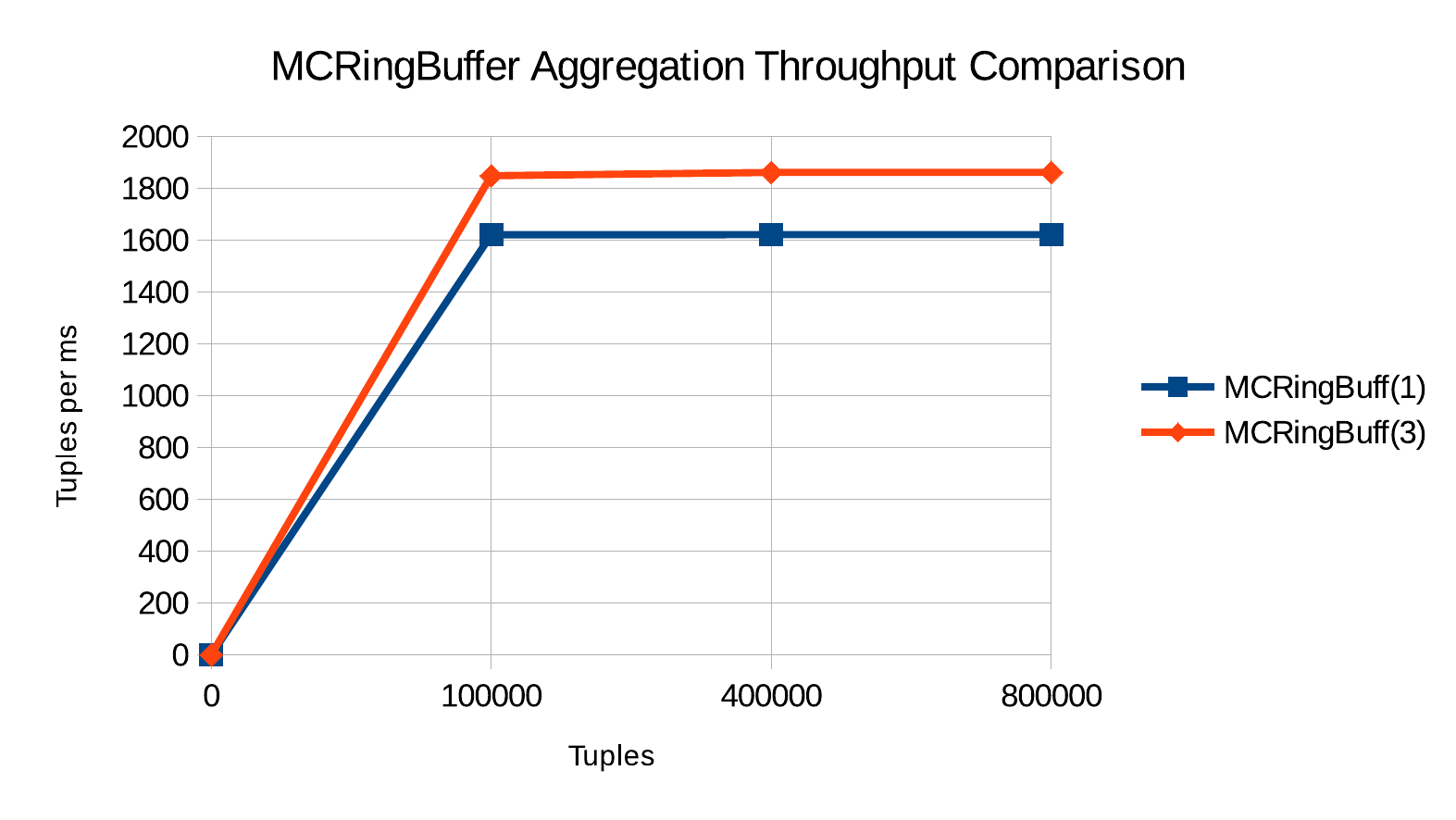}
\caption{MCR aggregation - throughput comparison \label{chag-MCRAggregationVelocityComparison}}
\end{figure}

\begin{figure}[b!]
\centering
\includegraphics[width=140mm]{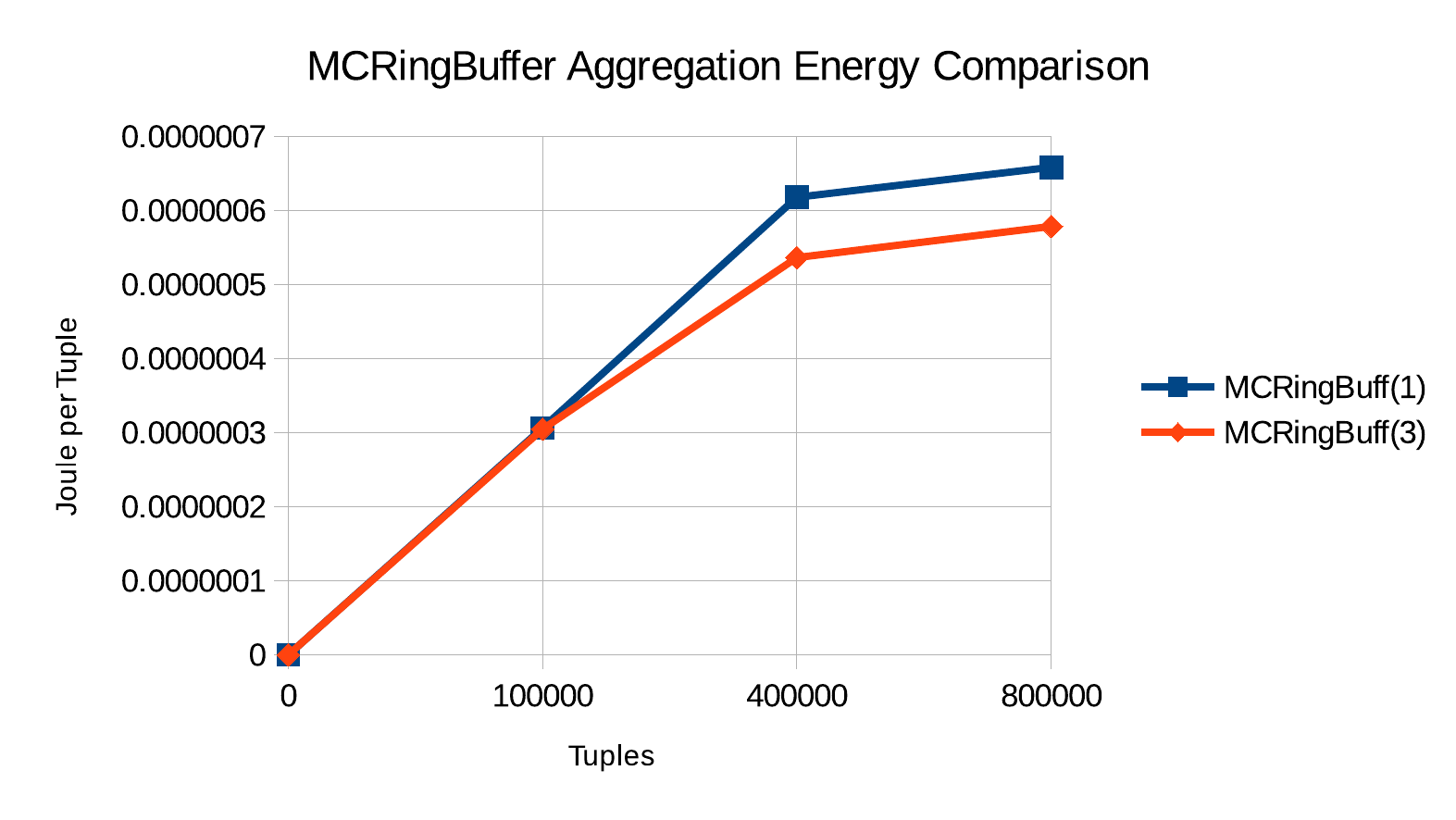}
\caption{MCR aggregation - energy comparison \label{chag-MCRAggregationEnergyComparison}}
\end{figure}

\subsubsection{Three producers variation}

In this variation as depicted by Figure \ref{chag-3_producers_aggregation}, three producers feed tuples to eight aggregators which communicate their windows to a final aggregator. The same approach as for the single producer variation concerning the placement of the data structures in memory holds (see Section~\ref{chag-1Prod}). Each producer feeds tuples to assigned consumers. In this way more pressure on the queues is achieved and the initial workload is equally shared among the SHAVES. Due to the fact that the producing phase of the process is faster than before, this approach is expected to provide better performance.

The pattern observed in Figure \ref{chag-3ProducerAggregationVelocityComparison} is the same as for the single producer variation in Figure \ref{chag-1ProducerVelocityComparison}. The difference lies in that the algorithms, as expected, have become faster now that three producers are used. MCR is more efficient by $75\%$ from BQ, by $7\%$ from Lamport and by $6\%$ from FF.


Energy wise, with the exception of BQ which exhibits $77\%$ higher energy needs from all other contenders, the rest of the algorithms exhibit similar behaviour (Fig. \ref{chag-3ProducerAggregationEnergyComparison}).

As for the single producer aggregation, BQ's performance is dependent on the size of the queues used (Fig. \ref{chag-BQ_BuffVariation_3Prod_Velocity} and \ref{chag-BQ_BuffVariation_3Prod_Energy}). When buffer capacity is doubled, throughput is reduced by $44\%$ while energy consumption is increased by $45\%$. On the contrary, when buffer size is reduced to half, throughput is increased by $39\%$ while energy consumption is decreased by $40\%$.

Figures \ref{chag-LamportAggregationVelocityComparison}, \ref{chag-LamportAggregationEnergyComparison}, \ref{chag-FFAggregationVelocityComparison}, \ref{chag-FFAggregationEnergyComparison}, \ref{chag-BQAggregationVelocityComparison}, \ref{chag-BQAggregationEnergyComparison}, \ref{chag-MCRAggregationVelocityComparison} and \ref{chag-MCRAggregationEnergyComparison} demonstrate the improvement gained through the second variation for every algorithm separately. The name of the algorithm followed by the number one, corresponds to the evaluation of the algorithm for the single producer aggregation. The name of the algorithm followed by the number three, corresponds to the evaluation of the algorithm for the aggregation where three producers are used.

BQ's performance and energy consumption do not have that much big of a difference for the two different scenarios.
When using three producers, although tuples may be communicated faster to and from the second stage aggregators at the beginning, the pace of the last aggregator stays the same as for the single producer version.

As a consequence aggregators may also wait longer for the last aggregator to consume everything from one batch until they stop busy waiting. Although initially there is an improvement in the stages up until the second stage aggregators, the application loses at the last stage of communication which also slows down the processing of incoming tuples by the aggregators, a fact which with its turn slows down the producers as well.

Table \ref{chag-table} provides the improvement gained by every algorithm from using three instead of a single producer for the aggregation. 

\begin{table}
\centering
\caption{Improvement from using three producers instead of one}
\label{chag-table}
\begin{tabular}{|c|l|c|l|c|l|c|l|}
\hline
           & Lamport & FF & BQ & MCR                                    \\ \hline
Throughput & 12\% & 12\% & 6\% & 3\% \\\hline
Energy     & 26\% & 25\% & 2\% & 10\% \\\hline
\end{tabular}
\end{table}

A remark worth taking into consideration, is that apart from BQ's behaviour, data retrieved from the investigation on aggregation resemble the pattern seen in data element evaluation, especially for data element size larger than 64 bytes (see Fig. \ref{chag-elementSizeVelocity} and \ref{chag-elementSizeEnergy}). This is because the evaluation on data element size, in a way simulates occasions where some work is performed between enqueues and dequeues. This is also what happens during the aggregation phase as well, where different aggregation processing algorithms are run from stage to stage corresponding to the work performed at the data element evaluation. This explains why MCR does not have the same advantage over the other algorithms as in other types of evaluation performed in Section~\ref{chag-ch7}.

\subsection{Discussion and Conclusion}
\label{chag-ch9}

\subsubsection{Research aim and main findings}
The emergence of multicore platforms destined for embedded systems
hosting computationally demanding applications, requires that
limitations of these systems are taken into consideration in order to
achieve optimal performance. Algorithmic design of concurrent data
structures used in such applications play a major role on performance
and energy consumption during the execution time, rendering their
investigation a necessity for the research community.

Although extensive research has been conducted on concurrent data
structures, their use and evaluation in data streaming applications
and specifically in the aggregation phase has been neglected. For this
reason four SPSC algorithms are used within the aggregation phase of a
data streaming application evaluated on the embedded platform Myriad 2
in order to research their behaviour and applicability under the new
circumstances.

Performance evaluation of the algorithms alone, corresponds to the
behavior observed in a data streaming aggregation context as well,
with the exception of BatchQueue, which shows that adopting data
structures that employ batch processing of big batches is not the way
to go. The smaller the batches the better the performance.

The winner of the investigation turns out to be MCRingBuffer as it
manages to synchronize its processes more efficiently than other
algorithms by making use of shared variables more sparsely. It needs
to be noted that batches were not used for MCRingBuffer, as they did
not seem to provide much of a difference on Myriad 2.

Furthermore the algorithms provide a much more efficient solution than
lock-based approaches which, along with interrupt handling, provide
the main ways of synchronization for embedded systems nowadays. The
performance gains through the use of lock-free approaches over
lock-based ones range from $83\%$ to $97\%$, a fact that is quite
encouraging, demonstrating the potential of these kinds of algorithms
on embedded devices.

%% file: uit-temp/ibrahim-uit-myriad2.tex
Recent research suggested that the energy consumption of future computing systems 
will be dominated by the cost of data movement~\cite{Dally11}. 
As predicted by Dally, accessing data from nearby memory in a 10nm chip 
is 75$\times$ more energy
efficient than accessing them from across the chip. 
Therefore, in order to construct energy efficient software systems, data structures and 
algorithms must support not only high parallelism but also fine-grained data locality. 
Moreover, the fine-grained data locality should be portable across platforms.

Concurrent trees are fundamental data structures that are widely used in different contexts such as load-balancing \cite{DellaS00, HaPT07, ShavitA96} and searching \cite{Afek:2012:CPC:2427873.2427875, BronsonCCO10, Brown:2011:NKS:2183536.2183551, Crain:2012:SBS:2145816.2145837, DiceSS2006, EllenFRB10}.
Most of the existing highly-concurrent search trees are not considering the fine-grained
data locality. The non-blocking concurrent search trees 
\cite{Brown:2011:NKS:2183536.2183551, EllenFRB10} and Software Transactional
Memory (STM) search trees
\cite{Afek:2012:CPC:2427873.2427875, BronsonCCO10,
Crain:2012:SBS:2145816.2145837, DiceSS2006} have been regarded
as the state-of-the-art concurrent search trees. They have been proven
to be scalable and highly-concurrent. 
However these trees are not designed for fine-grained data locality. 
Prominent concurrent search trees which are often included in several benchmark 
distributions such as the concurrent red-black
tree \cite{DiceSS2006} by Oracle Labs and the concurrent AVL tree
developed by Stanford \cite{BronsonCCO10} are not designed for data locality either. It is challenging to devise search trees that are portable, highly concurrent and fine-grained locality-aware. A platform-customized locality-aware search trees \cite{KimCSSNKLBD10, Sewall:2011aa} are not portable while there are big interests of concurrent data structures for unconventional platforms~\cite{Ha:2010aa, Ha:2012aa}. Concurrency control techniques such as transactional memory~\cite{Herlihy:1993aa,Ha:2009aa} and 
multi-word synchronization~\cite{Ha:2005aa,Ha:2003aa,Larsson:2004aa} do not take into account fine-grained locality while fine-grained locality-aware techniques such as van Emde Boas layout \cite{Prokop99,vanEmdeBoas:1975:POF:1382429.1382477} poorly support concurrency.

However, there are no studies discussing the effect of
portable fine-grained locality on energy efficiency
and performance (e.g., throughput) in concurrent search trees. 
This work aims to provide insights into whether it would be worthwhile 
to develop portable fine-grained locality-aware concurrent search trees, 
and what improvements and drawbacks can be expected from 
such solutions with respect to energy-efficiency and performance.

Evaluating the effect of portable fine-grained locality on energy efficiency
and performance in concurrent 
search trees is challenging, mainly because 
practical fine-grained locality-aware concurrent search trees for a 
platform are usually customized for the particular platform and therefore not portable. 
To the best of our knowledge, the DeltaTree and GreenBST is the 
only practical {\em portable} fine-grained locality-aware concurrent search
tree~\cite{UmarAH16_PPoPP, GBST-sigmetrics2015}. 

Please note that the names of our trees have been changed according to comments we 
have received since 2013. DeltaTree that was described in the previous deliverables (i.e., D2.1 \cite{HaTUTGRWA:2014} and D2.2 \cite{HaTUAGRT15}) and our technical report uploaded to Arxiv.org in 2013 \cite{deltatreeTR2013} is a non-blocking and locality-aware concurrent search tree. Our 
SIGMETRICS'15 poster~\cite{GBST-sigmetrics2015} presents DeltaTree, a {\em lock-based homogeneous B-link tree} with vEB-layout nodes, and it is the same tree as 
the Balanced $\Delta$Tree found in the previous deliverables. GreenBST is a {\em lock-based heterogeneous B-link tree} with vEB-layout nodes, which is a more improved tree than DeltaTree and is the same tree as the Heterogeneous $\Delta$Tree found in the previous deliverables.

We present the study on the 
effect of portable fine-grained locality on energy efficiency and performance in 
concurrent search trees. We found portable fine-grained locality-aware concurrent search tree can reduce the energy cost incurred by data movement within a system, 
and also achieve the best energy efficiency and performance 
over the evaluation platforms (cf. Table \ref{tbl:platforms}).

\begin{table}[!t]
\centering
\footnotesize
\begin{tabular}{| c || p{3.5cm} | p{3cm} | p{3cm} | p{3.2cm} |}
\hline
\bf Name & 	\bf HPC &  		\bf		ARM & 			\bf	MIC  & \bf Myriad2\\
\hline
\hline
\bf System &	Intel Haswell-EP &	Samsung Exynos5 Octa &	Intel Knights Corner & Movidius Myriad2\\
\hline
\bf Processors&	2x Intel Xeon E5-2650L v3 &	1x Samsung Exynos 5410		&1x Xeon Phi 31S1P & 1x Myriad2 SoC \\
\hline
\bf \# cores&	24 (without hyperthreading)&	$-$ 4x Cortex A15 cores \newline $-$ 4x Cortex A7 cores	& 57 (without hyperthreading) &	$-$ 1x LeonOS core \newline $-$ 1x LeonRT core \newline $-$ 12x Shave cores \\
\hline
\bf Core clock&	2.5 GHz& $-$ 1.6 GHz (A15 cores) \newline $-$ 1.2 GHz (A7 cores) &	1.1 GHz &  600 MHz\\
\hline
\bf L1 cache	& 32/32 KB I/D	& 32/32 KB I/D	& 32/32 KB I/D		& $-$ LeonOS (32/32 KB I/D) \newline  $-$ LeonRT (4/4 KB I/D) \newline $-$ Shave (2/1 KB I/D)\\
\hline
\bf L2 cache	& 256 KB	& $-$ 2 MB (shared, A15 cores) \newline $-$ 512 KB (shared, A7 cores)  & 512 KB & $-$ 256 KB (LeonOS) \newline  $-$ 32 KB (LeonRT) \newline $-$ 256 KB (shared, Shave)\\
\hline
\bf L3 cache	& 30 MB (shared)&	-	& - & 2MB "CMX" (shared) \\
\hline
\bf Interconnect&	8 GT/s Quick Path Interconnect (QPI)&	CoreLink Cache Coherent \newline Interconnect (CCI) 400&	5 GT/s Ring Bus Interconnect & 400 GB/sec Interconnect\\
\hline
\bf Memory&	64 GB DDR3&	2 GB LPDDR3&	6 GB GDDR5 & 128 MB LPDDR II\\
\hline
\bf OS&		Centos 7.1 (3.10.0-229 kernel)	&Ubuntu 14.04 (3.4.103 kernel)	& Xeon Phi uOS (2.6.38.8+mpss3.5) & RTEMS (MDK 15.02.0)\\
\hline
\bf Compiler &	GNU GCC 4.8.3 & GNU GCC 4.8.2 & Intel C Compiler 15.0.2 & Movidius MDK 15.02.0\\
\hline
\end{tabular}
\vspace{-5pt}
\caption{We use 4 different benchmark platforms to evaluate tree's energy efficiency and performance.}
\label{tbl:platforms}
\end{table}

\subsection{Benchmarks and Profiles of Concurrent Search Trees}

To evaluate our conceptual idea of a locality-aware concurrent search tree, we compare GreenBST 
energy consumption and throughput with three prominent concurrent search trees (cf. Table \ref{tbl:algos}). GreenBST is a concurrent B+tree based on the locality-aware concurrent search tree layout~\cite{GBST-sigmetrics2015} and is an improved tree than DeltaTree.
LFBST is the state-of-the-art fast concurrent non-blocking binary search tree.
B-link (CBTree) tree is a concurrent B+tree that has been around for decades but still popular due to its effectiveness and practicality. Moreover, B-link tree also still used as a backend in popular database systems such as 
PostgreSQL\footnote{\url{https://github.com/postgres/postgres/blob/master/src/backend/access/nbtree/README}}~\cite{pgsql}. Citrus is an example of a concurrent search tree that uses read-copy-update (RCU), which is normally found in operating system internals, to handle concurrent operations within its data structure.

\begin{table}[!t]
\footnotesize
\begin{center}
\begin{tabular}{ | c | c | c | p{3cm} | p{3cm} | c | c |}
\hline
\bf \# & \bf Algorithm & \bf Ref & \bf Description & \bf	Synchronization& \bf Code authors & \bf Data structure \\
\hline
1	&	GreenBST&  -	&	Locality aware concurrent search tree  & lock-based &  this report &	b+tree	\\
2	&	LFBST	&\cite{Natarajan:2014:FCL:2692916.2555256} & Improved non-blocking binary search tree & lock free	&	UT Dallas & binary tree\\
3	&	CBTree	&\cite{Lehman:1981:ELC:319628.319663}	& Concurrent B+tree (B-link tree\protect\footnotemark)&  lock-based & this report & b+tree \\				
4	&	Citrus	&  \cite{Arbel:2014:CUR:2611462.2611471}	&	RCU-based search tree  & lock-based & Technion &	binary tree	\\
\hline
\end{tabular}
\end{center}
\vspace{-5pt}
\caption{List of evaluated concurrent tree algorithms. These algorithms are sorted by synchronization type.}
\label{tbl:algos}
\end{table}

Besides energy efficiency and throughput, we also profile the cache and branching behavior of the tested trees to determine whether data-locality
optimization can be translated into better energy efficiency and higher throughput. The experimental benchmarks were conducted on Intel high performance computing (HPC) platform, an ARM embedded platform, 
an accelerator platform based on the Intel Xeon Phi architecture (MIC platform) (cf. Table  \ref{tbl:platforms}). 

We measured the energy efficiency (in operations/Joule) and throughput (in operations/second)  
of all the trees in this evaluation.
\textit{Energy efficiency} indicators in {\em operations/Joule} were calculated 
using the number of operations ($\mathit{rep}=\num[group-separator={,}]{5000000}$)  
divided by the total CPU and DRAM energy consumption for the whole operations.
The ARM and Myriad2 platforms were equipped with a built-in on-board power measurement system that 
was able to measure the energy of all CPU cores and memory (DRAM) continuously in real-time. 
For the Intel HPC platform, the Intel PCM library using built-in CPU counters was used to measure 
the CPU and DRAM energy.
Energy indicators on MIC platform were collected by polling the \texttt{/sys/class/micras/power}
interface every 50 milliseconds.
\textit{Throughput} indicators in {\em operations/second} were calculated 
using ($\mathit{rep}=\num[group-separator={,}]{5000000}$) operations
divided by the maximum time the threads need to finish the whole operations.

All trees were pre-filled with $init$ values to simulate trees 
that partially fit into the last level cache. 
We used  $\mathit{init}=\num[group-separator={,}]{8388607}$ 
for the benchmarks on the Intel HPC and MIC platforms and  
$\mathit{init}=\num[group-separator={,}]{4194303}$ for the benchmarks on the ARM platform.

Combination of
update rate $u=\{0, 50\}$ and 
selected number of threads 
were used for each run.
Update rate of 0 equals to 100\% search, 
while 50 update rate equals to 50\% search and 50\% insert/delete operations out of $\mathit{rep}$ operations.
All involved operations were using randomly generated
values $v \in (0, \mathit{init} \times 2], v \in \mathbb{N}$ as their parameter. The same range 
was also used to generate random numbers for the $\mathit{init}$ values. 

To ensure fair
comparisons, all of the tree benchmark programs were using a unified benchmark source 
code that was linked directly during compilation.
We also had set the $\UB$ values of the GreenBST and the CBTree's 
B+tree order to their respective values so that each GNode and each 
CBTree's pages were within the system's page size of 4KB.  
We used POSIX thread library for concurrency on the HPC, ARM, and MIC platforms and 
running threads were pinned to the available physical cores.

\subsubsection{Energy Evaluation}

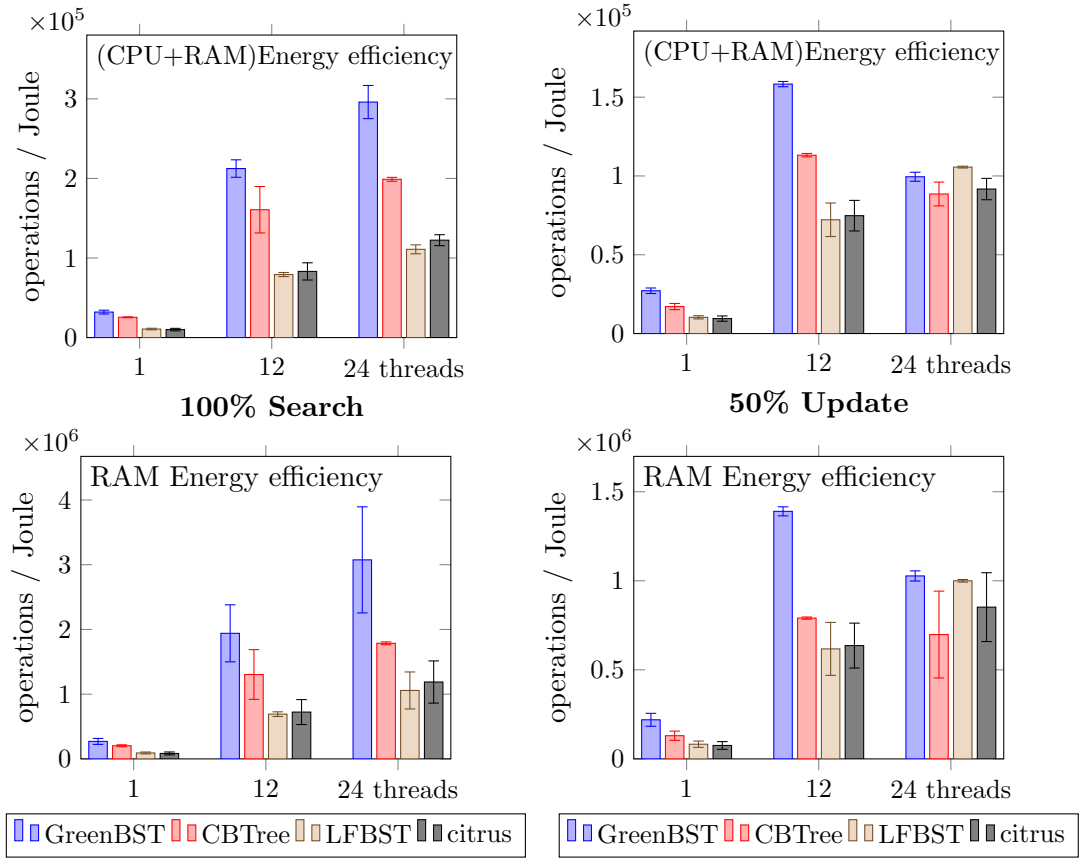
\begin{figure}[!t] 
{\footnotesize
\centering
\input{figs-uit/hpc-energy.tex}
}
\caption{Energy comparison using 
$2^{23}$ initial values on the  HPC platform. GreenBST is up to 50\% more
energy efficient than other trees in the 100\% search benchmark using 12 threads 
(namely, on single CPU).}\label{fig:ene-hpc}
\end{figure}

In the HPC platform (cf. Figure \ref{fig:ene-hpc}), GreenBST  is
up to 50\% more energy efficient than CBTree 
as seen in the 100\% searching case using 24 threads.
For updates, GreenBST can be 30\%
more efficient than CBTree in the 50\% update benchmark using 12 threads.

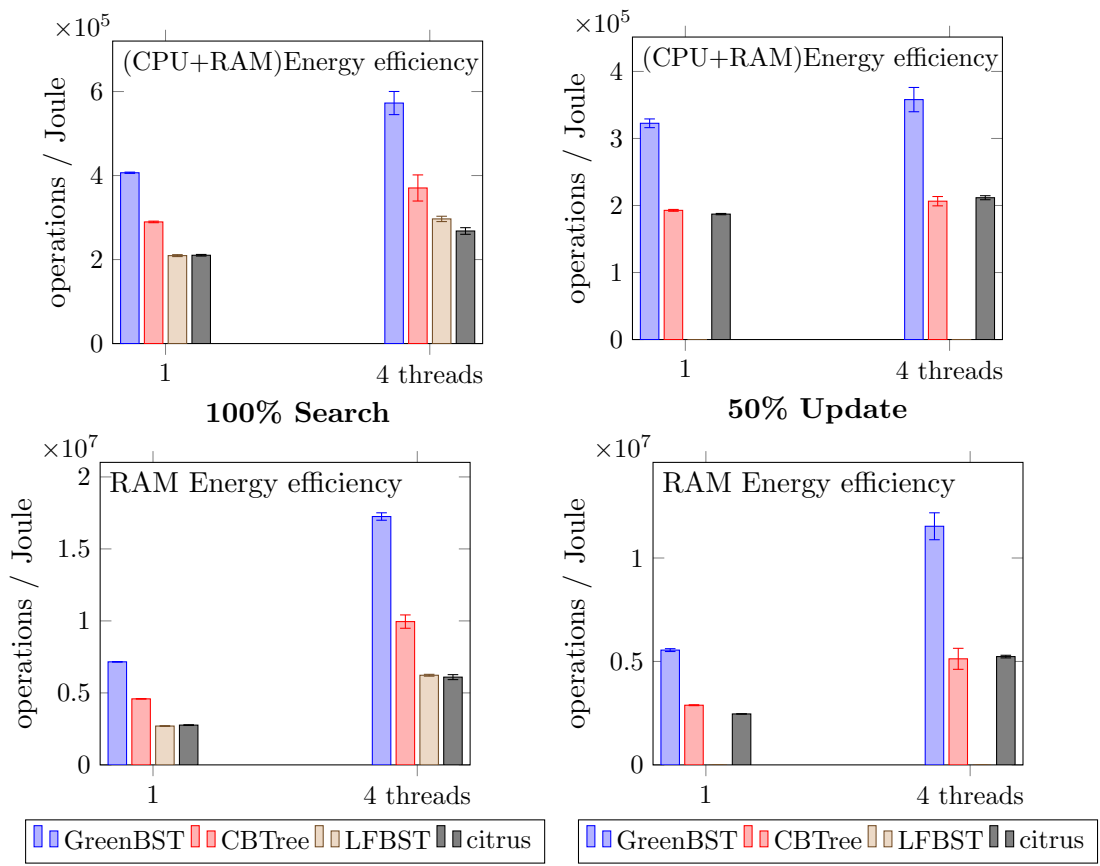
\begin{figure}[!t]
{\footnotesize
\centering
\input{figs-uit/arm-energy.tex}
}
\caption{Energy comparison using 
$2^{22}$ initial values on the ARM platform.  GreenBST is up to 100\% more
energy efficient than the other trees in the 50\% update operation using 4 threads.
Unfortunately, LFBST does not support the concurrent update
operations on the ARM 32-bit platform.}\label{fig:ene-arm}
\end{figure}

In the ARM platform (cf. Figure \ref{fig:ene-arm}), GreenBST is
100\% more energy efficient than CBTree 
in 50\% update benchmark using 4 threads. Also, GreenBST is 
75\% more efficient than the other trees in the 100\% search benchmark using 4 threads.

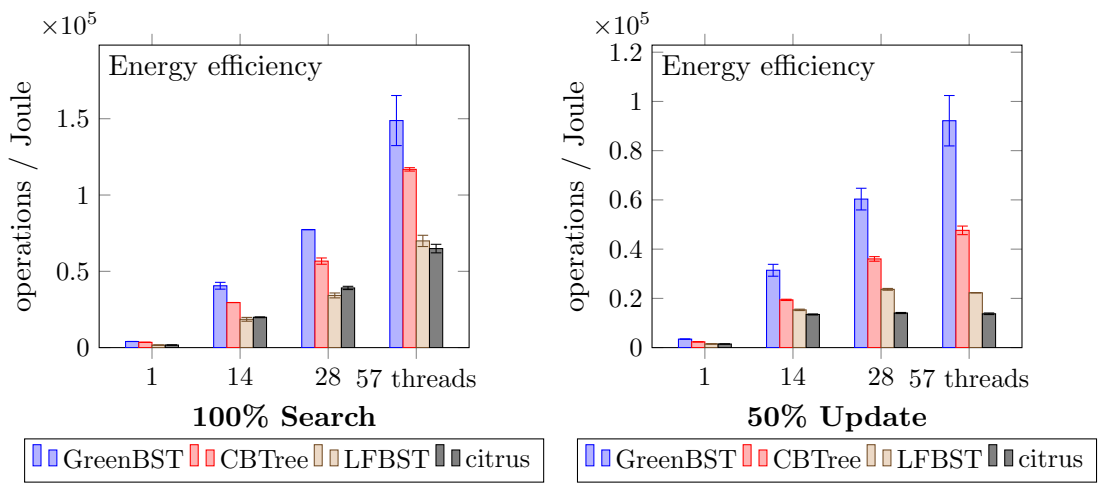
\begin{figure}[!t]
{\footnotesize
\centering
\input{figs-uit/mic-energy.tex}
}
\caption{Energy comparison using 
$2^{23}$ initial values on the MIC platform. GreenBST is up to 90\% more
energy efficient than the other trees in the 50\% update operation using 57 threads.}\label{fig:ene-mic}
\end{figure}

In the MIC platform (cf. Figure \ref{fig:ene-mic}), GreenBST is
up to 90\% more energy efficient than CBTree 
in the 50\% update benchmark using 57 threads. As for the search-only results, 
GreenBST is 20\% more efficient than CBTree in the 100\% search case
using 28 threads.

Energy results from the HPC, ARM, and MIC platforms highlight the advantage
in using the energy efficient and platform-independent locality-aware layout. 
GreenBST outperformed the other concurrent search trees on the HPC, ARM, and MIC
platforms that have different memory hierarchies.

\subsubsection{Throughput Evaluation}
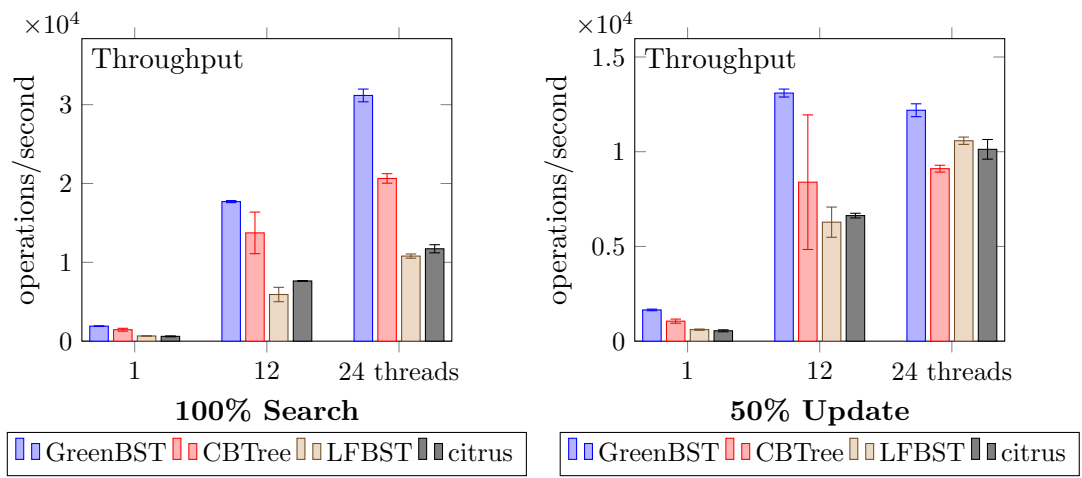
\begin{figure}[!t] 
{\footnotesize
\centering
\input{figs-uit/hpc-perf.tex}
}
\caption{Throughput comparison using 
$2^{23}$ initial values on the HPC platform. GreenBST is up to 50\% faster
 than the other trees in the 100\% search operation benchmark using 24 threads.}\label{fig:res-hpc}
\end{figure}

On the HPC platform (cf. Figure \ref{fig:res-hpc}), GreenBST throughput is
up to 50\% faster than CBTree 
in the 100\% search using 24 threads.
The 50\% update throughput of GreenBST is also 50\% faster than
CBTree when using 12 threads.

\begin{figure}[!t] 
{\footnotesize
\centering
\input{figs-uit/arm-perf.tex}
}
\caption{Throughput comparison using 
$2^{22}$ initial values on the ARM platform. GreenBST is up to 50\% faster
than other trees in 100\% search operation with 4 threads.
Unfortunately, LFBST does not support the concurrent update
operations on the ARM 32-bit platform.}\label{fig:res-arm}
\end{figure}
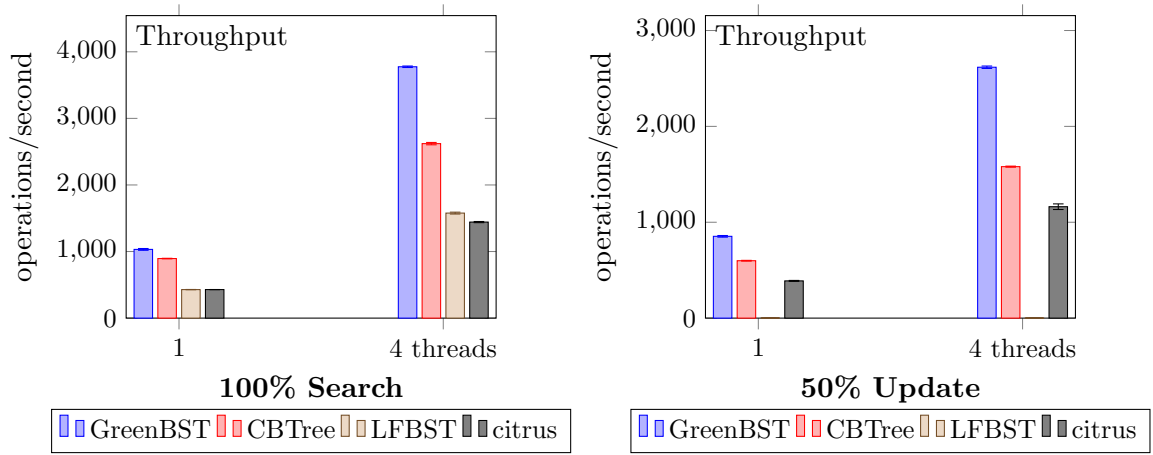

In the ARM platform (cf. Figure \ref{fig:res-arm}), GreenBST is
50\% faster than CBTree 
in the 100\% search case using 4 threads. GreenBST is also 50\% faster
than CBTree in the  50\% update case using 4 threads.

\begin{figure}[!t] 
{\footnotesize
\centering
\input{figs-uit/mic-perf.tex}
}
\caption{Throughput comparison using 
$2^{23}$ initial values on the MIC platform. GreenBST is up to 100\% faster
 than the other trees in 50\% update operation with 57 threads.}\label{fig:res-mic}
\end{figure}
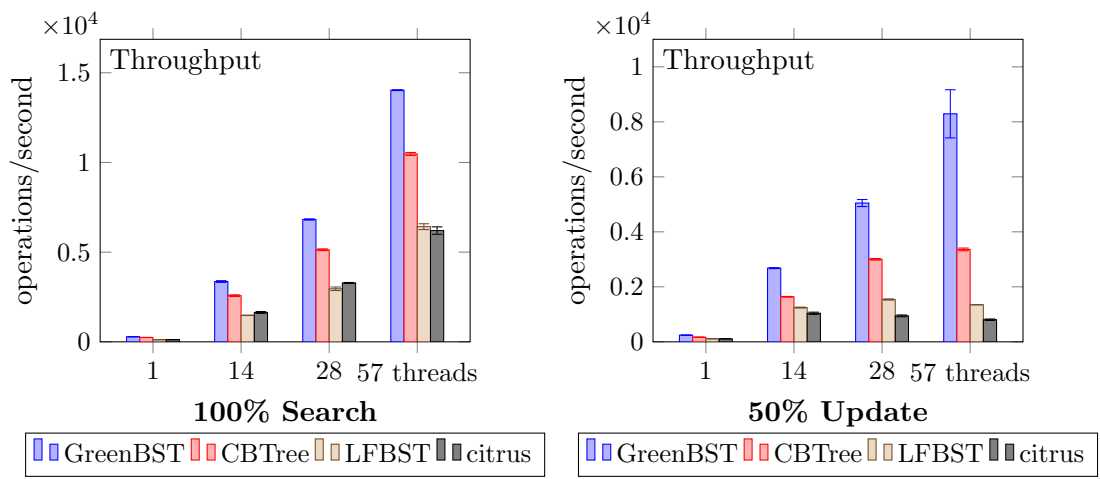

In the MIC platform (cf. Figure \ref{fig:res-mic}), GreenBST is
up to 100\% faster than the other trees 
in 50\% update case using 57 threads. GreenBST also
outperforms CBTree by 20\% in the 100\% search case using 57 threads.

As in the previous energy evaluation sector, GreenBST has managed to outperform 
CBTree, LFBST, and citrus tree. These benchmark results have proved the advantages
of the locality-aware layout on concurrent search tree. 
However, to provide insights 
into whether these energy efficiency and throughput advantages
are really caused by the higher degree of locality that GreenBST has, 
we also collect and evaluate several key
profiling results (i.e., cache miss and branch miss) of all the trees during the benchmarks.

\subsubsection{Profiling Results}

\begin{figure}[!t] 
{\footnotesize
\centering
\input{figs-uit/hpc-prof.tex}
}
\caption{Profiling result on the HPC platform.}
\label{ap:hpc-prof}
\end{figure}
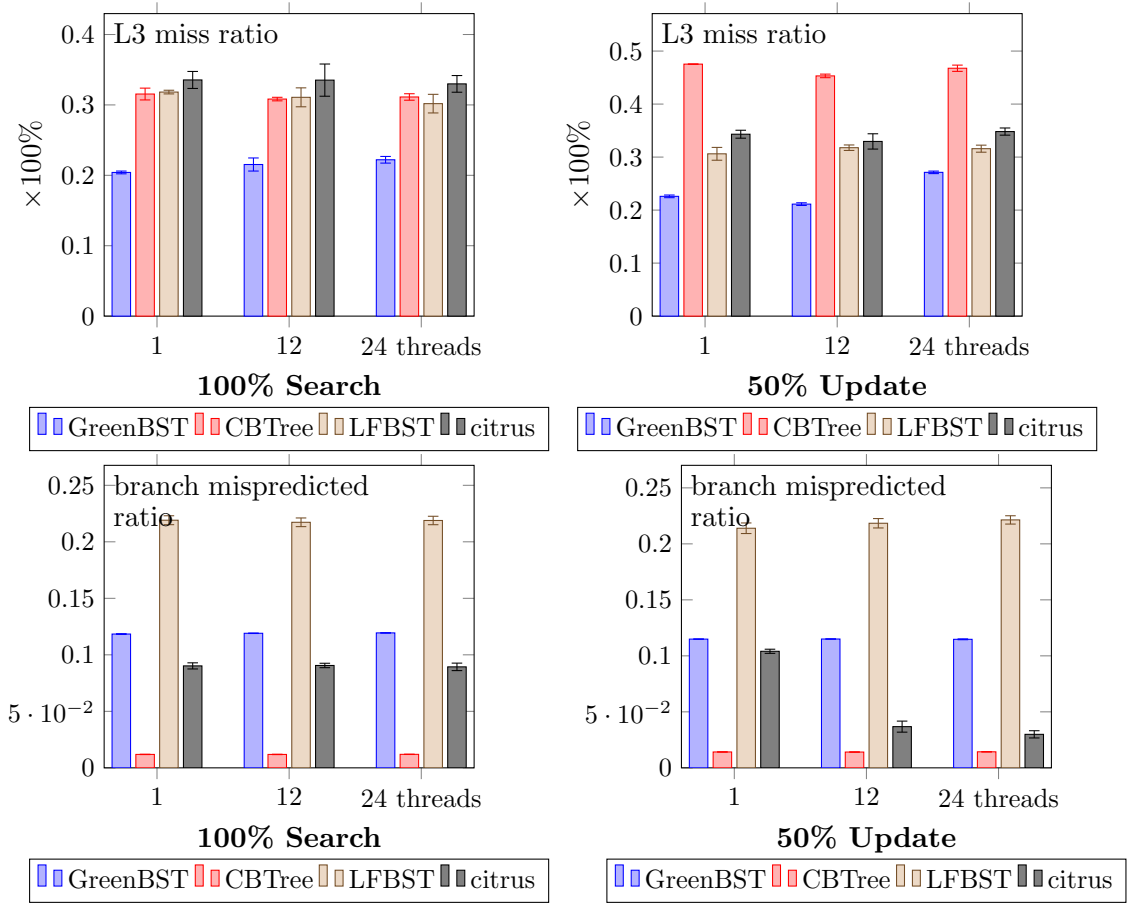

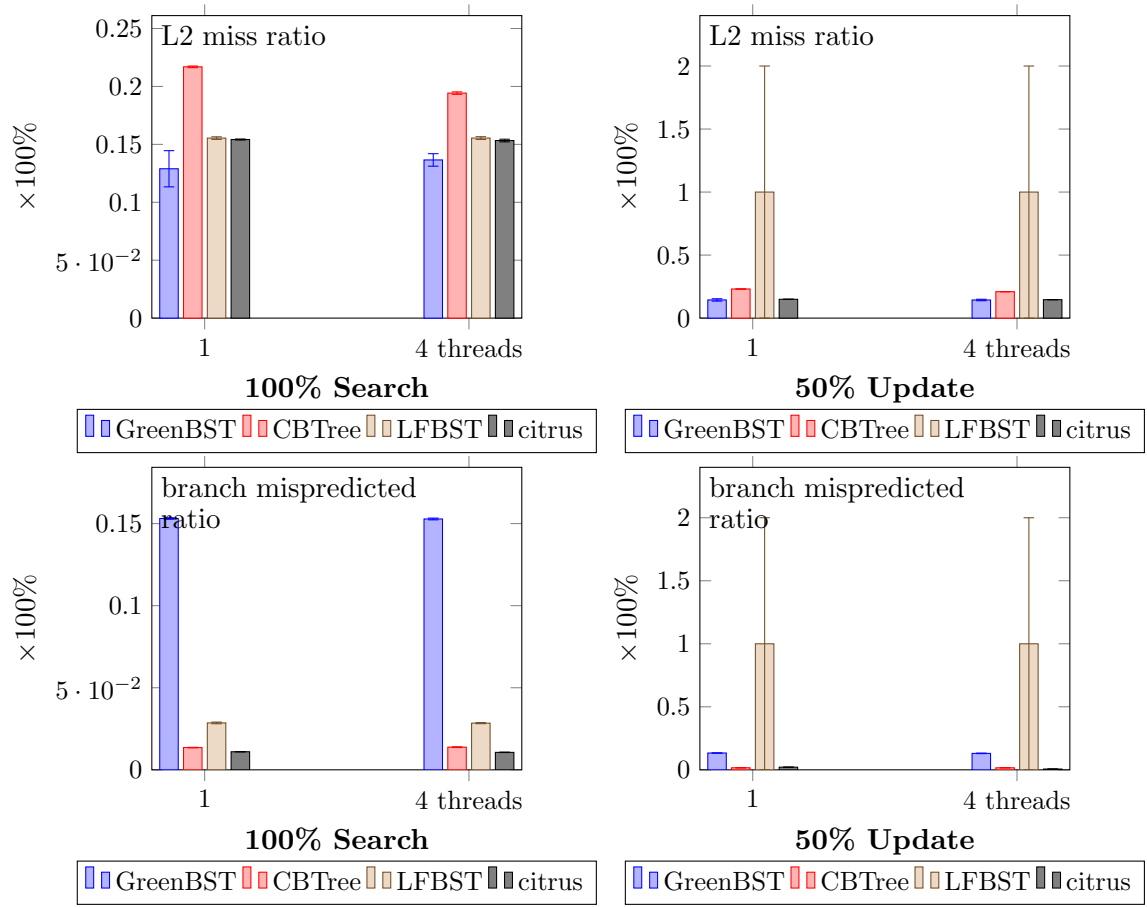
\begin{figure}[!t] 
{\footnotesize
\centering
\input{figs-uit/arm-prof.tex}
}
\caption{Profiling result on the ARM platform.}
\vspace{20pt}
\label{ap:arm-prof}
\end{figure}

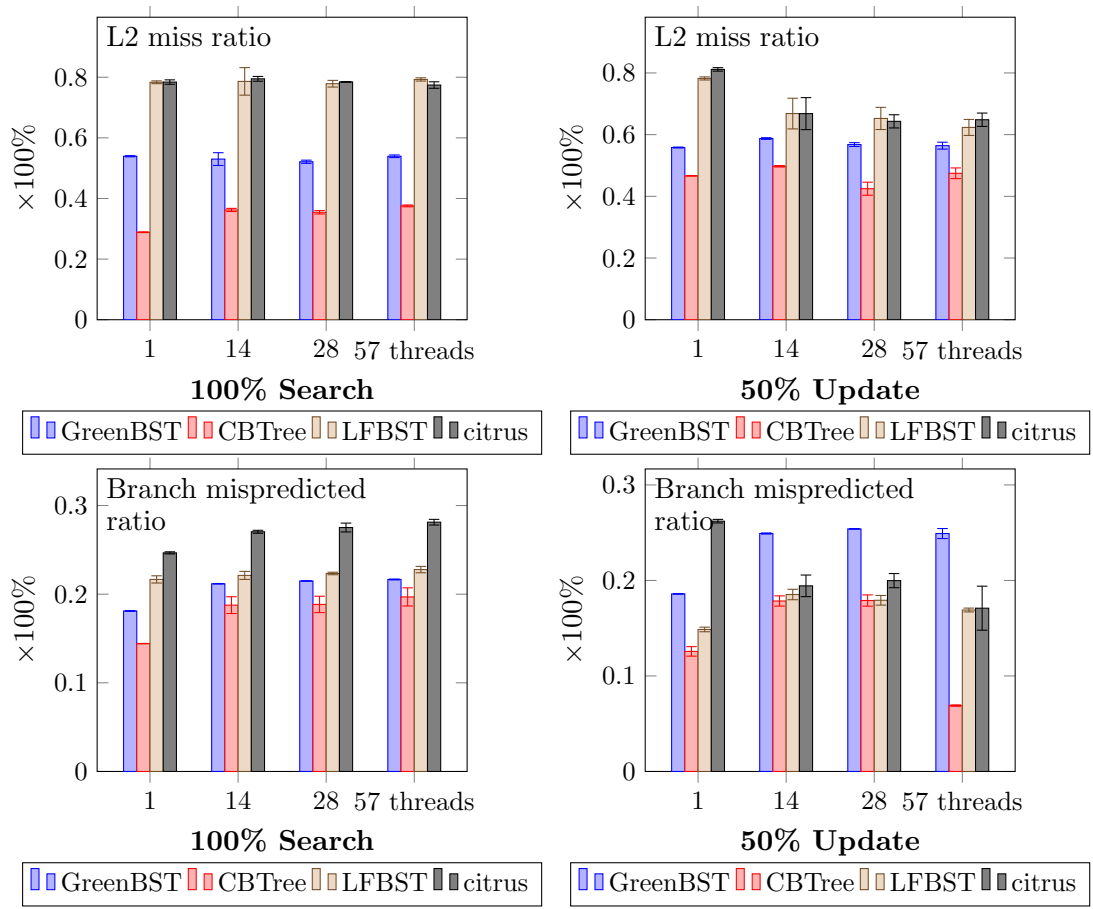
\begin{figure}[!t] 
{\footnotesize
\centering
\input{figs-uit/mic-prof.tex}
}
\caption{Profiling result on the MIC platform.}
\label{ap:mic-prof}
\end{figure}

To gain more insight into which factors that have caused GreenBST's good energy efficiency
and throughput, we extensively profiled all the trees in Table \ref{tbl:algos} when running the 
benchmarks on the HPM, ARM, and MIC platforms.

The profiling result on the HPC platform (cf. Figure \ref{ap:hpc-prof}) has revealed
that GreenBST has the lowest L3 cache miss ratio.
On the side note, GreenBST's branching performance is not as efficient as
CBTree, which is mainly because of the recursive locality-aware layout~\cite{GBST-sigmetrics2015}. However, it is obvious that 
GreenBST's efficient L3 cache performance eclipses 
its branching weakness, as GreenBST is the best performing tree compared to other trees.

On the ARM platform (cf. Figure \ref{ap:arm-prof}), the GreenBST's 
L2 cache miss ratio is the lowest compared to the other trees. Again,
the branching performance of GreenBST is 
considerably worse than the other trees, but it does not affect the GreenBST
energy efficiency and throughput because of the energy and time savings
improvement obtained from the lower data transfer.
 
Lastly, on the MIC platform (cf. Figure \ref{ap:mic-prof}), the GreenBST's  
L2 cache miss ratio is worse than CBTree's, however after careful inspection, 
the total data transferred 
is actually less than CBTree's that indicates GreenBST has 
a more efficient data re-use. 
Also, similar to other platforms' results, 
CBTree branch miss ratio is slightly lower than GreenBST's
because of the recursive tree layout adopted by GreenBST.

\paragraph*{Profiling Results on a Cycle-accurate Platform Simulator}
To obtain even more insights into whether fine-grained data locality is able to
lower data movements in a system, we tested the trees in Table \ref{tbl:algos}
in the cycle-accurate computer system simulator platform GEM5. We collect
the trees' load/write access to level 1, level 2, level 3 caches and DRAM 
when running the \textit{100\% search/1-thread} micro-benchmark using 4095 initial
value.

The simulation results (cf. Table \ref{tbl:gem5}) shows GreenBST's L3 and
DRAM load and store access are the lowest, despite having bigger 
L1 and L2 access compared to other trees. This indicates that most
of GreenBST's data load/store instructions are served within the CPU, which
implies that GreenBST can lower the number of 
data transfer between memory hierarchy. 

\begin{table}[!t]
\centering
\footnotesize
\begin{tabular}{| c | r | r | r | c |}
\hline
\bf \# of access & \bf GreenBST & \bf CBTree &  \bf	Citrus & \bf LFBST \\
\hline
\bf L1& $2451225$	& 	$7227496$	&	$2408209$ 	& $-$\\ \hline
\bf L2&	$49639$	&	$46907$	&	$81188$ 	& $-$ \\ \hline
\bf L3&	$1576$	&	$3002$	&	$13178$ 	& $-$ \\		\hline		
\bf DRAM&	$1394$	& 	$2142$	&	$3093$ 	& $-$ \\ \hline
\hline
\end{tabular}
\caption{Number of access as reported by the GEM5 system platform simulator. Unfortunately, LFBST is not able to run on the GEM5 simulator because it
was designed only for 64-bit platforms.}
\label{tbl:gem5}
\end{table}

\subsection{Locality-aware Concurrent Search Tree on Myriad2 platform}

We have implemented DeltaTree that works on the Myriad2 platform.
The main
idea of DeltaTree on Myriad2 platform is that we modify DeltaTree concurrency control while keeping the DeltaTree's original tree structure. The reason for the modification is because of the lack of support for atomic operations and limited number of usable hardware mutex on Myriad2. To circumvent these limitations, we utilize LeonRT as a lock manager for the shaves. With LeonRT acts as a lock manager, all shaves need to request a DeltaNode lock from LeonRT before it can lock the DeltaNode for
update and maintenance operation. Our locking technique implementation uses only
a shared array structure with $2\times\mathit{sv}$ size, where $\mathit{sv}$ is the number of active shaves. Advanced locking techniques \cite{HaPT07_JSS, KarlinLMO91, LimA94} can also be used. For low latency lock operations, we put this lock structure in the Myriad2's CMX memory. All other DeltaTree structures (e.g., the tree itself) are placed in the DDR memory.

\begin{figure}[!t] 
\centering 
\includegraphics[width=0.8\linewidth]{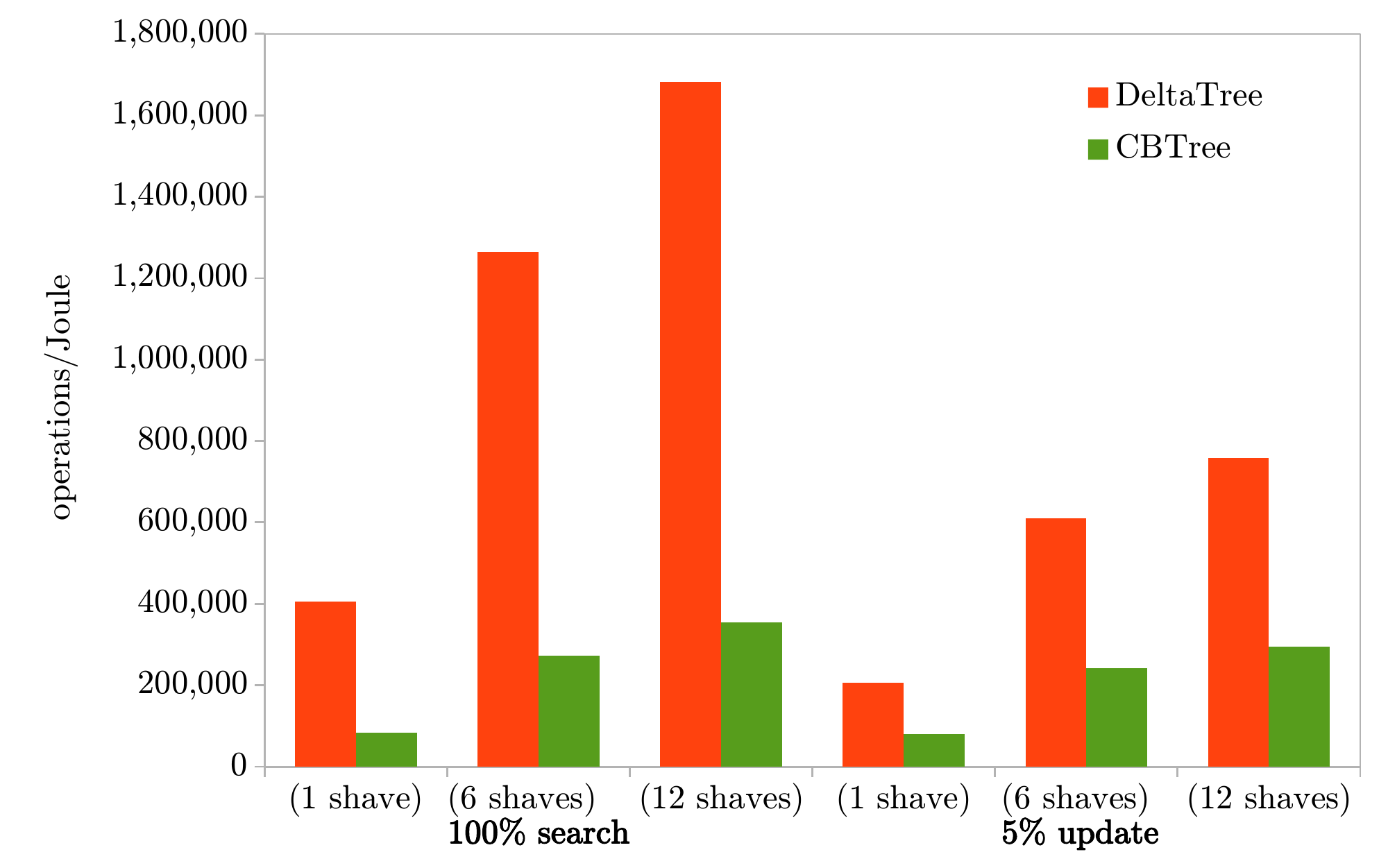}
\caption{Energy comparison using 
$2^{20}$ initial values on an Myriad2 platform. DeltaTree is up to 4.7$\times$ more
energy efficient than CBTree in 100\% search operation with 12 shaves.}\label{fig:ene-myriad2}
\end{figure}

\begin{figure}[!t]
\footnotesize 
\centering 
\resizebox{0.8\linewidth}{!}{
\input{figs-uit/perf-1048576.tex}
}
\caption{Throughput comparison using 
$2^{20}$ initial values on an Myriad2 platform. DeltaTree is up to 4$\times$ faster 
than CBTree in 100\% search operation with 12 shaves}\label{fig:res-myriad2}
\end{figure}
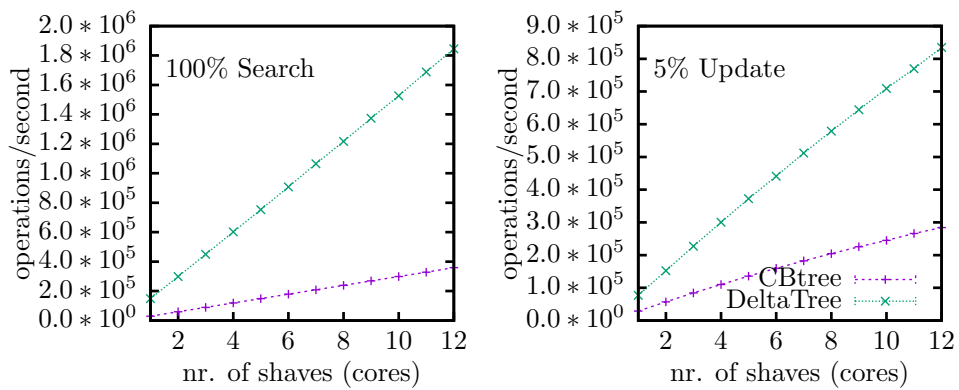

We tested our
DeltaTree implementation on Myriad2 against the concurrent B+tree (Blink tree) \cite{Lehman:1981:ELC:319628.319663}.
The Blink tree implementation (CBTree) also utlized the same locking technique and memory placement strategy as DeltaTree.
Figure \ref{fig:ene-myriad2} shows that the energy efficiency of DeltaTree 
is up to 4.7$\times$ better than CBTree in the 100\% search using 12 shaves on the Myriad2 platform. For the 5\% update case, DeltaTree is 150\% more 
energy efficient than CBTree when using 1, 6, and 12 shaves. In terms of 
throughput on the Myriad2 platform, Figure \ref{fig:res-myriad2} indicates that 
DeltaTree has up to 4$\times$ more throughput than CBTree in the 100\% search case
when using all available 12 shaves.

%% file: figs-uit/hpc-energy.tex
\pgfplotstableread[col sep = comma, header=false]{figs-uit/FINAL/results-energy-x86/GreenBST.x86.csv}\GTableE
\pgfplotstableread[col sep = comma, header=false]{figs-uit/FINAL/results-energy-x86/CBTree.x86.csv}\BTableE
\pgfplotstableread[col sep = comma, header=false]{figs-uit/FINAL/results-energy-x86/LFBST.x86.csv}\LTableE
\pgfplotstableread[col sep = comma, header=false]{figs-uit/FINAL/results-energy-x86/citrus.x86.csv}\CTableE
\pgfplotstableread[col sep=comma, header=false]{
1
12
24 threads
}\hpclabel

\pgfplotsset{tick scale binop=\times}
\foreach \st/\bg/\op in {2/6/100\% Search, -1/3/50\% Update}{%
\centering
\begin{tikzpicture}

  \begin{axis}[
  every y tick scale label/.append style={
    at={(0,1.05)},anchor=east,inner sep=0pt},
  small,
  ybar,
  ylabel=operations / Joule,
  xlabel={\bf\op},
  xtick=data,
  xticklabels from table={\hpclabel}{0},
  enlarge y limits={0.2,upper},
  enlarge x limits=0.2,
  bar width=7pt,
  ymin=0,
  x filter/.code={
        \ifnum\coordindex>\st
            \ifnum\coordindex<\bg
                \def\pgfmathresult{}
            \fi
	\fi }
 ]

\addplot+ [error bars/.cd, y dir=both,y explicit relative] table[y error expr = {(\thisrowno{6}+\thisrowno{8}+\thisrowno{14}+\thisrowno{16})/(\thisrowno{5}+\thisrowno{7}+\thisrowno{13}+\thisrowno{15})}, x expr=\coordindex+1, y expr = {\thisrowno{0}/(\thisrowno{5}+\thisrowno{7}+\thisrowno{13}+\thisrowno{15})}]{\GTableE};
\addplot+ [error bars/.cd, y dir=both,y explicit relative] table[y error expr = {(\thisrowno{6}+\thisrowno{8}+\thisrowno{14}+\thisrowno{16})/(\thisrowno{5}+\thisrowno{7}+\thisrowno{13}+\thisrowno{15})}, x expr=\coordindex+1, y expr = {\thisrowno{0}/(\thisrowno{5}+\thisrowno{7}+\thisrowno{13}+\thisrowno{15})}]{\BTableE};
\addplot+ [error bars/.cd, y dir=both,y explicit relative] table[y error expr = {(\thisrowno{6}+\thisrowno{8}+\thisrowno{14}+\thisrowno{16})/(\thisrowno{5}+\thisrowno{7}+\thisrowno{13}+\thisrowno{15})}, x expr=\coordindex+1, y expr = {\thisrowno{0}/(\thisrowno{5}+\thisrowno{7}+\thisrowno{13}+\thisrowno{15})}]{\LTableE};
\addplot+ [error bars/.cd, y dir=both,y explicit relative] table[y error expr = {(\thisrowno{6}+\thisrowno{8}+\thisrowno{14}+\thisrowno{16})/(\thisrowno{5}+\thisrowno{7}+\thisrowno{13}+\thisrowno{15})},x expr=\coordindex+1, y expr = {\thisrowno{0}/(\thisrowno{5}+\thisrowno{7}+\thisrowno{13}+\thisrowno{15})}]{\CTableE};

\node at (\pgfkeysvalueof{/pgfplots/xmin}, \pgfkeysvalueof{/pgfplots/ymax}) [anchor=north west] {\footnotesize (CPU+RAM)Energy efficiency};

\end{axis}
\end{tikzpicture}
\qquad
}

\foreach \st/\bg/\op in {2/6/100\% Search, -1/3/50\% Update}{%
\centering
\begin{tikzpicture}

  \begin{axis}[
  every y tick scale label/.append style={
    at={(0,1.05)},anchor=east,inner sep=0pt},
  legend columns=-1,
  legend style={anchor=north,at={(axis description cs:0.5,-0.18)},font=\footnotesize},
  legend entries={GreenBST,CBTree,LFBST,citrus},
  small,
  ybar,
  ylabel=operations / Joule,
  xlabel={\bf\op},
  xtick=data,
  xticklabels from table={\hpclabel}{0},
  enlarge y limits={0.2,upper},
  enlarge x limits=0.2,
  bar width=7pt,
  ymin=0,
  x filter/.code={
        \ifnum\coordindex>\st
            \ifnum\coordindex<\bg
                \def\pgfmathresult{}
            \fi
	\fi }
 ]

\addplot+ [error bars/.cd, y dir=both,y explicit relative] table[y error expr = {(\thisrowno{8}+\thisrowno{16})/(\thisrowno{7}+\thisrowno{15})}, x expr=\coordindex+1, y expr = {\thisrowno{0}/(\thisrowno{7}+\thisrowno{15})}]{\GTableE};
\addplot+ [error bars/.cd, y dir=both,y explicit relative] table[y error expr = {(\thisrowno{8}+\thisrowno{16})/(\thisrowno{7}+\thisrowno{15})}, x expr=\coordindex+1, y expr = {\thisrowno{0}/(\thisrowno{7}+\thisrowno{15})}]{\BTableE};
\addplot+ [error bars/.cd, y dir=both,y explicit relative] table[y error expr = {(\thisrowno{8}+\thisrowno{16})/(\thisrowno{7}+\thisrowno{15})}, x expr=\coordindex+1, y expr = {\thisrowno{0}/(\thisrowno{7}+\thisrowno{15})}]{\LTableE};
\addplot+ [error bars/.cd, y dir=both,y explicit relative] table[y error expr = {(\thisrowno{8}+\thisrowno{16})/(\thisrowno{7}+\thisrowno{15})}, x expr=\coordindex+1, y expr = {\thisrowno{0}/(\thisrowno{7}+\thisrowno{15})}]{\CTableE};

\node at (\pgfkeysvalueof{/pgfplots/xmin}, \pgfkeysvalueof{/pgfplots/ymax}) [anchor=north west] {\small RAM Energy efficiency};

\end{axis}
\end{tikzpicture}
}


%% file: figs-uit/arm-energy.tex
\pgfplotstableread[col sep = comma, header=false]{figs-uit/FINAL/results-energy-arm/GreenBST.arm.csv}\GTableE
\pgfplotstableread[col sep = comma, header=false]{figs-uit/FINAL/results-energy-arm/CBTree.arm.csv}\BTableE
\pgfplotstableread[col sep = comma, header=false]{figs-uit/FINAL/results-energy-arm/LFBST.arm.csv}\LTableE
\pgfplotstableread[col sep = comma, header=false]{figs-uit/FINAL/results-energy-arm/citrus.arm.csv}\CTableE

\pgfplotstableread[col sep=comma, header=false]{
1
4 threads
}\armlabel

\pgfplotsset{tick scale binop=\times}
\foreach \st/\bg/\op in {1/4/100\% Search, -1/2/50\% Update}{%
\centering
\begin{tikzpicture}

  \begin{axis}[
  every y tick scale label/.append style={
    at={(0,1.05)},anchor=east,inner sep=0pt},
  small,
  ybar,
  ylabel=operations / Joule,
  xlabel={\bf\op},
  xtick=data,
  xticklabels from table={\armlabel}{0},
  enlarge y limits={0.2,upper},
  enlarge x limits=0.2,
  bar width=7pt,
  ymin=0,
  x filter/.code={
        \ifnum\coordindex>\st
            \ifnum\coordindex<\bg
                \def\pgfmathresult{}
            \fi
	\fi }
 ]

\addplot+ [error bars/.cd, y dir=both,y explicit relative] table[y error expr = {(\thisrowno{4}+\thisrowno{6}+\thisrowno{8})/(\thisrowno{3}+\thisrowno{5}+\thisrowno{7})}, x expr=\coordindex+1, y expr = {\thisrowno{0}/(\thisrowno{3}+\thisrowno{5}+\thisrowno{7})}]{\GTableE};
\addplot+ [error bars/.cd, y dir=both,y explicit relative] table[y error expr = {(\thisrowno{4}+\thisrowno{6}+\thisrowno{8})/(\thisrowno{3}+\thisrowno{5}+\thisrowno{7})}, x expr=\coordindex+1, y expr = {\thisrowno{0}/(\thisrowno{3}+\thisrowno{5}+\thisrowno{7})}]{\BTableE};
\addplot+ [error bars/.cd, y dir=both,y explicit relative] table[y error expr = {(\thisrowno{4}+\thisrowno{6}+\thisrowno{8})/(\thisrowno{3}+\thisrowno{5}+\thisrowno{7})},x expr=\coordindex+1, y expr = {\thisrowno{0}/(\thisrowno{3}+\thisrowno{5}+\thisrowno{7})}]{\LTableE};
\addplot+ [error bars/.cd, y dir=both,y explicit relative] table[y error expr = {(\thisrowno{4}+\thisrowno{6}+\thisrowno{8})/(\thisrowno{3}+\thisrowno{5}+\thisrowno{7})}, x expr=\coordindex+1, y expr = {\thisrowno{0}/(\thisrowno{3}+\thisrowno{5}+\thisrowno{7})}]{\CTableE};

\node at (\pgfkeysvalueof{/pgfplots/xmin}, \pgfkeysvalueof{/pgfplots/ymax}) [anchor=north west] {\footnotesize (CPU+RAM)Energy efficiency};

\end{axis}
\end{tikzpicture}
\qquad
}

\foreach \st/\bg/\op in {1/4/100\% Search, -1/2/50\% Update}{%
\centering
\begin{tikzpicture}

  \begin{axis}[
  every y tick scale label/.append style={
    at={(0,1.05)},anchor=east,inner sep=0pt},
  legend columns=-1,
  legend style={anchor=north,at={(axis description cs:0.5,-0.18)},font=\footnotesize},
  legend entries={GreenBST,CBTree,LFBST,citrus},
  small,
  ybar,
  ylabel=operations / Joule,
  xlabel={\bf\op},
  xtick=data,
  xticklabels from table={\armlabel}{0},
  enlarge y limits={0.2,upper},
  enlarge x limits=0.2,
  bar width=7pt,
  ymin=0,
  x filter/.code={
        \ifnum\coordindex>\st
            \ifnum\coordindex<\bg
                \def\pgfmathresult{}
            \fi
	\fi }
 ]

\addplot+ [error bars/.cd, y dir=both,y explicit relative] table[y error expr = {\thisrowno{8}/\thisrowno{7}}, x expr=\coordindex+1, y expr = {\thisrowno{0}/\thisrowno{7}}]{\GTableE};
\addplot+ [error bars/.cd, y dir=both,y explicit relative] table[y error expr = {\thisrowno{8}/\thisrowno{7}}, x expr=\coordindex+1, y expr = {\thisrowno{0}/\thisrowno{7}}]{\BTableE};
\addplot+ [error bars/.cd, y dir=both,y explicit relative] table[y error expr = {\thisrowno{8}/\thisrowno{7}}, x expr=\coordindex+1, y expr = {\thisrowno{0}/\thisrowno{7}}]{\LTableE};
\addplot+ [error bars/.cd, y dir=both,y explicit relative] table[y error expr = {\thisrowno{8}/\thisrowno{7}}, x expr=\coordindex+1, y expr = {\thisrowno{0}/\thisrowno{7}}]{\CTableE};

\node at (\pgfkeysvalueof{/pgfplots/xmin}, \pgfkeysvalueof{/pgfplots/ymax}) [anchor=north west] {\small RAM Energy efficiency};

\end{axis}
\end{tikzpicture}
}


%% file: figs-uit/mic-energy.tex

\pgfplotstableread[col sep = comma, header=false]{figs-uit/FINAL/results-energy-mic/GreenBST.mic.csv}\GTableE
\pgfplotstableread[col sep = comma, header=false]{figs-uit/FINAL/results-energy-mic/CBTree.mic.csv}\BTableE
\pgfplotstableread[col sep = comma, header=false]{figs-uit/FINAL/results-energy-mic/LFBST.mic.csv}\LTableE
\pgfplotstableread[col sep = comma, header=false]{figs-uit/FINAL/results-energy-mic/citrus.mic.csv}\CTableE

\pgfplotstableread[col sep=comma, header=false]{
1 
14 
28 
57 threads
}\miclabel

\pgfplotsset{tick scale binop=\times}
\foreach \st/\bg/\op in {3/8/100\% Search, -1/4/50\% Update}{%
\centering
\begin{tikzpicture}

  \begin{axis}[
   every y tick scale label/.append style={
    at={(0,1.07)},anchor=east,inner sep=0pt},
  legend columns=-1,
  legend style={anchor=north,at={(axis description cs:0.5,-0.3)},font=\footnotesize},
  legend entries={GreenBST,CBTree,LFBST,citrus},
  small,
  ybar=0pt,
  ylabel=operations / Joule,
  xlabel={\bf\op},
  xtick=data,
  xticklabels from table={\miclabel}{0},
  enlarge y limits={0.2,upper},
  enlarge x limits=0.2,
  bar width=5pt,
  ymin=0,
  x filter/.code={
        \ifnum\coordindex>\st
            \ifnum\coordindex<\bg
                \def\pgfmathresult{}
            \fi
	\fi }
 ]

\addplot+ [error bars/.cd, y dir=both,y explicit relative] table[y error expr = {\thisrowno{4}/\thisrowno{3}}, x expr=\coordindex+1, y expr = {\thisrowno{0}/\thisrowno{3}}]{\GTableE};
\addplot+ [error bars/.cd, y dir=both,y explicit relative] table[y error expr = {\thisrowno{4}/\thisrowno{3}}, x expr=\coordindex+1, y expr = {\thisrowno{0}/\thisrowno{3}}]{\BTableE};
\addplot+ [error bars/.cd, y dir=both,y explicit relative] table[y error expr = {\thisrowno{4}/\thisrowno{3}}, x expr=\coordindex+1, y expr = {\thisrowno{0}/\thisrowno{3}}]{\LTableE};
\addplot+ [error bars/.cd, y dir=both,y explicit relative] table[y error expr = {\thisrowno{4}/\thisrowno{3}}, x expr=\coordindex+1, y expr = {\thisrowno{0}/\thisrowno{3}}]{\CTableE};

\node at (\pgfkeysvalueof{/pgfplots/xmin}, \pgfkeysvalueof{/pgfplots/ymax}) [anchor=north west] {\small Energy efficiency};

\end{axis}
\end{tikzpicture}
}


%% file: figs-uit/hpc-perf.tex
\pgfplotstableread[col sep = comma, header=false]{figs-uit/FINAL/results-x86/GreenBST.x86.csv}\GTable
\pgfplotstableread[col sep = comma, header=false]{figs-uit/FINAL/results-x86/CBTree.x86.csv}\BTable
\pgfplotstableread[col sep = comma, header=false]{figs-uit/FINAL/results-x86/LFBST.x86.csv}\LTable
\pgfplotstableread[col sep = comma, header=false]{figs-uit/FINAL/results-x86/citrus.x86.csv}\CTable

\pgfplotstableread[col sep=comma, header=false]{
1
12
24 threads
}\hpclabel

\pgfplotsset{tick scale binop=\times}
\foreach \st/\bg/\op in {2/6/100\% Search, -1/3/50\% Update}{%
\centering
\begin{tikzpicture}

  \begin{axis}[
   every y tick scale label/.append style={
    at={(0,1.05)},anchor=east,inner sep=0pt},
  legend columns=-1,
  legend style={anchor=north,at={(axis description cs:0.5,-0.3)},font=\footnotesize},
  legend entries={GreenBST,CBTree,LFBST,citrus},
  small,
  ybar,
  ylabel=operations/second,
  xlabel={\bf\op},
  xtick=data,
  xticklabels from table={\hpclabel}{0},
  enlarge y limits={0.2,upper},
  enlarge x limits=0.2,
  bar width=7pt,
  ymin=0,
  x filter/.code={
        \ifnum\coordindex>\st
            \ifnum\coordindex<\bg
                \def\pgfmathresult{}
            \fi
	\fi }
 ]

\addplot+ [error bars/.cd, y dir=both,y explicit relative] table[y error expr = {\thisrowno{4}/\thisrowno{3}}, x expr=\coordindex+1, y expr = {\thisrowno{0}/\thisrowno{3}}]{\GTable};
\addplot+ [error bars/.cd, y dir=both,y explicit relative] table[y error expr = {\thisrowno{4}/\thisrowno{3}}, x expr=\coordindex+1, y expr = {\thisrowno{0}/\thisrowno{3}}]{\BTable};
\addplot+ [error bars/.cd, y dir=both,y explicit relative] table[y error expr = {\thisrowno{4}/\thisrowno{3}}, x expr=\coordindex+1, y expr = {\thisrowno{0}/\thisrowno{3}}]{\LTable};
\addplot+ [error bars/.cd, y dir=both,y explicit relative] table[y error expr = {\thisrowno{4}/\thisrowno{3}}, x expr=\coordindex+1, y expr = {\thisrowno{0}/\thisrowno{3}}]{\CTable};

\node at (\pgfkeysvalueof{/pgfplots/xmin}, \pgfkeysvalueof{/pgfplots/ymax}) [anchor=north west] {\small Throughput};

\end{axis}
\end{tikzpicture}
}


%% file: figs-uit/arm-perf.tex
\pgfplotstableread[col sep = comma, header=false]{figs-uit/FINAL/results-arm/GreenBST.arm.csv}\GTable
\pgfplotstableread[col sep = comma, header=false]{figs-uit/FINAL/results-arm/CBTree.arm.csv}\BTable
\pgfplotstableread[col sep = comma, header=false]{figs-uit/FINAL/results-arm/LFBST.arm.csv}\LTable
\pgfplotstableread[col sep = comma, header=false]{figs-uit/FINAL/results-arm/citrus.arm.csv}\CTable

\pgfplotstableread[col sep=comma, header=false]{
1
4 threads
}\armlabel


\foreach \st/\bg/\op in {1/4/100\% Search, -1/2/50\% Update}{%
\centering
\begin{tikzpicture}

  \begin{axis}[
  legend columns=-1,
  legend style={anchor=north,at={(axis description cs:0.5,-0.3)},font=\footnotesize},
  legend entries={GreenBST,CBTree,LFBST,citrus},
  small,
  ybar,
  ylabel=operations/second,
  xlabel={\bf\op},
  xtick=data,
  xticklabels from table={\armlabel}{0},
  enlarge y limits={0.2,upper},
  enlarge x limits=0.2,
  bar width=7pt,
  ymin=0,
  x filter/.code={
        \ifnum\coordindex>\st
            \ifnum\coordindex<\bg
                \def\pgfmathresult{}
            \fi
	\fi }
 ]

\addplot+ [error bars/.cd, y dir=both,y explicit relative] table[y error expr = {\thisrowno{4}/\thisrowno{3}}, x expr=\coordindex+1, y expr = {\thisrowno{0}/\thisrowno{3}}]{\GTable};
\addplot+ [error bars/.cd, y dir=both,y explicit relative] table[y error expr = {\thisrowno{4}/\thisrowno{3}}, x expr=\coordindex+1, y expr = {\thisrowno{0}/\thisrowno{3}}]{\BTable};
\addplot+ [error bars/.cd, y dir=both,y explicit relative] table[y error expr = {\thisrowno{4}/\thisrowno{3}}, x expr=\coordindex+1, y expr = {\thisrowno{0}/\thisrowno{3}}]{\LTable};
\addplot+ [error bars/.cd, y dir=both,y explicit relative] table[y error expr = {\thisrowno{4}/\thisrowno{3}}, x expr=\coordindex+1, y expr = {\thisrowno{0}/\thisrowno{3}}]{\CTable};

\node at (\pgfkeysvalueof{/pgfplots/xmin}, \pgfkeysvalueof{/pgfplots/ymax}) [anchor=north west] {\small Throughput};

\end{axis}
\end{tikzpicture}
}


%% file: figs-uit/mic-perf.tex

\pgfplotstableread[col sep = comma, header=false]{figs-uit/FINAL/results-mic/GreenBST.mic.csv}\GTable
\pgfplotstableread[col sep = comma, header=false]{figs-uit/FINAL/results-mic/CBTree.mic.csv}\BTable
\pgfplotstableread[col sep = comma, header=false]{figs-uit/FINAL/results-mic/LFBST.mic.csv}\LTable
\pgfplotstableread[col sep = comma, header=false]{figs-uit/FINAL/results-mic/citrus.mic.csv}\CTable

\pgfplotstableread[col sep = comma, header=false]{figs-uit/FINAL/results-energy-mic/GreenBST.mic.csv}\GTableE
\pgfplotstableread[col sep = comma, header=false]{figs-uit/FINAL/results-energy-mic/CBTree.mic.csv}\BTableE
\pgfplotstableread[col sep = comma, header=false]{figs-uit/FINAL/results-energy-mic/LFBST.mic.csv}\LTableE
\pgfplotstableread[col sep = comma, header=false]{figs-uit/FINAL/results-energy-mic/citrus.mic.csv}\CTableE

\pgfplotstableread[col sep=comma, header=false]{
1 
14 
28 
57 threads
}\miclabel

\pgfplotsset{tick scale binop=\times}
\foreach \st/\bg/\op in {3/8/100\% Search, -1/4/50\% Update}{%
\centering
\begin{tikzpicture}

  \begin{axis}[
      every y tick scale label/.append style={
    at={(0,1.05)},anchor=east,inner sep=0pt},
  legend columns=-1,
  legend style={anchor=north,at={(axis description cs:0.5,-0.3)},font=\footnotesize},
  legend entries={GreenBST,CBTree,LFBST,citrus},
  small,
  ybar=0pt,
  ylabel=operations/second,
  xlabel={\bf\op},
  xtick=data,
  xticklabels from table={\miclabel}{0},
  enlarge y limits={0.2,upper},
  enlarge x limits=0.2,
  bar width=5pt,
  ymin=0,
  x filter/.code={
        \ifnum\coordindex>\st
            \ifnum\coordindex<\bg
                \def\pgfmathresult{}
            \fi
	\fi }
 ]

\addplot+ [error bars/.cd, y dir=both,y explicit relative] table[y error expr = {\thisrowno{4}/\thisrowno{3}}, x expr=\coordindex+1, y expr = {\thisrowno{0}/\thisrowno{3}}]{\GTable};
\addplot+ [error bars/.cd, y dir=both,y explicit relative] table[y error expr = {\thisrowno{4}/\thisrowno{3}}, x expr=\coordindex+1, y expr = {\thisrowno{0}/\thisrowno{3}}]{\BTable};
\addplot+ [error bars/.cd, y dir=both,y explicit relative] table[y error expr = {\thisrowno{4}/\thisrowno{3}}, x expr=\coordindex+1, y expr = {\thisrowno{0}/\thisrowno{3}}]{\LTable};
\addplot+ [error bars/.cd, y dir=both,y explicit relative] table[y error expr = {\thisrowno{4}/\thisrowno{3}}, x expr=\coordindex+1, y expr = {\thisrowno{0}/\thisrowno{3}}]{\CTable};

\node at (\pgfkeysvalueof{/pgfplots/xmin}, \pgfkeysvalueof{/pgfplots/ymax}) [anchor=north west] {\small Throughput};

\end{axis}
\end{tikzpicture}
}


%% file: figs-uit/hpc-prof.tex

\pgfplotstableread[col sep = comma, header=false]{figs-uit/FINAL/results-x86/GreenBST.x86.csv}\GTable
\pgfplotstableread[col sep = comma, header=false]{figs-uit/FINAL/results-x86/CBTree.x86.csv}\BTable
\pgfplotstableread[col sep = comma, header=false]{figs-uit/FINAL/results-x86/LFBST.x86.csv}\LTable
\pgfplotstableread[col sep = comma, header=false]{figs-uit/FINAL/results-x86/citrus.x86.csv}\CTable

\pgfplotstableread[col sep=comma, header=false]{
1 
12 
24 threads
}\hpclabel


\foreach \val/\err/\metric/\unit in {
						 11/12/L3 miss ratio/$\times 100 \%$,
						 15/16/branch mispredicted ratio/{}
						 }{%
\foreach \st/\bg/\op in {2/6/100\% Search, -1/3/50\% Update}{%
\begin{tikzpicture}
\centering
  \begin{axis}[
   legend columns=-1,
  legend style={anchor=north,at={(axis description cs:0.5,-0.3)},font=\footnotesize},
  legend entries={GreenBST,CBTree,LFBST,citrus},
  small,
  ybar,
  ylabel=\unit,
  xlabel=\bf\op,
  xtick=data,
  xticklabels from table={\hpclabel}{0},
  enlarge y limits={0.2,upper},
  enlarge x limits=0.2,
  bar width=7pt,
  ymin=0,
  x filter/.code={
        \ifnum\coordindex>\st
            \ifnum\coordindex<\bg
                \def\pgfmathresult{}
            \fi
	\fi }
 ]

\addplot+ [error bars/.cd, y dir=both,y explicit relative] table[y error expr = {\thisrowno{\err}/\thisrowno{\val}}, x expr=\coordindex+1, y index = \val]{\GTable};
\addplot+ [error bars/.cd, y dir=both,y explicit relative] table[y error expr = {\thisrowno{\err}/\thisrowno{\val}}, x expr=\coordindex+1, y index = \val]{\BTable};
\addplot+ [error bars/.cd, y dir=both,y explicit relative] table[y error expr = {\thisrowno{\err}/\thisrowno{\val}}, x expr=\coordindex+1, y index = \val]{\LTable};
\addplot+ [error bars/.cd, y dir=both,y explicit relative] table[y error expr = {\thisrowno{\err}/\thisrowno{\val}}, x expr=\coordindex+1, y index = \val]{\CTable};

\node at (\pgfkeysvalueof{/pgfplots/xmin}, \pgfkeysvalueof{/pgfplots/ymax}) [anchor=north west, text width=4cm] {\small\metric};

\end{axis}
\end{tikzpicture}
}

}


%% file: figs-uit/arm-prof.tex
\pgfplotstableread[col sep = comma, header=false]{figs-uit/FINAL/results-arm/GreenBST.arm.csv}\GTable
\pgfplotstableread[col sep = comma, header=false]{figs-uit/FINAL/results-arm/CBTree.arm.csv}\BTable
\pgfplotstableread[col sep = comma, header=false]{figs-uit/FINAL/results-arm/LFBST.arm.csv}\LTable
\pgfplotstableread[col sep = comma, header=false]{figs-uit/FINAL/results-arm/citrus.arm.csv}\CTable

\pgfplotstableread[col sep=comma, header=false]{
1
4 threads
}\armlabel

\foreach \val/\err/\metric/\unit in {
						 9/10/L2 miss ratio/$\times 100 \%$,
						 13/14/branch mispredicted ratio/$\times 100 \%$
						 }{%
\foreach \st/\bg/\op in {1/4/100\% Search, -1/2/50\% Update}{%
\centering
\begin{tikzpicture}

  \begin{axis}[
   legend columns=-1,
  legend style={anchor=north,at={(axis description cs:0.5,-0.3)},font=\footnotesize},
  legend entries={GreenBST,CBTree,LFBST,citrus},
  small,
  ybar,
  ylabel=\unit,
  xlabel=\bf\op,
  xtick=data,
  xticklabels from table={\armlabel}{0},
  enlarge y limits={0.2,upper},
  enlarge x limits=0.2,
  bar width=7pt,
  ymin=0,
  x filter/.code={
        \ifnum\coordindex>\st
            \ifnum\coordindex<\bg
                \def\pgfmathresult{}
            \fi
	\fi }
 ]

\addplot+ [error bars/.cd, y dir=both,y explicit relative] table[y error expr = {\thisrowno{\err}/\thisrowno{\val}}, x expr=\coordindex+1, y index = \val]{\GTable};
\addplot+ [error bars/.cd, y dir=both,y explicit relative] table[y error expr = {\thisrowno{\err}/\thisrowno{\val}}, x expr=\coordindex+1, y index = \val]{\BTable};
\addplot+ [error bars/.cd, y dir=both,y explicit relative] table[y error expr = {\thisrowno{\err}/\thisrowno{\val}}, x expr=\coordindex+1, y index = \val]{\LTable};
\addplot+ [error bars/.cd, y dir=both,y explicit relative] table[y error expr = {\thisrowno{\err}/\thisrowno{\val}}, x expr=\coordindex+1, y index = \val]{\CTable};

\node at (\pgfkeysvalueof{/pgfplots/xmin}, \pgfkeysvalueof{/pgfplots/ymax}) [anchor=north west, text width=4cm] {\small\metric};

\end{axis}
\end{tikzpicture}
}

}


%% file: figs-uit/mic-prof.tex
\pgfplotstableread[col sep = comma, header=false]{figs-uit/FINAL/results-mic/GreenBST.mic.csv}\GTable
\pgfplotstableread[col sep = comma, header=false]{figs-uit/FINAL/results-mic/CBTree.mic.csv}\BTable
\pgfplotstableread[col sep = comma, header=false]{figs-uit/FINAL/results-mic/LFBST.mic.csv}\LTable
\pgfplotstableread[col sep = comma, header=false]{figs-uit/FINAL/results-mic/citrus.mic.csv}\CTable

\pgfplotstableread[col sep=comma, header=false]{
1 
14 
28 
57 threads
}\miclabel


\foreach \val/\err/\metric/\unit in {
						 9/10/L2 miss ratio/$\times 100 \%$,
						 17/18/Branch mispredicted ratio/$\times 100 \%$
						 }{%
\foreach \st/\bg/\op in {3/8/100\% Search, -1/4/50\% Update}{%
\begin{tikzpicture}
\centering
  \begin{axis}[
   legend columns=-1,
  legend style={anchor=north,at={(axis description cs:0.5,-0.3)},font=\footnotesize},
  legend entries={GreenBST,CBTree,LFBST,citrus},
  small,
  ybar=0pt,
  ylabel=\unit,
  xlabel=\bf\op,
  xtick=data,
  xticklabels from table={\miclabel}{0},
  enlarge y limits={0.2,upper},
  enlarge x limits=0.2,
  bar width=5pt,
  ymin=0,
  x filter/.code={
        \ifnum\coordindex>\st
            \ifnum\coordindex<\bg
                \def\pgfmathresult{}
            \fi
	\fi }
 ]

\addplot+ [error bars/.cd, y dir=both,y explicit relative] table[y error expr = {\thisrowno{\err}/\thisrowno{\val}}, x expr=\coordindex+1, y index = \val]{\GTable};
\addplot+ [error bars/.cd, y dir=both,y explicit relative] table[y error expr = {\thisrowno{\err}/\thisrowno{\val}}, x expr=\coordindex+1, y index = \val]{\BTable};
\addplot+ [error bars/.cd, y dir=both,y explicit relative] table[y error expr = {\thisrowno{\err}/\thisrowno{\val}}, x expr=\coordindex+1, y index = \val]{\LTable};
\addplot+ [error bars/.cd, y dir=both,y explicit relative] table[y error expr = {\thisrowno{\err}/\thisrowno{\val}}, x expr=\coordindex+1, y index = \val]{\CTable};

\node at (\pgfkeysvalueof{/pgfplots/xmin}, \pgfkeysvalueof{/pgfplots/ymax}) [anchor=north west, text width=4cm] {\small\metric};

\end{axis}
\end{tikzpicture}
}

}


%% file: figs-uit/perf-1048576.tex
\begingroup
  \makeatletter
  \providecommand\color[2][]{%
    \GenericError{(gnuplot) \space\space\space\@spaces}{%
      Package color not loaded in conjunction with
      terminal option `colourtext'%
    }{See the gnuplot documentation for explanation.%
    }{Either use 'blacktext' in gnuplot or load the package
      color.sty in LaTeX.}%
    \renewcommand\color[2][]{}%
  }%
  \providecommand\includegraphics[2][]{%
    \GenericError{(gnuplot) \space\space\space\@spaces}{%
      Package graphicx or graphics not loaded%
    }{See the gnuplot documentation for explanation.%
    }{The gnuplot epslatex terminal needs graphicx.sty or graphics.sty.}%
    \renewcommand\includegraphics[2][]{}%
  }%
  \providecommand\rotatebox[2]{#2}%
  \@ifundefined{ifGPcolor}{%
    \newif\ifGPcolor
    \GPcolortrue
  }{}%
  \@ifundefined{ifGPblacktext}{%
    \newif\ifGPblacktext
    \GPblacktexttrue
  }{}%
  \let\gplgaddtomacro\g@addto@macro
  \gdef\gplbacktext{}%
  \gdef\gplfronttext{}%
  \makeatother
  \ifGPblacktext
    \def\colorrgb#1{}%
    \def\colorgray#1{}%
  \else
    \ifGPcolor
      \def\colorrgb#1{\color[rgb]{#1}}%
      \def\colorgray#1{\color[gray]{#1}}%
      \expandafter\def\csname LTw\endcsname{\color{white}}%
      \expandafter\def\csname LTb\endcsname{\color{black}}%
      \expandafter\def\csname LTa\endcsname{\color{black}}%
      \expandafter\def\csname LT0\endcsname{\color[rgb]{1,0,0}}%
      \expandafter\def\csname LT1\endcsname{\color[rgb]{0,1,0}}%
      \expandafter\def\csname LT2\endcsname{\color[rgb]{0,0,1}}%
      \expandafter\def\csname LT3\endcsname{\color[rgb]{1,0,1}}%
      \expandafter\def\csname LT4\endcsname{\color[rgb]{0,1,1}}%
      \expandafter\def\csname LT5\endcsname{\color[rgb]{1,1,0}}%
      \expandafter\def\csname LT6\endcsname{\color[rgb]{0,0,0}}%
      \expandafter\def\csname LT7\endcsname{\color[rgb]{1,0.3,0}}%
      \expandafter\def\csname LT8\endcsname{\color[rgb]{0.5,0.5,0.5}}%
    \else
      \def\colorrgb#1{\color{black}}%
      \def\colorgray#1{\color[gray]{#1}}%
      \expandafter\def\csname LTw\endcsname{\color{white}}%
      \expandafter\def\csname LTb\endcsname{\color{black}}%
      \expandafter\def\csname LTa\endcsname{\color{black}}%
      \expandafter\def\csname LT0\endcsname{\color{black}}%
      \expandafter\def\csname LT1\endcsname{\color{black}}%
      \expandafter\def\csname LT2\endcsname{\color{black}}%
      \expandafter\def\csname LT3\endcsname{\color{black}}%
      \expandafter\def\csname LT4\endcsname{\color{black}}%
      \expandafter\def\csname LT5\endcsname{\color{black}}%
      \expandafter\def\csname LT6\endcsname{\color{black}}%
      \expandafter\def\csname LT7\endcsname{\color{black}}%
      \expandafter\def\csname LT8\endcsname{\color{black}}%
    \fi
  \fi
    \setlength{\unitlength}{0.0500bp}%
    \ifx\gptboxheight\undefined%
      \newlength{\gptboxheight}%
      \newlength{\gptboxwidth}%
      \newsavebox{\gptboxtext}%
    \fi%
    \setlength{\fboxrule}{0.5pt}%
    \setlength{\fboxsep}{1pt}%
\begin{picture}(7200.00,2880.00)%
    \gplgaddtomacro\gplbacktext{%
      \csname LTb\endcsname%
      \put(976,512){\makebox(0,0)[r]{\strut{}$0.0*10^{0}$}}%
      \put(976,730){\makebox(0,0)[r]{\strut{}$2.0*10^{5}$}}%
      \put(976,947){\makebox(0,0)[r]{\strut{}$4.0*10^{5}$}}%
      \put(976,1165){\makebox(0,0)[r]{\strut{}$6.0*10^{5}$}}%
      \put(976,1382){\makebox(0,0)[r]{\strut{}$8.0*10^{5}$}}%
      \put(976,1600){\makebox(0,0)[r]{\strut{}$1.0*10^{6}$}}%
      \put(976,1817){\makebox(0,0)[r]{\strut{}$1.2*10^{6}$}}%
      \put(976,2035){\makebox(0,0)[r]{\strut{}$1.4*10^{6}$}}%
      \put(976,2252){\makebox(0,0)[r]{\strut{}$1.6*10^{6}$}}%
      \put(976,2470){\makebox(0,0)[r]{\strut{}$1.8*10^{6}$}}%
      \put(976,2687){\makebox(0,0)[r]{\strut{}$2.0*10^{6}$}}%
      \put(1276,352){\makebox(0,0){\strut{}$2$}}%
      \put(1683,352){\makebox(0,0){\strut{}$4$}}%
      \put(2090,352){\makebox(0,0){\strut{}$6$}}%
      \put(2497,352){\makebox(0,0){\strut{}$8$}}%
      \put(2904,352){\makebox(0,0){\strut{}$10$}}%
      \put(3311,352){\makebox(0,0){\strut{}$12$}}%
      \put(1184,2361){\makebox(0,0)[l]{\strut{}100\% Search}}%
    }%
    \gplgaddtomacro\gplfronttext{%
      \csname LTb\endcsname%
      \put(128,1599){\rotatebox{-270}{\makebox(0,0){\strut{}operations/second}}}%
      \put(2191,112){\makebox(0,0){\strut{}nr. of shaves (cores)}}%
    }%
    \gplgaddtomacro\gplbacktext{%
      \csname LTb\endcsname%
      \put(4576,512){\makebox(0,0)[r]{\strut{}$0.0*10^{0}$}}%
      \put(4576,754){\makebox(0,0)[r]{\strut{}$1.0*10^{5}$}}%
      \put(4576,995){\makebox(0,0)[r]{\strut{}$2.0*10^{5}$}}%
      \put(4576,1237){\makebox(0,0)[r]{\strut{}$3.0*10^{5}$}}%
      \put(4576,1479){\makebox(0,0)[r]{\strut{}$4.0*10^{5}$}}%
      \put(4576,1720){\makebox(0,0)[r]{\strut{}$5.0*10^{5}$}}%
      \put(4576,1962){\makebox(0,0)[r]{\strut{}$6.0*10^{5}$}}%
      \put(4576,2204){\makebox(0,0)[r]{\strut{}$7.0*10^{5}$}}%
      \put(4576,2445){\makebox(0,0)[r]{\strut{}$8.0*10^{5}$}}%
      \put(4576,2687){\makebox(0,0)[r]{\strut{}$9.0*10^{5}$}}%
      \put(4876,352){\makebox(0,0){\strut{}$2$}}%
      \put(5283,352){\makebox(0,0){\strut{}$4$}}%
      \put(5690,352){\makebox(0,0){\strut{}$6$}}%
      \put(6097,352){\makebox(0,0){\strut{}$8$}}%
      \put(6504,352){\makebox(0,0){\strut{}$10$}}%
      \put(6911,352){\makebox(0,0){\strut{}$12$}}%
      \put(4784,2361){\makebox(0,0)[l]{\strut{}5\% Update}}%
    }%
    \gplgaddtomacro\gplfronttext{%
      \csname LTb\endcsname%
      \put(3728,1599){\rotatebox{-270}{\makebox(0,0){\strut{}operations/second}}}%
      \put(5791,112){\makebox(0,0){\strut{}nr. of shaves (cores)}}%
      \csname LTb\endcsname%
      \put(6176,815){\makebox(0,0)[r]{\strut{}CBtree}}%
      \csname LTb\endcsname%
      \put(6176,655){\makebox(0,0)[r]{\strut{}DeltaTree}}%
    }%
    \gplbacktext
    \put(0,0){\includegraphics{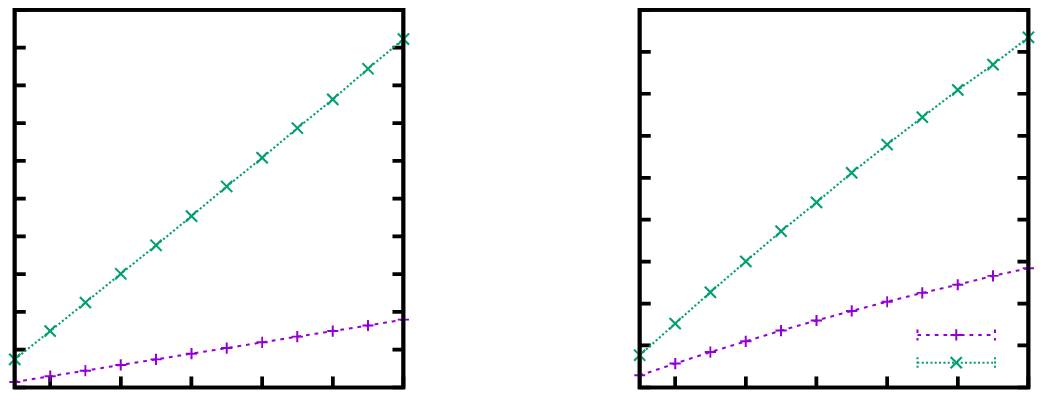}}%
    \gplfronttext
  \end{picture}%
\endgroup

%% file: conclusion.tex
In this work, we have reported our current results on the new energy/power models modeling the trade-off of energy efficiency and performance of data structures and algorithms; as well as the latest prototype of libraries and programming abstractions.

\begin{itemize}
\item We have improved the power model for Myriad1 from the power models presented in Deliverable 2.2.
\item  We have proposed a new energy complexity model for multithreaded algorithms. This new general and validated energy complexity model for parallel (multithreaded) algorithms abstracts away possible multicore platforms by their static and dynamic energy of a computational operation and data access, and derives the energy complexity of a given algorithm from its {\em work}, {\em span} and {\em I/O} complexity.
\item We have continued the modelling of the performance and the
energy consumption of \dss on a CPU platform and need even less measurements points than previously.
\item We have investigated on the optimization of streaming
applications on Myriad2, from three points of view, which are
performance, energy consumption, and space. 
\item We have implemented DeltaTree and a fast concurrent B-Tree on Myriad2 platform, and have shown that a specialized ultra low-power embedded platform such as Movidius Myriad2 can also benefit from the fine-grained locality data structures. 
\end{itemize}

In the next steps of this work package, WP2 will continue the works of Task 2.1-2.4. The future works are to continue develop novel concurrent data structures and novel adaptive memory access algorithms. Moreover, identifying the best configuration (e.g., auto-tuning) to run the algorithms is also considered in the next steps.

%% file: glossary.tex
\begin{flushleft}
\begin{tabular}{lp{12cm}}
\textbf{BRU}    &  Branch Repeat Unit (on SHAVE processor) \\
\textbf{CAS}    &  Compare-and-Swap instruction \\
\textbf{CMX}    &  Connection MatriX on-chip (shared) memory unit, 128KB (Movidius Myriad) \\
\textbf{CMU}    &  Compare-Move Unit (on SHAVE processor) \\
\textbf{Component} & 1. [hardware component] part of a chip's or motherboard's 
  circuitry; \ 2. [software component] encapsulated and annotated reusable
  software entity with contractually specified interface and
  explicit context dependences only, subject to third-party (software) composition.\\
\textbf{Composition}    & 1. [software composition] Binding a call to a 
  specific callee (e.g., implementation variant of a component) and allocating
  resources for its execution; \ 2. [task composition] Defining a macrotask and
  its use of execution resources 
  by internally scheduling its constituent tasks in serial,
  in parallel or a combination thereof. \\
  
\textbf{CPU}    &  Central (general-purpose) Processing Unit\\

\textbf{uncore}    &  including the ring interconnect, shared cache, integrated memory controller, home agent, power control unit, integrated I/O module, config Agent, caching agent and Intel QPI link interface \\ 
\textbf{CTH}    &  Chalmers University of Technology \\
\textbf{DAQ}    &  Data Acquisition Unit \\
\textbf{DCU}    &  Debug Control Unit (on SHAVE processor) \\
\textbf{DDR}    &  Double Data Rate Random Access Memory \\
\textbf{DMA}    &  Direct (remote) Memory Access \\
\textbf{DRAM}   &  Dynamic Random Access Memory \\
\textbf{DSP}    &  Digital Signal Processor \\
\textbf{DVFS}   &  Dynamic Voltage and Frequency Scaling \\
\textbf{ECC}    &  Error-Correcting Coding \\
\textbf{EXCESS} &  Execution Models for Energy-Efficient Computing Systems\\
\textbf{GPU}    &  Graphics Processing Unit\\
\textbf{HPC}    &  High Performance Computing\\
\textbf{IAU}    &  Integer Arithmetic Unit (on SHAVE processor) \\
\textbf{IDC}    &  Instruction Decoding Unit (on SHAVE processor) \\
\textbf{IRF}    &  Integer Register File (on SHAVE processor) \\
\textbf{LEON}    &  SPARCv8 RISC processor in the Myriad1 chip\\
\textbf{LIU}    &  Link\"oping University \\
\textbf{LLC}    &  Last-level cache\\
\textbf{LSU}    &  Load-Store Unit (on SHAVE processor) \\
\textbf{Microbenchmark} & Simple loop or kernel developed to measure one or few properties of the underlying architecture or system software\\
\textbf{PAPI}   &  Performance Application Programming Interface\\
\end{tabular}
\end{flushleft}

\newpage 

\begin{flushleft}
\begin{tabular}{lp{12cm}}
\textbf{PEPPHER} &  Performance Portability and Programmability for Heterogeneous Many-core Architectures. FP7 ICT project, 2010-2012, www.peppher.eu \\
\textbf{PEU}    &  Predicated Execution Unit (on SHAVE processor) \\
\textbf{Pinning} &  [thread pinning] Restricting the operating system's CPU scheduler in order to map a thread to a fixed CPU core \\
\textbf{QPI}    &  Quick Path Interconnect\\
\textbf{RAPL}   &  Running Average Power Limit energy consumption counters (Intel)\\
\textbf{RCL}   &  Remote Core Locking (synchronization algorithm)\\
\textbf{SAU}    &  Scalar Arithmetic Unit (on SHAVE processor) \\
\textbf{SHAVE}  &  Streaming Hybrid Architecture Vector Engine (Movidius) \\
\textbf{SoC}    &  System on Chip \\
\textbf{SRF}    &  Scalar Register File (on SHAVE processor) \\
\textbf{SRAM}   &  Static Random Access Memory \\
\textbf{TAS}    &  Test-and-Set instruction\\
\textbf{TMU}    &  Texture Management Unit (on SHAVE processor) \\
\textbf{USB}    &  Universal Serial Bus \\
\textbf{VAU}    &  Vector Arithmetic Unit (on SHAVE processor) \\
\textbf{Vdram}  &  DRAM Supply Voltage \\
\textbf{Vin}    &  Input voltage level  \\
\textbf{Vio}    &  Input/Output voltage level  \\
\textbf{VLIW}   &  Very Long Instruction Word (processor) \\
\textbf{VLLIW}  &  Variable Length VLIW (processor) \\
\textbf{VRF}    &  Vector Register File (on SHAVE processor) \\
\textbf{Wattsup}&  Watts Up .NET power meter \\
\textbf{WP1}   &  Work Package 1 (here: of EXCESS) \\
\textbf{WP2}   &  Work Package 2 (here: of EXCESS) \\
\end{tabular}
\end{flushleft}